\documentclass[12pt,a4paper,twoside]{report}
\usepackage[cp852]{inputenc}
\usepackage[dvips]{graphicx}
\usepackage{t1enc}
\usepackage{pgf}
\usepackage{hyperref}
\usepackage{supertabular}
\usepackage{caption}
\usepackage{subfigure}
\usepackage{thumbpdf}
\usepackage{amsmath}
\usepackage{rotating}
\usepackage{array}
\usepackage{sidecap}
\usepackage{xspace}
\usepackage[bottom]{footmisc}
\author{Lech Wiktor Piotrowski}

\let\Otemize =\itemize
\let\Onumerate =\enumerate
\let\Oescription =\description
\def\Nospacing{\itemsep=0pt\topsep=0pt\partopsep=0pt\parskip=0pt\parsep=0pt}

\makeatletter
\newcommand\ackname{Acknowledgements}
\if@titlepage
  \newenvironment{acknowledgements}{%
      \titlepage
      \null\vfil
      \@beginparpenalty\@lowpenalty
      \begin{center}%
        \bfseries \ackname
        \@endparpenalty\@M
      \end{center}}%
     {\par\vfil\null\endtitlepage}
\else
  
\fi
\makeatother

\title{Modelling of the \pin detector}
\date{Warsaw, 2011}
\author{Lech Wiktor Piotrowski}

\begin{document}

\newcommand{\dx}{\mathrm{d}x }
\newcommand{\dy}{\mathrm{d}y }
\newcommand{\m}{\mathrm{m}}

\newcommand{\pin}{``Pi~of~the~Sky''\xspace}
\def\captionlabeldelim{.}%

\renewcommand{\captionsize}{\small}
\renewcommand{\captionlabelfont}{\bfseries}

\begin{titlepage}

\vspace*{50pt}

\begin{center}
\large
\textit{
University of Warsaw \\
Faculty of Physics \\[80pt]
}

\huge \bf
Modelling of the \pin detector \\[40pt]

\large \bf
Lech Wiktor Piotrowski \\[80pt]

\end{center}

\begin{center}
\normalsize
	PhD thesis written under supervision\\
	of prof. dr hab. Aleksander Filip \.Zarnecki \\[60pt]
\end{center}

\begin{center}

\includegraphics[width=0.2\textwidth]{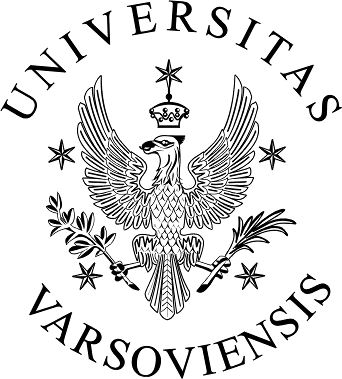} \\ [60pt]

\normalsize
Warsaw, 2011
\end{center}
\end{titlepage}

\thispagestyle{empty}
\cleardoublepage


\begin{abstract}
	The ultimate goal of the ``Pi of the Sky'' apparatus is observation of optical flashes of astronomical origin and other light sources variable on short timescales, down to tens of seconds. We search mainly for optical emission of Gamma Ray Bursts, but also for variable stars, novae, blazars, etc. This task requires an accurate measurement of the source's brightness (and it's variability), which is difficult as ``Pi of the Sky'' single camera has a large field of view of about $20^\circ \times 20^\circ$. This causes a significant deformation of a point spread function (PSF), reducing quality of brightness and position measurement with standard photometric and astrometric algorithms. Improvement requires a careful study and modelling of PSF, which is the main topic of the presented thesis. A dedicated laboratory setup has been created for obtaining isolated, high quality profiles, which in turn were used as the input for mathematical models. Two different models are shown: diffractive, simulating light propagation through lenses and effective, modelling the PSF shape in the image plane.
	
	The effective model, based on PSF parametrization with selected Zernike polynomials describes the data well and was used in photometry and astrometry analysis of the frames from the \pin prototype working in Chile. No improvement compared to standard algorithms was observed in brightness measurements, however more than factor of 2 improvement in astrometry accuracy was reached for bright stars. Additionally, the model was used to recalculate limits on the optical precursor to GRB080319B -- a limit higher by $0.75^\m$ compared to previous calculations has been obtained.

	The PSF model was also used to develop a dedicated tool to generate Monte Carlo samples of images corresponding to the \pin observations. The simulator allows for a detailed reproduction of the frame as seen by our cameras, taking into account PSF, electronics and mechanical fluctuations, etc. A comparison of photometry performed on real and simulated data resulted in very similar results, proving the simulator a worthy tool for future \pin hardware and software development.
\end{abstract}

\thispagestyle{empty}
\cleardoublepage



\tableofcontents



\chapter{Introduction}

The sky has fascinated mankind from the very beginning. It's main inhabitants - Sun, Moon and stars - inspired most ancient cultures to think about their place in the Universe and the passage of time, for behaviour of these celestial bodies was regular and predictable. However, as soon as man started to look into the sky with something more than a pure amazement, he found among the stars some strange ones, not following the rules of the most. These were planets. From time to time the wild blue yonder was raided by some other visitors, such as comets, or much more often -- meteors. Those dwellers of heavens shaped the knowledge about the world above our heads for millennia.

During passage of centuries, the sky, recognized for a long time as immutable and a rather peaceful place slowly occurred to be a home to many violent events. Among these were those more spectacular, such as supernovae or less destructive such as novae or flare stars. However, quite recently, it were the Gamma Ray Bursts that shocked the astronomers and introduced a set of new ideas to this branch of science.

GRBs are perhaps the most energetic phenomena known to man, emitting enormous amounts of energy mainly in gamma-rays, but also in other wavelengths of electromagnetic radiation. More than 4 decades of observation since discovery in 1969 by a pair of Vela satellites showed, that they are also one of the most remote type of objects, appearing always at cosmological distances. A few outbursts are present on the sky per week, but their position is totally random. Thus detection can be performed only by satellite gamma-ray detectors covering very large fractions of the sky. To perform simultaneous observations also in optical band, similar approach has to be introduced to on-ground optical observatories. Up to now most of the telescopes focused on observing a very small fraction of the sky with a big magnification and thus were nearly useless for such simultaneous observations. \pin introduces a completely new approach -- constant monitoring of a very large part of the sky, aimed at detection of bright transients coming from random directions.

This work was prepared as a part of the \pin project. Constant sky monitoring requires cameras with a very large field of view (for astronomical standards). Lenses required to obtain such parameters introduce large deformations to the image, increasing with the distance from the frame centre (optical axis). Study of these deformations, described by the point spread function (PSF) of the detector, is the main subject of this thesis.

The first chapter describes short time-scale variable phenomena on the sky, focusing on Gamma Ray Bursts -- the main field of interest of the \pin experiment. The general idea of the experiment is presented in the second chapter, along with the construction of the prototype and the concept of the final system. Analysis software, main scientific results and uncertainties considered in the data analysis are briefly discussed.

The fourth chapter shows results of the laboratory measurements of the point spread function as well as of some other characteristics of the detector. Attempts to reproduce these results with mathematical models are presented in the subsequent two chapters: in chapter \ref{chap_diff} via physical modelling of the light passage through lenses and in chapter \ref{chap_polynomials} via parametrization of the PSF shape on the frame.

The final outcome of this work are the applications of the resulting detector response model. The first being the development of a new algorithm for measuring stars (or other objects) brightness and position on the frame, as well as search for optical precursor to GRB080319B, presented in chapter \ref{chap_modelling_real_sky_data}. The other application, shown in chapter \ref{chap_simulator}, is the procedure of a realistic simulator of the \pin frame, featuring deformed stars shapes, signal fluctuations and numerous other features, which can be used for future hardware and software development\footnote{Simulation with Monte Carlo techniques becomes a common tool for improving our understanding of the complex detector systems.}.

All results presented in chapter \ref{chap_lab} to \ref{chap_simulator} were obtained by the author of this thesis.

\chapter{Short time-scale phenomena in the sky}

The first that we know of, to question the immutability of the sky and to write about an appearance of a new resident of the firmament was a Greek astronomer Hipparchus (although such events are reported to be observed by Chinese about 1500 BC). In II century BC Hipparchus noticed a ``New Star'' -- gr. Nova (sec. \ref{sec_nova}) -- that he was sure he had not observed before. That discovery, a shock to the Greek community, which though of a sky as of a constant sphere, probably pushed Hipparchus to the creation of one of the first star catalogues, to allow future generations to discover other new stars.

The apparent immutability of the heavens was also questioned by the Chinese. In 185 AD they observed a supernova (sec. \ref{sec_sn}) SN 185\cite{first_sn}, which remained in the sky for 8 months\footnote{The first documented Chinese discovery of a star, that could be a supernova comes from 352 BC\cite{sn352}. Some historical sources claim, that suspected supernova was first seen in 2241 BC.}. Although we hardly know anything about the astronomic observations in the earlier times, this supernova is believed to be the first recorded by the mankind. Following that other supernovae were rarely discovered, every few hundreds years or so, like SN~1006\cite{sn1006} -- the brightest one so far, SN~1054\cite{sn1054} (the one that formed Crab nebula), SN~1572\cite{sn1572} and SN~1604\cite{sn1604}, the last one being the latest seen in the Milky Way.

The SN 1572\footnote{Although it was a supernova, at this time it was called ``nova'', and ambiguity remained up to year 1930.} discovered by Tycho Brahe had forced reinvention of the Hipparchus idea, that the night sky beyond planets and Moon is not immutable. This had begun a revolution in European astronomical thinking. Shortly afterwards, in year 1584, an idea that other stars are other Suns placed far from the Earth was reintroduced by Giordano Bruno, and given physical basis in the next century by Isaac Newton, who explained absence of the gravitational pull on the Solar System by equal distribution of stars in the sky.

It was not long before the first discovery of a variability of a known star. In year 1638 Johannes Holwarda found that the star Omicron Ceti (later named Mira) had a 11 months pulsation cycle\cite{holwarda}. Next, variations of the luminosity of star Algol were observed by Geminiano Montanari in 1667. By 1786 ten variable stars were known and the number started to increase rapidly after 1850, induced by an improvement of observational techniques. Along with all those discoveries, certain understanding of the nature of celestial bodies was born and is still growing, boosted by invention of quantum physics in the first half of XX century.

Other milestones in observation of astronomical objects, including variable stars and other short time-scale phenomena, was introduction of photography in late XIX century and CCD imaging in late XX century. However, stepping into XX century, one cannot diminish development of other fields of observational astronomy, such as radio, X-ray and gamma-ray astronomy. They allowed discoveries of more exotic phenomena with time-scales much shorter that those previously known. Some of these have also been observed in optical wavelengths, some still remain to be confirmed.

Close future will perhaps bring advancements in observation techniques for astronomy of largest photon energies (considered to be a part of the so-called astroparticle physics) or even for other mediums, such as neutrinos. However, in this chapter, and generally in this work I will focus on the astronomical objects that are of interest to the \pin project, i.e. those that can be observed optically and with variability ranging from seconds to years.

\section{Gamma Ray Bursts}
\label{sec_grb}

The main inspiration for the \pin project, described in more detail in chap. \ref{chap_pi}, are the so-called Gamma Ray Bursts (GRBs) -- cosmic explosions radiating mostly in gamma-rays (thus the name), on timescales of milliseconds to tens of seconds\cite{grb}. Some of their energy is also radiated in optical band, as well as radio and X-rays -- otherwise they would not be of any interest to the optical experiments. They still remain one of the most fascinating and challenging subjects in modern astrophysics, and their mystery begun with their discovery.

\subsection{A brief history of GRBs}

In times when many, if not most of the discoveries in science are born from the joint effort of experiment and theory, and thus are in some way expected, surprises hold a special value and nearly always focus attention of scientific world for at least a short time. This was the case of Gamma Ray Bursts, which were the first extra-terrestrial (excluding Sun) sources of gamma rays discovered by man. More so, the discovery was completely unexpected.

\begin{figure}[h!]
\begin{center}
	\includegraphics[width=0.6\textwidth]{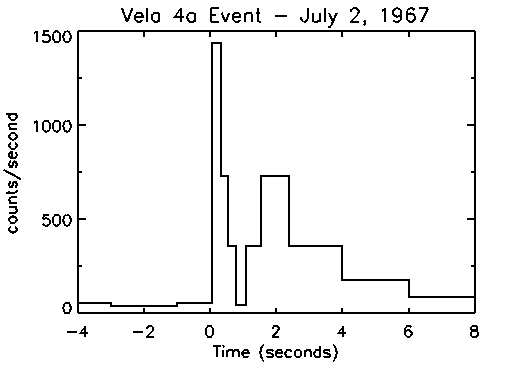}
\end{center}
\caption{Historical lightcurve of the first GRB ever detected. The detection was performed by the Vela satellites\cite{Astroph1973}.}
\label{vela_event}
\end{figure}

In year 1963 United States of America launched the first pair of military satellites Vela equipped with gamma-ray detectors. Their task was to monitor the obedience of the nuclear tests ban treaty by other countries. The treaty banned both the terrestrial and extra-terrestrial tests, therefore Vela satellites were placed on an orbit of about 120000 km above the Earth surface -- high enough to detect possible explosions on the dark side of the Moon. Four years passed till the first detection of gamma ray explosion on $2^{\mathrm{nd}}$ of July, 1967 (fig. \ref{vela_event}). The outburst contained no known signatures of a nuclear explosion. It's detection was possible with the 3rd generations of Vela satellites, launched in 1965, which were the first generation to store such events for further analysis.

However, $5^{\mathrm{th}}$ and $6^{\mathrm{th}}$ generation of Vela satellites, launched in 1969, were required to prove that the source of gamma ray bursts was not localized on Sun, Moon or the Earth's orbit. In the period 1969-1972 satellites detected 16 gamma ray bursts (GRBs) and results were published in Astrophysical Journal Letters in 1973 \cite{Astroph1973}. Additionally, data from Soviet Konus satellites confirming the detection of GRBs were published in 1974. These discoveries begun the era of GRB science, still one of the most intriguing branches of astrophysics.

The main tool for gamma ray bursts research up to 1991 was the InterPlanetary Network (IPN)\cite{ipn} formed in 1976\footnote{IPN is working up to this day, although spacecrafts forming it have changed.}. At the beginning it consisted of existing spacecrafts equipped with gamma ray detectors, already studying Sun and solar planets. The network allowed triangulation of GRB coordinates with uncertainty of a few arcminutes. This was enough to prove that GRBs are not counterparts of any of the known space sources of radiation, like, for example, X-ray sources. However IPN could not provide any insights to the distance between Earth and outbursts and thus the origin and nature of GRBs remained a speculation for the next 15 years.

Part of the scientific community shared an idea that GRBs origin at cosmological distances. However, very high intensity of the radiation brought most scientists to believe that the phenomenon occurs much closer, at Galactic distances or, at the beginning, even at the Solar System boundaries. The weight shifted across the years between supporters of both theories, often due to misleading observations.

One of the earliest hypothesis supporting Galactic origin was an idea of a neutron star bombarded by some kind of an object, probably a comet. A lack of any common pattern in GRBs lightcurves and an estimation of number of neutron stars and comets in the Milky Way made this a preferred explanation for some period of time. A significant argument for the supporters of cosmological distances came on the $5^{\mathrm{th}}$ of March 1979 with the discovery of a burst coincident in position with Supernova N49 in the Large Magellanic Cloud\cite{magel_cloud}. The distance of about $160$ ly from the Earth and a very big burst luminosity made the comet-neutron star explanation unacceptable. However, the supporters of the Galactic origin came with two explanations undermining the significance of this detection -- the position coincidence between GRB and Supernova could be purely accidental or the detected outburst was not a GRB, but an other class of phenomena.

Following years brought two strong evidences of short distances to GRBs. The first was a discovery of $30-60$ keV absorption lines in several dozen detections performed in years 1979-81. These lines were interpreted as a cyclotron frequencies of electrons in magnetic field of neutron stars. The calculated flux density of a magnetic field producing required effect was about $3\cdot 10^{12}$ gauss -- typical for Galactic neutron stars and far too small for extra-galactic sources. In year 1981 Bradley Schaefer from MIT discovered on an archival photography from 1928 an optical flash in the place of a burst from 1978. In 1984 two more such coincidences were found. That lead to an idea, that GRBs are a repetitive phenomenon with an average outburst frequency 1.1 per year, with source lifetime of more than 50 years. Thus GRBs could not be destructive for their sources, which means their energy is too low for extragalactic origin.

In 80ties there were just two arguments supporting extragalactic origin of GRBs, easily undermined and thus of much less significance. The first was an isotropic distribution of bursts on the celestial sphere, and thus no correlation with the Milky Way. However, such correlation would also not be observed if bursts were produced in the Galactic halo. The second argument was a shortage of weak GRBs.

If one considers the burst intensity $L\propto \frac{1}{R^2}$ (where $R$ is the distance towards the burst) and the number of GRBs $N\propto R^3$, assuming their constant density in the Universe, then we expect $N\propto L^{-\frac{3}{2}}$, which is a so-called power-law. Balloon experiments performed in 1982 showed a shortage of weak bursts and thus a significant deviation to this power law. If the phenomena took place only in much younger Universe, thus on cosmological distances, such a deviation would be expected. Per contra this could also be explained by a not high enough sensitivity of detectors, which could not detect longer bursts with smaller amplitude, expected from the larger distances.

Nonetheless the idea of Galaxy origin was also shattered. The spectral lines occurred to be an effect of deconvolution of the signal. The idea of GRBs as a repetitive phenomenon also lost its impact, for there were still no detections of optical flashes accompanying gamma ray bursts. Interestingly, GRB from the Large Magellanic Cloud truly turned out to be a different kind of phenomenon -- a so called Soft Gamma Repeater (SGR)\footnote{Soft Gamma Repeaters are sources of repeating gamma radiation outbursts. Such outbursts are believed to be produced on magnetastars -- neutron stars with huge magnetic fields.}.

\begin{figure}[h!]
\begin{center}
	\includegraphics[width=0.6\textwidth]{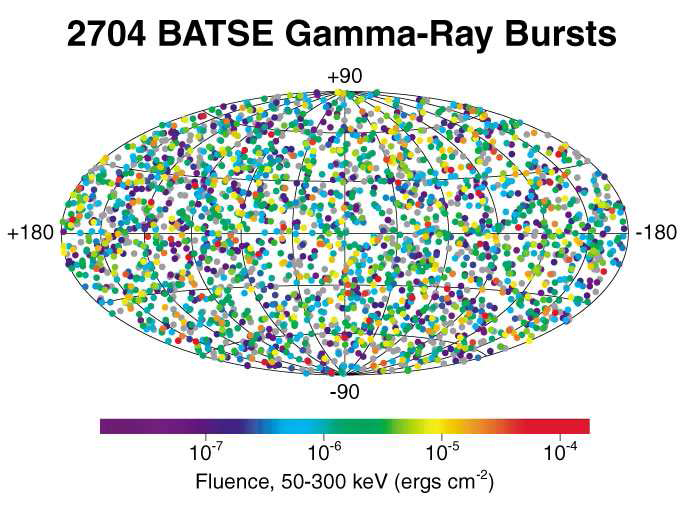}
\end{center}
\caption{Galactic coordinates of all GRBs detected by BATSE with their fluence indicated by colour.}
\label{batse_grbs}
\end{figure}

This uncertain situation held until 1991, when the satellite Compton Gamma Ray Observatory (CGRO) equipped with the BATSE detector\cite{batse} was launched. BATSE was an instrument dedicated to GRB observation and had much better sensitivity and angular resolution then any other experiment up to that day. It detected about 1 burst per day\footnote{BATSE detected nearly 3000 GRBs during its life period. It was discarded due to political reasons in year 2000.} with position uncertainty of $4^{\circ}-10^{\circ}$. This detection rate together with high enough sensitivity lead to a definite confirmation of the deviation from the power-law. That fact ultimately shifted weight towards the cosmological origin of gamma ray bursts. Also confirmed, with high statistics, was the isotropic distribution of bursts on the sky (see fig. \ref{batse_grbs}).

The definite confirmation of the cosmological distance hypothesis however, was yet to come. The main role in this event played the Dutch-Italian satellite BeppoSAX\cite{bepposax}, designed for spectral and variability measurements of X-Ray sources, as well as all-sky monitoring for transient phenomena. The monitoring was performed by a so called Phoswitch Detector System (PDS) a crystal scintillator with nearly whole-sky coverage, but poor angular resolution. PDS allowed the first detection of an X-Ray afterglow of a GRB on $28^{\mathrm{th}}$ of January 1997. But it was the Wide Field Camera instrument with $40^{\circ}\times40^{\circ}$ field of view and angular resolution of about 3 arcminutes that made a first measurement of distance to GRB possible. On $5^{\mathrm{th}}$ of August 1997 it precisely measured a position of a detected burst and sent it to the on-ground stations, which detected a radio afterglow and, 3 days later, an optical afterglow (Keck telescopes \cite{Keck}). Precise measurement of the optical afterglow resulted in determination of the redshift $z=0.835$. This meant that the distance to the burst was about $7\times 10^9$ ly -- definitely outside the Galaxy.

A very bright afterglow to the burst detected by BeppoSAX was observed on 28th of April 1998. The recorded spectrum and lightcurve were similar to those generated by supernovae. This lead to a hypothesis that GRBs longer than 10 s are produced during a collapse of a very massive (more than 20 solar masses) stars -- hypernovae.

It seems that the final proof was given by GRB030329\footnote{All gamma-ray bursts are named after the date of their detection. Each pair of numbers after the ``GRB'' corresponds respectively to year, month and day. If more than one GRB was detected on a single date, the distinction is made with consecutive letters of alphabet added at the end of the name.}, with a rapidly darkening afterglow, which started to brighten 24 hours after the explosion, having a maximum 1.6 days after the trigger. It's further decay was much more gentle than usual and after 9 days it became similar in shape and spectrum to that of supernovae. The interpretation is that the rapidly darkening part of the afterglow was generated by the GRB, while the following raise and gentle decay was created by the hypernova itself. 

If we assume that the hypernova is responsible for all bursts longer than 10s, why haven't it been seen in all the other GRBs? The answer is that the GRB030329 was close enough to the Earth (2 Gly), that the hypernova was bright enough to be seen, while for other, more distant bursts, we would only see the GRB and its afterglow.

For a long time optical counterparts were believed to be delayed temporally with respect to the gamma ray radiation. However, development in observational techniques as well as the increasing number of detections pushed the start of the optical emission further and further towards very beginning of the phenomenon. Important dates are the $23^{\mathrm{nd}}$ of January 1999, when small robotic telescope ROTSE discovered an afterglow, with peak brightness $9^{\rm{m}}$\footnote{Magnitudo $^{\rm{m}}$ stands for an apparent brightness of an astronomical object in relation to an object of reference: $m=-2.5\log{\frac{I}{I_r}}-m_r$, where $m_r$ is the magnitudo of the object of reference, $I_r$ its observed radiation flux and $I$ is the observed radiation flux of the discussed object. This unit has been normalized according to the Ptolemaeus scale, where the most bright stars seen with the naked eye had brightness of $1^{\rm{m}}$ and the darkest of $6^{\rm{m}}$. The Sun has a brightness of $-26^{\rm{m}}$, and Vega $0^{\rm{m}}$.}, 20 seconds after the trigger sent from BATSE. For years it remained the fastest observation and the most luminous optical counterpart observed, up to $19^{\mathrm{th}}$ of March 2008. On this day \pin experiment was the first one to detect GRB030819B in optical band simultaneously to the start of emission in gamma rays. The other 2 optical experiments to confirm this discovery were TORTORA\cite{tortora} and Raptor\cite{raptor}. The burst is so far the brightest observed by man in optical band (reaching $5.3^{\rm{m}}$), and the observation confirmed that the optical radiation is produced at the same stage of GRB explosion as gamma radiation, although due to different physical mechanism (sec. \ref{ssec_ourgrb}).

The last discovery mentioned, among numerous other, was possible due to the observations performed by a SWIFT\cite{swift} satellite detecting about 3 GRBs per week. Other orbital experiments currently performing research on this phenomenon are Fermi\cite{fermi}, Agile\cite{agile}, Suzaku\cite{suzaku}, Konus\cite{konus} and Integral\cite{integral}.

\subsection{Experimental knowledge about Gamma Ray Bursts}

So far a few thousand GRBs have been detected. Their distribution on the celestial sphere is isotropic and so far no convincing evidence for grouping in any subclass of bursts has been found. They occur outside the Galaxy, mostly on cosmological distances -- the most distant one occurred at $z=8.2$ -- 13 Gly from the Earth, becoming the second most distant object observed by men. The closest GRB had $z=0.1685$, meaning it was about 2 Gly from us. 

Hardly any regularity is observed in GRBs lightcurves. All exhibit a certain raise and decay, but these properties can be seen as well in a single or multiple peaks, which vary in shape and duration. The overall duration of this phenomena ranges from milliseconds to hundreds of seconds. Some of the bursts have a precursor -- a much smaller burst preceding the main explosion. So far precursor has only been seen in gamma rays. One property is common among all GRBs -- energy of each one is of the order of $10^{49}-10^{51}$ ergs.

\begin{figure}[b!]
\begin{center}
	\includegraphics[width=0.6\textwidth]{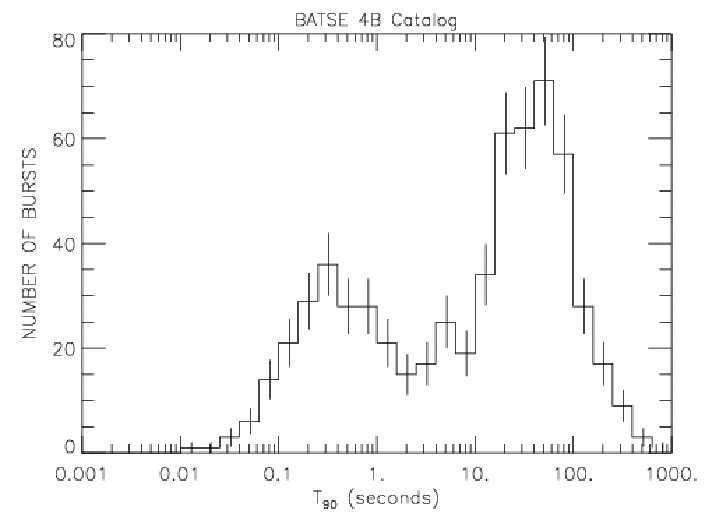}
\end{center}
\caption{Durations of gamma ray bursts detected by BATSE. $\rm{T_{90}}$ stands for the time in which $90\%$ of the burst energy was emitted. Visible is a definite division into two groups -- with $\mathrm{T_{90}}<2$ s and $\mathrm{T_{90}}>2$ s.}
\label{batse_division}
\end{figure}

BATSE observations\cite{batse_cat} showed that bursts can be divided into two subgroups -- long bursts, with duration of more than 2 seconds and average duration of about 30 seconds, and short bursts with average duration of 0.3 seconds (fig. \ref{batse_division}). Additionally, short bursts tend to have a much harder spectra\footnote{Hardness ratio of a GRB is usually defined as a ratio of two fluences in different energy bands, integrated in time over the duration of the burst.}. Long bursts seem to be more luminous and occur at larger distances.

There have been numerous attempts to provide other divisions, but so far none has been nearly as convincing as the described one. Perhaps the number of bursts observed with modern instruments is still too low -- in year 2010 SWIFT detected its $500^{\mathrm{th}}$ GRB, the number still much lower than about 2700 bursts detected by BATSE.

\subsection{Most popular models}

\begin{figure}
  \centering
  \includegraphics[width=0.3\textwidth]{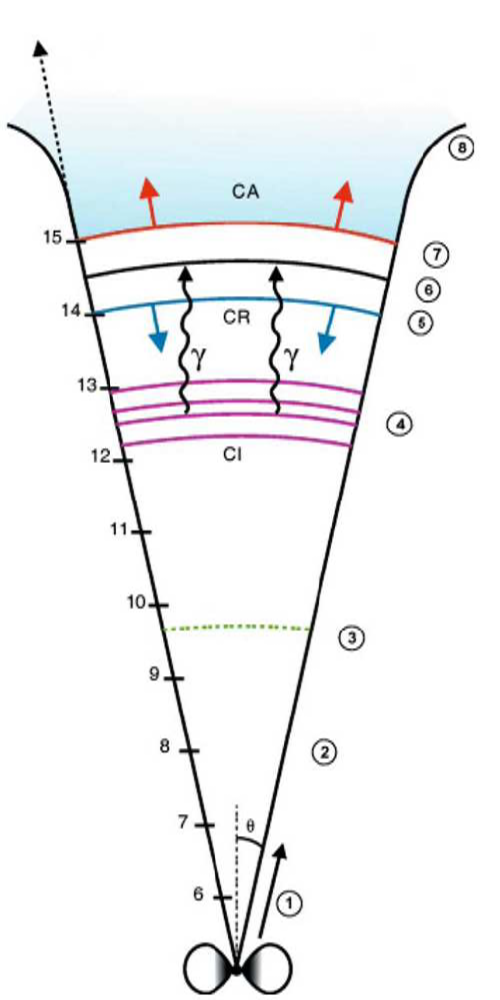}
  \caption{Fireball shock model of GRB radiation generation (image taken from \cite{fbs_model_image}). The outflow from the central engine is accelerated in phase 1 to 2, then in 3 it becomes transparent. The first stage of electromagnetic emission takes places when internal shocks (CI) occur in phase 4, then the outflow propagates into external medium, causing forward (6,7) and reverse shocks (5) producing electromagnetic radiation of lower energies. The jet becomes subrelativistic in phase 8.}
\label{fig_fbs_model}
\end{figure}

In general, all GRBs are believed to be created in the following way. At the beginning, a so-called central engine starts to emit 
vast amounts of matter (or energy). While the nature of the central engine is still a subject of speculations and may be different for various bursts, the generation of the radiation visible on the Earth is currently described by a fireball shock model (modified according to recent knowledge)\cite{fbs_model}, as seen on fig. \ref{fig_fbs_model}.

The outflow\footnote{The composition of the outflow is still unknown, although weight seems to shift from baryon-dominated to magnetic-dominated recently.} is emitted in relativistic packets (``shells'') with different Lorentz factors $\Gamma\gg100$. The difference in velocities of the packets cause internal shocks, which in turn lead to emission of radiation in so-called ``prompt emission phase''. It has been a recent discovery, that not only gamma rays are produced at this stage, but also optical radiation, although via different physical processes\cite{nature_grb}.

The second most important stage of emission is the collision of the outflow with interstellar medium (ISM) or wind from the GRB progenitor. The forward shock is responsible mainly for generation of the X-Ray radiation\cite{forward_shock}, while the reflected wave propagating back into the outflow causes so called reverse shocks, which are believed to be responsible for afterglow emission, mainly in optical and radio bands\cite{reverse_shock}. The overall energy emitted in this model ranges from $10^{49}$ to $10^{51}$ ergs.

Data from the SWIFT satellite exhibited another property of GRBs -- so-called X-ray flares, which are ``bumps'' of X-ray radiation appearing in the lightcurve after the start of gamma emission. Those flares are seen in a significant fraction of GRBs and in most cases are 10 times less fluent then the GRB emission. The important fact that arises from the discovery of flares is that they are most probably due to the late activity of the central engine\cite{xray_flares}.

As mentioned before, long GRBs exhibit some properties of supernovae, thus a hypothesis of a collapse of a very heavy supernova -- hypernova\footnote{Hypernova is now generally a name for supernova type Ib and Ic. There is an evidence showing that some of the type Ic supernovae are responsible for GRBs, although it is believed that both Ib and Ic can form a GRB, depending on the geometry of the explosion.} is the leading one. In general, after the physical mechanism of a large object collapsing, this family of models is called ``collapsar'' models\cite{grb_collapsar}. The nuclei of hypernova collapses forming a black hole\footnote{In most models, the black hole is formed in case of a GRB, while in a case of a common supernovae a neutron star is formed. However, there are some exotic GRB models which introduce a multi-stage collapse, first to a neutron star, than to a black hole or even a strange star\cite{grb_quark}.} surrounded by an accreting disk of matter. Due to the rotation, a stream of relativistic barions is ejected which collides with collapsing shell of the star, and the mechanism predicted by fireball shock model follows.

While attributes of supernovae have been seen for long bursts, no such correlation has been detected for short bursts, thus a central engine driving this type of phenomena is much more of a speculation. However, experimental data such as the hardness of short bursts spectra and their correlation with certain regions of the universe suggest a ``merger'' family of models, in which two compact objects, such as two neutron stars or a neutron star and a black hole revolving around each other, collide and merge, forming a central black hole with a disk of accreting matter\cite{grb_merger}. Following mechanism of producing radiation is in general driven by the fireball shock model.

Some of the short bursts observed are probably stronger outbursts of the so-called Soft Gamma Repeaters (SGR), which in most cases emit a short pulse of gamma radiation from time to time, but in most cases much weaker than in case of GRBs. However, there have been certain observations of strong pulses coming from SGRs, which, in case of unknown repeaters may be detected as short GRBs.

In most cases, both the collapsar and the merger type models include the spinning of the central engine, which is probably the main reason behind the outflow being emitted in jets. When the jest axis is directed at Earth, the electromagnetic emission from ultrarelativistic jets, reaching Lorentz factor of up to $\Gamma\simeq1000$, can be seen from cosmological distances, making GRBs one of the most distant objects detected by man. This would not be possible if the emission was isotropic, like we assume in case of supernovae.

The nature of jets is not yet well understood. The high brightness of ``the naked eye burst'' GRB080319B -- the optically most luminous GRB detected so far -- reaching $5.3^m$\cite{nature_grb} is attributed to the two-component jet structure. The inner jet with higher Lorentz factor and the opening angle of only $0.4^{\circ}$ was directed at Earth resulting in higher apparent brightness of the optical emission. it is possible that more GRBs exhibit two jet structure, however, in most cases we only see the outer, wider and less energetic jet, thus the higher magnitudo of other GRBs.

\section{Other phenomena of interest}

\subsection{Novae}
\label{sec_nova}

\begin{figure}[b!]
\begin{center}
	\includegraphics[width=0.6\textwidth]{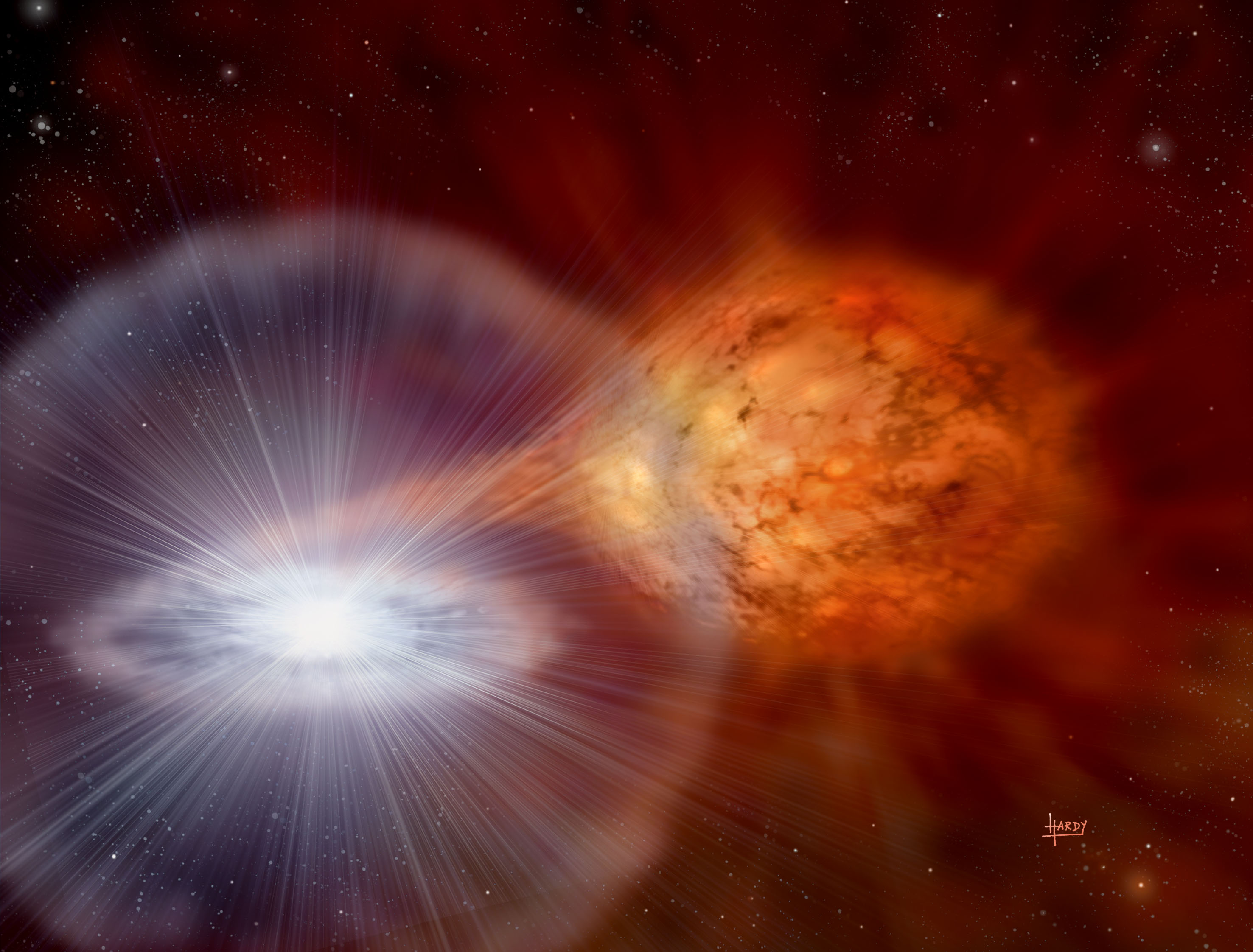}
\end{center}
\caption{An artist view of a white dwarf during a nova explosion (by David. A. Hardy).}
\label{nova_star}
\end{figure}

Novae are perhaps the most famous representatives of so-called cataclysmic variables -- a binary systems consisting of a white dwarf and its companion, a donor star, from which mass is transferred to the dwarf, forming an accretion disk. They exhibit increases of brightness from 6 to 18 magnitudo and thus were among the first variable stars to be discovered. Emission takes place in optical and ultraviolet, as well as X-ray and gamma rays\cite{fermi_nova}.

The increase has the two-step structure. The first step during which brightness is increased by $6^{\rm{m}}-13^{\rm{m}}$ takes approximately 1 to 2 days and is often followed by a few hours to few days plateau and a sudden brightening by about $2^{\rm{m}}$. After reaching its peak brightness, star begins to darken -- the time of darkening by $2^{\rm{m}}$ indicates the type of nova, from ``very fast'' (less than 2 days) to very slow (between 151 and 250 days).

Spectroscopic observations show that during the explosion the star loses about $0.001\%$ of its mass, releasing about $10^{38}-10^{39}$ J. Such energies are very likely to be released only in nuclear fusion, which is believed to be the mechanism of nova outburst. The fusion is ignited by hydrogen from the accretion disk flowing onto the surface of the white dwarf (see fig. \ref{nova_star}). Due to high gravity, hydrogen is compressed and heated to very high temperatures. When it reaches about $20\cdot 10^6$ K, the thermonuclear reaction occurs, and the hydrogen is burned in the CNO cycle. Remaining gas on the stars surface is blown away by the explosion, producing light.

All novae are believed to be a repetitive phenomenon, for only about $0.01\%$ solar mass is ejected in the explosion, which is relatively small compared to the white dwarf mass. Then the accretion and flowing of the hydrogen on the stars surface probably repeats\footnote{If the white dwarf exceeds its Chandrasekhar limit due to long accretion, the interior may become a subject to a carbon fusion, destroying the star in a type Ia supernova explosion.}. However, the time between outbursts is probably too large to be observed for most classical novae. There is a separate class of cataclysmic variables called ``recurrent novae'', exploding much more frequently (10 to 80 years), but not as luminous, and brightening by $4^{\rm{m}}$ to $9^{\rm{m}}$ only due to the fact, that the amount of matter participating in the nuclear fusion is about 10 times smaller than in classical novae.

\subsection{Supernovae}
\label{sec_sn}

\begin{figure}[b!]
\begin{center}
	\includegraphics[width=0.49\textwidth]{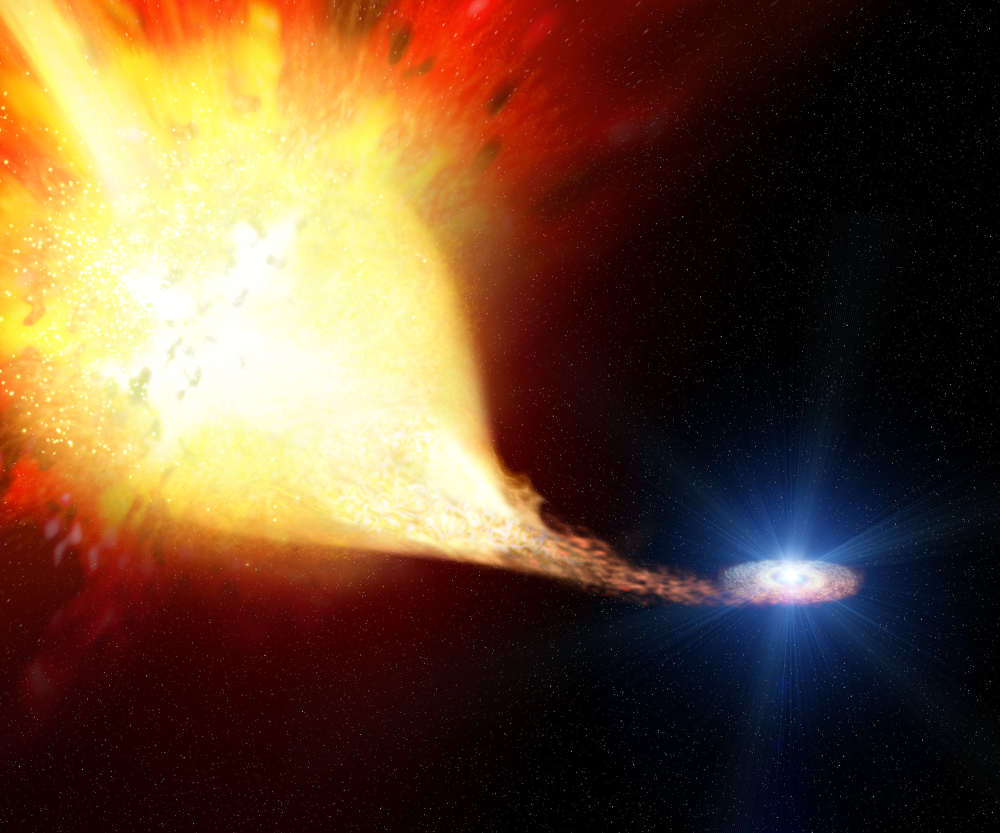}
	\includegraphics[width=0.4082\textwidth]{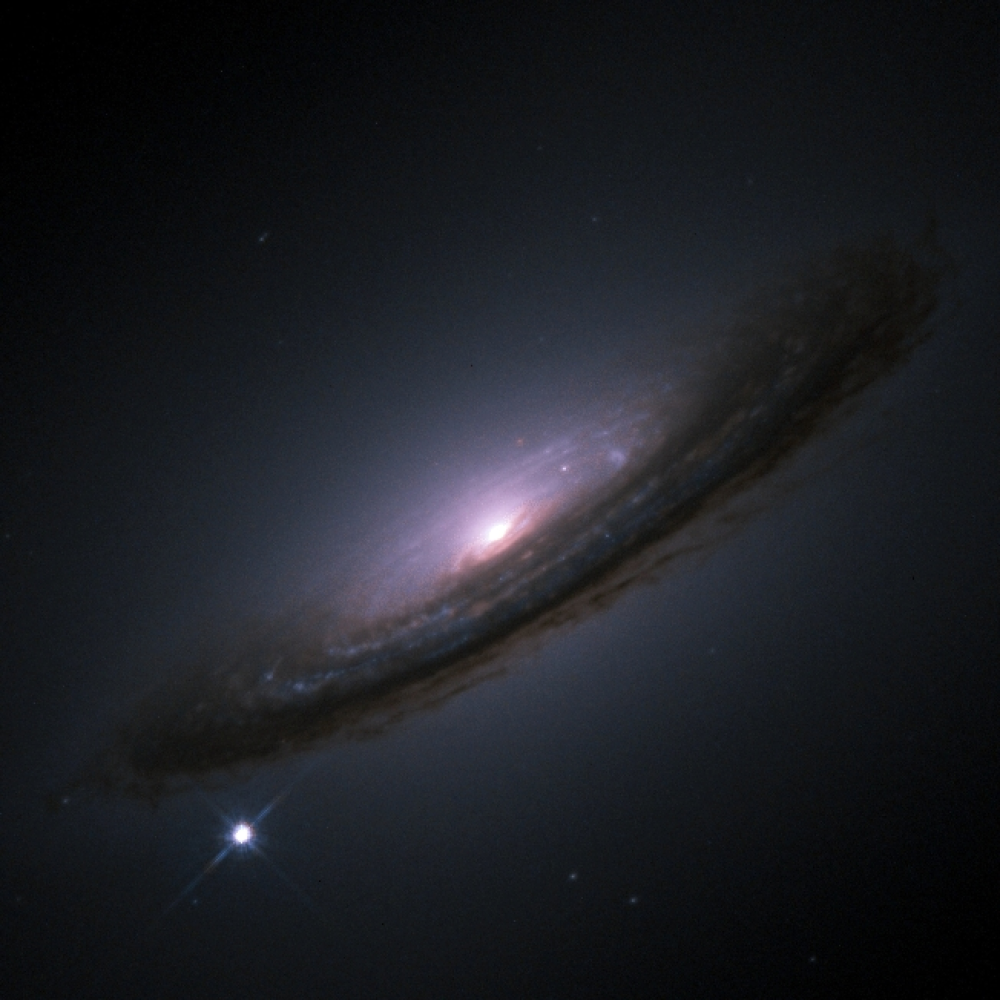}
\end{center}
\caption{On the left: artist's impression on supernova SN 1993J, which later occurred to be mixed type II/Ib ( (c) by ESA ); on the right: Hubble telescope reconstruction of supernova SN 1994D, type Ia ( (c) by Hubble ).}
\label{sn_images}
\end{figure}

Up to year 1930 there was no distinction between novae and supernovae, for both types of explosions appeared as new stars to observes and no methods to compare amounts of energy released were available. In that year, a new class of stars -- supernovae -- was proposed by Baade and Zwicky \cite{sn_naming}. They analysed a nova-like event S~Andromedae, which occurred in 1885 in the Andromeda galaxy. The distance to the galaxy, measured in 1917 by Ritchey, turned out to be much bigger than previously suspected and thus S~Andromedae had to release a much larger energy than typical nova.

Supernovae are extremely bright explosions, often outshining the entire galaxy they reside in. They are catastrophic in their nature, completely destroying the star, but releasing a shock wave into surrounding interstellar medium, creating a so-called supernova remnant (SNR). The phenomenon can be triggered by two different mechanisms. The first happens for massive stars, which burn out nuclear fuel in their core. They can no longer sustain a gravitational pressure and collapse into a neutron star or a black hole. Vast amounts of gravitational potential energy are released in this process, creating a shock wave and heating outer layers of the star. The second scenario is characteristic for white dwarfs -- if the amount of mass accreted from a companion star is large enough for the dwarf to come close to its Chandrasekhar limit, a carbon fusion begins, which becomes a so-called runaway fusion and drives the supernova explosion.

At the first glance, supernovae seem likely to be categorized by their triggering mechanisms -- ignition or collapse of the core. However, they were categorized mainly according to their spectroscopic properties (and historical reasons). Type I (consisting mainly of subtypes: a, b, c, distinguished by other spectroscopic features) lacks a Balmer series (line of hydrogen) in their spectrum, in contradiction to the type II (consisting mainly of subtypes P and L, distinguished by features of a lightcurve). Type Ia is a white dwarf reaching its Chandrasekhar limit and igniting its core, while types Ib,c and type II are probably triggered by collapses.

Supernova of different types and subtypes differ in the mechanisms of converting their mass and gravitational potential into released energy, but the estimated energy output is on the level of $10^{51}$ ergs\footnote{Similar as in the Gamma Ray Bursts (sec. \ref{sec_grb}), but released in a much longer period of time.}. Most of the supernova explosions are triggered by a collapse of the core (about $80\%$). The frequency of supernovae events in the galaxy of size of the Milky Way is estimated to be 1 per 50 years, although the last supernova in Galaxy was observed in year 1604. However, about 150 such explosions are yearly discovered in other galaxies.

\subsection{Flare stars}

\begin{figure}[b!]
\begin{center}
	\includegraphics[width=0.6\textwidth]{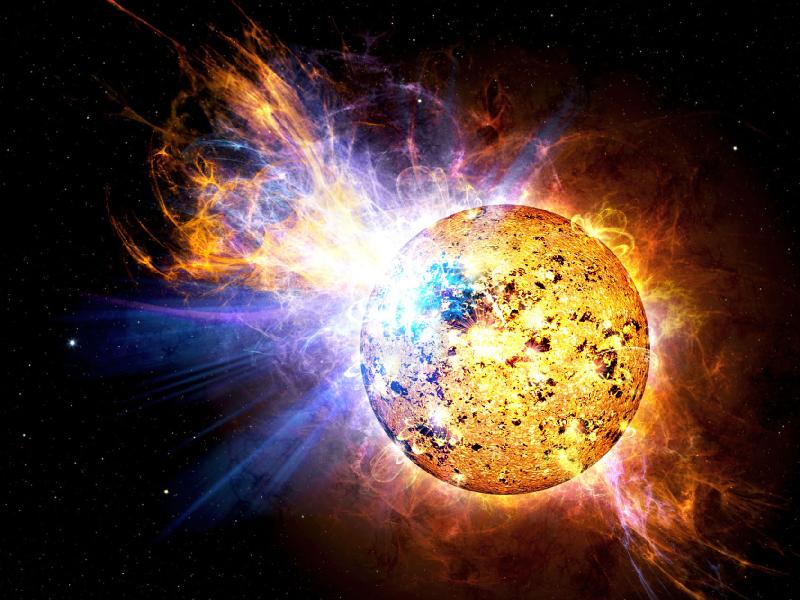}
\end{center}
\caption{An artist view of flare star EV Lacertae during an explosion ( (c) NASA ).}
\label{flare_star}
\end{figure}

Flare stars (also called UV Ceti variables) were discovered much later than novae or supernovae, although they exhibit as violent, or even more violent behaviour than the former. That is probably due to the fact, that most of them appear as a brightening of already visible star, not a new star, and that they are rather a rare phenomenon. Most of the flare stars observed are closer then 50 ly from Earth -- that is due to the fact, that flaring occurs in red dwarfs, which are very dim\footnote{Recent research indicates that flaring may also occur in brown dwarfs.}. However, some were seen as far as 1000 ly away. 

Their brightness can increase by two magnitudo just in seconds or minutes and drop to normal brightness within minutes to hours. The increase takes place in nearly whole spectrum, ranging from radio to X-ray, and the event is highly unpredictable. The flaring takes place due to the similar mechanism as solar flares -- magnetic reconnection. Flux patterns of plasma in the star's atmosphere are typically aligned with the star's magnetic field. From time to time the field rearranges, dropping to a lower energy state. Realignment of the plasma causes it to collide with itself, heating up even from 3000 K to 10000 K, and thus resulting in a star brightening.

Such flares are much brighter than solar flares for two connected reasons. First of all, red dwarf mass is only few tenths of the Sun's mass, thus the material can be ejected much more energetically due to lower gravity. Additionally, red dwarfs are much dimmer than Sun, thus the flaring, which can cover up to $20\%$ of their surface, is much more noticeable.

The flare star closest to Earth is Proxima Centauri, however, the first to be discovered were V1396 Cygni and AT Microscopii in 1924. The most famous one is UV Ceti discovered in 1948, from which the whole group took name (UV Ceti variables). 

\subsection{Other variable stars}

Supernovae, novae and flare stars are of special interest to the \pin project, for they may often appear as a completely new objects on the sky and thus are not subject to regular observation but rather to detection. However, an important amount of astronomical knowledge is derived from study of other variable stars. This study can be successfully performed by an experiment such as ours.

Probably most stars are variable in some way -- even Sun has a 11 year solar cycle, during which its energy output varies by about $0.1\%$. There are currently two main categories of variables, distinguished by the mechanism causing the fluctuations in their light flux.

Intrinsic variables change their actual luminosity due to some phenomenon happening in the star. Among this category are flare stars, novae and supernovae mentioned before, belonging to the subtype of eruptive or cataclysmic (or even catastrophic) variables. There are also stars which, for example, change their radius due to varying temperature and density in their interiors, such as pulsating variables.

In case of extrinsic variables external properties cause the variability in the amount of light that can reach an observer. The first subgroup are eclipsing variables, normally binary systems where one member covers another during their orbiting. The second group -- rotating variables -- change their apparent brightness due to the different parts of the star being visible to the observer and releasing different amounts of light.

\subsection{Blazars}

\begin{figure}[h!]
\begin{center}
	\includegraphics[width=0.58\textwidth]{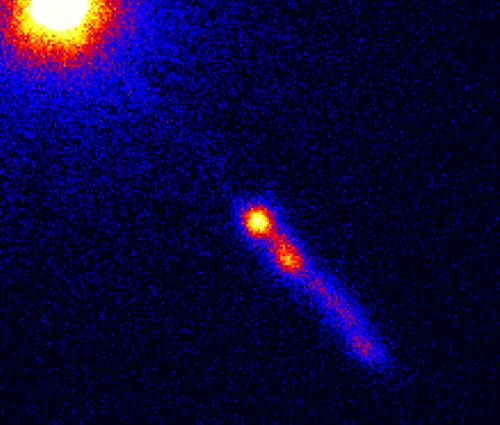}
\end{center}
\caption{The first discovered quasar and blazar -- 3C 273. Image by Chandra X-Ray observatory.}
\label{blazar_img}
\end{figure}

There was a number of objects observed by optical astronomers counted as irregular variable stars -- objects that changed their brightness on a wide range of periods but without any apparent pattern. With development in radio astronomy, especially with increasing the angular resolution of radio telescopes and discovery of quasars in late 1950s it became apparent, that those stars were in fact different objects -- blazars.

It is believed, that in the centre of most or even all massive galaxies supermassive black holes exist. Especially in case of giant elliptical galaxies, they form  active galactic nuclei (AGNs), which are powered by matter falling onto the black hole. Falling matter forms an accretion disk, which emits vast amounts of energy in form of particles. However, the property that decides that an AGN is a blazar are highly relativistic jets (0.95-0.99~c) emitted along the angular momentum of the accretion or the spin of the black hole direction. Plasma in the jets interacts with itself and magnetic field emitting radiation. The spectrum ranges from radio to gamma-rays. For some cases TeV photons were also observed. If a jet is pointing in the direction of Earth, the object is called a blazar.

The important distinction between blazars and other AGNs comes from the velocities of emitted jets. Effects of special relativity, so-called Lorentz boost causes a jet to be a few to few hundred times brighter for an observer than in its rest frame. Additionally, the time scale of the jet variability is much shorter when seen from Earth.

Blazars are the most luminous persistent sources in the Universe known to man, highly polarized ($3-20\%$) in radio and optical frequencies, showing an irregular variability on timescales from days to years. Their brightness and variability makes some of them objects well suited for observation by small, high time-resolution systems such as \pin.

\subsection{Near-Earth objects}

There is a number of visitors to the night's sky that are not of stellar or quasi-stellar nature. Comets and meteors were observed since the birth of astronomy. Nowadays we know, that there are many more celestial bodies much closer to Earth than stars, such as asteroids and planetoids. Careful analysis of their lightcurves can reveal their shape and other properties. However, a pure detection of such an object and deriving its trajectory may prove more important in the future than analysis of its properties -- an answer to an unending threat towards Earth.

There is, however, a group of objects that became a subject for studies very recently. These are artificial satellites of Earth, both machines still functioning and working for man and cosmic debris -- mostly leftovers of past, when man could not imagine, that polluting our planet's orbit may turn out to be a big problem in the future.

The study of satellites is important due to an increasing demand for placing new functioning objects on the orbit. Finding a free place for them is problematic, for trajectories of most cosmic debris (as well as some spy satellites) are unknown. Additionally, debris is a large threat for objects lifted from the planet's surface to the orbit.

Fortunately, most if not all artificial satellites reflect Sun's light and thus are visible from the Earth. Careful studies may result in catalogues of cosmic debris with their parameters. However, this job may only be performed by joint efforts of experiments with a wide field of view and thus large detection capabilities. One of such experiments is \pin.

\chapter{The \pin experiment}
\label{chap_pi}

Progress in astronomical instrumentation constantly increases our understanding of the Universe and reveals new, often unsuspected, mysteries. Unraveling some of those is possible by gathering and careful study of already existing data. However, some new discoveries call for a completely new scientific approach. This was the case with gamma ray bursts -- the main inspiration for the \pin project.

\section{General concept}

The instruments that accidentally discovered GRBs had a non-scientific purpose -- monitoring of large parts of the sky -- a purpose strange to astronomers at these times (sec. \ref{sec_grb}). This idea became more popular in this field of science along with the understanding of GRB phenomenon. Following 30 years of studies lead to construction of the satellite instrument BATSE, then SWIFT and Fermi -- satellites capable of constantly observing very large parts of the sky with high time resolution and thus detecting hundreds of bursts in few years time.

Up to now a few thousands of GRBs were detected by dedicated gamma-ray satellite experiments\cite{GCN}. Only a fraction of these bursts were also seen in other wavelengths, and there were only a few simultaneous observations of optical and gamma-ray emissions. Just one of those measurements had good enough optical time resolution to claim that the first stage of optical and gamma emissions took place at the same time.

The main reason for such a difference in the number of gamma-ray and optical detections of GRBs is the difference in methodology of experiments. Satellite experiments are aimed at the detection of GRBs. They constantly monitor a large fraction of the sky and perform a real-time analysis of data, to detect a signal. The idea that stands behind the vast majority of optical cameras and telescopes is opposite -- they are dedicated to detailed observation of specific astronomical sources, thus having very high sensibility (a big ``range''), a very small field of view and nearly no automatic detection capabilities. There are also systems dedicated to detection and analysis of variable sources, which have a bigger field of view, such as ASAS\cite{ASAS}. However, unlike satellite experiments, they do not monitor a large fraction of the sky, but perform a survey, taking a photo of a specific part of the sky every few hours or days. This limits their ability to measure object's variability on a scale of minutes or shorter and nearly rules out detection of short bursts such as GRBs.

The need for optical monitoring of a large part of the sky is clear, for studying the emission in visible wavelengths from its very beginning can unravel crucial information about this phenomenon (sec. \ref{ssec_ourgrb}). Moreover, if there are other optical transients in the sky not yet discovered, happening on timescales of seconds, is there a better way to find them?

\section{Final system design}

\begin{figure}[b!]
\begin{center}
	\includegraphics[width=0.4\textwidth]{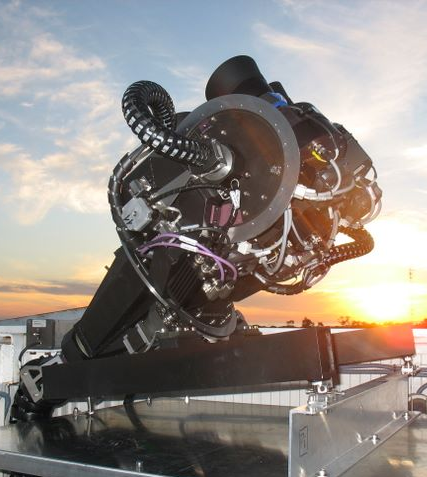}
\end{center}
\caption{\pin final system mount with four cameras installed at the INTA test site near Huelva, Spain.}
\label{fig_full_mount}
\end{figure}

The \pin project was designed to answer the need for optical monitoring of the large part of the sky. The name of the project is derived from the concept of watching $\pi$ steradians of the celestial sphere, almost the whole part available to astronomical observations from a single point on the Earth\footnote{The whole sky visible from a flat surface would be $2\pi$ steradians, but a large part of the sky close to the horizon is filled with intense ambient light, which is the main factor making this area not suitable for scientific observations.}. Currently we assume that the full system is going to cover 1.5 steradians of the sky. This area will hopefully increase in the future. Additionally, a high time resolution (in optical terms) and self-triggering capabilities are going to be the main features of the experiment.

The field of view of 1.5 steradians is obtained by using 12 cameras, each covering $20^{\circ} \times 20^{\circ}$ of the sky, using Canon EF lenses with $f=85$ mm and $f/d=1.2$. Cameras are of a unique construction and design prepared by \pin project members. Exception is a commercial CCD sensor, with roughly $2000\times 2000$ pixels\footnote{Term ``pixel'' in this work is used in two meanings: pixel as a (in this case) square area on the frame or CCD sensor or as a distance equal to the side of such square.} and a pixel size $15\times15$ $\rm{\mu m}^2$ (corresponding to an angular size of 36 arcseconds). This setup, with 10~s exposures and 2~s readout time gives estimated range of $11.5^{\rm{m}}$ on a single frame and $13-14^{\rm{m}}$ on 20 stacked frames. Parameters such as cooling, readout gain, etc. can be controlled by USB 2.0 or Ethernet interfaces. Cooling is performed with a two-stage Peltier stack and allows reaching $40$ K below ambient temperature. 

Additionally, a custom shutter has been developed. 10 s exposures with 2 s readout result in 2000 to 3000 frames per night, 200000 to 300000 exposures after 4 months of operation, the maximal lifetime of best commercial shutters. \pin shutter, sustaining $10^7$ cycles is enough for a few years of uninterrupted operation.

Four cameras are installed on each of three paralactic mounts (fig. \ref{fig_full_mount}). Two modes of observation are possible -- so called ``deep'' and ``wide''. In the ``deep'' mode axes of all cameras on a single mount are parallel, so they observe the same field of view. Proper summing of the frames in such a configuration increases the range of the instrument\footnote{The simplest way to increase the apparatus range using multiple images of the same field is to make an average of such images. This way the signal to noise ratio is increased, for the random noise is not additive, while the light from interesting sources is. Of course, more sophisticated algorithms of such averaging can be used, which take into account slightly different position and orientation of a source on different images.}. In case of detecting an interesting source (i.e. GRB optical counterpart), the flash recognition algorithm can automatically send a command to the mount to switch cameras into the ``deep'' mode. In the ``wide'' mode, which is the standard mode of operation resulting in the field of view of 1.5 steradians, cameras on a single mount cover neighbouring parts of the sky.

\begin{figure}[t!]
\begin{center}
	\begin{displaymath}
\begin{array}{ccc}
 	\reflectbox{\includegraphics[width=0.4\textwidth]{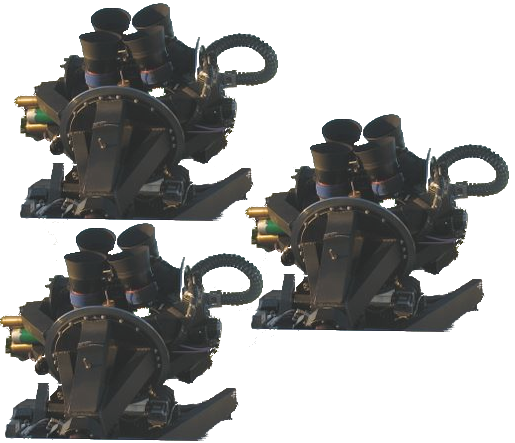}} & \raisebox{3cm}{$\underleftrightarrow{\sim 100\ \rm km}$} &	\includegraphics[width=0.4\textwidth]{pi_of_the_sky/3_mounts_clean.png}
\end{array}
	\end{displaymath}

\end{center}
\caption{A scheme of the final \pin system -- two sites, each with 12 cameras installed on 3 mechanical mounts. Sites are distant by about 100 km, to allow filtering out near-earth objects flashes with the parallax measurement.}
\label{fig_pi_final_scheme}
\end{figure}

\begin{table}[b]
\begin{center}
\begin{tabular}{cc}
\hline
\bf Site separation [km] & \bf Orbit [km] \\
\hline
30 & 92000 \\
100 & 293000 \\
200 & 579000 \\
\hline
\end{tabular}
\caption{Three exemplary sites separation distances and corresponding maximal orbits of satellites that can be rejected by parallax, assuming position difference threshold for automatic event rejection equal to twice the pixel size (72 arcseconds).}
\label{tab_parallax}
\end{center}
\end{table}

Two sites are being prepared for the final \pin system, aiming to observe the same part of the sky from two locations (see fig. \ref{fig_pi_final_scheme}). With distance of more than 30 km (probably about 100 to 200 km) such configuration can be efficiently used to eliminate flashes caused by artificial sources -- mostly satellites and cosmic debris reflecting the light of the Sun using parallax. Depending on their linear and rotational speed, construction and distance to Earth, part of such reflexes are similar in characteristics (both shape and duration) to optical flashes of astrophysical origin (sec. \ref{ssec_uncor_flash}) and parallax is the only way of undoubtful rejection. Table \ref{tab_parallax} shows maximal orbit radius of an artificial satellite which can be rejected with parallax for specific site separation distances. Separation distance of 30 km should allow for rejection of most near-earth objects such as communication satellites, however 100 km separation would be needed to reject very high orbit satellites, such as Vela, which should be sufficient for recognition of all near-Earth objects in the \pin range. The 200 km separation distance extends the parallax usability beyond Moon's orbit.

\begin{figure}[t!]
\begin{center}
	\includegraphics[width=0.89\textwidth]{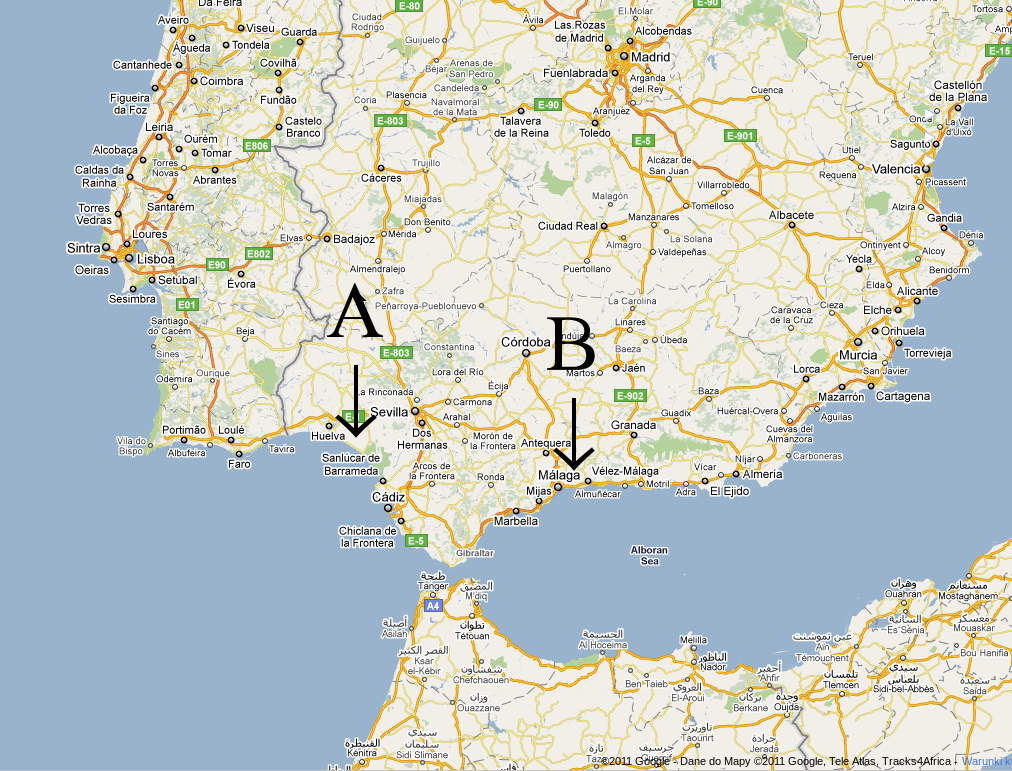}
\end{center}
\caption{Locations of the existing and a considered sites of the final \pin system. The existing site A lies near Huelva, while the second, B, near Malaga in Spain.}
\label{fig_spain_map}
\end{figure}

\begin{figure}[b!]
\begin{center}
	\includegraphics[width=0.4\textwidth]{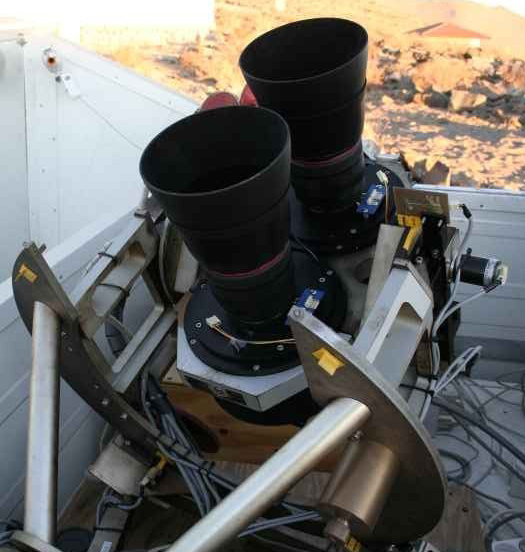}
\end{center}
\caption{The \pin prototype at the in Las Campanas Observatory on the Atacama desert in Chile.}
\label{fig_prototype}
\end{figure}

The whole setup has been designed to work fully autonomously, requiring a direct human assistance as rarely as possible -- hopefully regular maintenance only once or twice a year, or in case of serious emergency. All elements -- cameras, mounts, domes, etc. can be controlled via Internet, thus allowing remote managing of minor disturbances. The flash detection is done on-line through a multilevel trigger, nearly instantly generating information about detected interesting events. Simultaneously, selected parts of data are gathered for further, offline analysis. The system starts and stops automatically, changing the modes of observation according to specified rules and scripts. All collected data are a subject to data reduction algorithms and the final results are stored in the dedicated database.

The installation of the first mount of the final systems with four cameras took place in October 2010. The observation site is located near Huelva in Spain in the site of INTA\cite{inta}. For the second site, CSIC\cite{csic} research station near Malaga is considered. The distance between the two sites is about 240 km, see fig. \ref{fig_spain_map}. Nearly all of the algorithms developed for the \pin project have been successfully tested with the prototype system working for 6 years in Chile.

\section{The prototype}

The prototype apparatus of the \pin project (fig. \ref{fig_prototype}) was installed in the Las Campanas Observatory\cite{lco} on the Atacama desert in Chile in the site of the Carnegie Institution of Washington, in June 2004. Thanks to the hospitality of the ASAS group\cite{ASAS} it operated in the ASAS dome till December 2009. This site was chosen mainly because of conditions, which are among the best on the Earth for observational astronomy -- very low humidity, an unclouded sky most of the year and low sky background -- due to the climate of this place and its high elevation (2380 m above sea). Recently the system was moved to the San Pedro de Atacama observatory in Chile, about 1300 km north of LCO, and started gathering data there.

Apparatus consists of a paralactic mount with two cameras having the field of view, number of pixels and optics similar to those being installed in the full system. Cameras watch the same part of the sky and work in a coincidence mode. The distance between them is far to small to make use of the parallax, but this mode allows rejecting false flashes caused by noise fluctuations and cosmic rays -- most significant background source for flash detection algorithms.

Since the overall field of view of the prototype is $20^{\circ} \times 20^{\circ}$, the standard mode of operation is to follow the centre of FoV of the SWIFT. The mode is changed into following INTEGRAL satellite FoV or observing selected sources, when the SWIFT FoV is below horizon. At the beginning and by the end of the night a scan of the whole sky is performed, taking 3 images of each field. Contrary to the normal mode, which is dedicated to detecting optical transients on scale of seconds to minutes, scan data is suitable for off-line analysis of sources variable on timescales of hours to months. All the modes of operation can be interrupted by a GCN alert -- information about a GRB in progress -- which cause the mount to point to the burst coordinates and start follow-up observations. Afterwards, the system reverts to the previous task.

A very high level of autonomy and reliability has been reached through years of testing the prototype. Currently system has very sophisticated self-monitoring capabilities, automatically detecting and solving problems and sending an appropriate e-mail or sms to people involved. No human interference has been required during normal night-shifts since about middle of 2009.

There are generally two types of data-analysis software running on the prototype system. On-line flash detection algorithm constantly analyses stream of collected frames (with $11^{\rm m}-12^{\rm{m}}$ range, depending on the conditions) and gives a real time information about detected transients (sec. \ref{ssec_flash_recog}). During the day-time, data is being analyzed in detail and catalogued to a database, on which off-line algorithms searching for variability on longer time-scales are being run (sec. \ref{ssec_other_recog}). The data analysis software is explained in more detail below.

\section{Analysis software}

\subsection{Flash recognition system}
\label{ssec_flash_recog}

A common approach in astronomy is to gather data during observation time and perform a sophisticated data-reduction later. This approach, although appropriate for dedicated studies of some objects of special interest, is not suited for a project like \pin. Single \pin camera gathers 16-24 GB of data per night. Full system is going to produce 512-768 GB per night, amount impossible to store for a longer time, not mentioning a detailed analysis. However, only a very small fraction of this is interesting for us and has to be saved. The question is, how to select this fraction?

In its approach to flash recognition, the \pin project is similar to large particle physics experiments, which analyse huge amounts of data in search for very few interesting events. In such a case, only interesting events including signatures of information required for detailed analysis are kept, the rest being discarded as ``background''. This task is performed by so-called triggers -- multilevel selection algorithms that receive huge number of events and reduce it by applying cuts on consecutive levels. They start with very fast and simple rules, eliminating the most obvious and numerous background and finish with sophisticated methods dedicated to analysis of the few events remaining. Same logic is utilized by our project, the only difference is the nature of the data stream -- series of images of the sky.

At the input to the selection algorithm, the frame, after dark subtraction, is modified by a kernel-based transform, similar to a discrete Laplace transform in electromagnetism. In ideal case this would make all real light sources pixel-like, remaining dark pixels a background. In the real case it sharpens the image and vastly reduces effects caused by background light gradients, mainly due to the Moon and clouds. Following steps of the algorithm are a simple recognition of pixels above estimated background level -- $\rm T_n$ (after removing saturated pixels), and veto on constant objects $\rm T_v$, which eliminates mainly stars -- pixels above background that were already present on previous frames. These two simple cuts gives a list of new candidates for flashes on the frame, still most of them not of astrophysical origin.

Following is a number of cuts that eliminate temporal errors of a CCD (like ``hot'' and ``black'' pixels). Then a list of pixels, which may be attributed to a single flash is transformed into a list of objects. The last of simple, pixel-based cuts applied removes stacks of close objects, mostly caused by moonlight illuminating dense, moving clouds.

A border line between the very fast, ``level 1'' trigger and the more sophisticated ``level 2'' trigger is the ``coincidence''. Lists of objects remaining after the cuts from the two cameras (observing the same field of view) are compared and only those flashes that are present on both are left for further analysis. In case of the prototype, this cut eliminates mostly events caused by cosmic radiation and random noise fluctuations. In the final system, due to parallax, coincidence will also remove flashes caused by near-earth objects, like satellites and planes. That will vastly reduce or even discard the need for the whole level 2 algorithm\footnote{An alternative to coincidence of flashes on two cameras is a requirement for flash to appear on two consecutive frames on a single camera. Although this approach requires an optical transient to be longer in time, it was used when there were problems with one camera in the prototype system. Currently, prior to the installation of the new apparatus of the final system in the second site, it is used in flash detection algorithm in INTA.}.

Level 2 cuts are designed to remove satellites, planes and fluctuating stars. This is mainly performed by discarding events whose position coincides with position of a satellite or a star found in dedicated catalogues. An event is rejected also if flashes observed on consecutive frames can be grouped into a track or have an elongated shape. The reconstruction of tracks require a number of frames after detecting the flash to be stored an analysed, which causes the time between the flash and a final confirmation of the detection to be quite long. Fortunately, as mentioned before, level 2 cuts should not be needed in the final system.

\begin{figure}[tb!]
\begin{center}
	\includegraphics[width=0.5\textwidth]{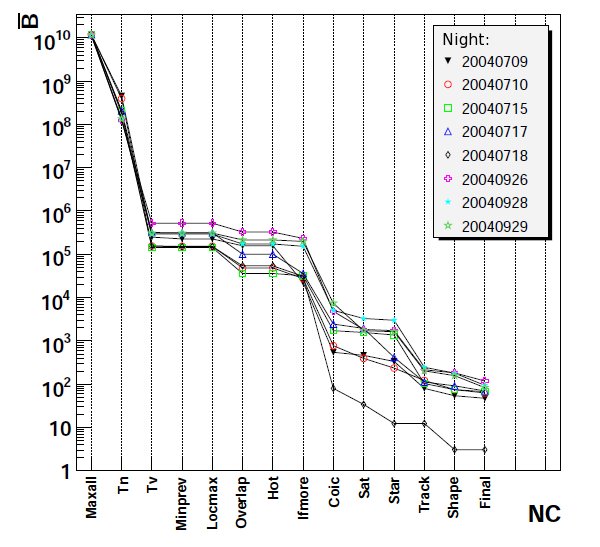}
\end{center}
\caption{Number of events $\mathrm{\bar{B}}$ considered in the on-line flash detection algorithm after consecutive levels of the trigger NC\cite{lwp_mgr}. Different markers represent different nights. The analysis was performed in 2004, at the beginning of the prototype operation. Later studies allowed to reduce the number of final events by an order of magnitude.}
\label{fig_background_simul}
\end{figure}

The number of pixels considered on the input of the flash recognition algorithm in a single frame mode exceeds $10^{10}$ per night for a pair of cameras (fig. \ref{fig_background_simul}). Simple, pixel-based cuts reduce number of flash candidates to the order of $10^5-10^6$ and coincidence reduces it to the order of slightly more than $10^2$ events per night. In the time of writing this work, the number of final events on which an ``eye analysis'' has to be performed is approximately few to over a dozen per night.

However, more than a dozen events per night for a pair of cameras means even as much as 300 hundred events for human examination in the full system. That's why a third level algorithm is being worked on, which includes even more sophisticated methods such as Hugh transform for finding tracks of meteor and plane like events\cite{msok_general}.

For each of the final events some basic astronomical data has to be calculated, like its brightness and position in celestial coordinates. These tasks are performed on-line by special photometric and astrometric algorithms which are much faster but less accurate than off-line objects cataloguing procedures.

\subsection{Astrometry and photometry}

Astrometry is a term used to describe a way of finding the celestial coordinates of an astronomical object from its image. The first step is determining the position of the object on the frame, which is not a straightforward procedure. Each optical imaging system is characterized by a point spread function (PSF) -- a response of the detector (in this case lenses + CCD sensor) to a point source of light\footnote{Objects such as stars or most of the artificial satellites of Earth, due to their large distance and an angular resolution of 36 arcseconds are point-sources to \pin}. Thus a point-source that normally would be contained in a single pixel is spread over several pixels, mainly due to diffraction and optical aberrations. That fact makes the calculation of the source's position often a complicated task and a number of algorithms analysing data stored in multiple pixels were invented. Fortunately, it also allows precision of coordinates determination to be a small fraction of a pixel, whereas for a single pixel the uncertainty is $\frac{1}{\sqrt{12}}$.

There is a wide variety of astrometric methods. Starting with simple or more sophisticated pixel-based, where the centre of an object is calculated from a set of chosen pixels as a, for example, centre of mass. This type of procedure is widely used in ``Pi of the Sky''. Often slower, but giving more precise results is fitting of a known point spread function to the object -- one of the main topics of this thesis (chap. \ref{chap_polynomials}).

\begin{figure}[tb!]
\begin{center}
	\includegraphics[width=0.5\textwidth]{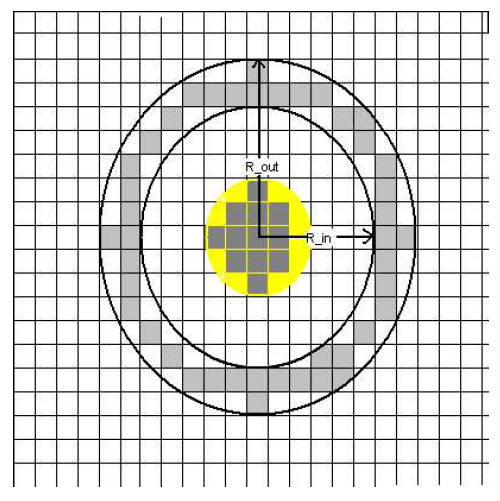}
\end{center}
\caption{An aperture used in the fast photometry algorithm\cite{msok_phd}. Pixels in the yellow circle are summed as signal, while a median of pixels in the black ring (radius R\_in to R\_out) is used as a background estimator.}
\label{fig_fast_photometry}
\end{figure}

Astrometry procedures often include also measuring of the object's brightness, which is, in general, the aim of a photometric algorithm. These also range from simple, pixel based algorithms, many of them described as ``aperture photometry''\footnote{The name ``aperture photometry'' comes from a defined configuration of pixels attributed to an object in this algorithm, called an aperture.}, up to those mentioned above, based on the detailed PSF modelling, called ``profile photometry''. Aperture photometry is often very fast, but not necessarily precise. One, simple variation of this algorithm, involving calculation of a cluster of pixels belonging to a star and subtracting background, estimated as a median of pixels in a ring around the star (fig. \ref{fig_fast_photometry}), is used as a ``fast photometry algorithm'' in \pin, for quick determination of a detected flash brightness and coordinates. 

More sophisticated astrometric algorithm, adapted from ASAS\cite{ASAS}, is used for cataloguing stars on stacked frames. In general, it calculates signal within several apertures of different sizes and returns an average. This algorithm is currently used in the analysis of variable stars in the \pin project, although modifications are being worked on, involving results presented in this thesis.

The coordinates of an object on the frame have to be recalculated then to celestial coordinates. This involves an iterative procedure of determining the pointing direction of the apparatus. For this purpose the number of bright reference stars is chosen from the frame, their positions (and brightnesses) calculated based on their coordinates on the frame and an approximate detector's orientation. This data is then compared with star catalogues and, as the final result, celestial coordinates of the frame centre and its orientation as well as coordinates of all recognized objects are found.

\subsection{Novae, flare stars and other variable stars recognition}
\label{ssec_other_recog}

Although some very rapid and big changes in brightness of known objects may be detected by the on-line flash recognition algorithm, most variable stars are recognized through off-line algorithms. The step preceding all off-line variability searches is cataloguing of the stars, i.e. performing astrometry and photometry for each object on single or 20 stacked frames in case of normal measurements and 3 stacked frames in case of images from the sky scan. Results for each object are then put into a database, along with information about conditions during the measurement, temperature of the detector, total number of objects detected, etc. A database created in this way is then processed by algorithms searching for novae-like events, flare-like events and those determining variability period.

Novae search algorithms look for new stars in the database. For each new object found in the night being analysed, a number of cuts is applied, to eliminate false events, such as fluctuations coming from very bright stars, hot pixels, vibrations of the mount, etc. Remaining objects are a subject to an examination by a member of the project. This method is proper for finding novae, supernovae (although no supernovae during the \pin work period was in our range), blazars, some flare stars or other variable stars with large amplitude, that are normally below \pin range.

Flare star recognition method is similar to the novae recognition. The difference is, that all existing objects in the database are examined in search for a sudden increase in magnitudo compared to previous measurements. Stars satisfying this criteria are a subject to consecutive quality cuts, same or similar to those in the novae search. Resulting lightcurves have to be examined by man.

The last one is the algorithm determining the periods of variability for variable stars in the database. Based on this algorithm, periods of already known variables can be determined, as well as possible periods for stars suspected to be variable. Also calculated is the uncertainty of estimated period, amplitude of magnitudo variations and the parameter reflecting quality of the fit -- this information is a key to finding new variable stars.

\section{Scientific results}
\subsection{Variable stars and quasi stellar sources}

\begin{figure}[b!]
\begin{center}
	\includegraphics[width=0.7\textwidth]{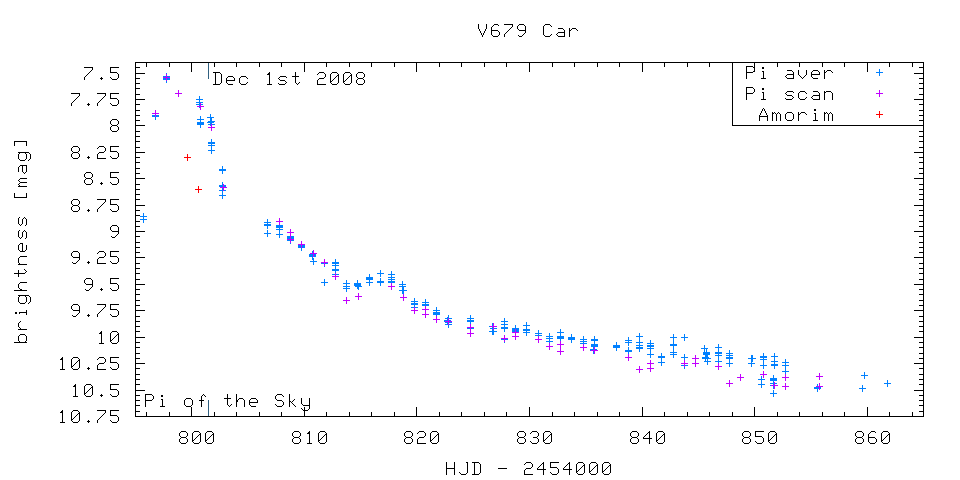}
\end{center}
\caption{V679 Carinae nova explosion, automatically discovered by the \pin project with the offline novae detection algorithm\cite{nova_v679}. ``HJD'' stands for Heliocentric Julian Day.}
\label{fig_nova_pi}
\end{figure}

During 2004-2009 working period, the novae detection algorithm of the \pin prototype system automatically detected 3 novae stars. The nova  V679 Carinae (fig. \ref{fig_nova_pi}) was discovered on 1st of December 2008 and was observed for many days later\cite{nova_v679}. \pin released an AAVSO\cite{AAVSO} notice requesting followup observations which resulted in precise astrometry and spectroscopic observations by the SMARTS\cite{smarts} 1.5m telescope. The star occurred to be a classical nova of the ``Fe II'' type.

On $16^{\mathrm{th}}$ of September 2007 outburst was detected in the place of 1RXS~J023238.8-371812 object. Followup observations showed redshift zero and a blue spectrum, making this object possible SU UMa-type or a WZ Sge-type outburst\cite{nova_1rxs}. On $15^{\mathrm{th}}$ of December 2007, \pin discovered brightening coincident with a $20.88^{\rm m}$ star GSC2.3~S55U020591, which reached nearly $11.5^{\rm m}$. Followup spectroscopic observations with Gunma Astronomical Observatory\cite{gunma} 1.5-m telescope and a low resolution spectrograph, showed a very blue continuum and other features characteristic for a WZ Sge-type dwarf nova\cite{nova_gsc}. Several other novae have been automatically detected by \pin, but we were not the first to discover them.

\begin{figure}[t!]
\begin{center}
	\includegraphics[width=0.6\textwidth]{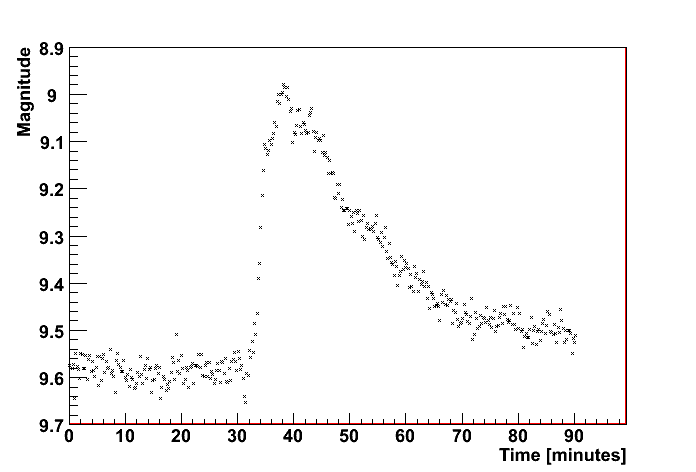}
\end{center}
\caption{Flare star GJ3331A automatically discovered by an off-line flares detection algorithm\cite{flare_gj3331a}.}
\label{fig_flare_pi}
\end{figure}

The first outburst detected by \pin was a flare star CN Leo explosion on $2^{\mathrm{nd}}$ of April 2005\cite{cn_leo}. It was so violent that it has been found by an on-line flash recognition system. Off-line flare detection algorithm discovered two additional flare stars. On $28^{\mathrm{th}}$ of November 2006 a GJ3331A brightened from $9.58^{\rm m}$ to $9.03^{\rm m}$, reaching its peak brightness in less than 10 minutes\cite{flare_gj3331a}. Following was a decay monitored for more than an hour (fig.~\ref{fig_flare_pi}).

During the night of $18^{\mathrm{th}}$ July 2008 another outburst was detected, connected to a USNO1050.00569325 star\cite{flare_2}. It's characteristics, like significant proper motion (~0".1/year) and 2MASS J-K color of 0.85 indicate that the object is an M dwarf, suggesting a flare explosion.

\begin{figure}[b!]
\begin{center}
	\includegraphics[width=0.3\textwidth]{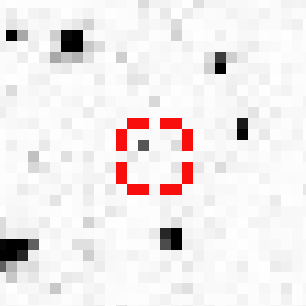}
\end{center}
\caption{Image of the blazar 3C 454.3 taken by the \pin prototype. Blazar, being on the edge of the \pin range, is the dark pixel in the top left corner of the marker.}
\label{fig_blazar}
\end{figure}

Additionally one blazar was observed by the \pin project (fig. \ref{fig_blazar}). It has been automatically detected by the novae detection algorithm in the database, unfortunately during its testing period. Thus no regular observations have been performed and its unprecedented bright peak has been missed. However, it would be possibly seen by the full \pin system and showed \pin potential at discovering and observing bright blazars.

In addition to observations described above, a catalogue of 725 variable stars with periods from 0.1 to 10 days, based on 2 years data (2004-2005) has been published\cite{var_cat_2004}. The classification was performed with the procedures developed for ``Pi of the Sky'', inheriting from light-curve shape analysis methods described in the GCVS catalogue. For 15 stars with previously unknown period, the period has been found and some stars were classified to a different type than claimed in the GCVS catalogue. Preliminary information on the variable stars catalogue based on measurements from years 2006-2007 has also been published\cite{var_cat_2006}, but the data are still in analysis.

\subsection{Uncorrelated flashes}
\label{ssec_uncor_flash}

\begin{figure}[t!]
\begin{center}
$
\begin{array}{c|c|c}
	\includegraphics[width=0.31\textwidth]{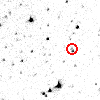} &
	\includegraphics[width=0.31\textwidth]{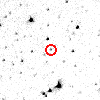} &
	\includegraphics[width=0.31\textwidth]{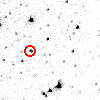} \\
\end{array}
$
\end{center}
\caption{A false \pin trigger generated by a rarely flashing satellite (frame 0, central plot). Satellite is also visible on frames -5 (left) and 6 (right), each with 10 s exposure and 2 s readout, thus it blinks approximately once per minute.}
\label{fig_satellite_serie}
\end{figure}

Up to now more than 250 uncorrelated flashes have been detected by the \pin on-line flash recognition algorithm. For a number of these optical transients their positions have been very close to some identified sources or archival GRB positions, however this is not an indicative information. Possibly most of flashes are generated by rotating artificial satellites reflecting the sunlight. Some satellites are observed as a track segment on the picture, blink often enough to be attributed to a track (fig. \ref{fig_satellite_serie}), have movement-distorted shapes or are simply in a catalogue. However, there are some that cannot be attached to any of these groups and those will be distinguishable from an optical transient of astrophysical origin only with the parallax, in the final system. 

\begin{figure}[b!]
\begin{center}
$
\begin{array}{c|c|c}
	\includegraphics[width=0.31\textwidth]{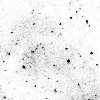} &
	\includegraphics[width=0.31\textwidth]{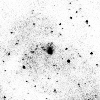} &
	\includegraphics[width=0.31\textwidth]{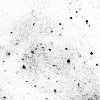} \\
\end{array}
$
\end{center}
\caption{An uncorrelated bright optical transient visible on 2 frames in their centre, detected by \pin 2005.05.31, 4:33:17. On the left -- a frame preceding the flash, followed by frames with the flash.}
\label{fig_flash_serie}
\end{figure}

Seven among the uncorrelated flashes were visible on more than one frame (fig. \ref{fig_flash_serie}). There are no known objects orbiting Earth that have small enough angular velocity to be seen in the same place (with \pin accuracy) and reflect enough light to be in the range of our experiment. Thus such events are probably of an astrophysical origin. With the full system an instant alert will be released in case of discovery of such an event, requesting follow-up, deeper range observations.

\subsection{GRBs}

In the time of writing this thesis, nearly 800 gamma ray bursts were detected by satellites after the \pin experiment start-up (tab. \ref{tab_grb_statistics}). As expected, about half of them happened during the day-time in Las Campanas Observatory. About $17\%$ of GRBs were out of the reach of our cameras, being either fully on north hemisphere or just below horizon. For nearly $18\%$ of bursts the apparatus was off, but this includes long period of about 10 months prior to the installation at the new site.

\begin{table}[t]
\begin{center}
\begin{tabular}{|c|c|c|c|c|}
\hline
 & \bf all det. GRBs & \bf apparatus off & \bf north hemisph. & \bf daytime \\
\hline \hline
\bf Events & 786 & 138 & 49 & 401 \\
\hline
\bf Fract. of all & 1 & 0.1756 & 0.0623 & 0.5102 \\
\hline \hline
 & \bf below horizon & \bf clouds & \bf outside FoV & \bf inside FoV \\
\hline \hline
\bf Events & 82 & 19 & 92 & 5 \\
\hline
\bf Fract. of all & 0.1043 & 0.0242 & 0.117 & 0.0064 \\
\hline
\end{tabular}
\caption{Statistics of all gamma ray bursts detected from the star of the \pin prototype on $1^{\mathrm{st}}$ of July 2004 up to $5^{\mathrm{th}}$ of October 2010. Only 5 out of almost 800 bursts were in the prototype's field of view and one of these bursts was detected by \pin. Up to date statistics can be found on the \pin project web page: http://grb.fuw.edu.pl/pi.}
\label{tab_grb_statistics}
\end{center}
\end{table}

More than a dozen of bursts could not be seen due to dense clouds, while nearly $12\%$ were initially outside of the apparatus field of view -- delay caused by a rotation towards proper coordinates is in most cases too long for us to spot the transient. However, 5 bursts were in the \pin prototype FoV from the very beginning. For 4 of them limits during and before the burst have been set -- still a very rare data in the study of this phenomenon. One burst have been autonomously detected by the flash recognition algorithm and it proved to be a very significant discovery.

\subsection{GRB080319B}
\label{ssec_ourgrb}

\begin{figure}[t!]
\begin{center}
	\includegraphics[width=0.35\textwidth]{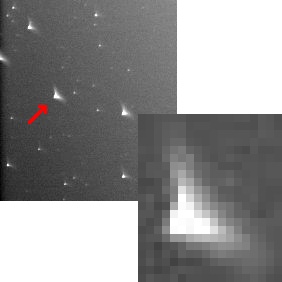}
\end{center}
\caption{The photograph of GRB080319B in its peak brightness, taken by the \pin prototype. For the first few frames, before the detector repositioned following GCN alert, the gamma ray burst was localized in the corner of the apparatus field of view, thus a large deformation of its shape.}
\label{fig_our_grb_part}
\end{figure}

The discovery of the GRB080319B optical counterpart by the \pin project, as well as TORTORA and Raptor experiments was a coincidence of an idea of full sky monitoring and luck. The burst was preceded by an another, which happened roughly half an hour earlier, in coordinates distant by about $10^{\circ}$. After receiving the GCN alert, \pin rotated to the GRB080319A and started the follow-up observations, remaining in the same position for half an hour. Field of view of $20^{\circ} \times 20^{\circ}$ was enough to see, in the corner of the frame, the second burst (fig. \ref{fig_our_grb_part}). Thus it was a mixture of luck, two explosions having quite close celestial coordinates, and idea of full sky monitoring -- without such a big field of view (which is very big in astronomical standards, even though it was still just a FoV of one camera), no such detection would be possible.

\begin{figure}[b!]
\begin{center}
	\includegraphics[width=0.65\textwidth]{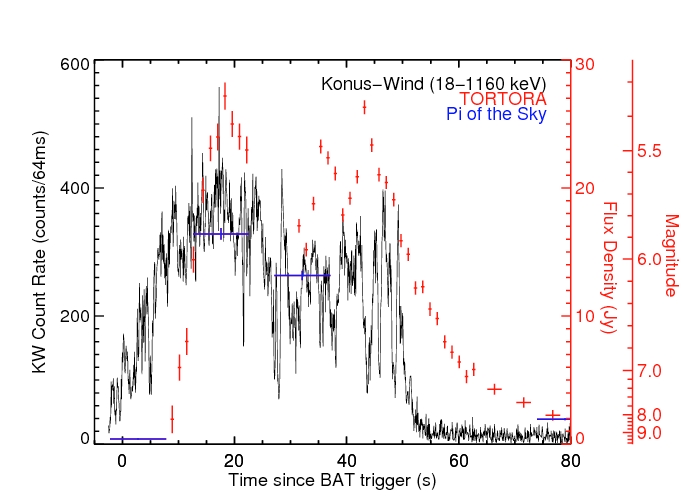}
\end{center}
\caption{Lightcurve of the initial emission from GRB080319B. Superimposed are points from \pin and TORTORA optical experiments, and from gamma-ray detector Konus on the WIND satellite. The first point, taken by the \pin prototype shows that the emission in the optical band and gamma rays started simultaneously\cite{nature_grb}.}
\label{fig_our_grb_nature_lc}
\end{figure}

The first, 10 s exposure of the \pin prototype on which the burst is visible, coincides for the first time in history with the start of emission in gamma rays. There were few previous simultaneous observations, but the time resolution was far to low to connect observed optical emission with the start of emission in higher energies. Additionally, a little bit later the burst became visible to TORTORA experiment, which, with its very high time resolution registered optical peaks similar to these visible in the gamma band (fig. \ref{fig_our_grb_nature_lc})\cite{nature_grb}. However, the existence of a correlation between lightcurves is still a subject of discussion.

The burst was unusually bright in X-Ray and the most bright ever observed in optical wavelengths, reaching peak brightness of $5.3^{\rm m}$. Measured distance was $z=0.937$, {i.e.} $7.5\cdot 10^9$ ly, meaning it is the brightest object observed so far by man. It's unusual brightness and some other lightcurve features can be explained by emission taking place in a two component jet -- the narrower, more luminous component directed at Earth, which may be a very rare occurrence.

\begin{figure}[t!]
\begin{center}
	\includegraphics[width=0.6\textwidth]{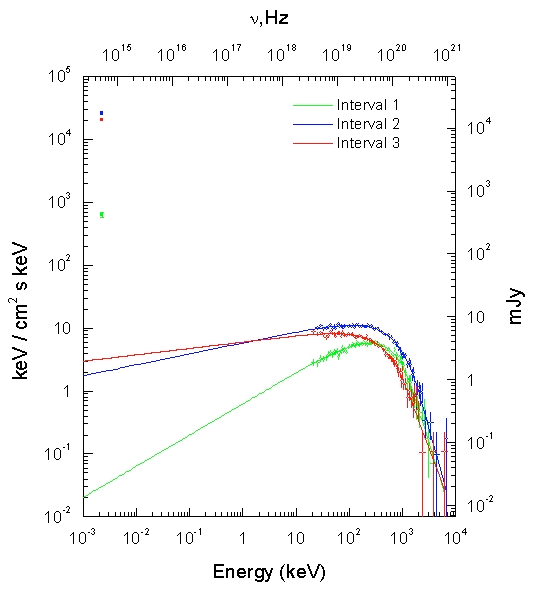}
\end{center}
\caption{GRB080319B spectrum in gamma rays and its extrapolation to optical energies in time intervals corresponding to the first three images of the burst by \pin. Three colourful points in the top left corner of the plot show optical flux for these intervals measured by \pin, much higher than the extrapolation. This data was used to prove, that initial optical emission is not a low-energy tail of the gamma ray emission.}
\label{fig_our_grb_nature_spectrum}
\end{figure}

The discovery of GRB080319B not only showed that there exists a simultaneous emission in optical and gamma bands, which was a hypothesis not accepted by the whole scientific community, but additionally provided scientific data to study that emission mechanisms. Before this observation, most models predicted that the optical emission, if observed from the very beginning of the burst, could be due to a low-energy tail of the high energy emission. However, the extrapolation of the measured gamma-ray spectrum to the optical energies (fig. \ref{fig_our_grb_nature_spectrum}) showed that in case of GBR080319B the optical flux exceeded the possible low-energy tail by 4 orders of magnitude. Thus radiation in these two different bands has to be produced by a different physical mechanism -- the most natural explanation is synchrotron emission for optical and synchrotron self-Compton (SSC) mechanism for gamma rays.

\pin experiment was fortunate to register GRB080319B from its very beginning. However, on the first 3 exposures, before the detector repositioned following GCN alert, the burst was in the very corner of the frame, and due to optical aberrations appeared very deformed (fig. \ref{fig_our_grb_part}). Standard photometric procedures are optimized for symmetric point spread functions, thus such a deformation can introduce additional uncertainties to measurements of brightness and position.

\section{Uncertainties}

\subsection{Photometry uncertainties}

The quality of brightness measurement for a given object strongly depends on the signal to noise ratio. In an ideal detector with no noise, no nonuniformities, stable and linear gain, photometry uncertainty would be determined by the Poisson statistics of photons and should be a monotonic function increasing with magnitudo. However, the real case is much more complicated. First of all, a saturation threshold makes objects below certain magnitudo level too bright for the detector. Additionally, there are multiply factors causing signal to fluctuate, like gain variations, tracking instability of the detector or changes in observation conditions.

\begin{figure}[t!]
\begin{center}
	\includegraphics[width=0.479\textwidth]{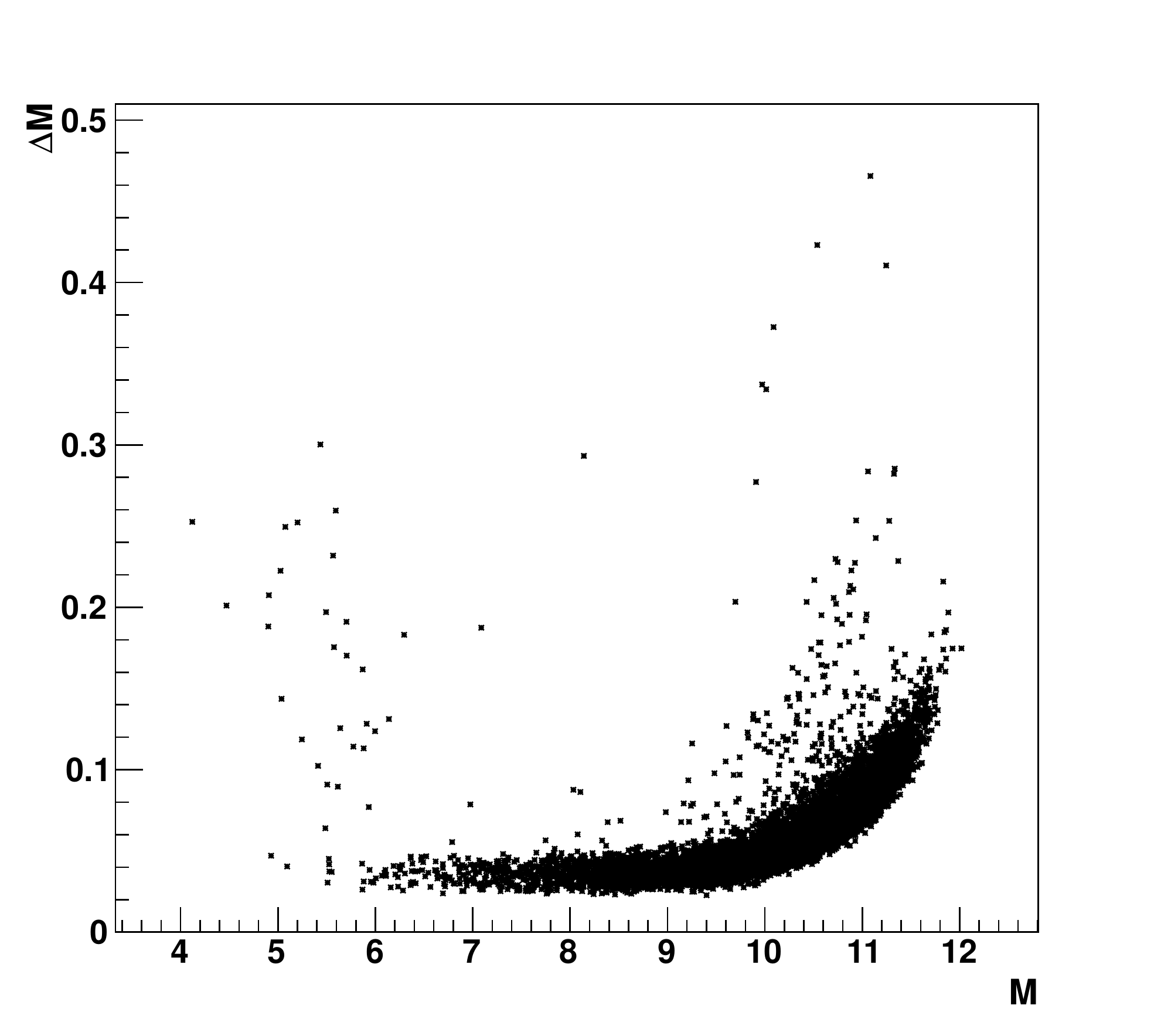}
	\includegraphics[width=0.479\textwidth]{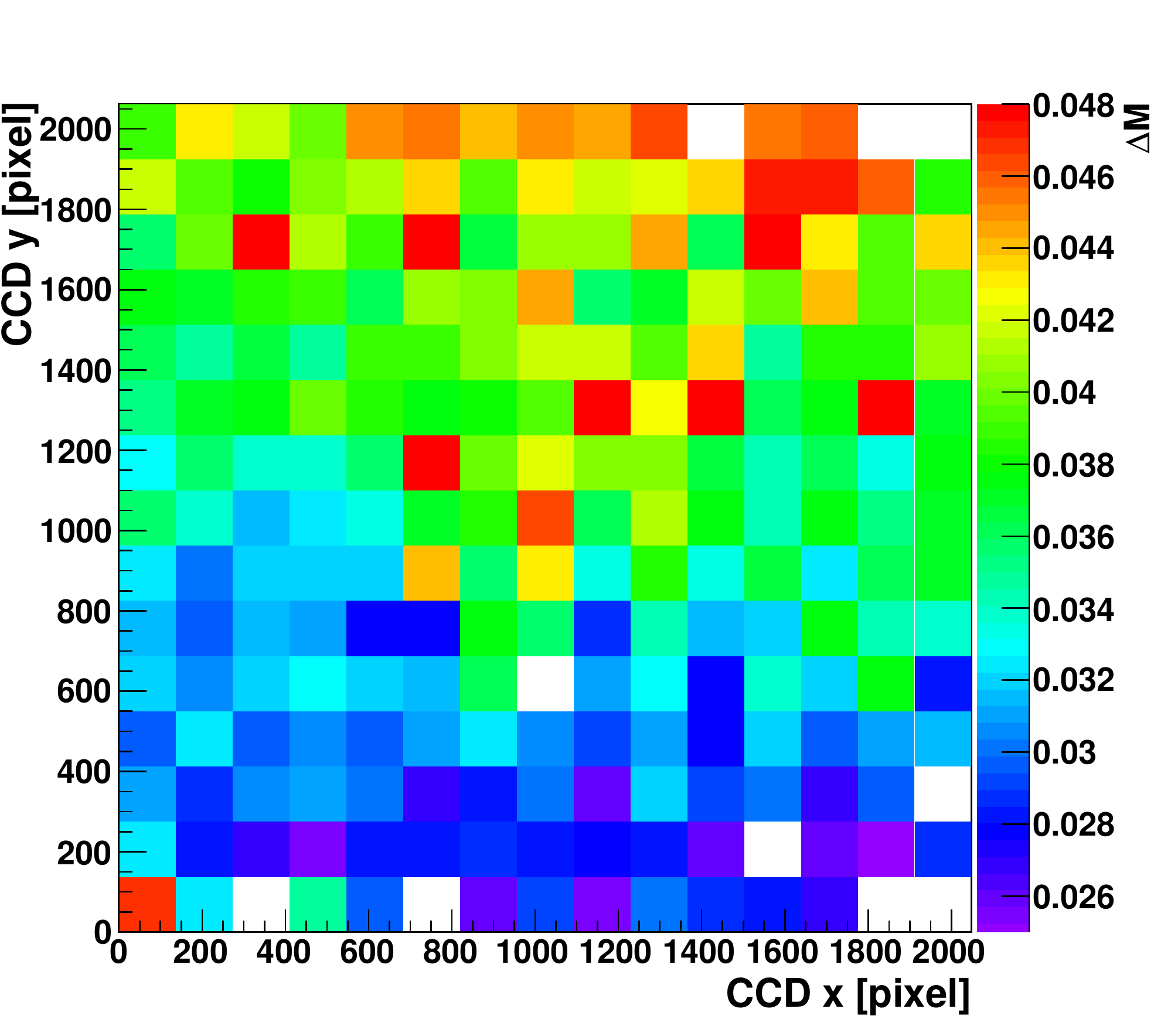}
\end{center}
\caption{Magnitudo uncertainty $\Delta$M distributions, as a function of the star magnitudo M (left) and as a function of the star position on the frame (right) for stars with $6<\mathrm{M}<9$.}
\label{fig_mag_uncert}
\end{figure}

All these factors result in the dependence of magnitudo uncertainty on magnitudo which is shown in fig. \ref{fig_mag_uncert} (left). The brightness of stars below $\sim 6^{\rm m}$ is very poorly measured. This is due to gain nonlinearity at highest charge values and pixel saturation. For stars above $\sim 9.8^{\rm m}$ a fast increase of uncertainty with magnitudo is observed, which continues up to detector maximal range of around $12^{\rm m}$. However, for stars between $\sim 6^{\rm m}$ and $\sim 9.8^{\rm m}$ the uncertainty in brightness determination does not depend on magnitudo. This behaviour is probably induced by a combination of many factors, such as a non-Poissonian signal fluctuations (for example, fluctuations of the CCD amplifier gain) and the photometric algorithm, which may be working best in this range.

Moreover, the photometry quality strongly depends on the position on the frame (fig. \ref{fig_mag_uncert}, right). The problem could be caused by a nonuniform illumination of the CCD, but this should be compensated to large extent by proper frame reduction (division by a ``flat'' frame) and photometric algorithm, which takes into account local background. The large vertical gradient of the magnitudo uncertainty observed in fig. \ref{fig_mag_uncert} (right) indicates that part of the fluctuations of the stars brightness could be caused by a vertical charge transfer procedure. However, slight differences along the horizontal axis are also visible, which may be caused by different shapes of the PSF on different parts of the frame (optical axis not exactly perpendicular to the CCD surface).

\subsection{Astrometry uncertainties}

To find celestial coordinates of an object visible on the frame several steps must be performed. First, a position on the CCD has to be found. Then it has to be transformed into celestial coordinates, which requires determining pointing direction of the apparatus.

\begin{figure}[t!]
\begin{center}
	\includegraphics[width=0.489\textwidth]{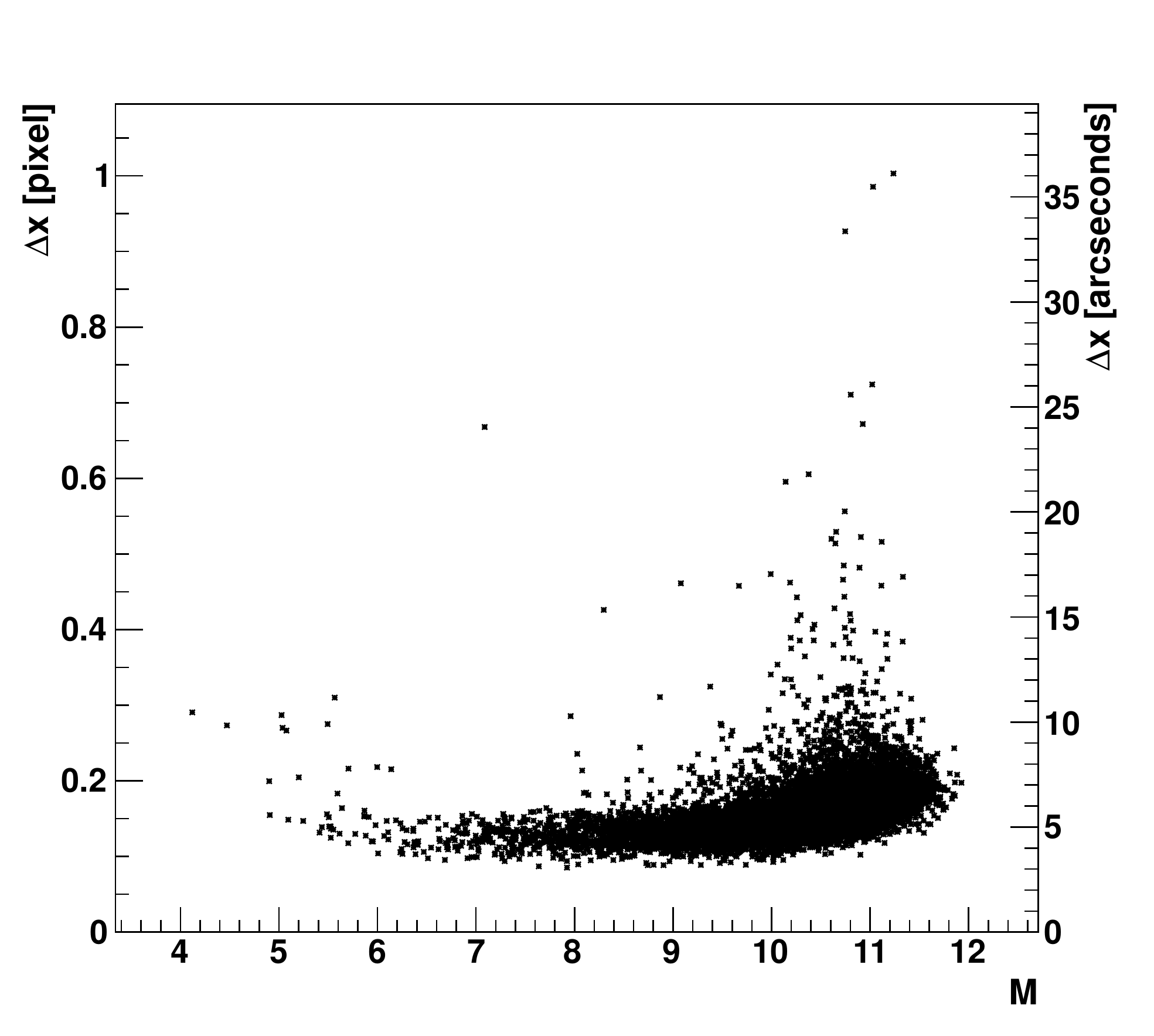}
	\includegraphics[width=0.489\textwidth]{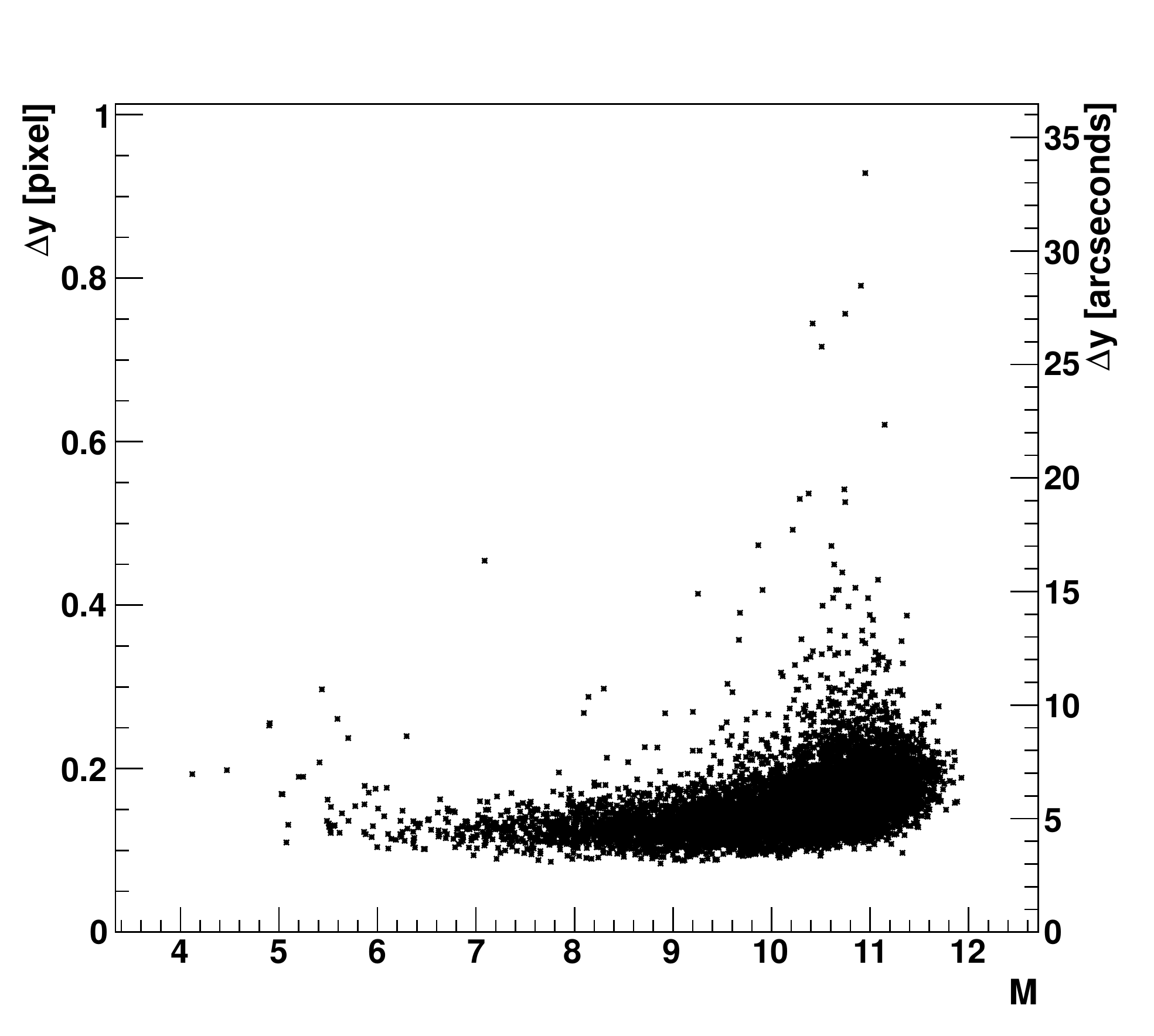}
\end{center}
\caption{Horizontal (left) and vertical (right) position uncertainty vs measured magnitudo M. The dependence is clearly non monotonic, the uncertainty largest for saturated stars and those close to the detector's maximal range.}
\label{fig_pos_uncert}
\end{figure}

The uncertainty of the latter step is rather small, for the calculation involves comparing a great number of stars with a catalogue. In the first step, similarly to the photometric algorithm, uncertainty depends on signal to noise ratio. Thus it should increase with the stars magnitudo. However, in case of the \pin project, the dependence is not monotonic. As seen in fig. \ref{fig_pos_uncert}, the lower edge of the uncertainty band drops with increasing magnitudo up to $9^{\rm m}-10^{\rm m}$, and then starts to increase. Additionally, saturation threshold does not influence astrometry as strong as photometry. 

\begin{figure}[b!]
\begin{center}
	\includegraphics[width=0.489\textwidth]{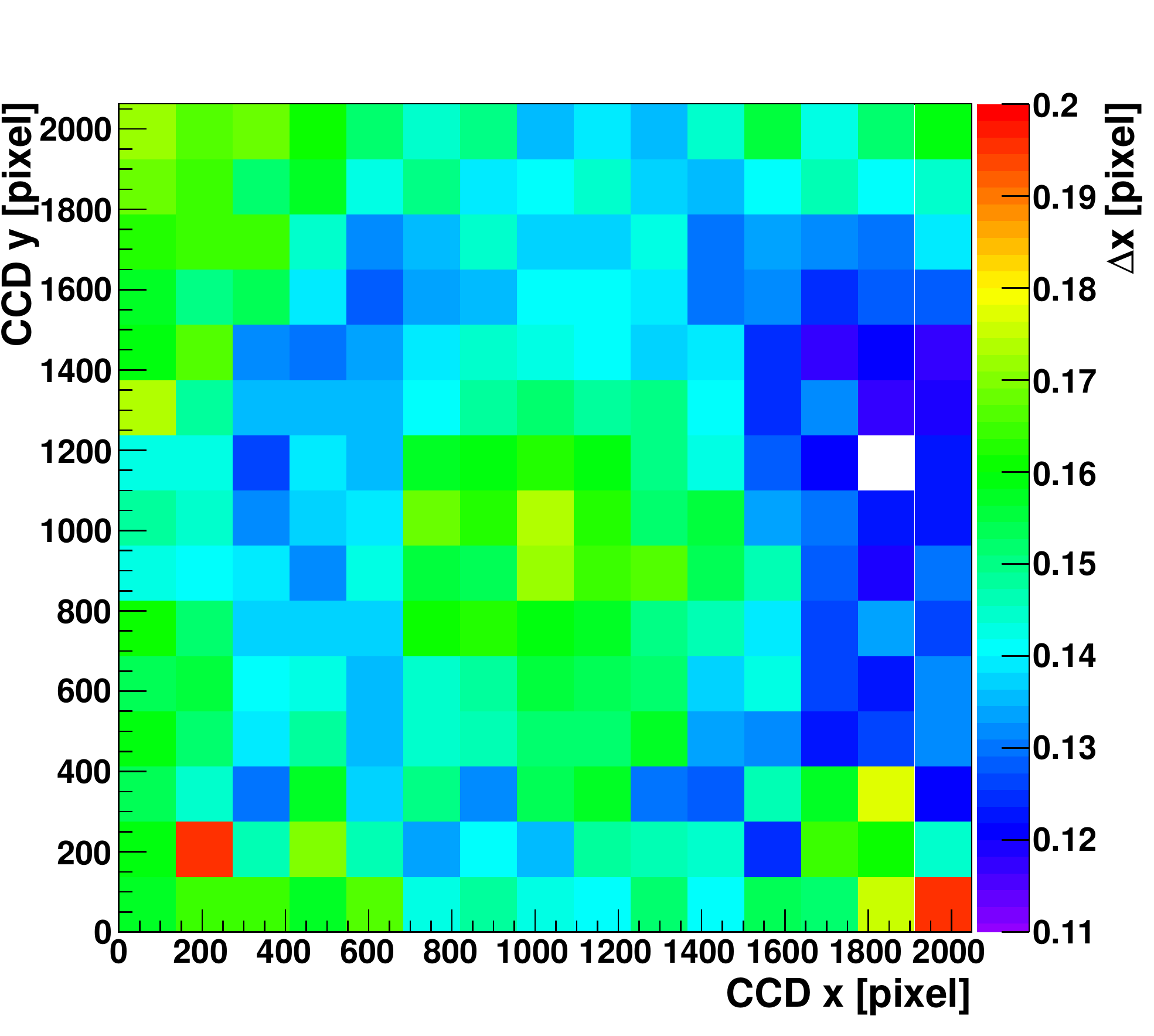}
	\includegraphics[width=0.489\textwidth]{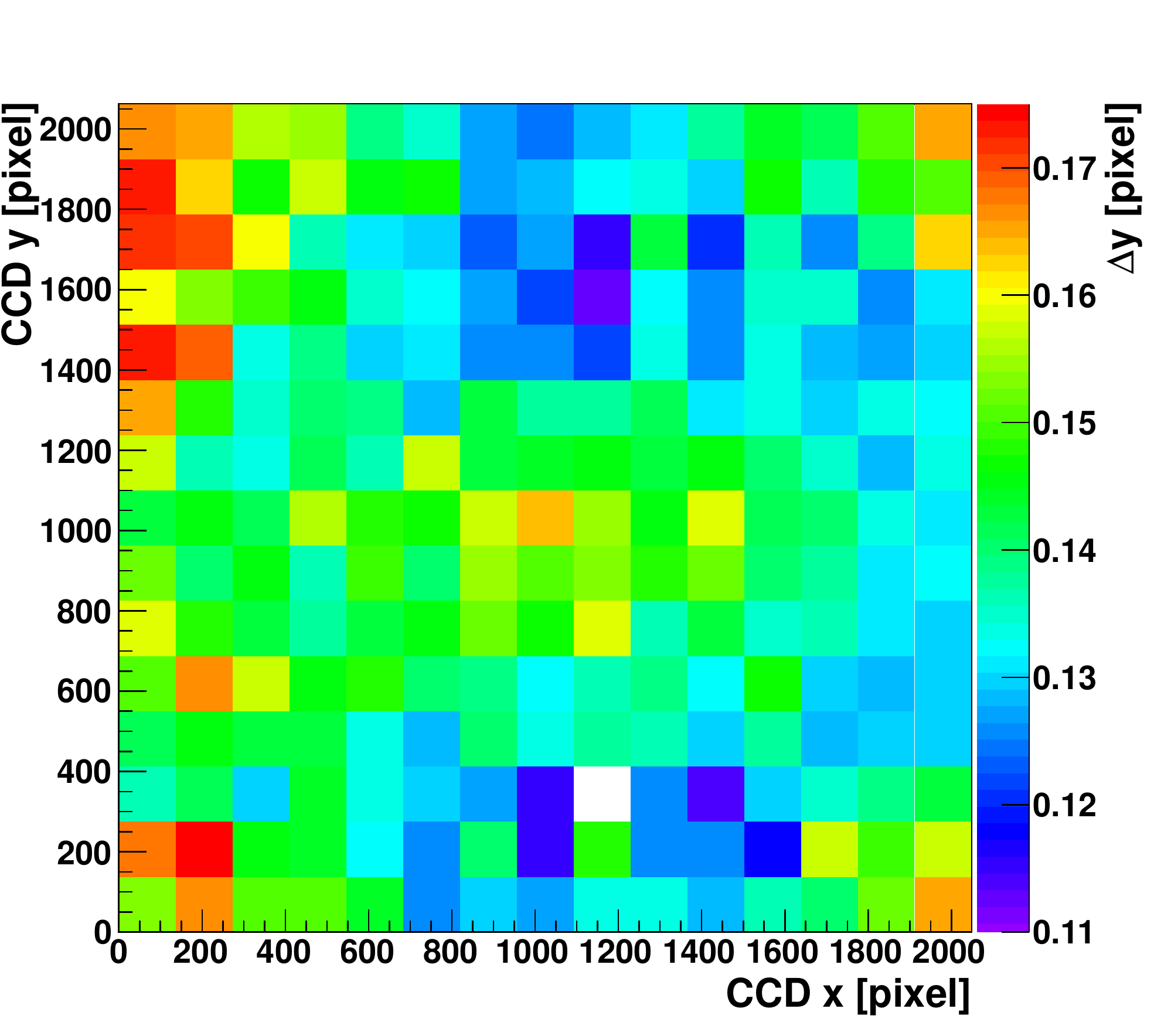}
\end{center}
\caption{Average position uncertainty vs position on the frame for stars with catalogue magnitudo $6^{\mathrm{m}}-10.5^{\mathrm{m}}$. The dependence shows rotational symmetry, which is consistent with PSF change along radial axis in the axially symmetric lenses.}
\label{fig_pos_uncert_xy}
\end{figure}

The dependence of the position uncertainty on the star position on the frame does not show an indicative vertical gradient (fig. \ref{fig_pos_uncert_xy}). In both horizontal and vertical directions uncertainties show an approximate rotational symmetry. However, the radial function is not monotonic, showing a broad minimum at about 800 pixels from the frame centre. It is also clear that the rotational symmetry is only approximate, as there are also changes along the azimuthal coordinate. The observed symmetry is a strong indication that the uncertainty in astrometry (and probably in photometry too) is enlarged by dependence of the PSF on the position on the frame. The PSF should have a significant rotational symmetry in the axially symmetric lenses, but there could be some deviations due to, for example, slight misalignment of the lenses and CCD axis.

\subsection{How to improve photometry and astrometry?}
\label{ssec_phot_ast_imp}

Analysis of the photometry and astrometry uncertainties show, that they depend on parameters such as position on the frame. It seems very likely that this dependence is caused by a change of the PSF on the frame -- a fact known, but never carefully studied in an experiment with such a big field of view as ``Pi of the Sky''. Using rotationally symmetrical apertures or PSFs for strongly deformed, definitely not rotationally symmetrical, images of stellar objects is a very risky approach. Therefore analysis of shape of the PSF and its dependence on the position on the frame is well justified, if not for the hope of improving photometry and astrometry then for better understanding of the \pin detector and future detectors with such FoVs.

\begin{figure}[b!]
\begin{center}
	\includegraphics[width=0.5\textwidth]{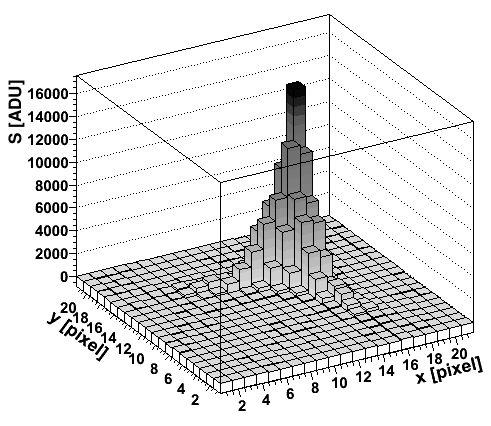}
\end{center}
\caption{An image of a single star, shown as a ``lego'' plot. The star is observed close to the frame corner. The signal S is spread over less than 30 pixels, thus making its shape difficult to extract.}
\label{star_points}
\end{figure}

\subsubsection{PSF parametrisation idea}

To analyse the shape of the PSF and its variation with the position on the frame, the shape itself has to be determined first. Images of the stars observed on frames are convolution of the PSF and the CCD pixel response. As the pixel size is quite large for \pin camera, most of single star images even at the very corner of the frame, consist of less than 30 pixels, thus giving less than 30 data points for shape analysis (fig. \ref{star_points}). While it is possible to fit a scale and position of a well known profile to this amount of data, it would be very hard to derive a profile's shape itself. This derivation requires a higher resolution, much better than pixel size. Assuming that a star is placed in a slightly different position on each frame, relative to pixel centre, a series of such images can result in sub-pixel resolution. Additionally, one can assume that the PSF is invariant under the rotation around the frame's centre and take into account all the stars within a certain distance from it. This operation performed on a long series of frames should give an average profile with a good enough resolution. However, it turned out that the obtained profiles were not satisfactory, especially for positions far from the frame centre.

There are many reasons why this method is not sufficient for shape derivation purposes. The most important is that star images used in the analysis need to be properly superimposed. The task is most difficult for peripheral positions, where the centre of the profile cannot be precisely defined, due to the lack of knowledge of the profile's shape. Additional blur comes also from the fact, that instead of considering stars at a fixed radius one has to sum the profiles from an annulus around the frame centre. Other uncertainties enter because we are observing stars with different spectral types (which can affect PSF shape), due to blur caused by cameras' vibrations when following the sky movement, or simply, because of atmospheric turbulence. 

Additionally, the behaviour of the CCD sensor itself may affect PSF, for example due to different sensitivities to different wavelengths. All these uncertainties can be eliminated or at least vastly reduced, when the data for PSF parametrisation is obtained from laboratory measurements -- using an immutable source with a known spectra and in controlled conditions.

\chapter{Cameras laboratory measurements}
\label{chap_lab}

Astrophysical objects that are of the interest for the ``Pi of the Sky'' experiment are so far away, that from the detector's point of view, can be treated as point sources. For the corresponding measurement in the laboratory the apparatus should consist of a CCD camera imaging a source which can be considered point-like -- an object for which an image on the sensor is much smaller than a pixel size. This condition can be fulfilled by a small enough artificial source of light placed far from the camera. 

Additionally, a high resolution point spread function measurement requires precise control of the relative position of the light source image on a CCD pixel. This could be accomplished by moving the camera or the light source. Considering camera shape, dimensions and weight, the latter occurs to be an easier solution. However, camera movement is the easiest way to perform a big change of the image position on CCD, when the PSF is to be measured in a different parts of the sensor. Based on these assumptions, a dedicated laboratory equipment has been prepared. 

\section{Laboratory setup description}

\begin{figure}[tb]
\begin{center}
	\includegraphics[width=0.2\textwidth,angle=90]{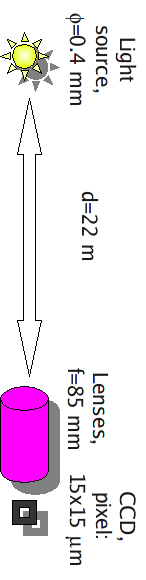}
\end{center}
\caption{The schematic layout of the setup used for laboratory measurements.}
\label{setup}
\end{figure}

The experimental setup consisted of a LED diode (red, green, yellow, blue or white) placed behind a pinhole of $0.1$ mm diameter at a distance of $22$ m from a CCD camera\footnote{The length of the corridor in the Particle Division building.}. A pixel size of the CCD sensor was $15 \times 15\ \rm{\mu m}$, and the camera was equipped with CANON lenses with focusing length of $85$ mm (fig. \ref{setup}) the same as in the \pin prototype and the final system. The expected geometrical diameter of the image, neglecting diffraction and assuming perfect optics is 1.5 $\mathrm{\mu m}$ (0.1 pixel).

\begin{figure}[tb]
\begin{center}
	\includegraphics[width=0.4\textwidth]{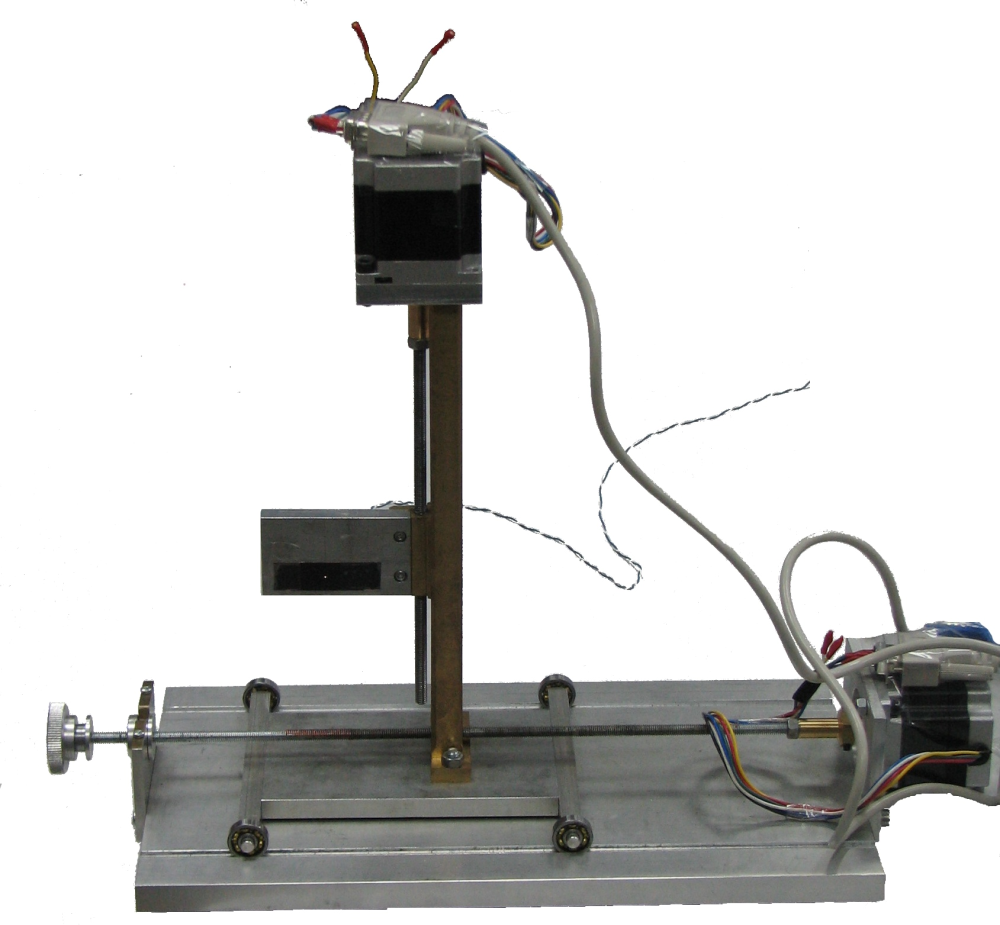}
\end{center}
\caption{The mechanical mount with two step motors used for precise positioning of the light source (LED diode) in the laboratory measurements}
\label{fig_mymount}
\end{figure}

The diode was placed in a mechanic mount, driven by two step motors, that allowed a precise movement in vertical and horizontal axis (as shown in fig. \ref{fig_mymount}). A constantly shining source was much too bright for PSF measurements, even with the lowest possible voltage in the diode working range. Thus a pulse generator was used as a power source, so that the diode image brightness on the CCD could be adjusted by changing the pulse length. Starting exposures, driving the step motors and triggering the generator pulse were controlled by a computer with self-written, dedicated software and scripts. The whole setup was placed in the basement corridor in the Soltan's Institute for Nuclear Studies' building on Ho\.za 69 in Warsaw. 

\section{Laboratory setup tests}

The main differences between a real star and the diode acting as one are the finite distance to the camera, artificial triggering of the light emission and the mechanical mounting. Thus, before performing PSF measurements a proper focusing of the lenses had to be found, as well as mechanical and optical stability had to be tested. 

\begin{figure}[tb!]
\begin{center}
\includegraphics[width=0.49\textwidth]{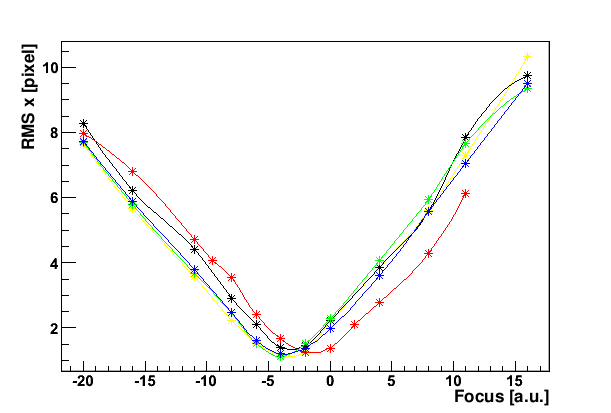}
\includegraphics[width=0.49\textwidth]{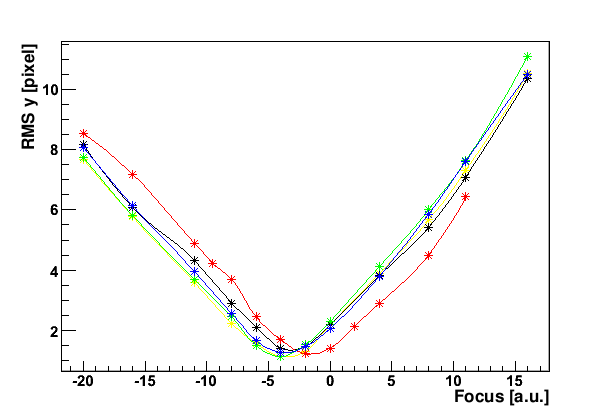}\\
\includegraphics[width=0.49\textwidth]{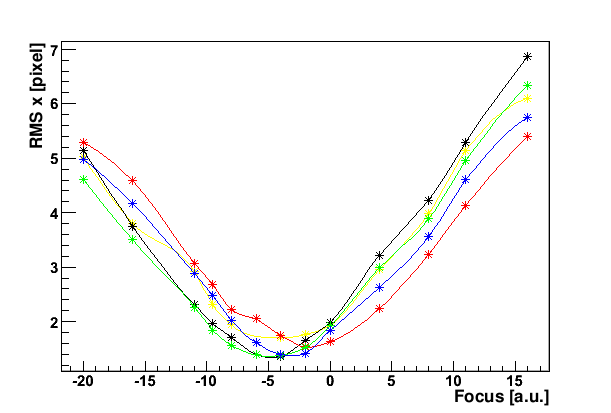}
\includegraphics[width=0.49\textwidth]{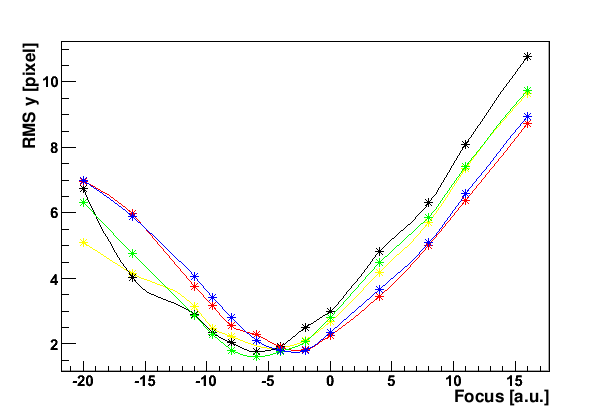}
\end{center}
\caption{Size of the diode image on the CCD as a function of the focus setting. Left column: horizontal RMS; right column: vertical RMS; top row: for the centre of the CCD sensor; bottom row: for the edge of the CCD sensor. Line colours represent diode colours, black standing for the white diode.}
\label{focusing}
\end{figure}

\subsection{Apparatus stability}

\subsubsection{Cameras focusing}
\label{sssec_cameras_focusing}

The first task was to find the focus of the lenses for a light source distant by 22 m, assuming that the star is best focused when its PSF is smallest. As an estimator of PSF size we chose root mean square (RMS) of a histogram of diode image pixels in horizontal and vertical axis. Measurements were performed for 5 diode colors and 13 focus settings.

As shown in fig. \ref{focusing}, there is a non-monotonic chromatic dependence in the focusing function, red and green colours having the most distant minima. Perhaps lenses, to compensate chromatic difference in focusing, try to achieve the minimum in the centre of the spectrum. Thus edges of the spectrum  -- red and blue -- have closer best focus values than the centre of the spectrum -- green.

There are nearly no differences in the horizontal and vertical RMS for the diode of the same colour in the central part of the CCD, where the PSF is nearly symmetrical. Such differences are expected and visible for the deformed PSF close to the edge of the frame, but minima of the focusing functions remain independent on the choice of the axis. There is, however, a difference in focusing minimum between the central and the edge PSF, which is smallest for the red diode. Because of the red diode giving the smallest side effect and the fact that the sensor is most sensitive in the red band we decided to use this colour best focusing $fs=1.4$ m as the general best focusing setting\footnote{The best focus setting of lenses for an object distant by 22 m is 1.4 m here due to the fact, that the \pin cameras are constructed in a different way than standard digital cameras, to which the scale on the lenses refers. The distance between the CCD and the lenses in our case is larger, to allow more freedom in focusing at real stars.}.

\subsubsection{Illumination stability}

\begin{figure}[tb]
\begin{center}
\includegraphics[width=0.49\textwidth]{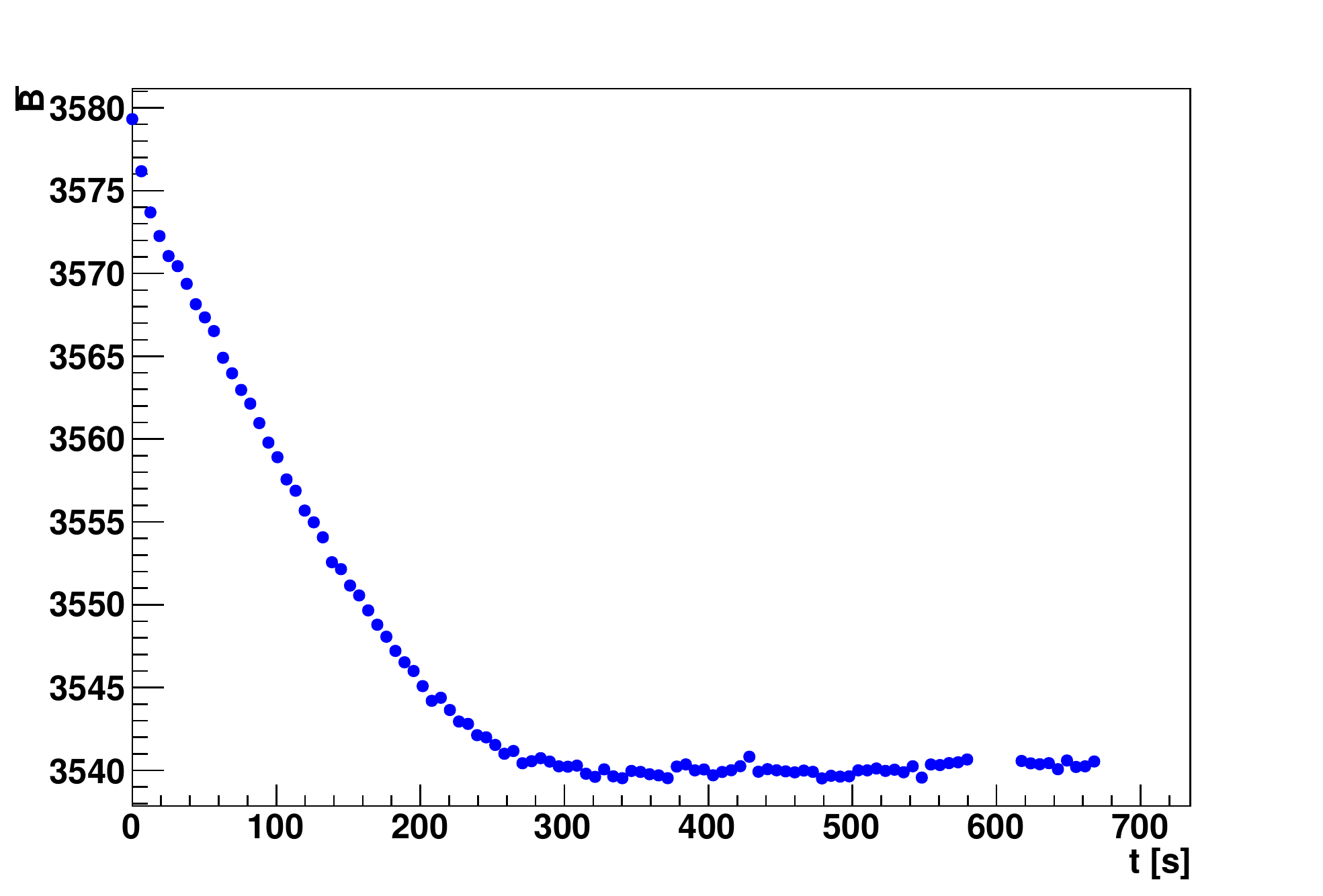}
\includegraphics[width=0.49\textwidth]{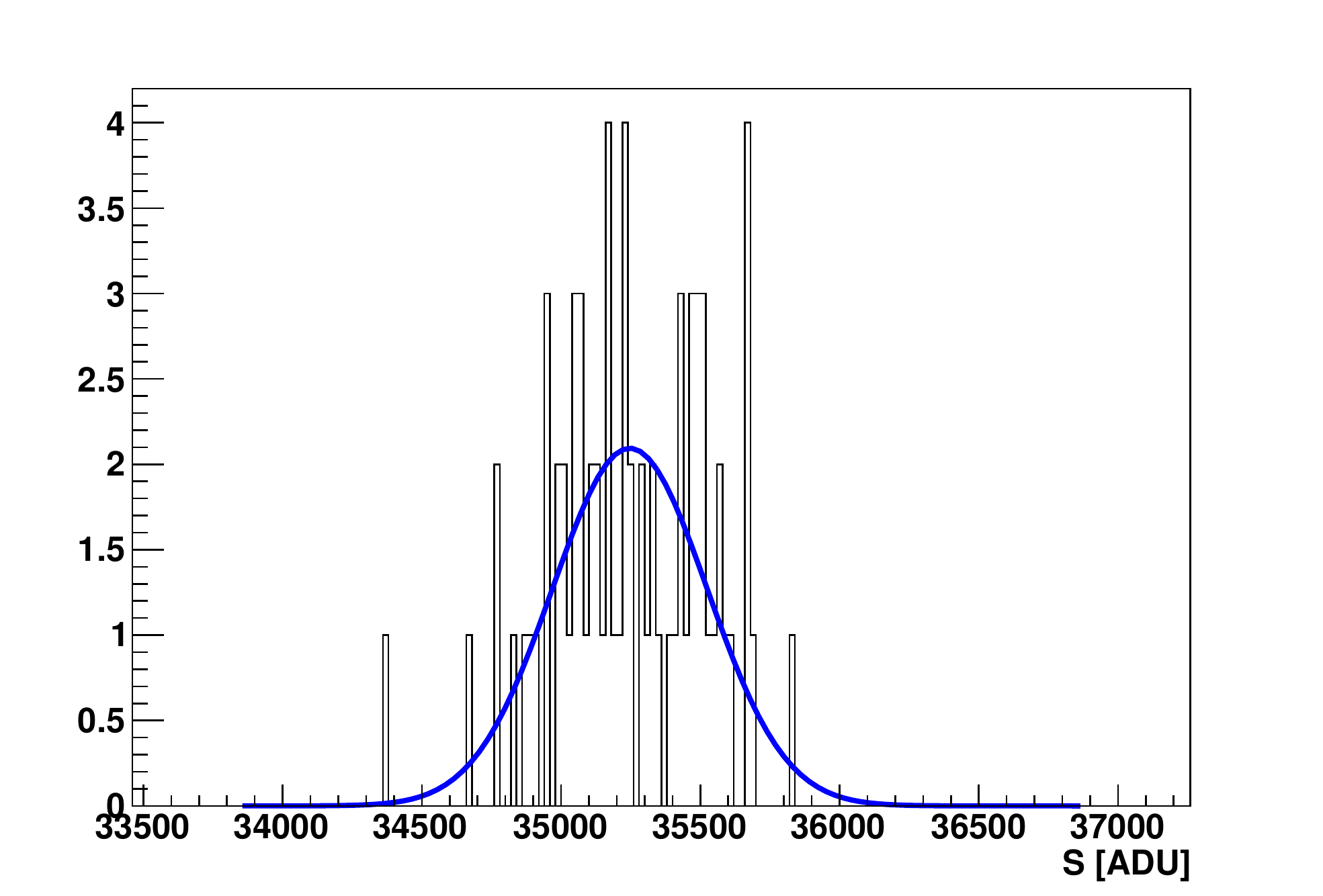}
\end{center}
\caption{Illumination stability. On the left: corridor background $\bar{\mathrm B}$ calculated as an average pixel value for frames as a function of time t after switching off corridor lights (uncertainties are smaller than the marker size); on the right: histogram of the diode signal S calculated in 100x100 pixels region around the diode, after subtracting background (signal level without the diode flash).}
\label{diode_stability}
\end{figure}

Reconstruction of high resolution PSF from single frames (sec. \ref{psf_reconstruction}) requires a diode with a stable light intensity as well as stable measurement of the signal. The latter involves estimating behaviour of the background coming from ambient light and other (however dim) light sources. 

The background does not need to be constant during the whole measurement serie, and such a condition would be difficult to achieve. However, it should not change significantly between so called ``corridor frames'' -- background frames taken without the diode pulse, subtracted from the frames with the diode flashing. The background becomes stable about 300 s after switching off corridor lights\footnote{The main factor responsible for the background not being stable immediately after switching off lights is probably deexcitation of some fluorescent signs in the corridor.}. After this time there is no visible trend in environmental light intensity, larger than frame to frame fluctuations (fig. \ref{diode_stability}, left).

The amount of light emitted during a single diode flash depends upon the shape of the current pulse coming from the generator, the diode response to such a pulse and Poisson statistics of photons. According to fig. \ref{diode_stability} (right), the fluctuations of the source light intensity (after subtracting background) are on the level of $0.8\%$. Nearly $70\%$ of those fluctuations can be explained by photon statistics. The results may vary slightly depending on the pulse length and other conditions, however the observed fluctuations are in general negligible compared to other uncertainties introduced during further data analysis. 

\subsubsection{Position stability}

\begin{figure}[t!]
\begin{center}
\includegraphics[width=0.49\textwidth]{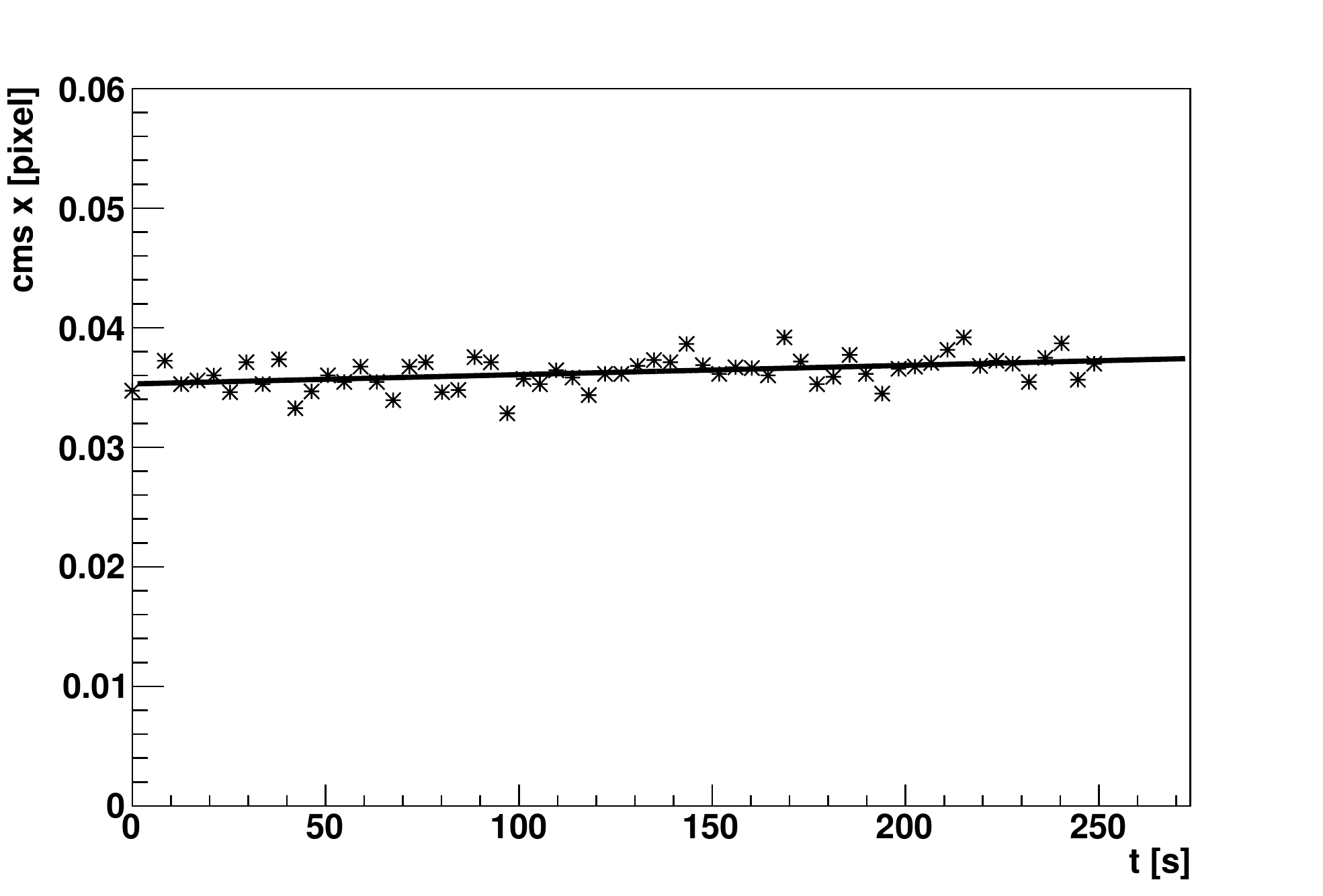}
\includegraphics[width=0.49\textwidth]{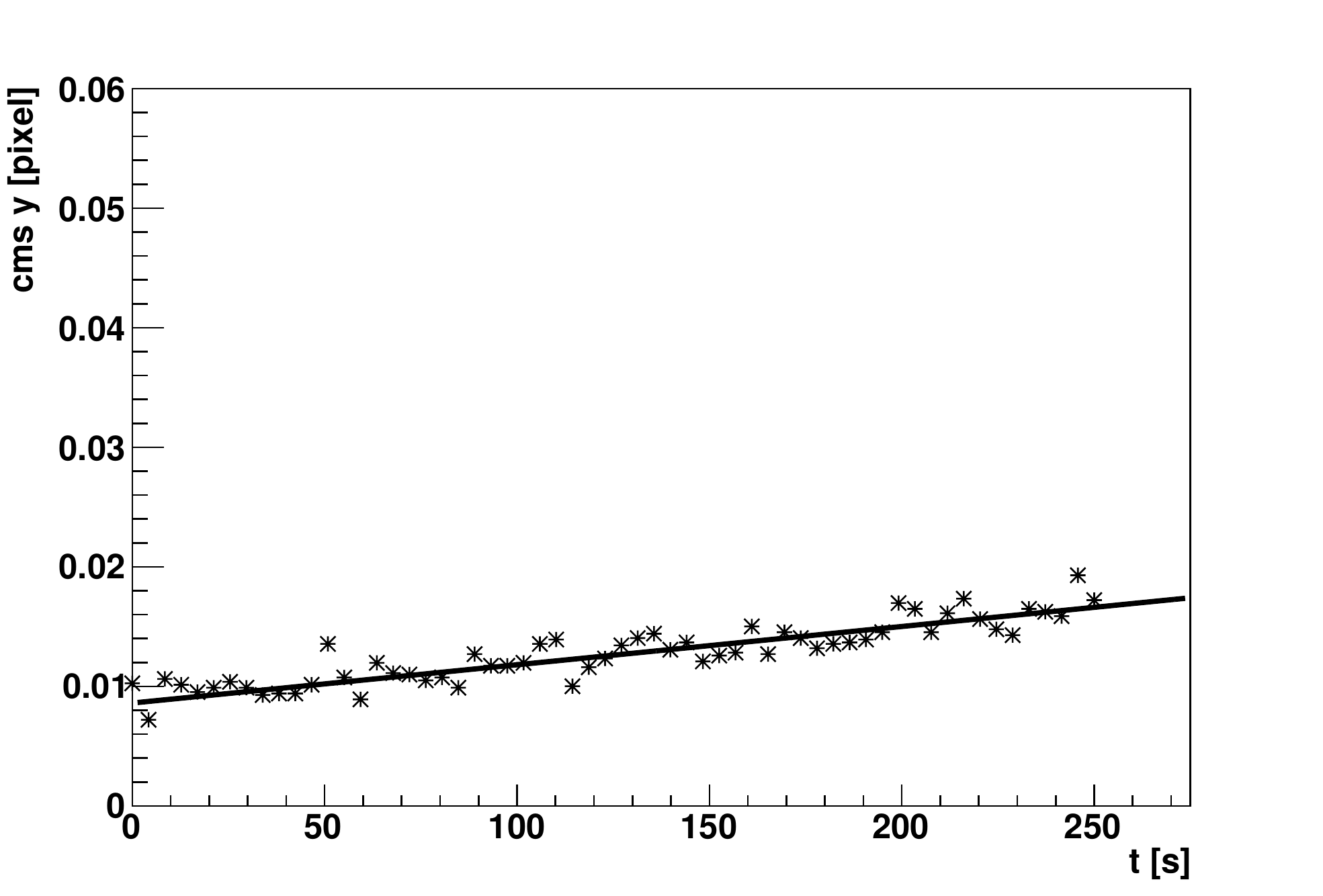}\\
\includegraphics[width=0.49\textwidth]{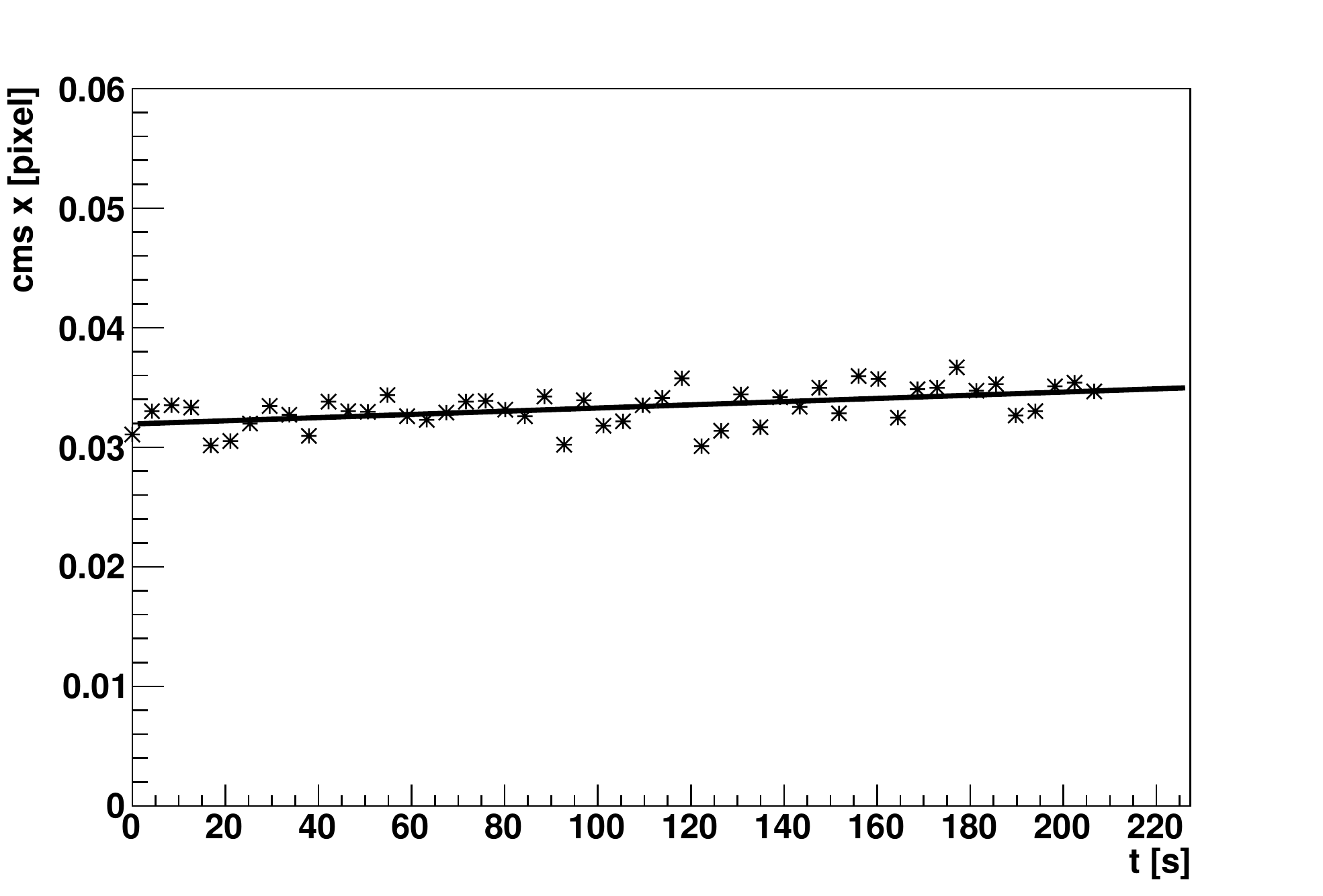}
\includegraphics[width=0.49\textwidth]{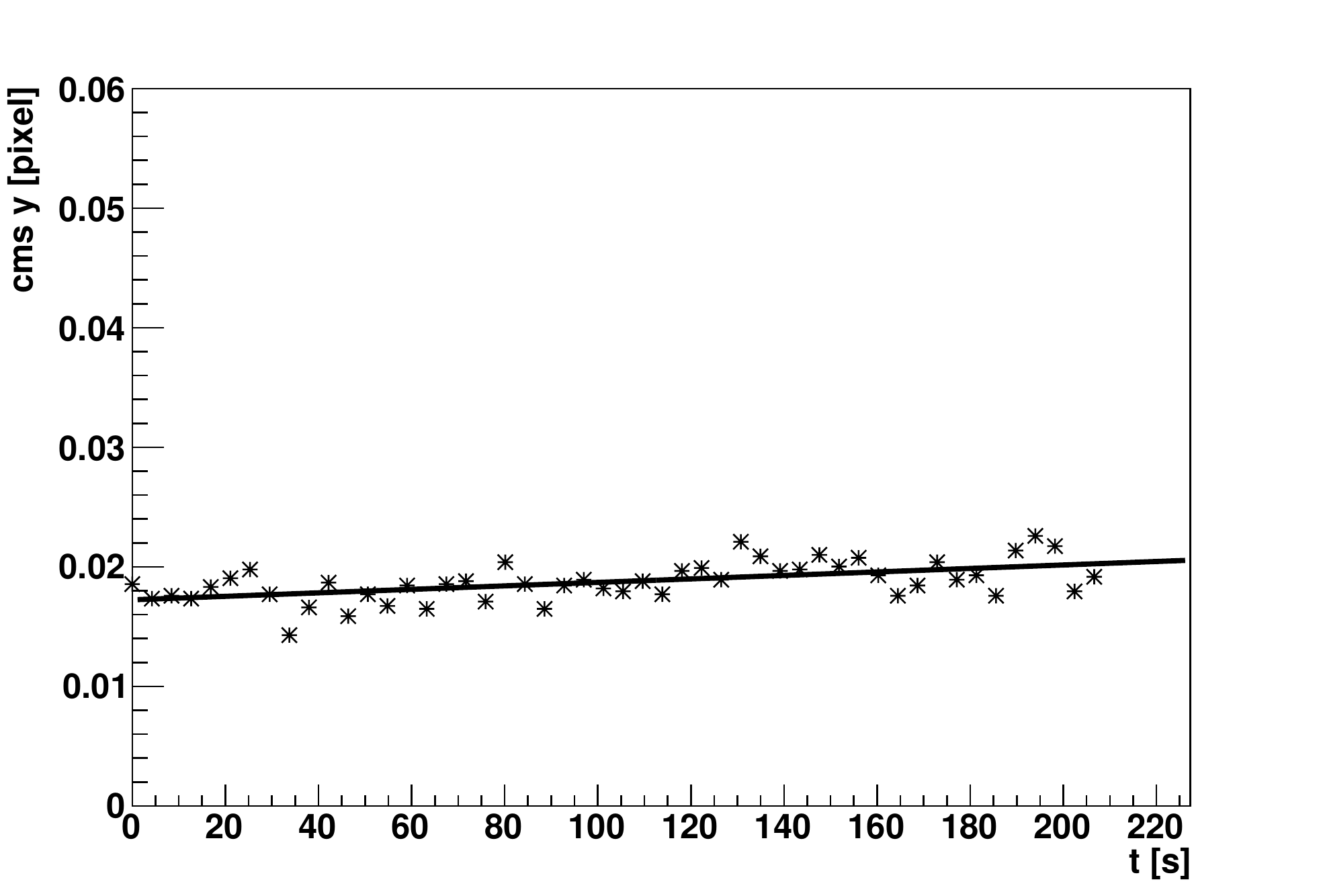}
\end{center}
\caption{Light source position stability (cms x, cms y) over time t. Top row: constantly opened shutter; bottom row: normal shutter mode. Results of a linear fit to the data points, indicated in the plot, are given in tab. \ref{tab_pos_movement}.}
\label{fig_pos_stability}
\end{figure}

Two series of measurements were performed to analyze the stability of the light source position in time. One serie was taken with mechanical shutter constantly opened (thus without vibrations induced by opening and closing of the shutter), second in normal (opening and closing) shutter mode. Fig. \ref{fig_pos_stability} shows slight change of the position of the light source image on the CCD, which is in general lower than $0.002$ pixels per minute (tab. \ref{tab_pos_movement}). These results rule out the possibility, that vibrations due to the shutter movement are the main reason of the position instability. The question is, why vertical shift in position is bigger for opened shutter? Frames with opened shutter were taken earlier, shortly after setting up the camera and the diode mount. Elements of the setup were probably still in a phase of mechanical stabilization due to thermal changes or gravitation, the latter hypothesis being favored by the observed difference of shift in vertical and horizontal axis. Measurements taken with normal shutter mode were taken after the period of the initial instability. Increased instability in the time period just after setting up the machinery was also observed during other data taking series.

\begin{table}[b]
\begin{center}
\begin{tabular}{|c|c|c|}\hline
 & horizontal [pixel per minute]& vertical [pixel per minute]\\
\hline
opened shutter & $(4.66 \pm 1.31)\cdot 10^{-4}$ & $(18.01 \pm 1.24)\cdot 10^{-4}$ \\
closed shutter & $(8.02 \pm 1.91)\cdot 10^{-4}$ & $(8.74 \pm 1.94)\cdot 10^{-4}$ \\
\hline
\end{tabular}
\caption{Estimated light source movement ratio -- results of a linear fit shown in fig. \ref{fig_pos_stability}.}
\label{tab_pos_movement}
\end{center}
\end{table}

Although we were unable to remove the position instability during the data taking, the expected shift of position is lower than $0.01$ pixel on a 100 frames measurement serie (about 6 minutes) and is therefore negligible. 



\subsection{Sample results}

\begin{figure}[h!]
\begin{center}
\subfigure[``Corridor frame'']{
\includegraphics[width=0.4\textwidth]{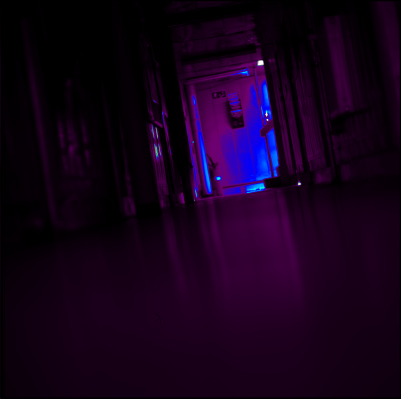}
\label{corridor}
}
\subfigure[Saturated diode flash with subtracted ``corridor frame'']{
\includegraphics[width=0.4\textwidth]{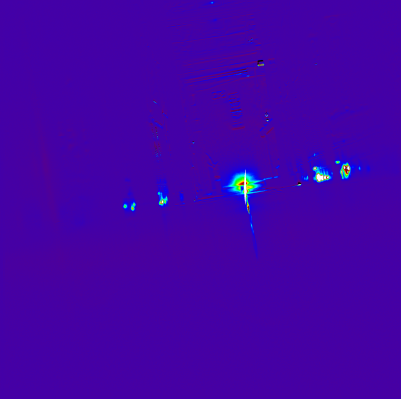}
\label{corr_flash}
}
\subfigure[Saturated diode flash magnified]{
\includegraphics[width=0.4\textwidth]{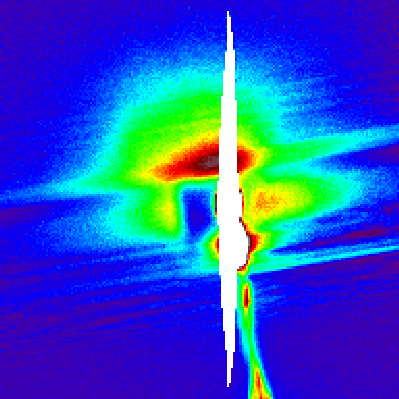}
\label{flash_mag}
}
\subfigure[Not saturated diode flash]{
\includegraphics[width=0.4\textwidth]{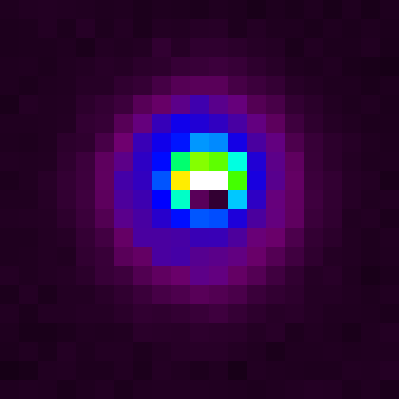}
\label{clean_flash}
}
\end{center}
\caption{Photographs taken using CCD camera during the tuning of the laboratory equipment.}
\label{initial_images}
\end{figure}

Fig. \ref{initial_images} shows pictures taken during tuning of the laboratory equipment. Fig. \ref{corridor} is a sample ``corridor frame'' -- frame without the diode blink - used for subtracting background from the images of the light source. The details of the corridor are quite well visible, although their absolute brightness is small. This shows the capabilities of the camera, especially very high signal to noise ratio. Fig. \ref{corr_flash} shows a saturated diode flash with subtracted ``corridor frame''. Such a bright blink produced many reflexes visible in the fig. \ref{flash_mag} -- around the actual diode image. Such photographs allowed removing ``light leaks'' in the setup, which resulted in very clean diode images, such as the one in fig.~\ref{clean_flash}. After performing laboratory setup tuning and stability tests, the main stage of data-taking could be started.

\section{Inter-pixel measurements}

The way the CCD sensor is designed causes a single-pixel light sensitivity to be not spatially uniform. That is mainly due to electrodes placed across the pixels and channel stops separating the sensor's columns. The non-uniformity can be measured with a source of light focused in a spot smaller than the pixel size\cite{subpixel}. The geometrical size of the spot in the described apparatus setup is smaller than the pixel size, but the PSF causes the light to be spread over several pixels. The way to restrain the PSF is to put a circular aperture in the front of the lenses. The smaller the aperture, the smaller the illuminated lenses area and the smaller the PSF. However, a small aperture causes the spot size to be diffraction limited (tab. \ref{diff_size}).

\begin{table}[h!]
\begin{center}
\begin{tabular}{|c|c|c|}
\hline
Aperture diameter [mm] & red light $\Delta _{CCD}$ [pixel] & blue light $\Delta _{CCD}$ [pixel] \\
\hline \hline
40 & 0.1 & 0.07 \\
\hline
20 & 0.2 & 0.14 \\
\hline
10 & 0.4 & 0.27 \\
\hline
\end{tabular}
\caption{Theoretical diffraction limited size of the light spot on CCD for different aperture diameters.}
\label{diff_size}
\end{center}
\end{table}

As a compromise between PSF size and the diffraction size an aperture of $20$ mm diameter had been chosen for the inter-pixel measurements.

\subsection{Pixel light intensity response}

The CCD sensor with accompanying electronics is a an analogue-digital converter of light intensity illuminating the camera into so-called ADU counts. As with most such devices there are some issues concerning the conversion process. Perhaps the most visible is the saturation, which is observed when the potential well is flooded with photoelectrons beyond its capacity, which causes the charge to escape to neighbouring wells, resulting in the distortion of the picture of the object and the surrounding area (so-called blooming, see fig. \ref{flash_mag}). Retrieving proper data from such a distorted image is close to impossible, that is why objects saturating the detector are rather not being analysed. 

However, there are some less obvious concerns like the dependence of the output signal in ADU counts on the intensity of light illuminating the CCD pixel. In order to determine this dependence measurements of single pixel signal for different diode light intensities were performed for 9 pixels close to the sensor centre. The light intensity was adjusted by triggering different number of diode pulses of the same length.

\begin{figure}[b!]
\begin{center}
\includegraphics[width=0.7\textwidth]{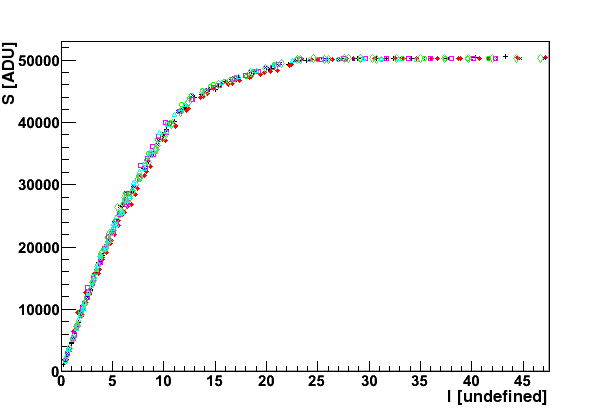}
\end{center}
\caption{Dependence of the average CCD signal S on the source intensity I for 9 selected pixels close to the centre of the chip (different markers).}
\label{pixel_linearity}
\end{figure}

Fig. \ref{pixel_linearity} shows the results of the described measurement. We observe that the dependence of signal on the light intensity is clearly not linear. While the nonlinearity broadens the magnitude range of permissible objects, it requires an application of a nonlinear correction for each pixel in order to determine the real signal. However, we also see that the dependence is very close to linear up to the output signal of about 20000 ADU, and, to simplify the analysis, this range was chosen for further measurements.

Sensor response depends on the mechanism of converting photons into measurable signal. This mechanism can cause the function describing pixel response to depend on the pixel placement on the CCD sensor. Fortunately, measurements taken at random positions far from the sensor centre showed that the linearity range remains the same, while the main change is in the saturation threshold, which is not relevant for the presented study.

\subsection{Pixel response function}
\label{ssec_prf}

\begin{figure}[b!]
\begin{center}
\includegraphics[width=0.328\textwidth]{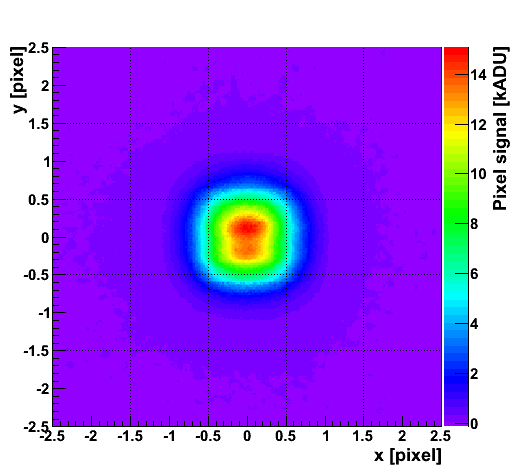}
\includegraphics[width=0.328\textwidth]{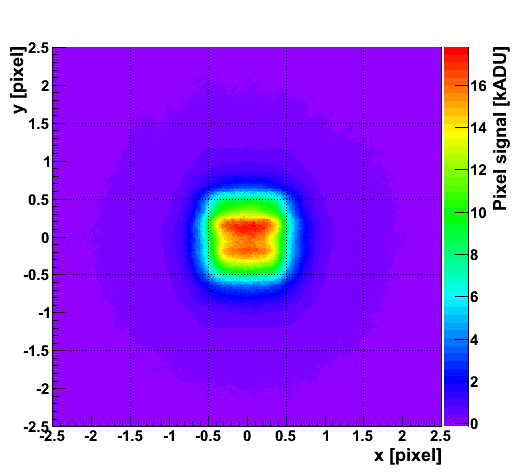}
\includegraphics[width=0.328\textwidth]{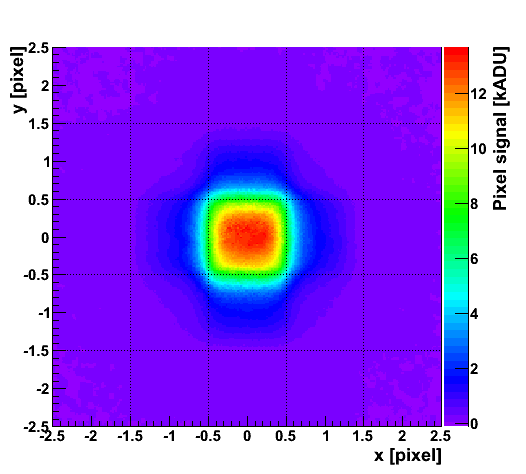}
\end{center}
\caption{Measurements used to extract pixel response function for the blue (left), red (centre) and white (right) diodes.}
\label{prf}
\end{figure}

Pixel response function (PRF) describes a single pixel signal value as a function of the spot position relative to the pixel edge. In an ideal case of infinitely small spot size and uniform pixel response, the PRF should have constant value inside the pixel and zero value outside. In the real case, it dependents on the pixel sensitivity and on the finite spot size, so that the spot may be only partially contained inside the pixel. The latter causes a narrow but not-negligible transition of PRF from high to low values close to the pixel border (fig. \ref{prf}). 

However, the function is also non-zero for spot fully outside the pixel. That may be caused by a finite PSF size or diffraction of the spot, restrained by setup parameters, but still not negligible. The more interesting possibility is that it is caused by a charge diffusion between pixels -- illuminating a single pixel causes some charge to be accumulated in a neighbouring pixels as well. In that case, the PRF ``tails'' contribute also to the observed PSF shape.

\subsection{Pixel sensitivity function}

\begin{figure}[t!]
\begin{center}
\includegraphics[width=0.328\textwidth]{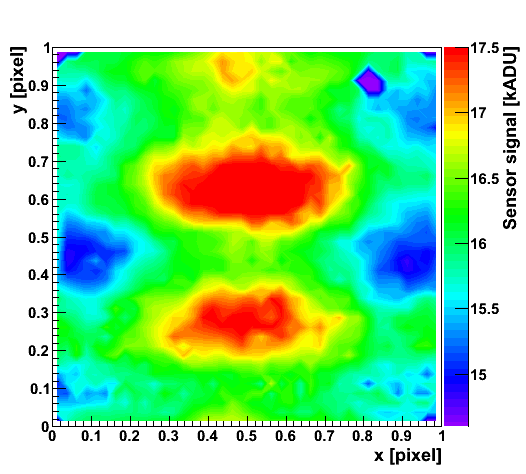}
\includegraphics[width=0.328\textwidth]{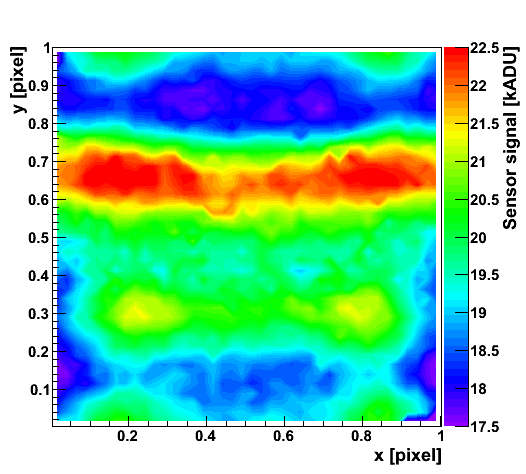}
\includegraphics[width=0.328\textwidth]{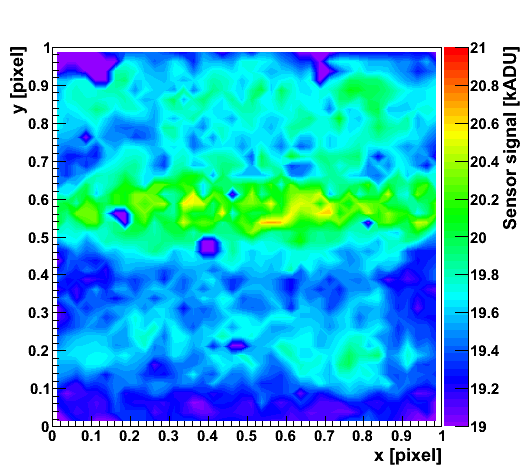}
\end{center}
\caption{Measurements used to extract pixel sensitivity function for the blue (left), red (centre) and white (right) diodes.}
\label{pixel_sens_func}
\end{figure}

Pixel sensitivity function is defined similarly to pixel response function, however an overall CCD signal is studied as a function of the spot position instead of the single pixel signal. Changes in pixel sensitivity are the main factor responsible for signal changes caused by the image movement across the CCD. With the knowledge of the pixel sensitivity function and the position of the source's centre on the pixel one should be able to compensate for this effect, performing more precise measurement of brightness.

The overall signal was estimated by a sum of $3\times 3$ pixels around the spot centre. Results of the measurement are shown in fig. \ref{pixel_sens_func}. The maximal observed changes in signal due to the pixel sensitivity non-uniformity for a red diode are more than $30\%$, and for a blue diode more than $20\%$. However, for a normal PSF image (taken without aperture reducing PSF size), which is spread over more pixels, the amplitude of corresponding fluctuations is lower (around $6\%$).

A visible vertical structure similar for both red and blue diodes is probably caused by the electrodes. The horizontal structure is clearly colour dependent, probably due to different penetration depths of light of different wavelengths. Both structures are much less visible for the white diode, for which the effect is averaged all over the white light spectrum, and thus much smaller.

\section{PSF measurements}

\subsection{PSF reconstruction}
\label{psf_reconstruction}

A high resolution profile for selected coordinates on the frame was obtained using multiply images of a diode. Each exposure was taken for a specific position of the diode's centre, the full set of images was covering $10\times 10$ points inside a single pixel. Additionally, 5 images with and without a light pulse from a diode were taken in each position, for noise reduction purposes. All the images were superimposed, taking into account coordinates of each image.

\begin{figure}[tb!]
\begin{center}
\subfigure[Distribution of the centre of mass for positions uniformly distributed in horizontal direction]{
\includegraphics[width=0.46\textwidth]{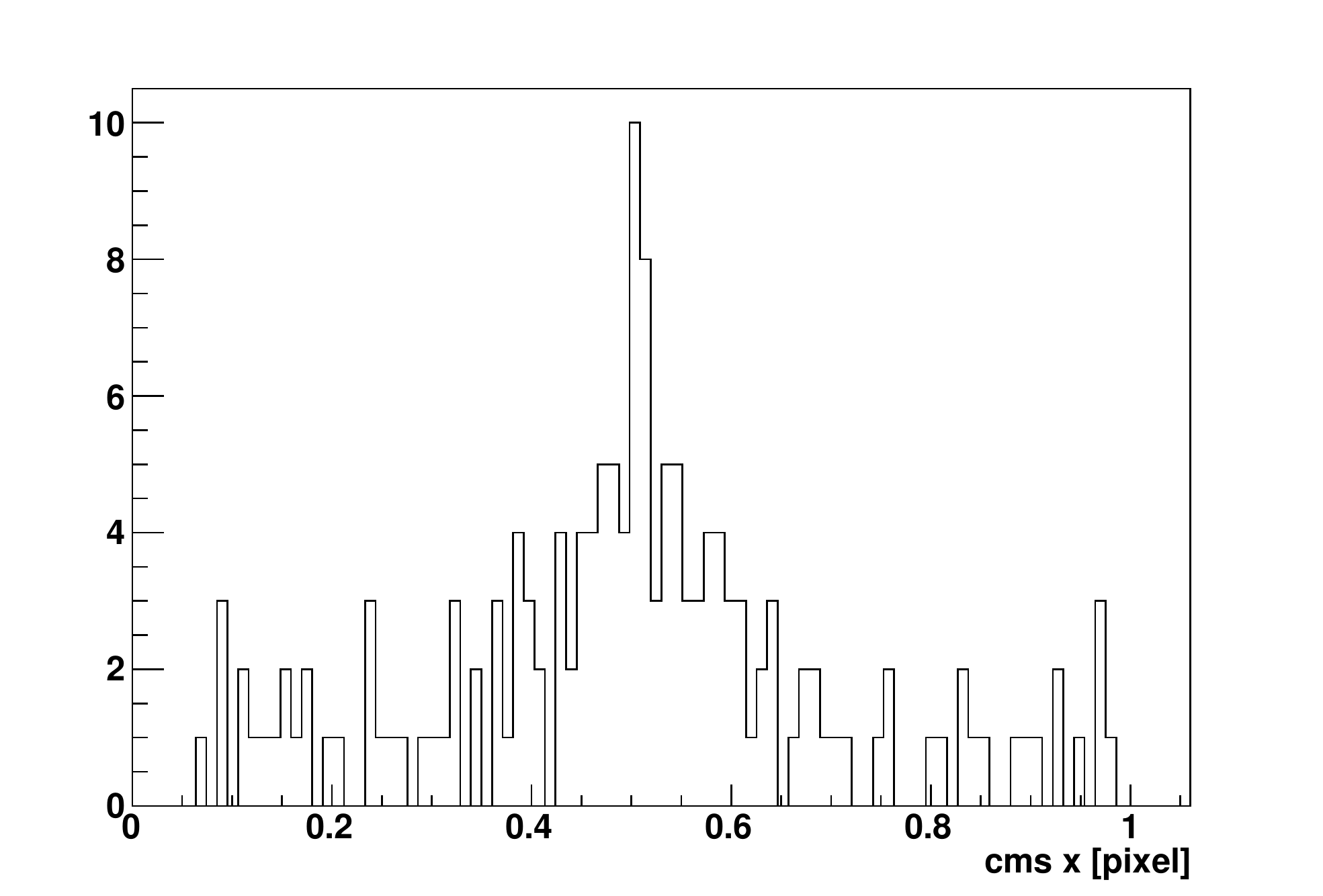}
\label{cms_distribution}
}
\hspace{4mm}
\subfigure[Measured (points) and simulated (line) dependence of the calculated cms x as a function of the real spot position x. Voigt PSF is assumed for simulation.]{
\includegraphics[width=0.46\textwidth]{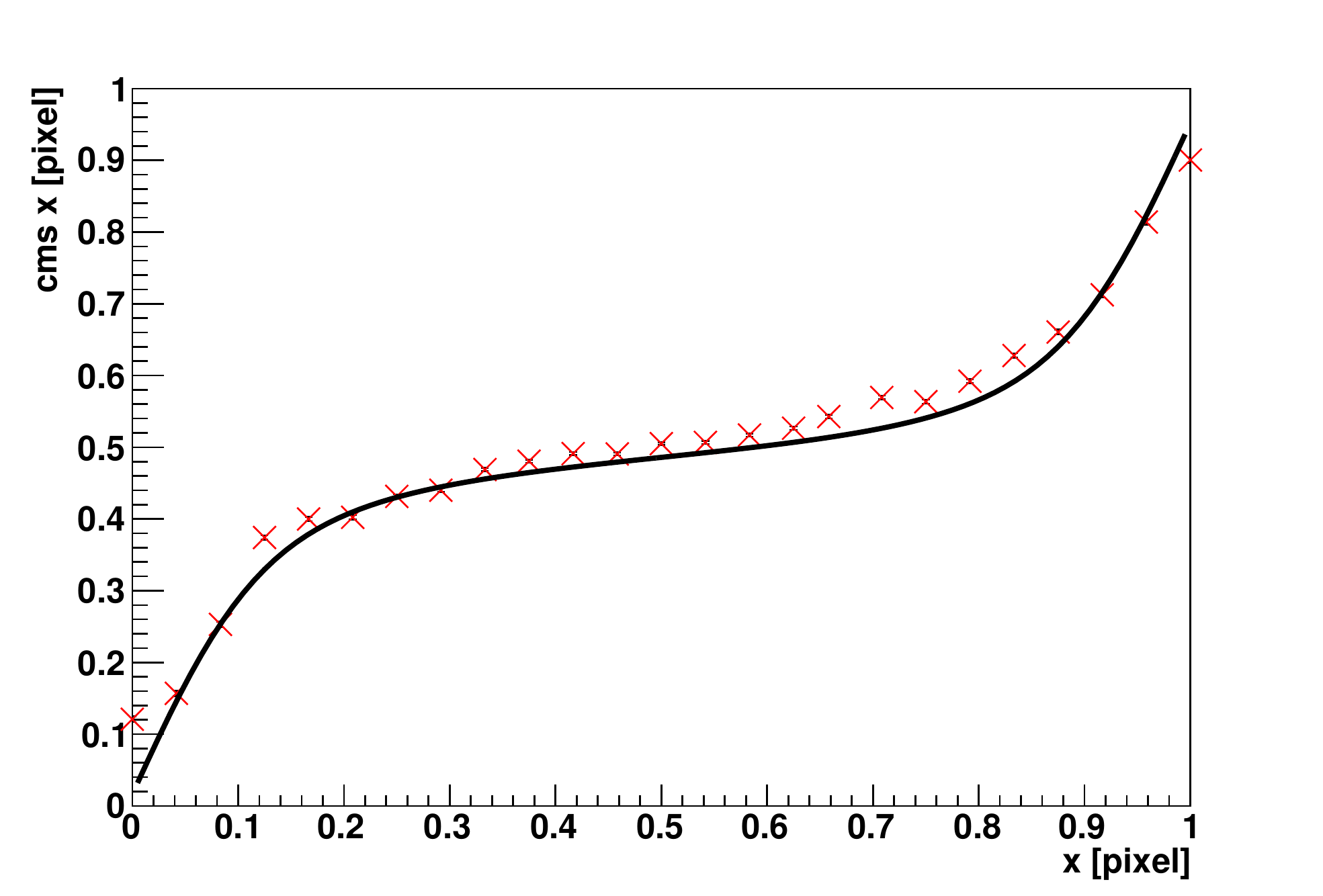}
\label{cms_simulation}
}
\subfigure[Corrected position $\mathrm{X(x)}$ -- the integral of the centre of mass position cms x]{
\includegraphics[width=0.46\textwidth]{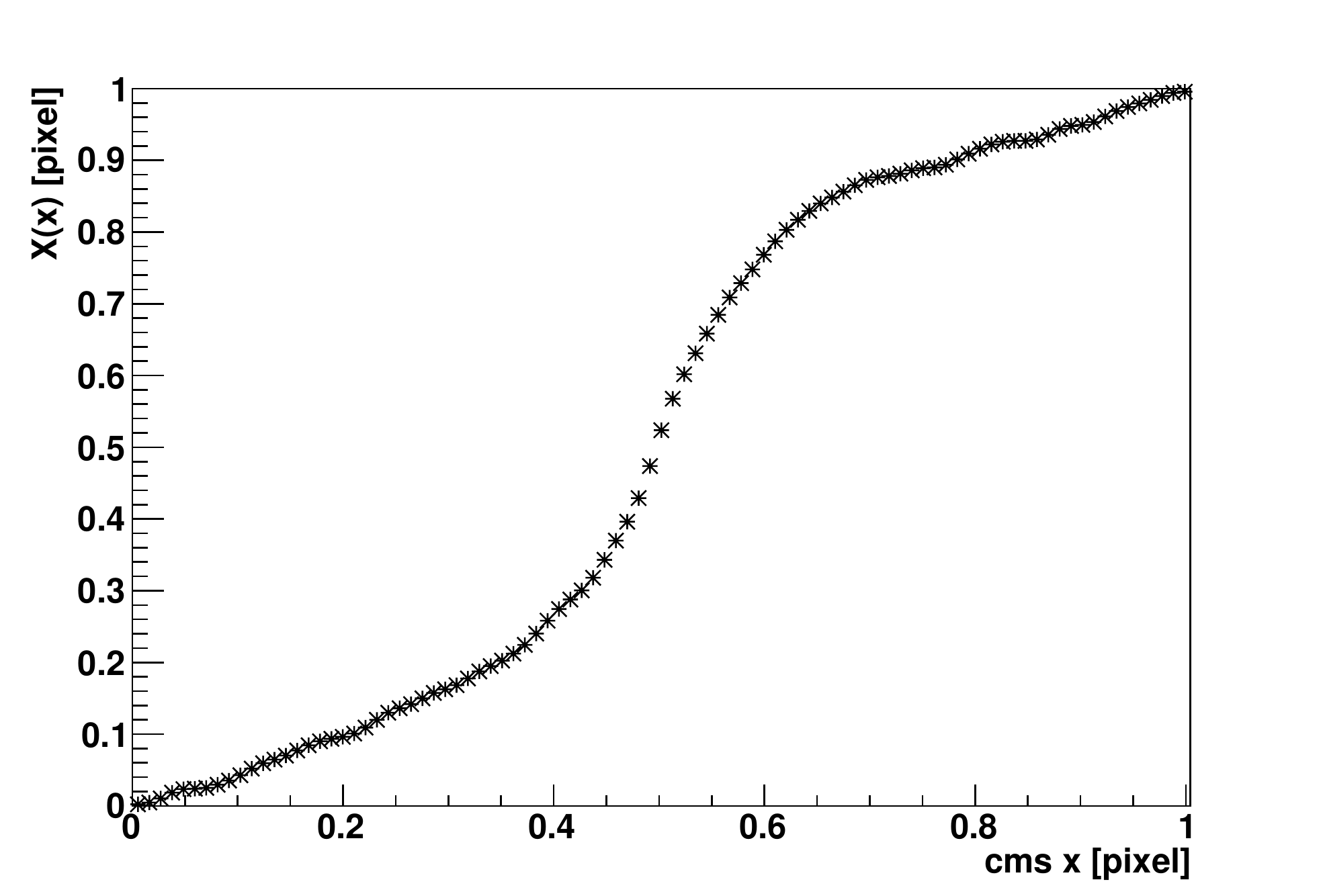}
\label{cms_distribuant}
}

\end{center}
\caption{Reconstruction of the spot horizontal position relative to the pixel edge, with the centre of mass method.}
\label{cms}
\end{figure}

The procedure of superposition requires determining the proper position of the single image relative to CCD pixels. The task is not straightforward, for estimators of the position like centre of mass\footnote{Mass in this case is, of course, signal stored in the pixels of the CCD.} (cms) tend to group around the brightest pixel centre (fig. \ref{cms_distribution}). This is caused by a convolution of a CCD pixel structure and a PSF shape analysed in a finite region. To simplify the discussion, let's consider a 1D cross-section of the PSF along the X axis and a cms in this direction calculated in three neighbouring pixels, the central one being the brightest one. In this case:

\begin{equation}
\mathrm{cms(x)}=S(-1,0)*(-0.5)+S(0,1)*0.5+S(1,2)*1.5
\label{cms_eq}
\end{equation}

where $S(a,b)=\int _a ^b PSF(x)$, a signal stored in a pixel between $a$ and $b$ coordinates, depends on the relative position of the PSF centre to the brightest pixel centre. Shown in fig. \ref{cms_simulation} is calculated cms as a function of the diode position for one of the single runs with the diode moving along the X axis (Y coordinate being fixed close to the pixel centre). The curve is a fit of eq. \ref{cms_eq} to these points, assuming Voigt profile\cite{voigt} shape of the PSF. The $\chi^2/\rm{NDF}$ of the fit is 102, so apparently Voigt profile does not properly describe the PSF cross section in horizontal axis\footnote{Theoretically, an ideal Fresnel diffraction shape without any aberrations is described by Bessel functions. However, in the aberrated case of \pin cameras, Bessel functions are not expected to describe PSF any better than the Voigt profile.}, but the plot shows that the interpretation of the cms grouping effect is correct.

To obtain a real position of the diode image centre, the grouping effect of the cms has to be removed. For an unbiased estimator of the spot position the expected distribution (assuming uniform distribution of the true position) should be flat, unlike the cms~x distribution on the fig. \ref{cms_distribution}. Therefore, cms has to be transformed in such a way as to result in a flat distribution. This can be obtained by integrating the $\mathrm{cms}$ distribution normalized to 1. The corrected estimator of position, $X$, is then:

\begin{equation}
	\mathrm{X(x)}=\int_0^x P(x')\dx'
\end{equation} 

where $\rm x$ is the position returned by the cms calculation and $P(x')$ is the probability density of $\mathrm{cms}(x')$. Calculated values of the corrected position, $\rm X(x)$, are shown as a function of the cms~x value in fig. \ref{cms_distribuant}. Corrected position for any measurement can be obtained by interpolating reconstructed $\rm X(x)$. The process of calculating such an integral involves some additional, technical steps like smoothing the histogram and fitting the absolute position of the diode image on the CCD\footnote{The relative position of the diode image on subsequent measurements is well known by the position of the step motors. However, determining the absolute position is a more complicated task, especially for non-symmetrical PSFs far from the centre of the frame.}.

\subsection{Measured PSFs}
\label{ssec_measured_psfs}

\begin{figure}[t]
\begin{center}
\includegraphics[width=0.4\textwidth]{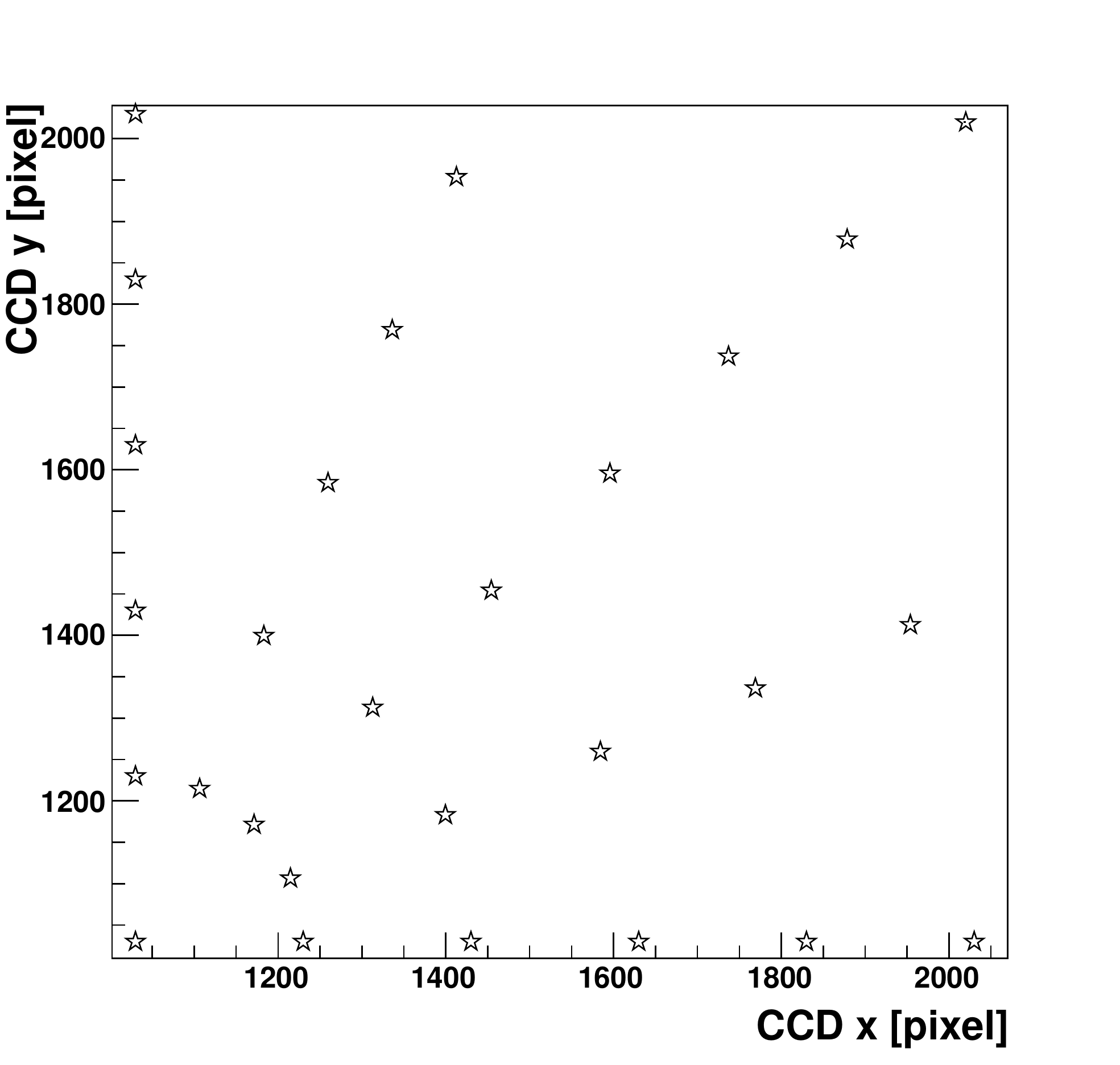}
\end{center}
\caption{Positions of the PSF measurements on the CCD surface in sensor coordinates. The measurements were performed for 5 angles (for $0^\circ$, $22.5^\circ$, $45^\circ$, $77.5^\circ$ and $90^\circ$ from the horizontal axis), and 5 or 8 distances from the frame's centre. For each angle, distances between considered positions were about 200 pixels. One quarter of the CCD chip was only considered.}
\label{meas_pos}
\end{figure}

PSF measurements and reconstruction were performed for a white diode for 5 angles and 6 distances from the frame centre, covering 1/4 of the CCD, as shown on fig. \ref{meas_pos}, each star representing a place on a CCD in which PSF was reconstructed.

\begin{figure}[tb]
\begin{center}
\includegraphics[width=0.6\textwidth]{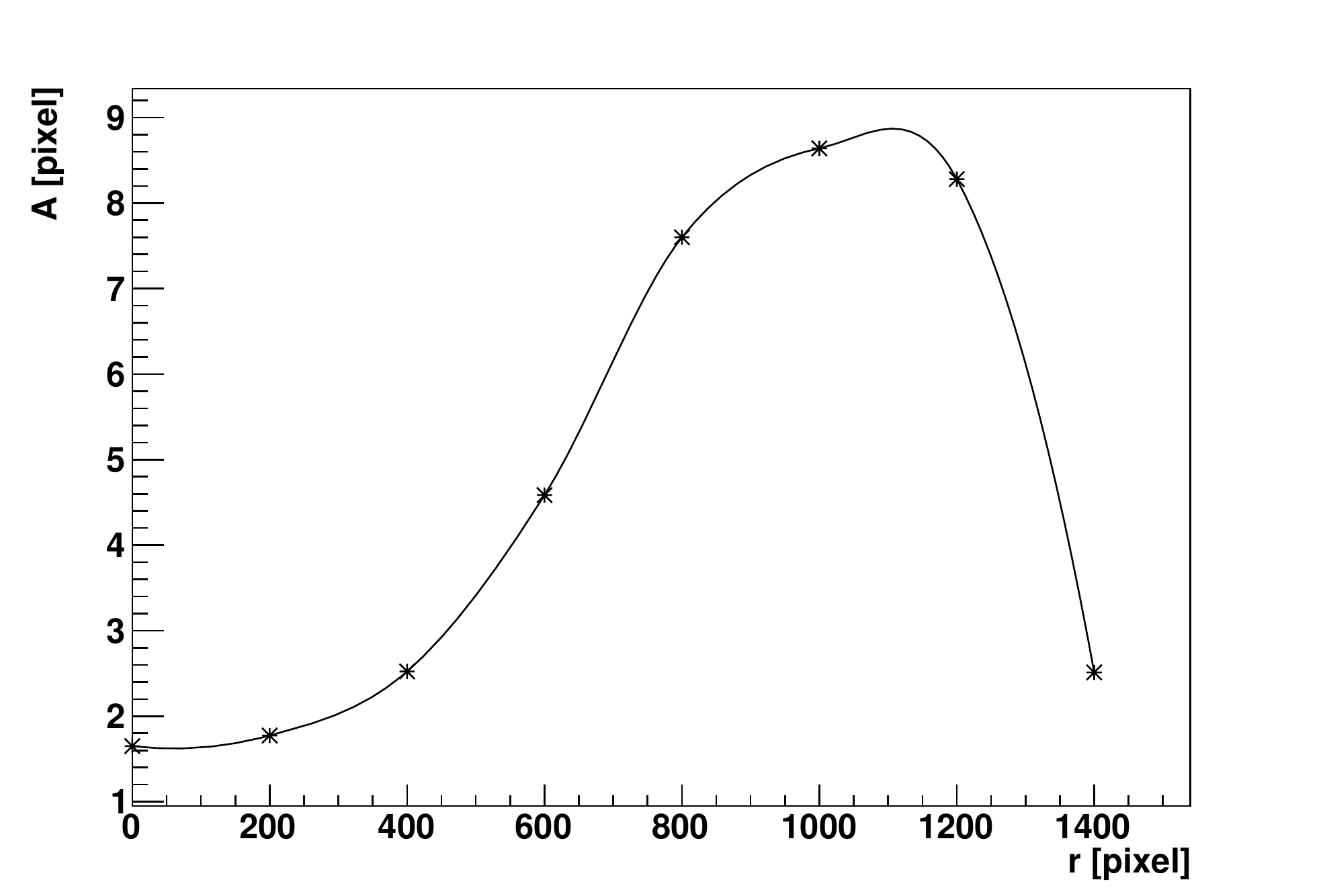}
\end{center}
\caption{The area A above half of the maximum height for PSFs measured along chip diagonal, as a function of the distance from the frame centre r.}
\label{fig_fwhm}
\end{figure}

\begin{figure}[b!]
\begin{center}
\subfigure[0 pixels]{
\includegraphics[width=0.328\textwidth]{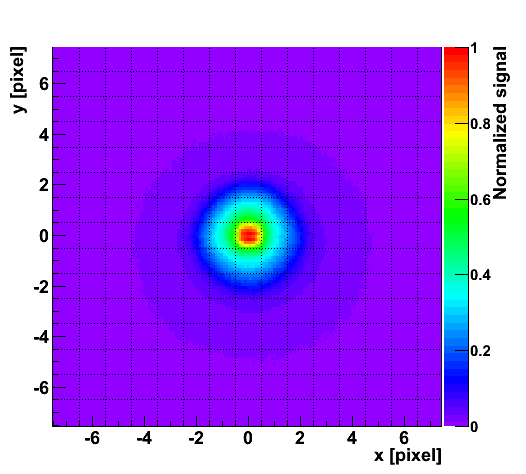}
}
\subfigure[200 pixels]{
\includegraphics[width=0.328\textwidth]{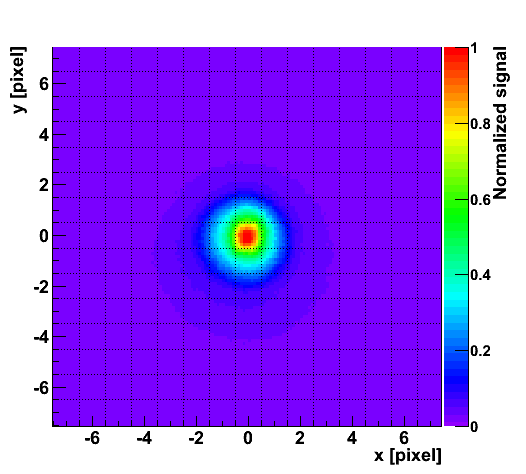}
}
\subfigure[400 pixels]{
\includegraphics[width=0.328\textwidth]{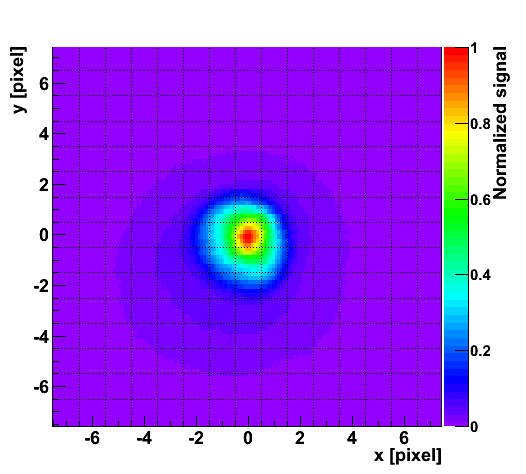}
}
\subfigure[600 pixels]{
\includegraphics[width=0.328\textwidth]{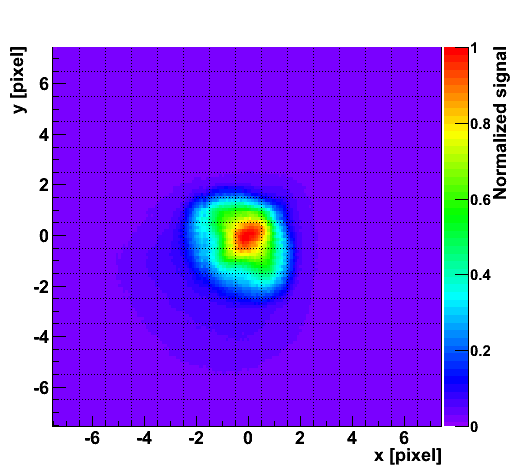}
}
\subfigure[800 pixels]{
\includegraphics[width=0.328\textwidth]{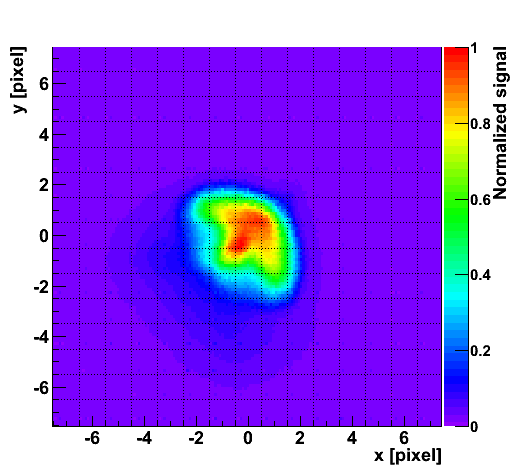}
}
\subfigure[1000 pixels]{
\includegraphics[width=0.328\textwidth]{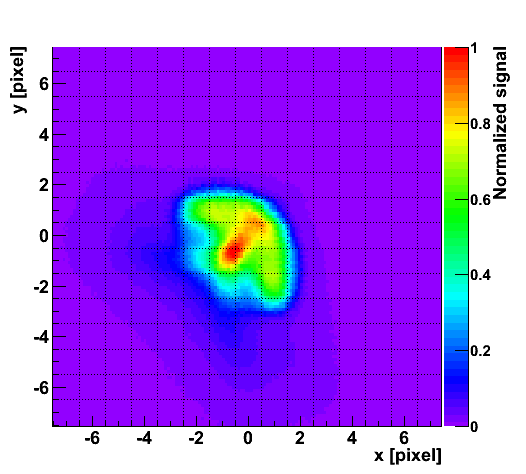}
}
\subfigure[1200 pixels]{
\includegraphics[width=0.328\textwidth]{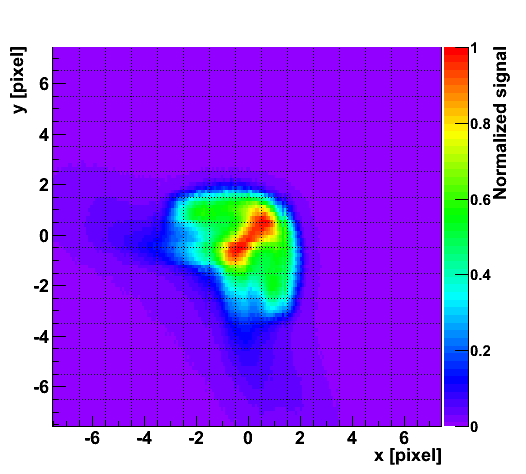}
}
\subfigure[1400 pixels]{
\includegraphics[width=0.328\textwidth]{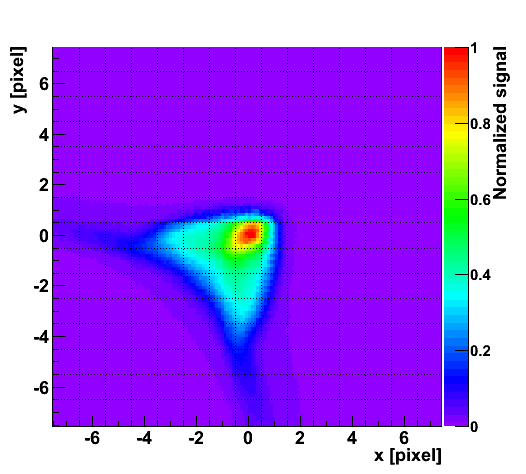}
}
\end{center}
\caption{PSFs of the white diode measured along the diagonal, for 0, 200, 400, 600, 800, 1000, 1200, 1400 pixels from the frame centre, as indicated below the plots.}
\label{diode_psf_white}
\end{figure}

A significant deformation develops with the distance from the frame centre, causing not only the shape of the profile to change, but also the area containing the signal to grow. The area covered by PSF profile at half-maximum, as shown in fig. \ref{fig_fwhm}, increases from slightly more than 1.5 pixel for the central profile to nearly 9 pixels for the profile 1000 pixels from the frame centre. Even ignoring the shape deformation, such a growth may be a non-negligible factor increasing uncertainties of aperture photometry. For profile photometry the situation is even more difficult, because the PSF shape changes dramatically with radius, as shown on fig. \ref{diode_psf_white} (for measurements at angle of $45^\circ$).

\begin{figure}[h!]
\begin{center}
\subfigure[800 pixels]{
\includegraphics[width=0.328\textwidth]{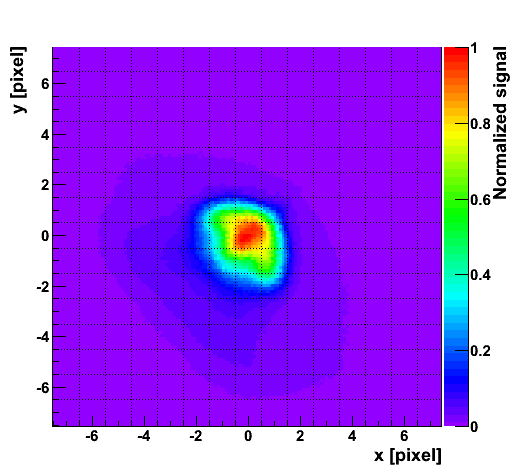}
}
\subfigure[1000 pixels]{
\includegraphics[width=0.328\textwidth]{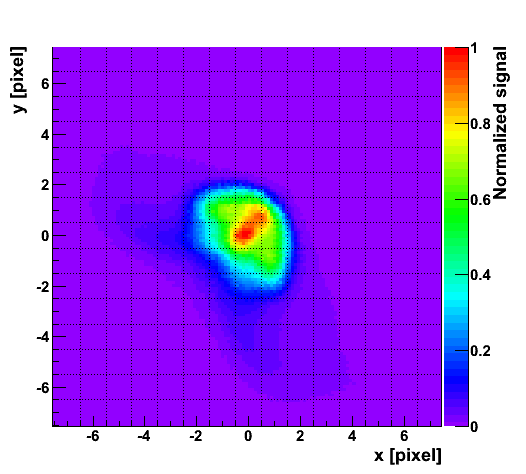}
}
\subfigure[1200 pixels]{
\includegraphics[width=0.328\textwidth]{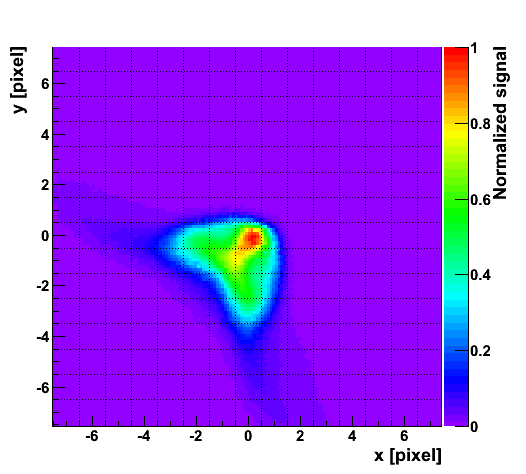}
}
\subfigure[1400 pixels]{
\includegraphics[width=0.328\textwidth]{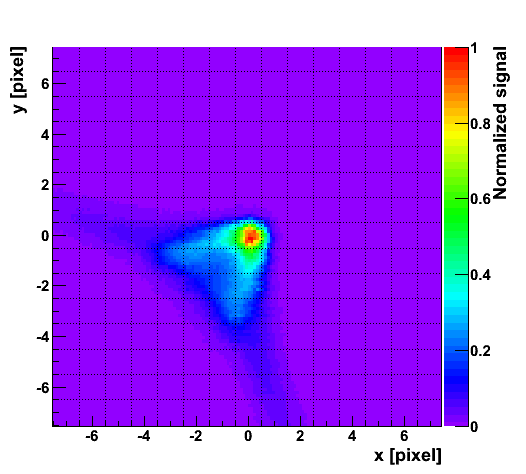}
}
\end{center}
\caption{PSFs of the blue diode measured along the diagonal, for 800, 1000, 1200, 1400 pixels from the frame centre.}
\label{diode_psf_blue}
\end{figure}

\begin{figure}[h!]
\begin{center}
\subfigure[800 pixels]{
\includegraphics[width=0.328\textwidth]{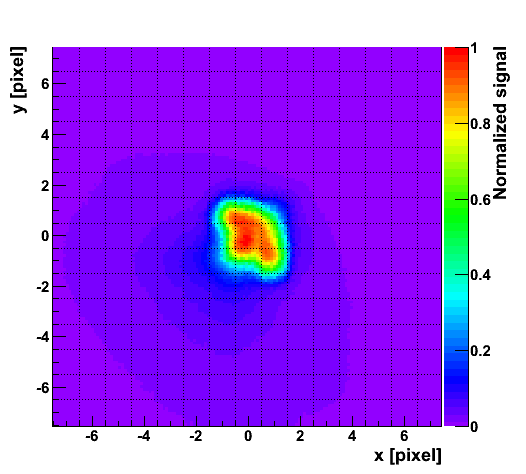}
}
\subfigure[1000 pixels]{
\includegraphics[width=0.328\textwidth]{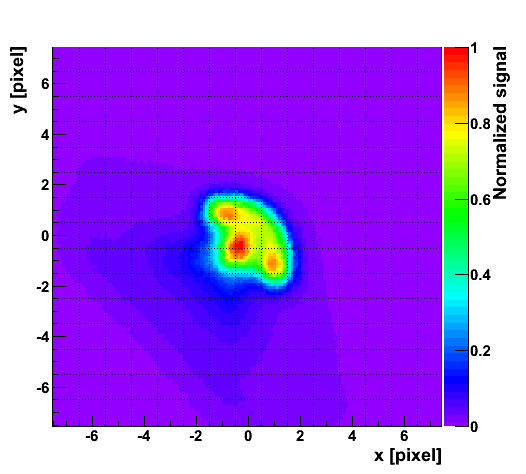}
}
\subfigure[1200 pixels]{
\includegraphics[width=0.328\textwidth]{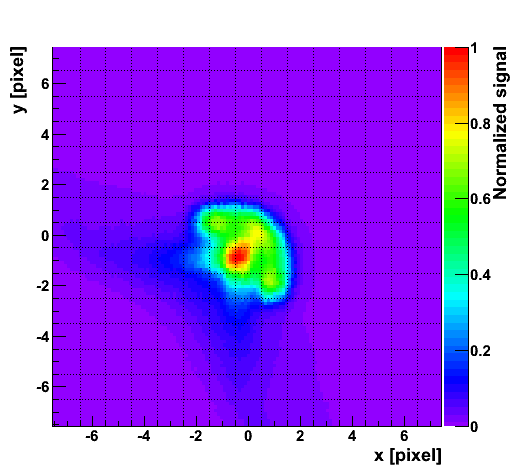}
}
\subfigure[1400 pixels]{
\includegraphics[width=0.328\textwidth]{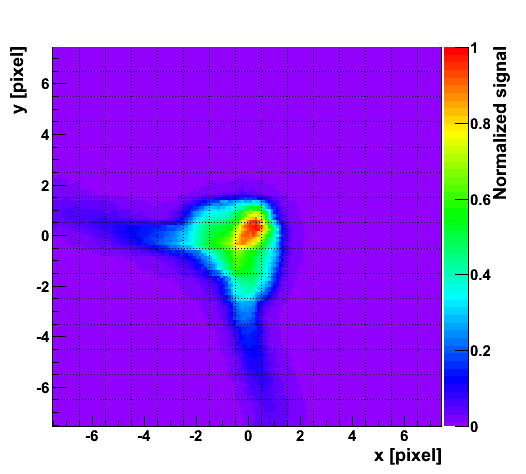}
}
\end{center}
\caption{PSFs of the red diode measured along the diagonal, for 800, 1000, 1200, 1400 pixels from the frame centre.}
\label{diode_psf_red}
\end{figure}

PSF reconstruction for the white diode is important from the point of view of studying a general image shape for real stars, which are hardly monochromatic. However, parametrization in terms of diffraction model (chap. \ref{chap_diff}) is much easier when a single wavelength is considered. Thus a similar PSF reconstruction was performed for the blue (fig. \ref{diode_psf_blue}) and the red diode (fig. \ref{diode_psf_red}), for 800, 1000, 1200 and 1400 pixels from the frame centre along its diagonal. It can be noticed, that PSFs of the white diode contain superimposed shape features of PSFs of the blue and the red diode, which is expected. However, polychromatic PSFs tend to be larger due to the fact, that different wavelengths are focused in a slightly different positions on the CCD.

\begin{figure}[h!]
\begin{center}
\begin{tabular}{cccc}
& $fs=1.2$ m & $fs=1.4$ m & $fs=1.6$ m \\
\begin{sideways}\begin{minipage}[c]{15mm}\begin{center}800\end{center}\end{minipage}\end{sideways}&
\includegraphics[width=0.3\textwidth]{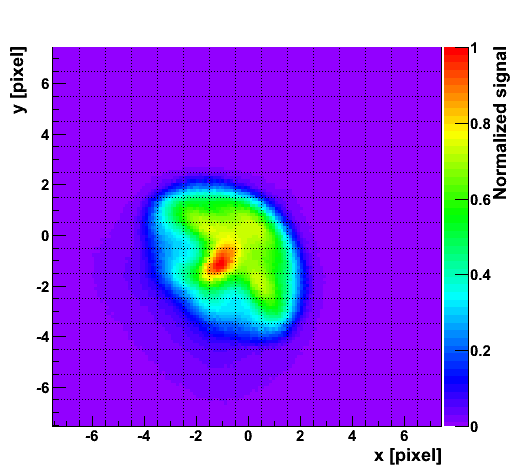} &
\includegraphics[width=0.3\textwidth]{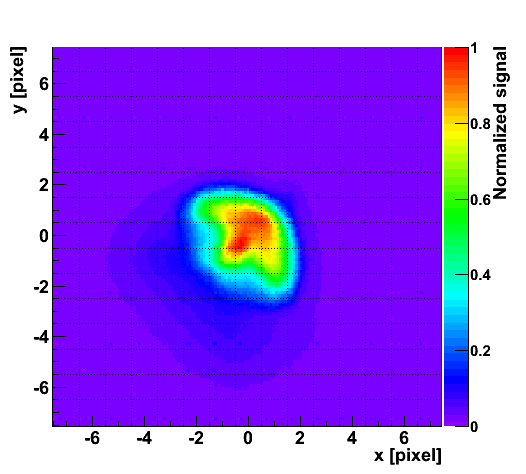} &
\includegraphics[width=0.3\textwidth]{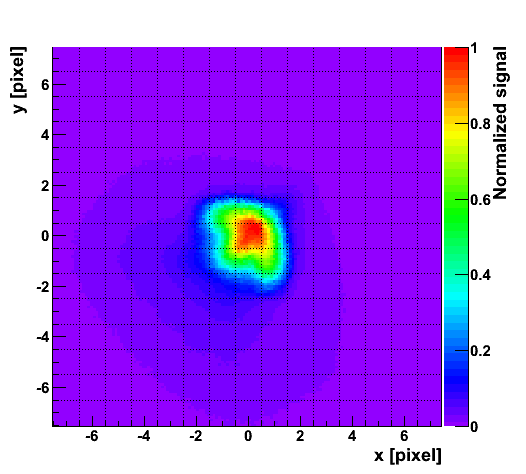}\\
\begin{sideways}1000\end{sideways} &
\includegraphics[width=0.3\textwidth]{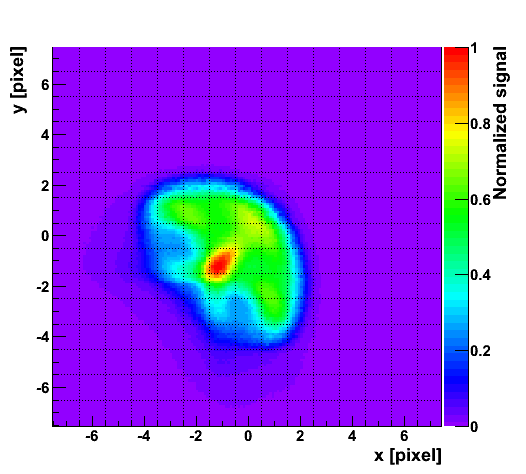} &
\includegraphics[width=0.3\textwidth]{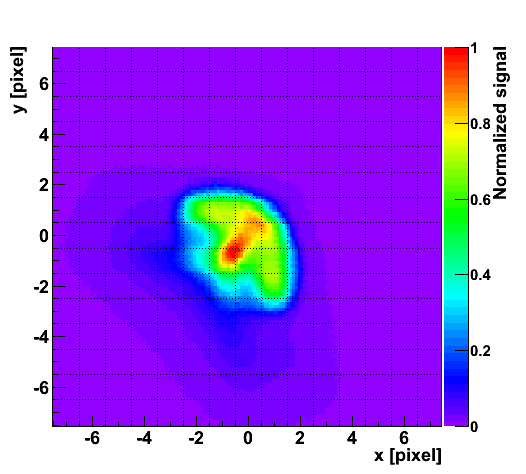} &
\includegraphics[width=0.3\textwidth]{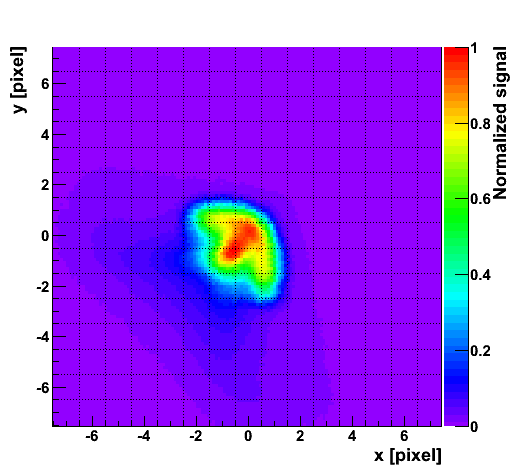}\\
\begin{sideways}1200\end{sideways} &
\includegraphics[width=0.3\textwidth]{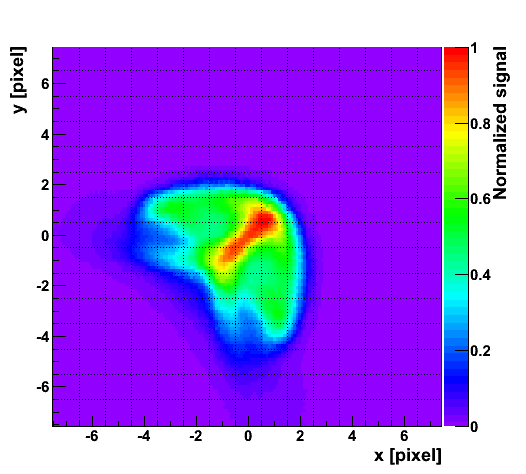} & 
\includegraphics[width=0.3\textwidth]{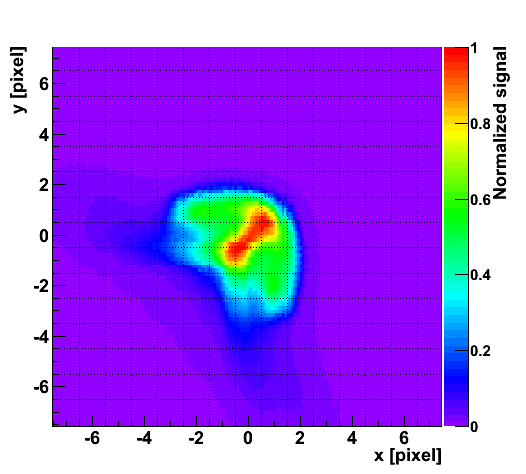} &
\includegraphics[width=0.3\textwidth]{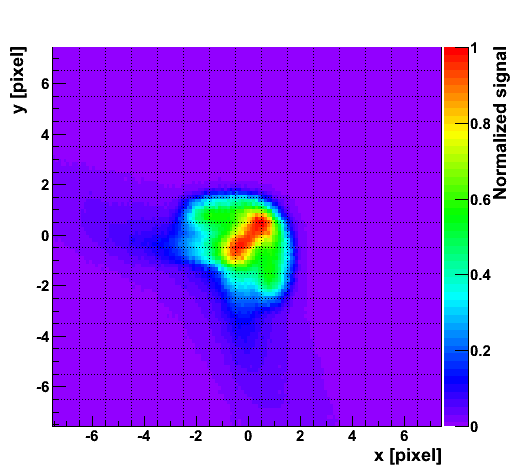}\\
\begin{sideways}1400\end{sideways} &
\includegraphics[width=0.3\textwidth]{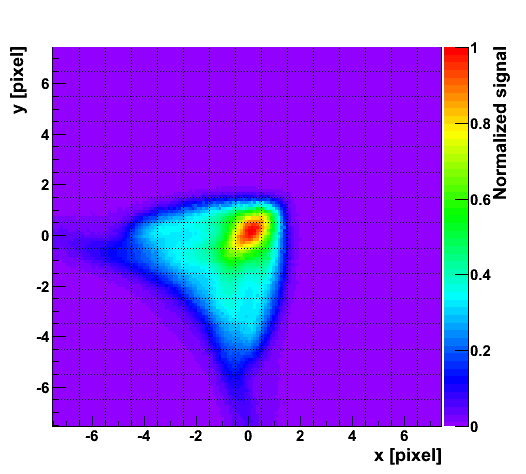} &
\includegraphics[width=0.3\textwidth]{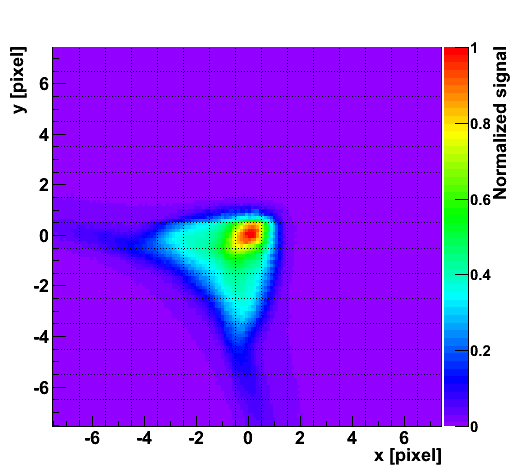} &
\includegraphics[width=0.3\textwidth]{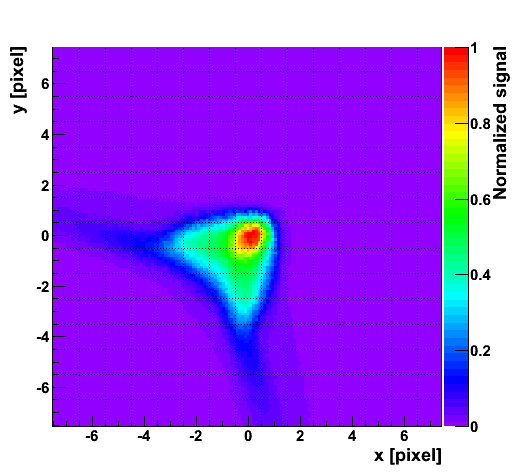} \\
\end{tabular}
\end{center}
\caption{PSFs of the white diode measured for different focus settings. Measurements were done for $fs=1.2$, $1.4$ and $1.6$ m (left, central and right columns respectively), and for 800, 1000, 1200 and 1400 pixels from the frame centre (from top to bottom).}
\label{diode_psf_focus}
\end{figure}

High quality photographic lenses of the same type should have PSFs in the same place close to identical, assuming the same focus is set. However, as mentioned before, focusing for the real stars is different than the focusing for the diode, and it is close to impossible to maintain exactly the same focus for different cameras. To study the possible influence of the focus setting on the PSF shape, measurements for three different focusing were repeated for the white diode, for 800, 1000, 1200 and 1400 pixels from the frame centre along its diagonal (fig. \ref{diode_psf_focus}).

The area covered by the PSF changes visibly with the focusing setting, but the general shape remains similar (blue and green parts). The very centre undergoes bigger changes, especially for 800 and 1000 pixels from the frame centre (red part). PSF seems to be best focused for the last setting, which was not the best focus for the central profile. This confirms the observation discussed in section \ref{sssec_cameras_focusing} that the best focus changes with the distance from the frame centre.

\chapter{Diffraction model of PSF}
\label{chap_diff}

A detailed knowledge about the point spread function for specific coordinates can be obtained directly from measurements, by reconstructing a high resolution profile of the point-like light source. However, to model the detector's response, a description of the PSF all over the frame is needed. For it is close to impossible to measure profiles for all positions on the CCD, the only viable solution is to create a mathematical model characterizing the PSF dependence on coordinates on the frame, based on obtained data.

Understanding the detector's response is inevitably connected with understanding how the image is formed on the CCD. While the conversion of the photons to electrons, charge processing and digitization are well explored for digital cameras, the PSF formation for such a wide-field experiment has not been described in literature, according to the author's knowledge. In this chapter we try to describe the image projected on the CCD as the result of a propagation of light through camera's lenses. Describing results of such a passage, if done properly and with high enough precision, should result in a proper PSF model.

\section{Diffraction based PSF}

In the case of a simple optical system, such as in the \pin project, the obstacle on the path of the light are lenses, whose size is much bigger than the wave length. This allows an assumption that the field inside their opening is similar to the field of an undisturbed light. Such an assumption validates use of the scalar diffraction theory, which should be an adequate approximation in the described circumstances.

In the basic case of image generation, the main calculation performed is the transformation of the front of the electromagnetic wave coming through an aperture (or lenses) into intensity of light on the screen -- the image. In this application, a Rayleigh-Sommerfeld equation can be used as the most general formula in the scalar diffraction theory:

\begin{figure}[tb!]
\begin{center}
	\includegraphics[width=0.47\textwidth]{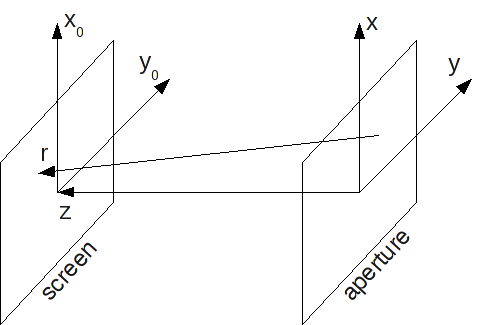}
\end{center}
\caption{Scheme of the aperture and screen coordinate systems, as used in the Rayleigh-Sommerfeld formula.}
\label{fig_diff_setup}
\end{figure}

\begin{equation}
PSF_L(x_0, y_0, z) = \left|\frac{1}{i\lambda}\iint\limits_{-\infty}^{\infty}U(x,y,0)\frac{ze^{ikr}}{r^2}\dx \dy\right|^2
\end{equation}

where $(x_0, y_0)$ are the Cartesian coordinates in the image plane, distant by $z$ from the aperture, $(x,y)$ are corresponding coordinates in the aperture plane, $r$ is the distance between chosen $(x,y,0)$ and $(x_0, y_0, z)$ (as shown in fig. \ref{fig_diff_setup}), $\lambda$ is the wavelength, $k$ is the wavenumber and $U(x,y,0)$ is the amplitude of the wavefront on the aperture\footnote{The amplitude of the wavefront should be constant on the aperture in the perfect case, but in reality can be altered by factors such as lenses transmission.}.

Calculating intensity of light in each point of the image requires integrating the wavefront all over the aperture plane. However, in the scalar theory it can be assumed that the wave is nonexistent outside the aperture $A$ in the aperture plane. This leads to the Kirchhoff approximation:

\begin{equation}
PSF_L(x_0, y_0, z) = \left|\frac{1}{i\lambda}\iint\limits_{A}U(x,y,0)\frac{ze^{ikr}}{r^2}\dx \dy\right|^2
\label{eq_kirchhoff_diff}
\end{equation}

In the case of many analysed optical systems, the size of the aperture $L_1$ and the screen $L_2$ are much smaller than the distance between them: $|z|\gg L_1+L_2$, or in a more detailed formulation: $\frac{L_1^2}{z\lambda}\ge1$. Then $z$ becomes dominant over the $x$ and $y$ and the Fresnel approximation becomes valid:

\begin{equation}
PSF_L(x_0, y_0, z) = \left|\frac{e^{ikz}}{i\lambda z}\iint\limits_{A}U(x,y,0)e^{\frac{ik}{2z}((x_0-x)^2+(y_0-y)^2)}\dx \dy\right|^2
\label{eq_fresnel_diff}
\end{equation}

The Fresnel formula is in fact a paraxial approximation, dealing only with small angles between optical axis and the image. However, it still takes into account curvature of the wavefront, which can be neglected for bigger distances, where $z\gg\frac{kL_1^2}{2}$. This is the core of the Fraunhofer approximation:

\begin{equation}
PSF_L(x_0, y_0, z) = \left|\frac{e^{ikz}}{i\lambda z}e^{\frac{ik}{2z}(x_0^2+y_0^2)}\iint\limits_{A}U(x,y,0)e^{\frac{ik}{z}(x_0x+y_0y)}\dx \dy\right|^2
\end{equation}

The Fresnel diffraction formula (eq. \ref{eq_fresnel_diff}) (as well as the Fraunhofer formula) can be easily transformed into an equation involving a Fourier transform:

$$PSF_L(x_0, y_0, z) = \left|\frac{e^{ikz}}{i\lambda z} e^{i\frac{\pi}{\lambda z}(x^2+y^2)} \mathcal{F}\{U(x,y,0)e^{i\frac{\pi}{\lambda z}(x_0^2+y_0^2)}\}\right|^2$$

which can be calculated using Fast Fourier Transform (FFT) algorithms, greatly reducing computing time. However, this is not the case for ``Pi of the Sky'', where the wide-angle lenses have $\frac{f}{d}=1.2$ ($f$ stands for the focusing length and $d$ for the lenses diameter). Thus both the aperture and the screen sizes are of the order of the size of the focusing length, i.e. $z \sim L_1,\ L_2$. This makes use of the Fraunhofer and Fresnel approximations impossible and forces us to use the Kirchhoff formulation (eq. \ref{eq_kirchhoff_diff}).

The approximation which can be introduced in the \pin case is an assumption of planar wave reaching the lenses, which is applicable for sources of light with almost infinite distance from the observer, such as stars. Thus in the perfect case of an empty opening, $U(x,y,0)$ could be treated as having a constant value all over the aperture.

\section{Monochromatic optical aberrations}

In the case of perfect lenses, a point-source of light should be visible as a point on the screen. However, the passage of light through an opening of a finite size diffracts the wave and causes spread of the image over finite area. The point spread function of a point-source visible through an aperture has been described by Airy and can be easily obtained from the eq. \ref{eq_kirchhoff_diff}. However, Kirchhoff formula in such form describes only light passing through a perfect optical system in which aperture is a ``source'' of a perfect, spherical wave. This is not the case in real optical systems, where introduction of such elements as mirrors or lenses aberrates the sphericity. Thus, in the case of a monochromatic wave, such deviations are called monochromatic optical aberrations.

To obtain an aberrated PSF from diffraction formulas, a deviation $W(x, y)$ from sphericity $e^{ikr}$ of the wavefront has to be introduced into eq. \ref{eq_kirchhoff_diff}:

\begin{equation}
PSF_L(x_0, y_0, z) = \left|\frac{1}{i\lambda}\iint\limits_{A}U(x,y,0)\frac{ze^{ikr}e^{W(x,y)}}{r^2}\dx \dy\right|^2
\label{eq_aber_diff}
\end{equation}

$W(x,y)$ may be expressed in multiple forms. Perhaps the most referred to is the sum of Seidel aberrations\cite{seidel} such as defocus, coma, astigmatism, spherical aberration, etc. This formulation is often used by optical engineers, mainly due to a quite convenient way of transforming into optical systems parameters.

The Zernike concept is generally more popular in the light wavefront analysis applications, mainly due to its mathematical properties. It introduces a set of polynomials, with names corresponding to aberrations in the Seidel formulation, which however are orthogonal on the whole aperture, making this formulation well suited for the wavefront fitting purposes. In this approach the wavefront can be expressed as:

\begin{equation}
W(\rho, \phi) = \sum_{n\geq m} Z^m_n(\rho, \phi) + \sum_{n\geq m} Z^{-m}_n(\rho, \phi)
\label{eq_wavefront_zernike}
\end{equation}

where $Z^m_n$ and $Z^{-m}_n$ stand for even and odd Zernike polynomials respectively and $n,\ m$ are non-negative integers, denoting the order and type of the aberration\cite{zernike_coef}. The wavefront distortion function $W$ was expressed in polar coordinates, more natural for radial aperture, where radial position $\rho$ is normalized to the aperture radius, $0 < \rho <1$. Each polynomial can be written as:

\begin{equation}
	Z^m_n(\rho, \phi) = R^m_n(\rho)\cos(m\phi),\ Z^{-m}_n(\rho, \phi) = R^m_n(\rho)\sin(m\phi)
\label{eq_zernike}
\end{equation}

the radial part being:

$$
	R^m_n = \sum^{(n-m)/2}_{k=0}\frac{(-1)^k(n-k)!}{k!((n+m)/2-k)!((n-m)/2-k)!}\rho^{n-2k}
$$

In fact, Zernike polynomials are compatible with Seidel aberrations if only aberrations up to $3^{\rm rd}$ order are used\cite{zernike_seidel_compat}. A few terms presenting aberrations used later in this work are shown in the tab. \ref{tab_aberr}.

\begin{table}
	\begin{center}
	\begin{tabular}{|c|c|c|c|c|}
		\hline
		\bf aberration & \bf n & \bf m & \bf polynomial & \bf wavefront \\
		\hline \hline
		astigmatism & 2 & 2 & $\rho^2 \cos{2\phi}$ & \includegraphics[width=0.1\textwidth]{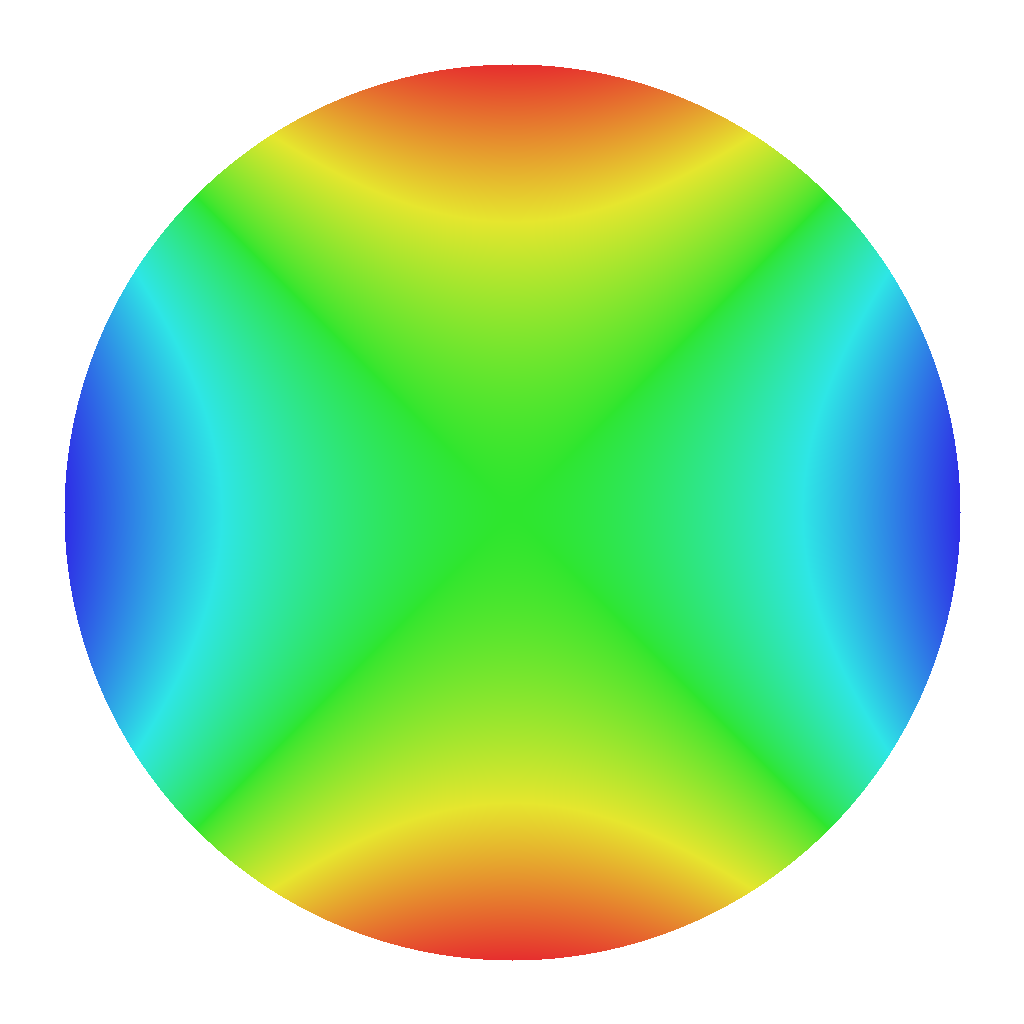} \\
		coma & 3 & 1 & $(3\rho^2-2)\rho\cos{\phi}$ & \includegraphics[width=0.1\textwidth]{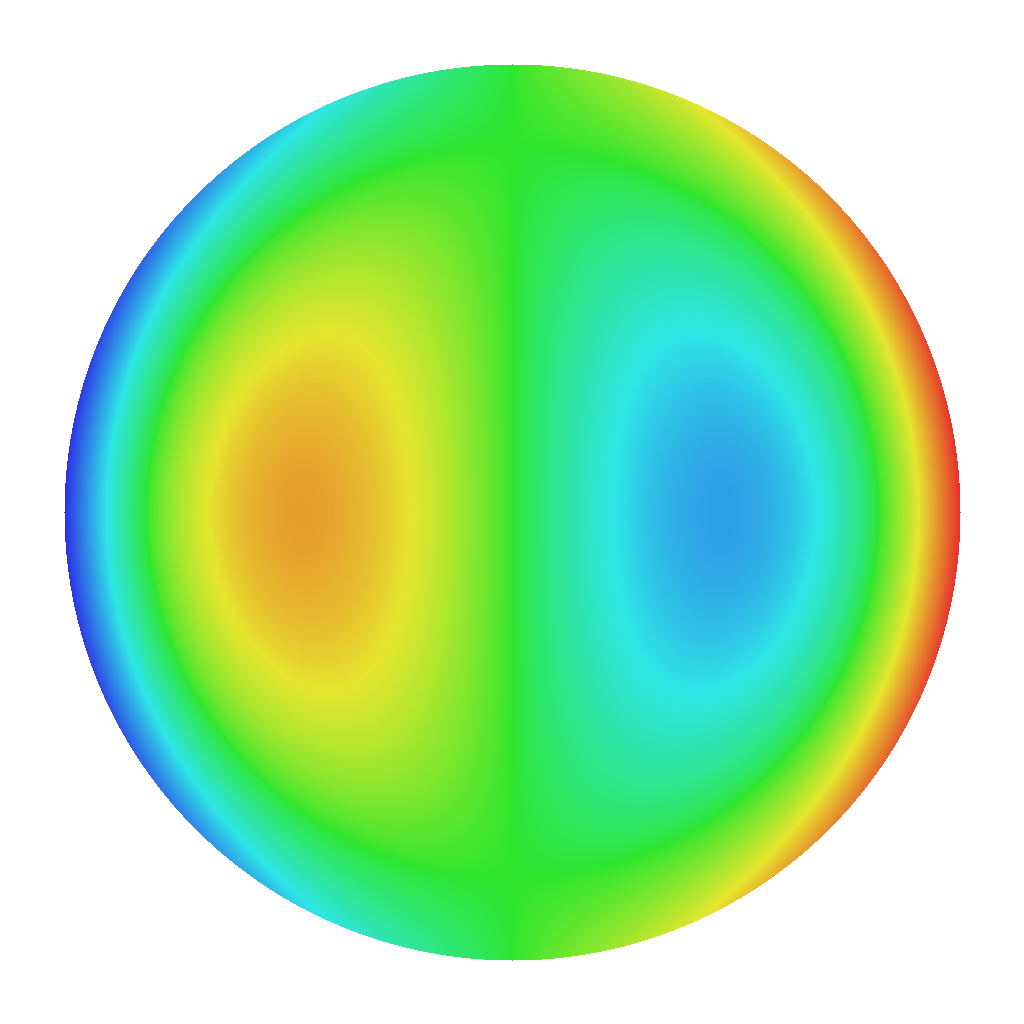}\\
		spherical aberration & 2 & 0 & $6\rho^4 - 6\rho^2 +1$ & \includegraphics[width=0.1\textwidth]{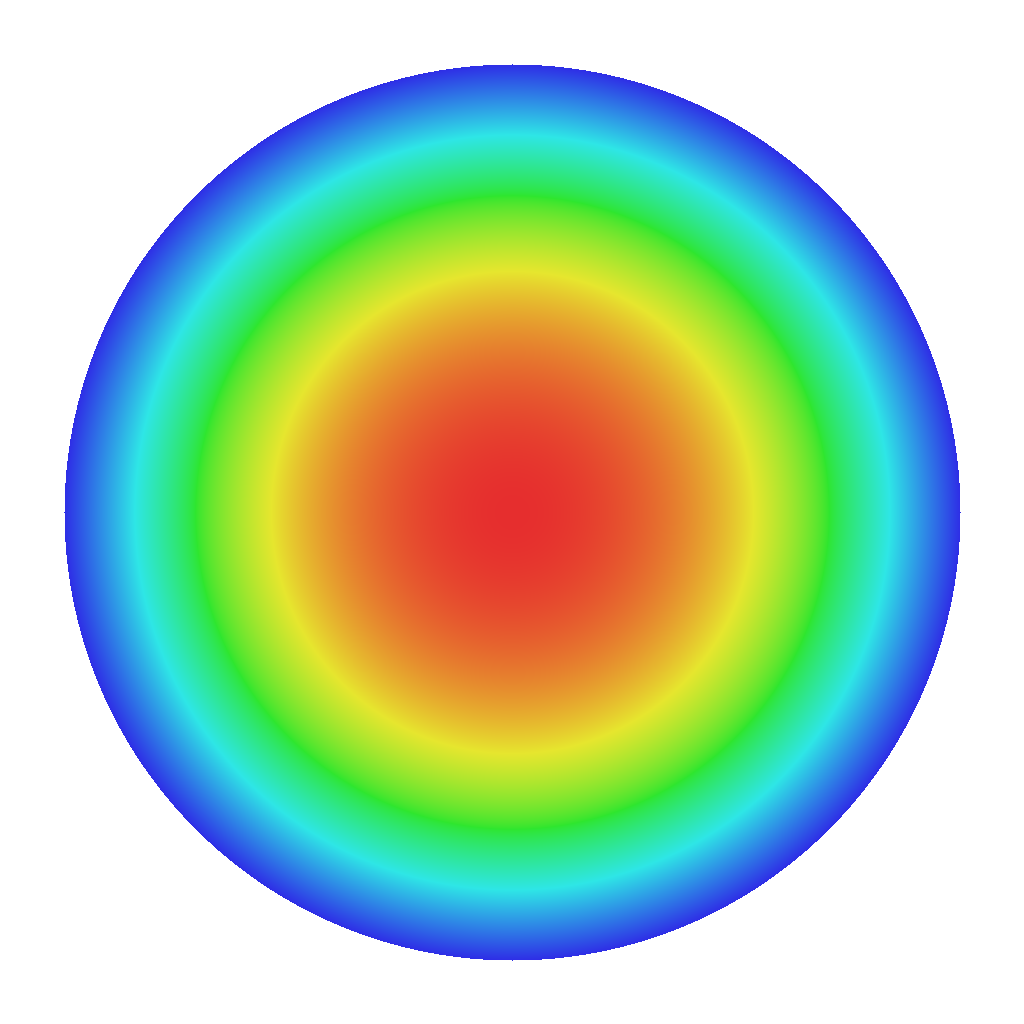} \\
		trefoil & 3 & 3 & $\rho^3\cos{3\phi}$ & \includegraphics[width=0.1\textwidth]{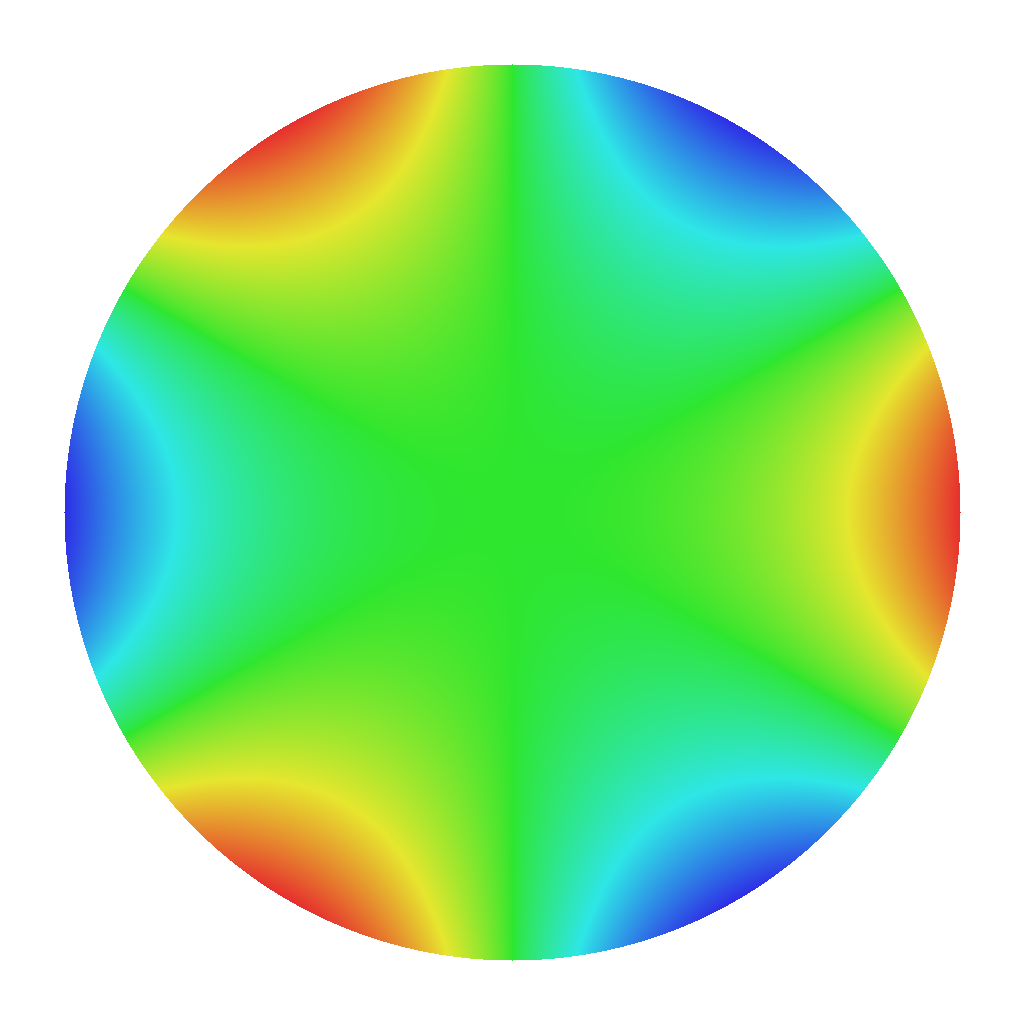} \\
		secondary coma & 5 & 1 & $(10\rho^5-12\rho^3+3\rho)\cos{\phi}$ & \includegraphics[width=0.1\textwidth]{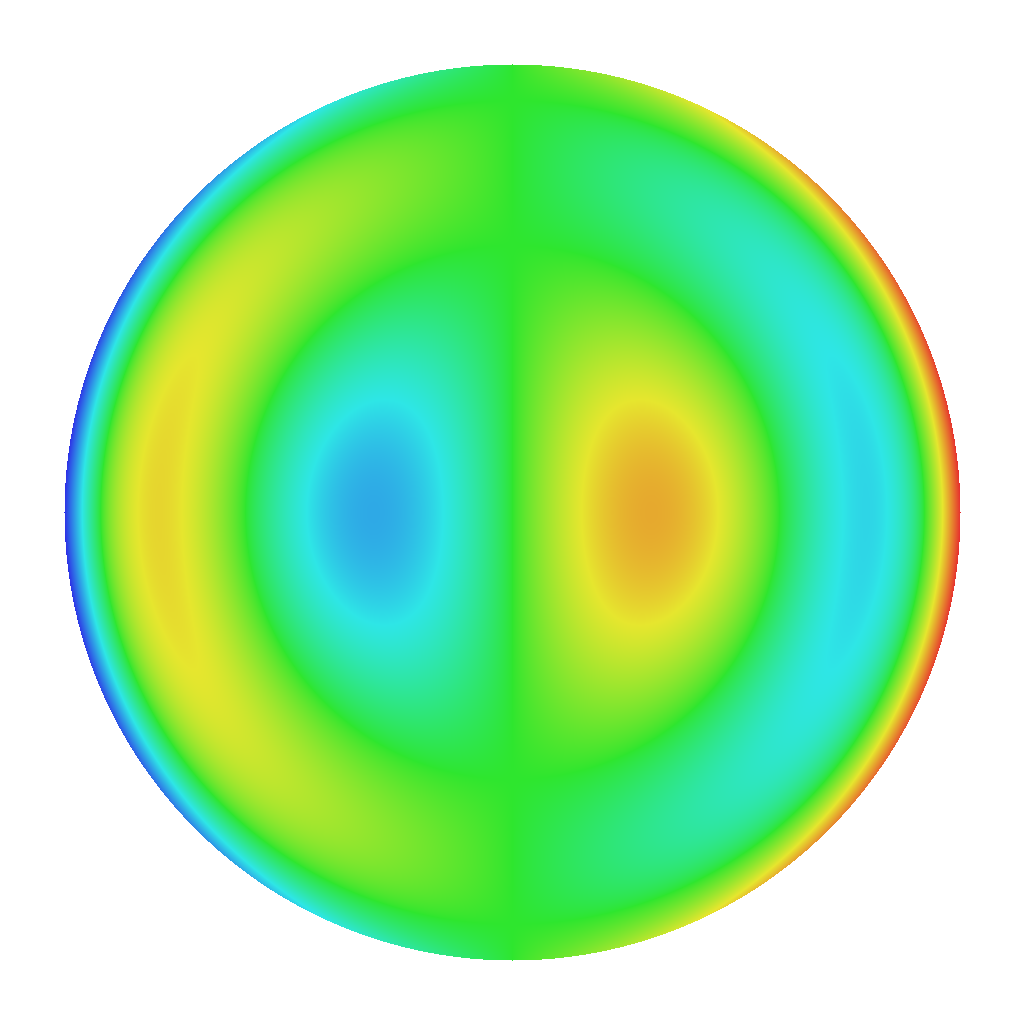} \\
		secondary spherical aberration & 6 & 0 & $20\rho^6-30\rho^4+12\rho^2-1$ & \includegraphics[width=0.1\textwidth]{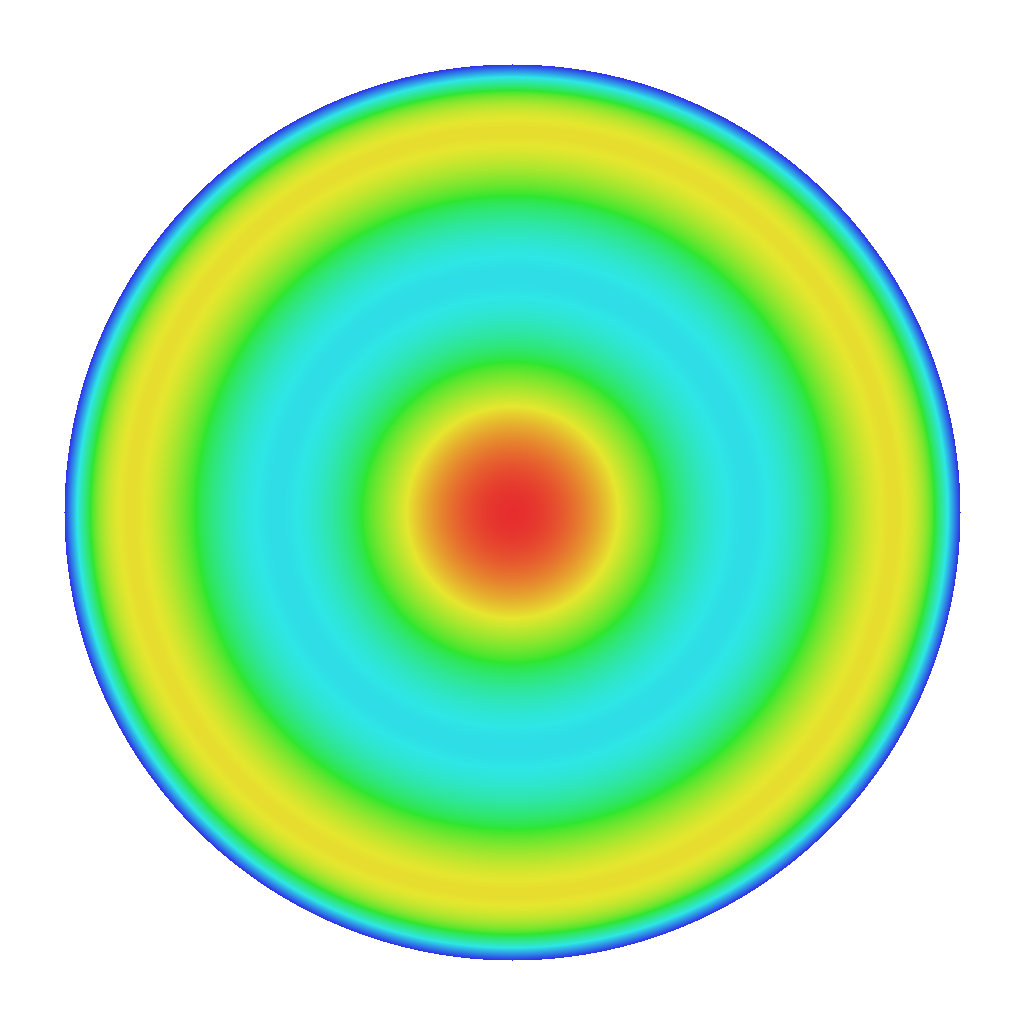} \\
		\hline
	\end{tabular}
	\end{center}
\caption{Monochromatic optical aberrations and their Zernike polynomials used in the diffractive PSF modelling.}
\label{tab_aberr}
\end{table}

\section{Modelling method}

A starting point for obtaining the model of the point spread function in the \pin project is to find a proper parametrisation for profiles reconstructed from laboratory measurement. Thus the eq. \ref{eq_aber_diff} was transformed into cylindrical coordinates, with astigmatism, coma, secondary coma, spherical aberration, secondary spherical aberration and trefoil aberrations used\footnote{These polynomials were chosen as those capable of providing significantly diverging contributions into the PSF, while still forming a rather small set.}:

\begin{equation}
	PSF_L(\rho_0, \phi_0, z) = \left|\frac{1}{i\lambda}\iint\limits_{A}U(\rho,\phi,0)\frac{ze^{ikr}e^{W(\rho,\phi)}}{r^2}\rho\mathrm{d}\rho \mathrm{d}\phi\right|^2
\label{eq_aber_final}
\end{equation}

where:

\begin{equation}
\begin{split}	
W(\rho,\phi) & = A(\rho^2 \cos{2\phi})+C((3\rho^2-2)\rho\cos{\phi})+S(6\rho^4 - 6\rho^2 +1)\\
& +T(\rho^3\cos{3\phi})+C'((10\rho^5-12\rho^3+3\rho)\cos{\phi})\\
&+S'(20\rho^6-30\rho^4+12\rho^2-1)
\end{split}
\label{eq_wavefront}
\end{equation}

$A,\ C,\ S,\ T,\ C',\ S'$ stand for coefficients of polynomials representing respectively astigmatism, coma, spherical aberration, trefoil, secondary coma and secondary spherical aberration. This is the maximal set of aberrations used in the modelling, however it can be easily extended to a bigger set. Polynomials used commonly for presenting deviation from the optical axis (tilt) or defocus are not needed here, since they are naturally present in the Kirchhoff formulation via coordinates of the image.

There are no known methods of analytically performing integration of eq. \ref{eq_aber_final}, especially in the presence of aberrations. Thus numerical integration has to be performed, which in this case is a complicated task. Due to the large wavenumber $k$ in the term $e^{ikr}$ (for any optical wavelength), the formula consist of a very fast oscillatory integral. Common adaptive integration methods fail at attempt of performing such a calculation. Other methods have to be used, perhaps one of the most obvious and giving good results is an integration using Gauss integration points in a two dimensional space. 

The quality of results strongly depends on the number of intervals used in the integration. Lenses as wide as in the \pin case (few centimeters) require hundreds to thousands intervals to obtain satisfactory result, the number growing with the distance of the calculated PSF from the frame centre. 

Altogether this makes the PSF computing a very time consuming task, reaching seconds on modern computers (for example of PSF with 250 points). Finding proper model parameters with fitting procedures often requires thousands of such calculations, making the modelling close to impossible.

Fortunately, a work by Miks et al.\cite{rapid_osc_int} makes the task of integrating rapidly oscillating functions in the diffraction equations much less time consuming. They obtain an accurate approximation (based on a Taylor series expansion of the function in the exponent), which can be formed as:

\begin{equation}
\begin{split}
\iint_S f(x,y)e^{ikg(x,y)}\mathrm{d}x\mathrm{d}y \simeq
4 \sum_n f(x_n, y_n)e^{ikg(x_n, y_n)}\frac{\sin{X_n}}{X_n}\frac{\sin{Y_n}}{Y_n}\Delta x_n \Delta y_n,\\
X_n=k\frac{\partial g(x_n, y_n)}{\partial x}\Delta x_n,\ Y_n=k\frac{\partial g(x_n, y_n)}{\partial y}\Delta y_n
\end{split}
\label{eq_int_approx}
\end{equation}

This formula allows obtaining better results and is a few times faster than previously mentioned methods -- it was used for the purpose of this work.

At this point we have to take into account, that the eq. \ref{eq_aber_final} describes only the PSF coming out of the optical system -- $PSF_L$, while the image is the convolution of the $PSF_L$ and the CCD response $PRF$:

\begin{equation}
	PSF(\rho_0', \phi_0', z) = \iint PRF(\rho_0, \phi_0, \rho_0', \phi_0')\cdot PSF_L(\rho_0, \phi_0, z) \mathrm{d}\rho_0 \mathrm{d}\phi_0
\label{eq_psf_conv_diff}
\end{equation} 

where $(\rho_0', \phi_0')$ are polar coordinates of the specific point of the final $PSF$ (corresponding to CCD pixel centres) in the image plane (the result of the convolution), while $(\rho_0, \phi_0)$ are polar coordinates of the current point of integration in the image plane (same as in eq. \ref{eq_aber_final}).

\section{PSF modelling results}

For modelling purposes, it was assumed that the camera's CCD-lenses setup can be approximated by an aperture-screen setup, with aperture of 2.5 cm radius (the exiting window of the lenses), and the basic distance between screen and aperture of 8.5 cm (the focusing length of the lenses). At this stage, it was assumed that the amplitude of the wave $U(x_0,y_0,0)=1$ all over the aperture.

The eq. \ref{eq_aber_final} approximated by eq. \ref{eq_int_approx} and convoluted as in eq. \ref{eq_psf_conv_diff} was used to model PSFs reconstructed in laboratory measurements (fig. \ref{diode_psf_blue} and \ref{diode_psf_red}). In general case the convolution would require integrating over the infinite space. However, the rapid drop of the PRF close to the pixel edge allows restricting the integration space. An area of $1.6 \times 1.6$ square pixels around the pixel centre was used, the integration replaced by a simple sum of $PRF$ and $PSF_L$ products on the uniformly distributed net of $20 \times 20$ point in this space.

\begin{figure}[tb!]
\begin{center}
\subfigure[simulated PSF]{
	\includegraphics[width=0.47\textwidth]{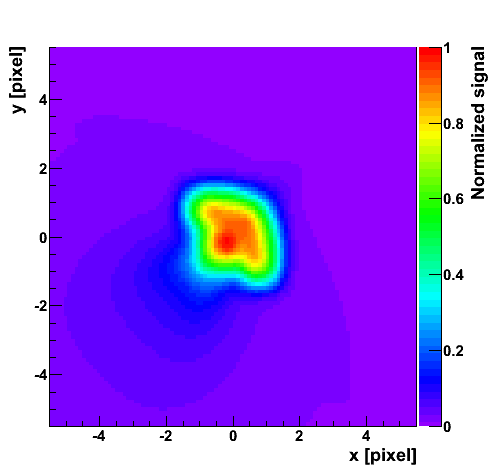}
}
\subfigure[real PSF]{
	\includegraphics[width=0.47\textwidth]{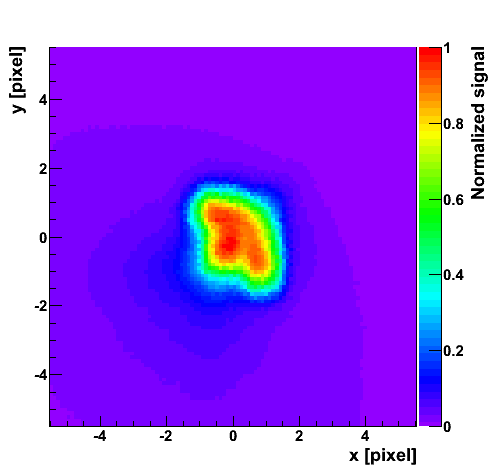}
}
\end{center}
\caption{The PSF description obtained from the diffraction model (left) and the reconstructed PSF 800 pixels from the frame centre (right). The overall shape given by the model is the same as the measurements, showing that the diffractive approach is proper in general, but it fails to reproduce details.}
\label{fig_diff_fit_psf_800}
\end{figure}

Model was obtained by fitting, with the least $\chi^2$ method, coordinates of an image $(x,y,z)$ (slight deviations from values obtained from the experiment were allowed) and aberration coefficients of formula \ref{eq_wavefront} to reconstructed profiles. The fit was performed starting in the centre of the frame and moving to the expected image position. The first reason for using such a method was that for large angles the deformation of the PSF caused by the plain deviation from the optical axis needs to be compensated by coma aberration and the proper coefficient could not be guessed. The second reason is a result of an observation, that the fitted model may, probably due to some simplifications in the model, describe the PSF better in a position different than it was actually registered.

Figure \ref{fig_diff_fit_psf_800} shows the fit result for the red diode PSF measured 800 pixels from the frame centre. Model is very similar to the measured point-source profile, showing, that the diffractive approach is, in general, the proper method of PSF modelling. Unfortunately, the obtained model fails to reproduce exact shape of the PSF centre (shown in red) and, even more, the tails of the profile (shown in blue).

The other problem in this approach is that the obtained results do not necessary represent the global minimum of $\chi^2$. During calculation we noticed, that many local $\chi^2$ minima exist in the parameter space and it is not guaranteed that the chosen ``path'' for parameters was the correct one.

\begin{figure}[tb!]
\begin{center}
\subfigure[simulated red PSF]{
	\includegraphics[width=0.47\textwidth]{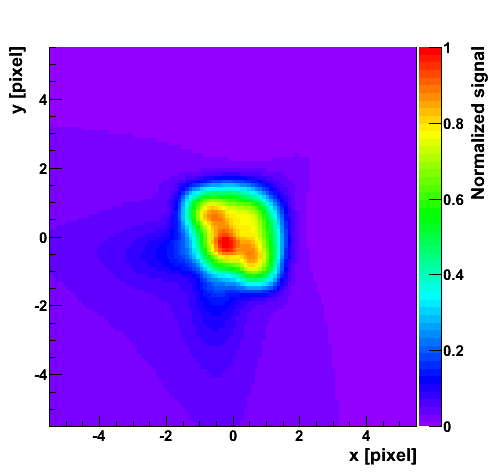}
}
\subfigure[measured red PSF]{
	\includegraphics[width=0.47\textwidth]{diffraction/red_800.png}
}
\subfigure[simulated blue PSF]{
	\includegraphics[width=0.47\textwidth]{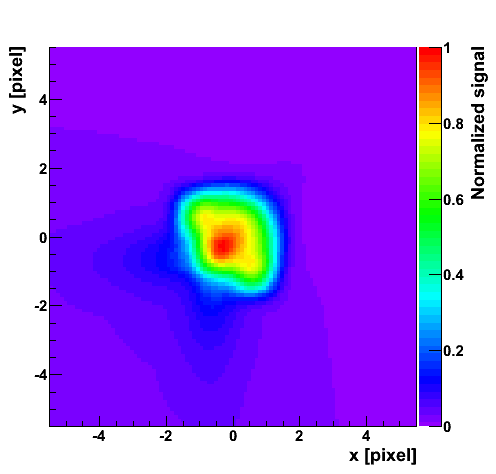}
}
\subfigure[measured blue PSF]{
	\includegraphics[width=0.47\textwidth]{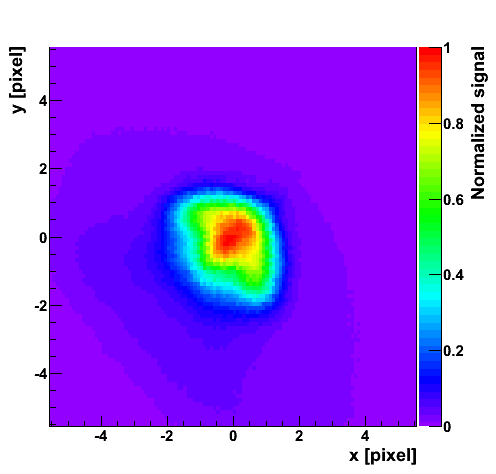}
}
\end{center}
\caption{The PSF description in the diffraction model for red (a) and blue (c) diode, obtained from the simultaneous fit to the red and blue PSF profile measurements results 800 pixels from the frame centre, shown in (b) and (d) respectively.}
\label{fig_diff_fit_psf_800_multi}
\end{figure}

To constrain the fit, an attempt to simultaneously minimize red and blue PSFs was performed, based on an assumption, that the aberrations coefficients do not depend on the wavelength. However, results of that procedure were worse for the red PSF, 800 pixels from the frame centre than in the single wavelength attempt (fig. \ref{fig_diff_fit_psf_800_multi}). Fits for larger distances were even more diverse from profiles reconstructed from laboratory measurements.

\begin{figure}[tb!]
\begin{center}
\subfigure[uniform transmission]{
	\includegraphics[width=0.457\textwidth]{diffraction/best_800_red.png}
}
\subfigure[nonuniform transmission]{
	\includegraphics[width=0.457\textwidth]{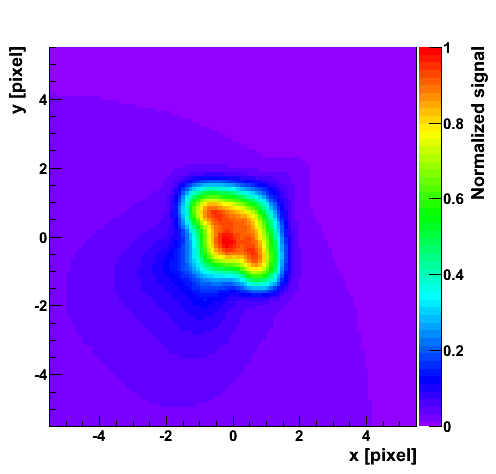}
}
\end{center}
\caption{Comparison of the PSF description obtained from the model with uniform lenses transmission (left) and with transmission changing with radial coordinate $\rho$ on the aperture as $0.73+0.27\rho^2$ (right).}
\label{fig_trans_comp}
\end{figure}

To improve agreement between the model and the data we also considered a fit where the $U(x_0, y_0, 0)$ function was modified in a final stage of model tuning to include the transmission function of the lenses. This however had a small impact on the overall shape, as shown in fig. \ref{fig_trans_comp}. The number of intervals used in the integration also had no impact on the fitting procedure, for increasing accuracy did not alter parameters significantly. The inconsistency is definitely not due to approximations used in the calculation procedure.

There are a few reasons why the diffractive approach failed to reproduce PSFs in high enough detail. The first is the time of computation -- although much shorter than in standard approach, thanks to approximation shown in eq. \ref{eq_int_approx}, still very long for a general search for the shape of the PSF, especially in a situation of multiple local minima. Additionally, fitting procedures used tend to find minima close to the initial parameters of the fit, while the initial parameters here (except for the image coordinates) were a pure guess. The procedure of starting in the frame centre, with aberrations coefficients close to zero and gradually changing them while proceeding to the final image position is also not guaranteed to give proper results. More advanced minimization algorithm, more likely to find global minimum, should perhaps be used, such as a genetic algorithm or simulated annealing. However, such algorithms require much longer computation times. 

It is also possible that the introduced model is too simple to reproduce profiles in sufficient detail. A simple aperture-screen setup is based on an approximation of thin lenses, while real photographic lenses definitely do not meet this assumption. In our model aberrations are added only in the aperture plane, while in reality they are formed in multiple planes.

It has to be stated, that although the approach failed, it shows promising results for further study. A number of dedicated measurements could be performed, such as direct measurements of the wavefront, that could aid the modelling. Such a study could result in a possibility of modelling PSFs taking into account the spectral type of the observed source, which in some special cases allowing long computing time, could provide interesting results.

\chapter{Polynomial model of PSF}
\label{chap_polynomials}

The point spread function describing the star image visible on the frame is a diffractive image of the source convoluted with the CCD pixel structure. Thus the most proper, in scientific terms, way to describe the PSF is to find parameters of the lenses and CCD and perform a diffraction calculations, as described in chapter \ref{chap_diff}. However, such an approach can only be used for an idealized, simplified version of an experimental setup. Moreover, in this case calculations of a diffractive pattern are a significantly time-consuming task. This makes the main aim of this work -- an application of the ``diffractive approach'' to data analysis and simulation -- a troublesome undertaking. 

A common way to simplify the use of physical models in practice is to introduce an effective approach. In the case of PSF description a parametrization in the image plane can be sought -- directly deriving it from the shape of reconstructed, high resolution point-like source's profiles. The integration of the wavefront is skipped, this way leaving only simple one-time function calculation for each of the PSF points. However, no simple function describes aberrated PSF in the image plane (except for some very simple cases). Therefore, to obtain an efficient effective profile a model has to be developed and for this task a proper mathematical basis has to be chosen.

\section{Model basis}

Deriving the PSF description in the image plane is a task similar to finding a mathematical description of complicated shapes of an aberrated wavefront. Thus similar basis of functions can be used in both tasks -- Zernike polynomials. However, in the case of the image plane they need to be modified slightly.

First modification is due to the fact that the PSF is a function described on an infinite plane, while the wavefront described by the Zernike polynomials was bound to a finite, circular aperture. Therefore, a transformation of a PSF's radial coordinate $r$ to argument of Zernike polynomial, $u(r)$, where we assume that $u(0)=0$ and $u(\infty)=1$, has to be performed. Thus set of modified Zernike polynomials is now defined as:

\begin{equation}
Z^m_n:=Z^m_n(u, \phi),\ u=1-e^{-\frac{r}{\lambda}}
\label{eq_zer_repar}
\end{equation} 

where $u$ and $\phi$ are standard radial and azimuthal coordinates of Zernike polynomials and $\lambda$ is a parameter modifying the transformation. Zernike polynomials are defined as in eq. \ref{eq_zernike}.

Additional modification is needed, for the real PSF has a maximal value around its centre and asymptotically drops to zero in infinity, while the wavefront has a sharp cut-off at the border of the aperture. The asymptotic behaviour is introduced to the PSF using a gaussian-like modification of the Zernike polynomials:

\begin{equation}
PSF_L(r, \phi) = e^{-0.5\cdot \sum_{m,n}Z^m_n(u, \phi)\cdot r^p}
\label{eq_psf_lenses}
\end{equation} 

where $PSF_L$ stands for a PSF generated by lenses only (no convolution with CCD) and $p$ is a factor modifying the asymptotic behaviour. Additionally, we assume that the PSF is symmetric, thus only symmetric terms are allowed in Zernike polynomials. The obtained set of functions is not orthogonal, in contrast to the standard Zernike polynomials, but still is well suited for fitting purposes. However, one has to keep in mind that in this approach hardly any physical meaning can be attributed to the parameters of the fit.

Similarly to the diffractive approach, PSF visible as an image on the frame is a convolution of a point spread function generated by lenses (eq. \ref{eq_psf_lenses}) and CCD pixel structure:

\begin{equation}
	PSF(r', \phi') = \int \int PRF(r, \phi, r', \phi')\cdot PSF_L(r, \phi) dr d\phi
\label{eq_psf_conv}
\end{equation} 
 
where $PRF(r, \phi, r', \phi')$ stands for pixel response function.

In the general case convolution of $PSF_L$ and $PRF$, given in eq. \ref{eq_psf_conv}, should be integrated over the whole CCD. In the real case the integration is performed numerically on interpolated PRF values (similarly as in the diffractive approach -- see sec. \ref{ssec_prf}) -- summed are products of $20\times 20$ $PSF_L$ points and bilineary interpolated $PRF$ points in the range of $1.6\times 1.6$ pixels around a pixel centre, which corresponds to the PRF size. The net of points is distributed and weighting for points chosen according to gaussian quadrature integration rules.

\section{Choosing optimal parameters for the basis}

On of the main aims of this work is to find a universal formula describing point spread function all over the frame. Due to limited time, the function was measured only for a few chosen positions (sec. \ref{ssec_measured_psfs}). Therefore deriving such a universal formula requires finding a local model for each of the measured PSFs first and then finding an interpolation method between these models.

Fitting eq. \ref{eq_psf_conv} to a reconstructed white diode profile gives very satisfactory results, especially when many (for example first 20) symmetrical Zernike polynomials are used as a basis. However, the bigger the basis the more complicated the mathematical description -- every extra polynomial means one extra dimension in the parameter space for the interpolation, not mentioning extra time needed for calculations. Moreover, some polynomials of a small significance may lead to distorted results in interpolated PSFs. Thus a proper basis -- set of Zernike polynomials that give significant contribution to the shape modelling needs to be chosen.

\subsection{Parameter reduction method}

The first conclusion of PSF fitting was that $p$ and $\lambda$ values should be fixed. These parameters are highly correlated with other free parameters and cause fits to be unstable. Moreover, the asymptotic behaviour adjusted by them is also altered by the circular Zernike Polynomials, i.e. terms 0, 4, 12 and 24. Thus $p=2$ and $\lambda=1$ values were chosen which are standard for normal distribution, simplify the computation of the eq. \ref{eq_psf_lenses} and proved to work well for all fits.

To choose optimal terms of Zernike Polynomials, a $\chi ^2$ fit of eq. \ref{eq_psf_conv} to measured profiles was performed in a special way. For the first iteration of the fit, the initial value of parameter 0 (weight for first Zernike polynomial $Z_0^0$) was set to 1 and all the other to a value close to 0\footnote{For all parameters set to 0 (or 1 in case of $Z_0^0$) a fit is highly unstable. The fit was more stable, when other parameters values were not initially equal to 0 (or 1), but close to it.}. The fit was performed with only the first parameter free, then repeated with only the second parameter free and so on until all parameters were tested. The parameter giving the best $\chi^2$ was chosen for the next iteration, in which the procedure of freeing consecutive parameters was repeated, with the parameter chosen in the first iteration permanently set as free. In the third iteration parameters chosen in the first and second were permanently free and so on, until the iteration in which all parameters were free was performed. This way not only parameters having the largest impact on the fit were chosen, but also their order, important for the future model reconstruction in different circumstances.

\begin{figure}[p!]
\begin{center}
$
\begin{array}{ccc}
	\begin{sideways}\begin{minipage}[c]{15mm}\begin{center}200\end{center}\end{minipage}\end{sideways} &
	\includegraphics[width=0.32\textwidth]{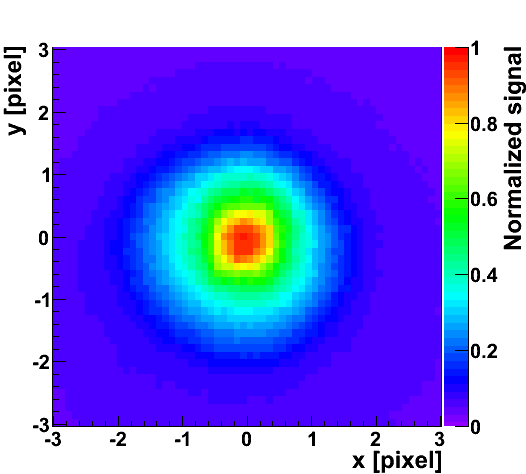} &
	\includegraphics[width=0.32\textwidth]{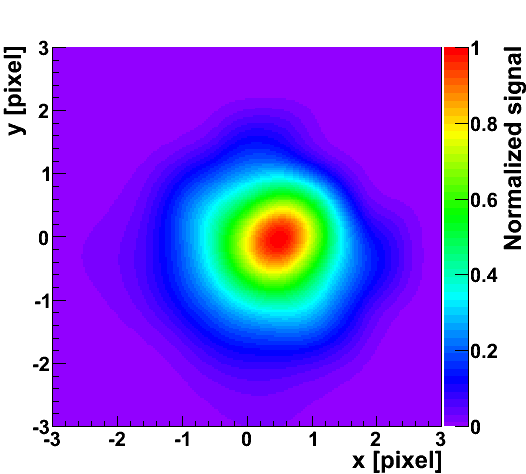} \\
	\begin{sideways}\begin{minipage}[c]{15mm}\begin{center}400\end{center}\end{minipage}\end{sideways} &
	\includegraphics[width=0.32\textwidth]{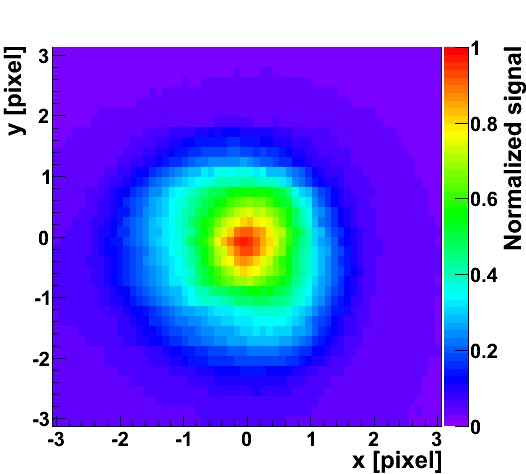} &
	\includegraphics[width=0.32\textwidth]{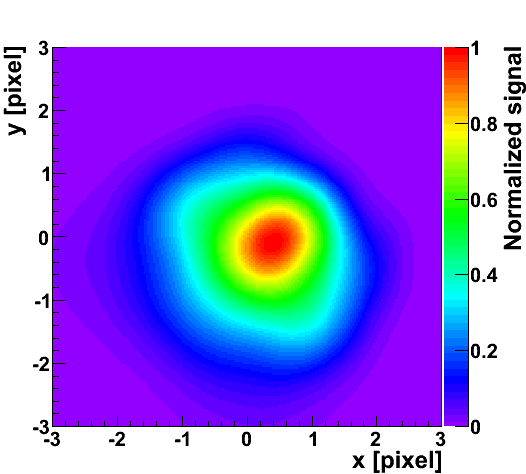} \\
	\begin{sideways}\begin{minipage}[c]{15mm}\begin{center}600\end{center}\end{minipage}\end{sideways} &
	\includegraphics[width=0.32\textwidth]{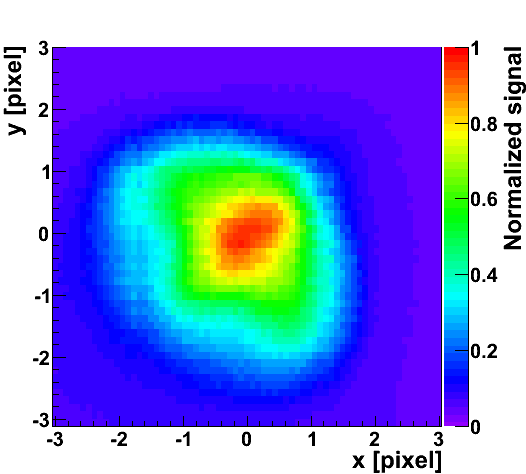} &
	\includegraphics[width=0.32\textwidth]{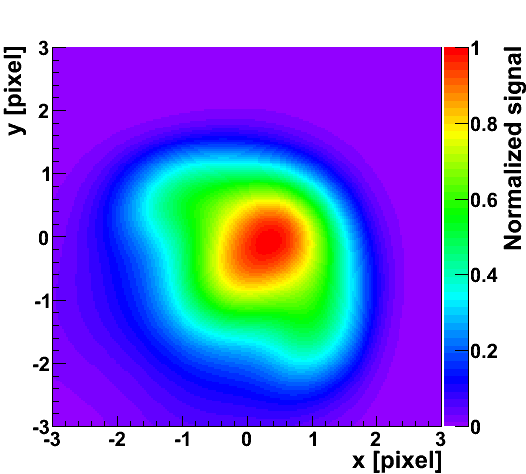} \\
	\begin{sideways}\begin{minipage}[c]{15mm}\begin{center}800\end{center}\end{minipage}\end{sideways} &
	\includegraphics[width=0.32\textwidth]{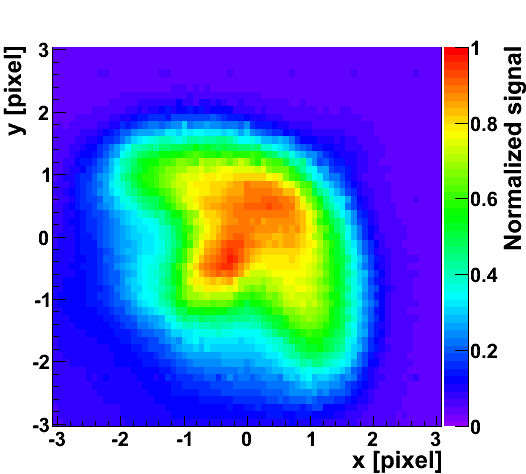} &
	\includegraphics[width=0.32\textwidth]{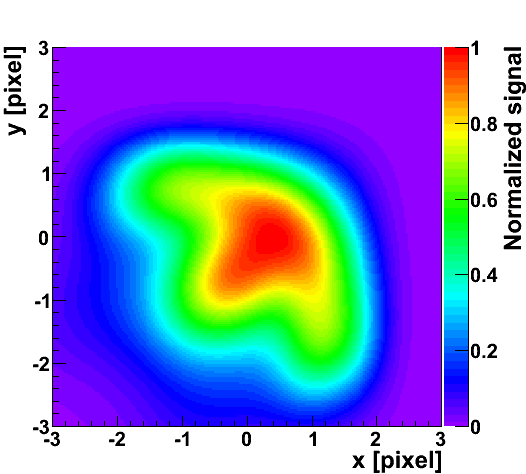} \\
\end{array}
$
\end{center}
\caption{Measured PSFs (left column) and their polynomial models (right column), resulting from the parameters searching and fitting procedure, for PSF measured 0 to 1400 pixels from the frame centre, as indicated in the plot. Continued on the next page.}
\label{fig_all_par_fit}
\end{figure}

\begin{figure}[t!]
\ContinuedFloat
\begin{center}
$
\begin{array}{ccc}
	\begin{sideways}\begin{minipage}[c]{15mm}\begin{center}1000\end{center}\end{minipage}\end{sideways} &
	\includegraphics[width=0.32\textwidth]{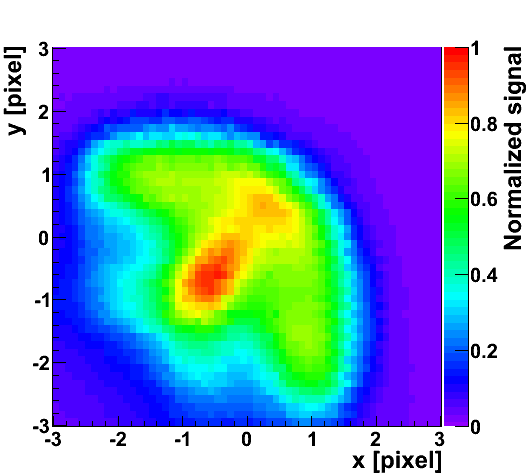} &
	\includegraphics[width=0.32\textwidth]{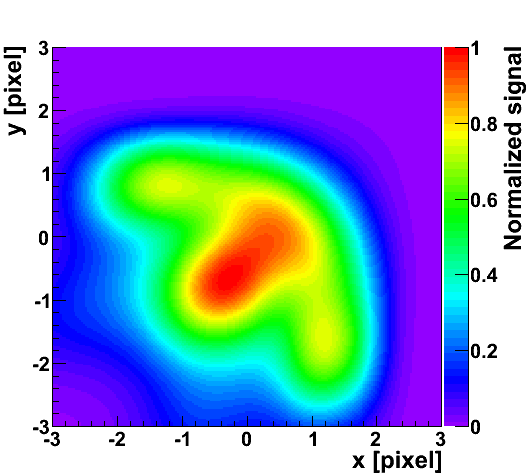} \\
	\begin{sideways}\begin{minipage}[c]{15mm}\begin{center}1200\end{center}\end{minipage}\end{sideways} &
	\includegraphics[width=0.32\textwidth]{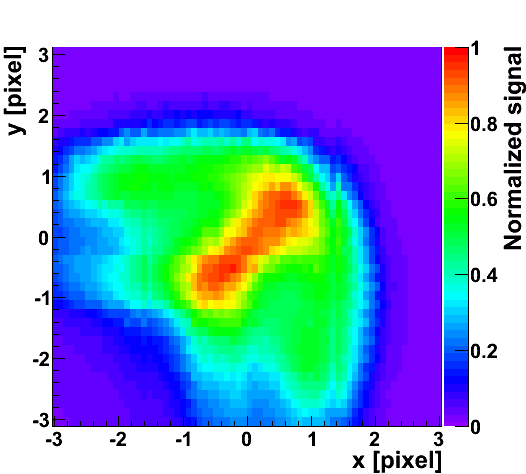} &
	\includegraphics[width=0.32\textwidth]{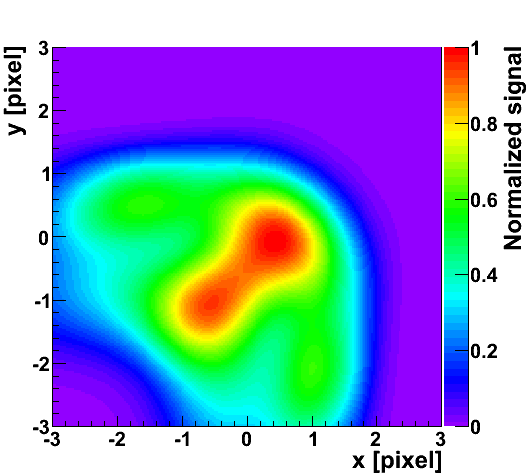} \\
	\begin{sideways}\begin{minipage}[c]{15mm}\begin{center}1400\end{center}\end{minipage}\end{sideways} &
	\includegraphics[width=0.32\textwidth]{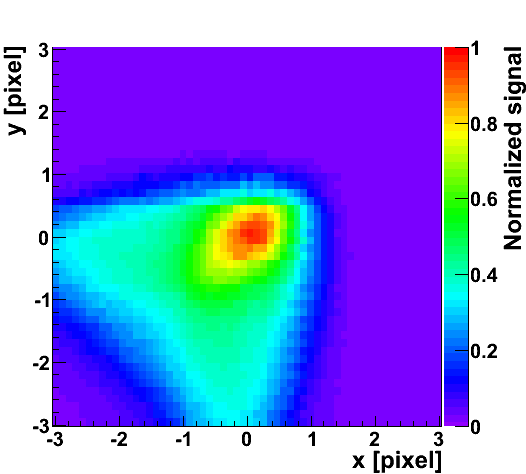} &
	\includegraphics[width=0.32\textwidth]{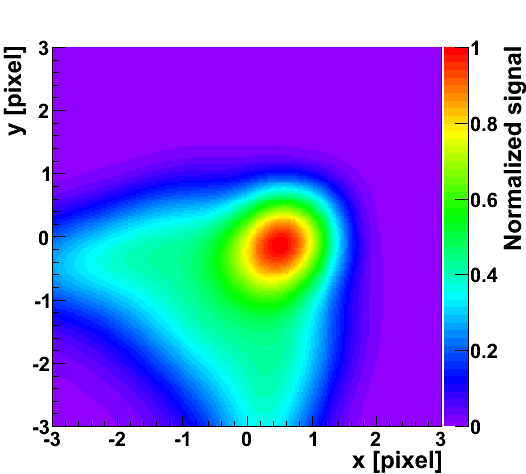} \\
\end{array}
$
\end{center}
\caption{Continued...}
\label{fig_all_par_fit}
\end{figure}

\subsubsection{Basis formed of the best parameters for all the diagonal PSFs}

\begin{figure}[tb!]
\begin{center}
	\includegraphics[width=0.99\textwidth]{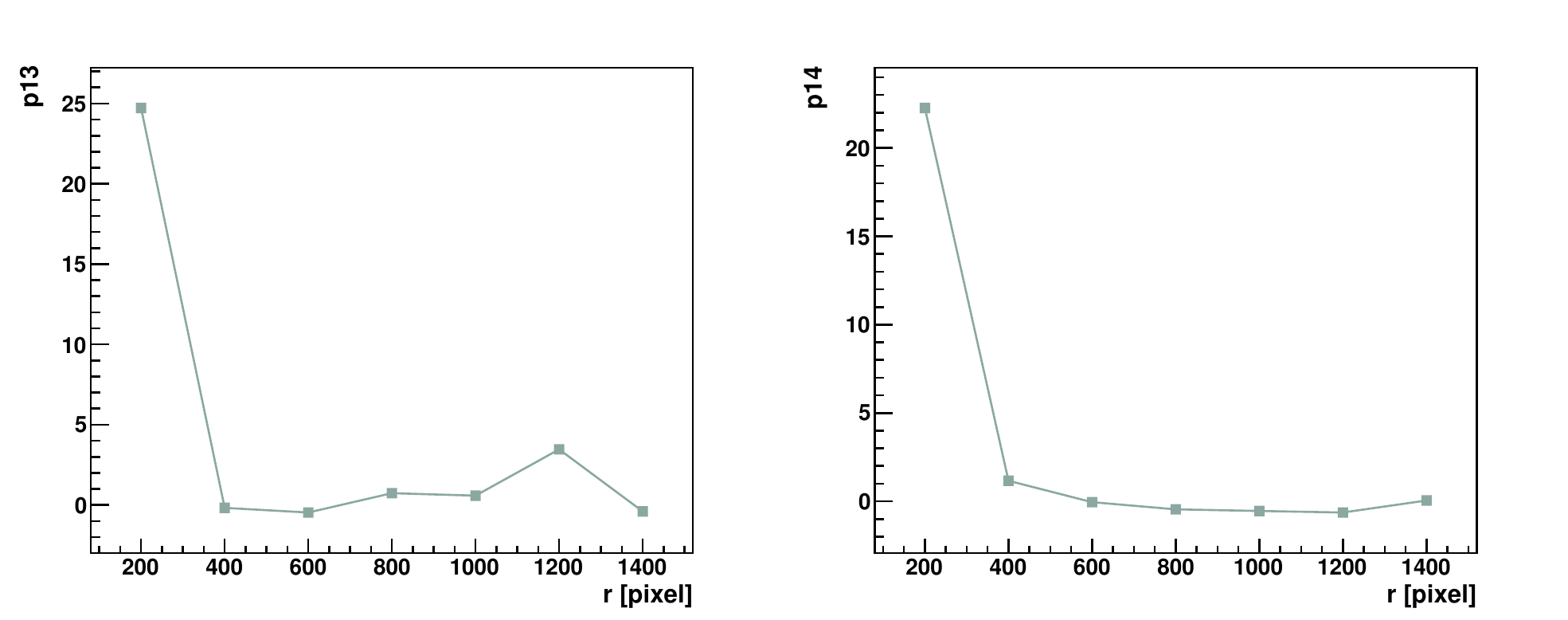}
\end{center}
\caption{Coefficients of Zernike polynomial terms 13 and 14 as a function of the distance from the frame centre r -- results of the best parameters searching and fitting procedure. No simple dependence of the parameter value on the distance from the frame centre is visible. Results for other parameters are included in appendix \ref{fig_app_first_par_val}.}
\label{fig_all_par_fit_dist}
\end{figure}

The overall fitting procedure could be performed to a single reconstructed PSF as well as to multiple profiles simultaneously, final $\chi^2$ being the sum of $\chi^2$ for single profiles. The fit to multiple profiles at once was the first attempt to find the optimal basis -- performed for all the diagonal PSFs. While the results of modelling for $6\times 6$ pixel area around the profile maximum were very satisfactory (fig. \ref{fig_all_par_fit}), no simple dependence of polynomial terms on the distance from the frame centre could be obtained (fig. \ref{fig_all_par_fit_dist}).

The attempts to obtain a simpler dependence were numerous. Mostly they focused on manually restraining the parameter space, such as removing parameters that were less important for some modelled PSFs or fitting separately PSFs at 0 to 600 pixels from the frame centre and 600 to 1400 pixels from the frame centre. Additionally, a constrain that parameters must follow a parabolic or cubic dependence on the distance from the frame centre was tested. The latter never gave simple dependence on the distance, the former resulted in unsatisfactory shapes.

However, one thing is apparent in most of the described fits -- parameters values resulting from a fit to PSF at 0 and 200 pixels from the frame centre differ significantly from values for other distances in nearly all cases. Moreover, the fitting procedure attempts to build the circularity of strongly circular PSFs -- such as these 0 and 200 pixels from the frame centre -- with non-circular terms. Those two factors led to a slight modification of the procedure -- 3 circular terms were always fitted at the beginning.

The results of parameter reduction with 3 circular parameters forced into the fit improved vastly the circularity of PSF at 0 and 200 pixels from the frame centre and described well all the profiles with overall number of 17 polynomial terms. The deviation of parameters for PSF at 0 pixels from the frame centre from results at other distances was vastly reduced in most cases, but still the parameter vs distance dependency could not be described by any simple function. The dependence of two selected fit parameters on the distance from the frame centre is presented in fig. \ref{fig_fin_par_val}.

\subsubsection{Basis formed of the least and the most deformed PSFs}

The need to include 3 circular profiles led to an idea, that the basis should include terms characteristic for the least and the most deformed PSF, i.e. 0 and 1200 pixels from the frame centre. Thus choosing of the best parameters was performed for these two profiles independently, results joined into a single, 15 parameter basis. Unfortunately, this approach also failed to describe parameters vs distance dependency in any simple way.

\subsection{Comparison of the selected basis}

\begin{figure}[b!]
\begin{center}
	\includegraphics[width=0.99\textwidth]{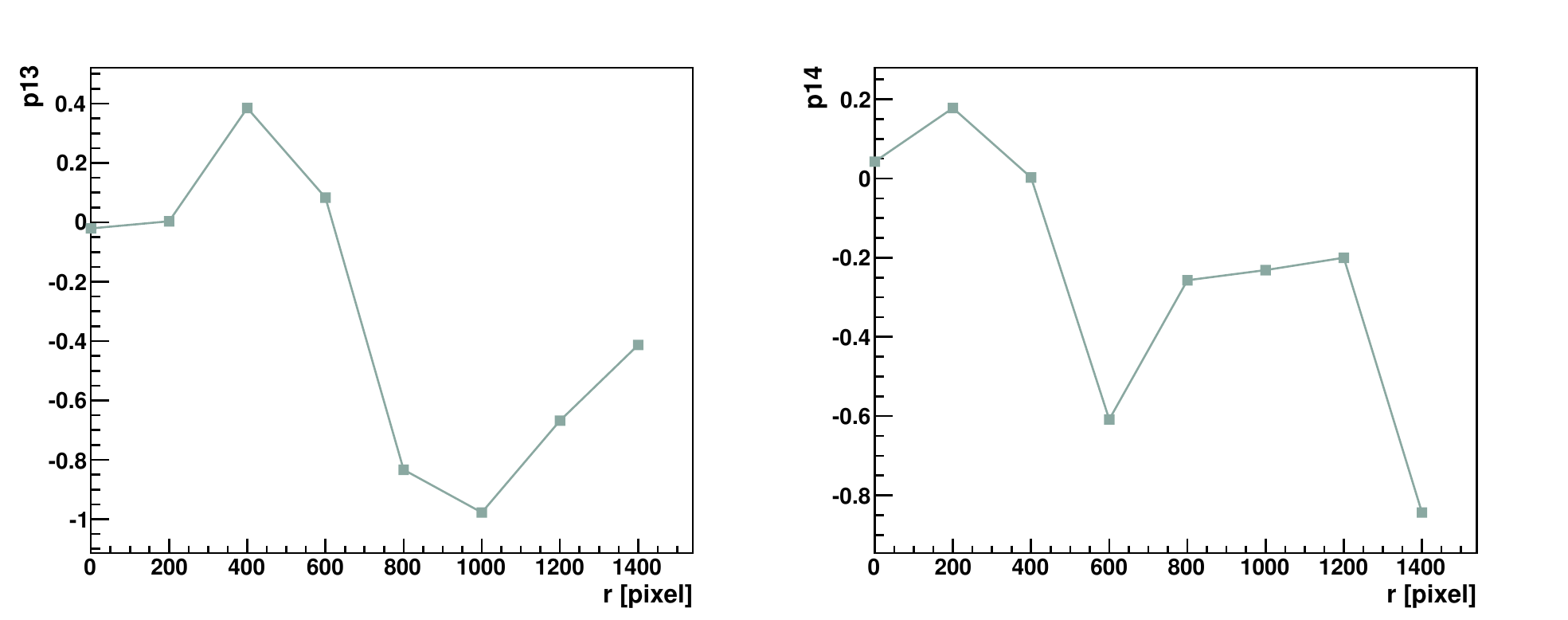}
\end{center}
\caption{Coefficients of Zernike polynomial terms 13 and 14 as a function of the distance from the frame centre r for the finally chosen basis of 17 polynomial terms -- 3 circular and 14 giving best $\chi^2$ for all measured profiles. Results for other polynomial terms are included in appendix \ref{fig_app_fin_par_val}.}
\label{fig_fin_par_val}
\end{figure}

Both basis, the first formed of 3 circular polynomials and those leading to best $\chi^2$ for all PSFs and the second formed by the best terms for the least and the most deformed PSFs, describe measured profiles comparably well and have comparably complicated parameter dependence on the distance from the frame centre. We did not succeed to implement the dependence in a simple and consistent mathematical model, due to multiple extrema not overlapping in the parameter space (see fig. \ref{fig_fin_par_val}) and a large number of parameters required to properly model measured profiles.

\begin{figure}[h]
\begin{center}
	\includegraphics[width=0.6\textwidth]{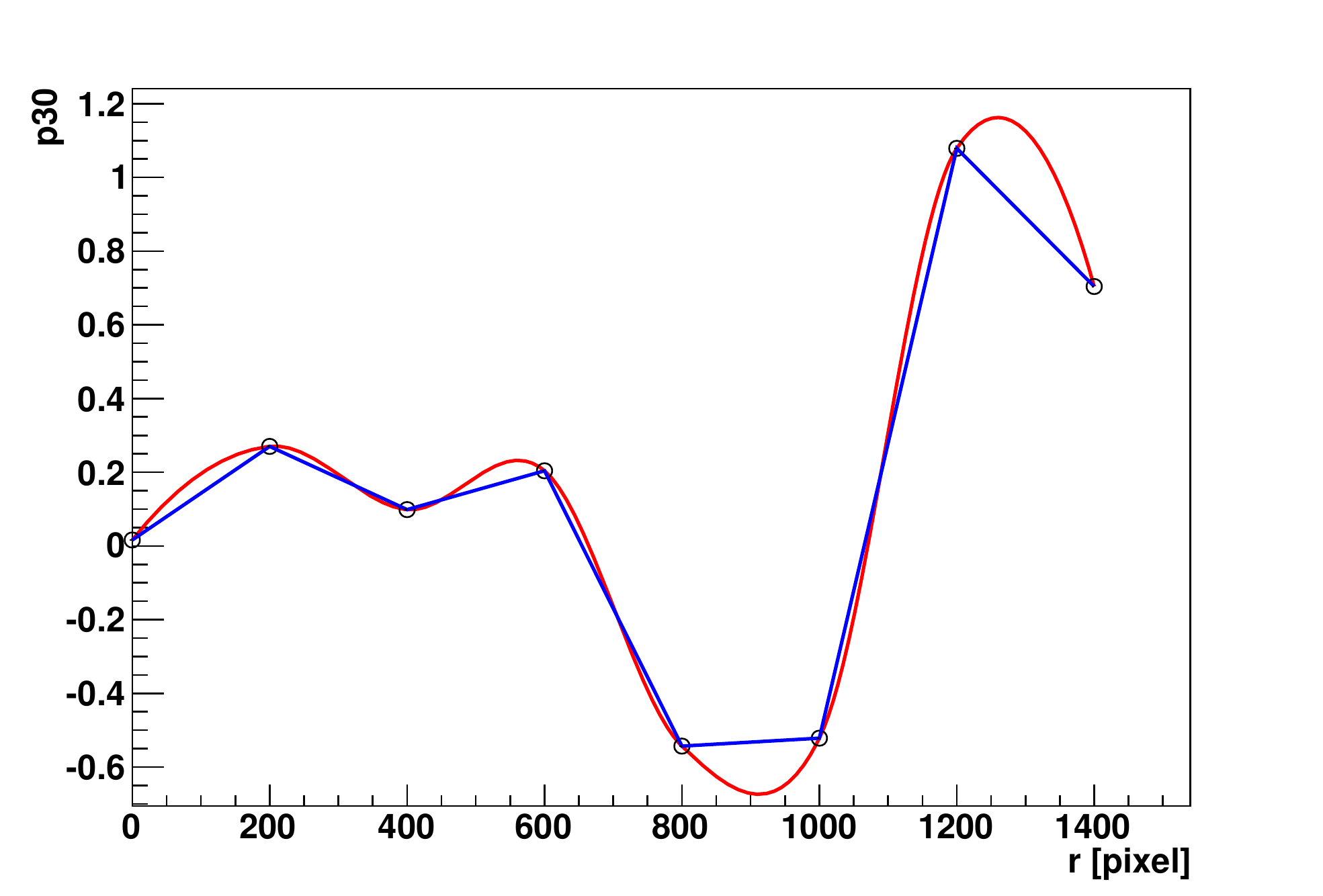}
\end{center}
\caption{Fitted weights for the Zernike polynomial term 30 as a function of the distance from the frame centre r (points). Also indicated are results of linear (blue line) and spline (red line) interpolation. Spline interpolation in this case leads to values of parameters far from these obtained from the fit, thus linear interpolation is safer.}
\label{fig_interp_comp}
\end{figure}

Therefore, for the effective approach, a linear interpolation in distance between parameters was chosen. The linear interpolation is not a perfect choice, for in some cases it leads to a sudden change in a parameter value, but does not cause a suspicious behaviour -- false extrema -- as in all attempted spline interpolations (example shown in fig. \ref{fig_interp_comp}).

\begin{figure}[b!]
\begin{center}
\subfigure[Fitted vs measured distance]{
	\includegraphics[width=0.479\textwidth]{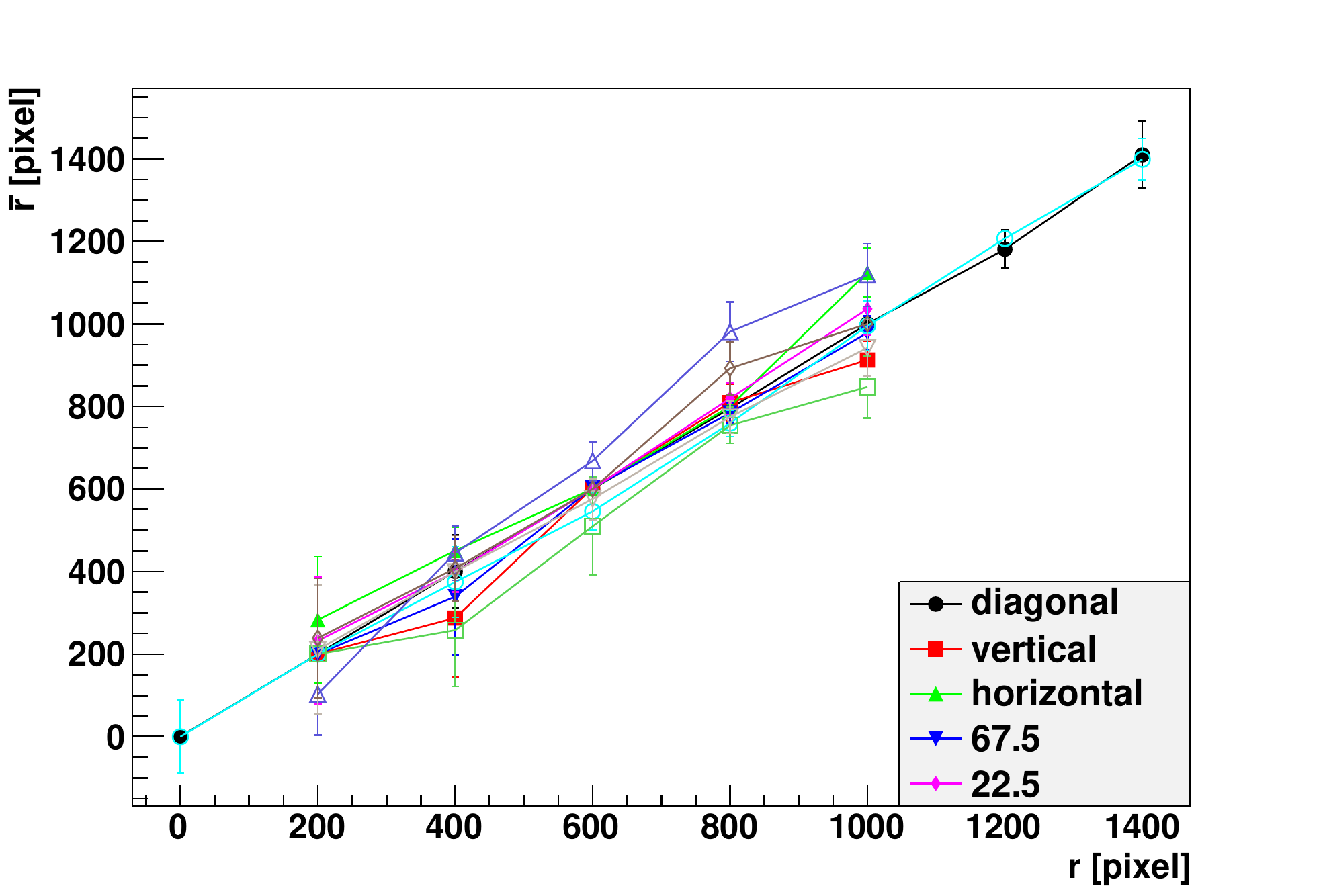}
\label{fig_distance}
}
\subfigure[Deviation of the fitted from measured distance]{
	\includegraphics[width=0.479\textwidth]{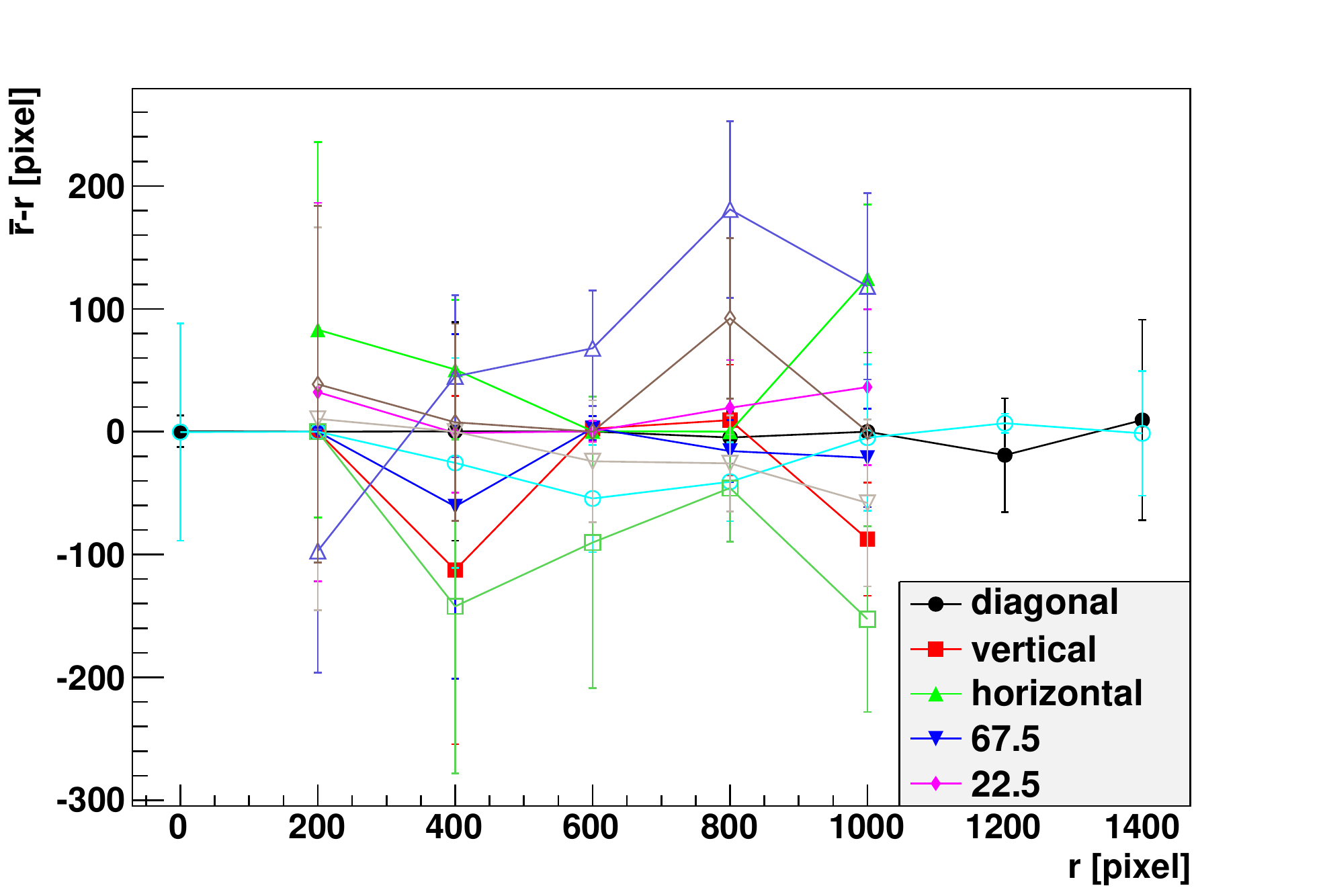}
\label{fig_distance_dev}
}
\end{center}
\caption{Consistency check of the obtained parametrizations -- distance $\bar{\mathrm r}$ resulting from the fit of model PSF to the PSF profile measured at the distance r from the frame centre. Empty markers denote least and most deformed PSFs base, filled the best base for all PSFs with 3 circular terms.}
\end{figure}

To test the selected interpolation method, a dedicated fitting procedure was performed. All the measured PSFs were fitted with one of the obtained parametrizations, but with radial coordinate on the frame, determining the shape, set free, although initialized to a proper value. For the comparison purposes, the test was performed for two parametrizations -- the first with 3 circular and other best terms for all profiles, and the second formed from the best parameters of the least and the most deformed profiles. The dependence of the PSF position determined from the fit (based on the shape only) on the actual distance from the frame centre is shown in fig. \ref{fig_distance} and the difference between the fitted and the measured position in fig. \ref{fig_distance_dev}.


The test shows quite good behaviour for diagonal profiles -- the deviation from the measured distance from the frame centre does not exceed 40 pixels for the first parametrization and 60 for the second. For other PSFs the result is worse, reaching almost 200 pixels deviation for some PSFs in the second parametrization and 100 for the first. This is probably due to the fact, that the diagonal parametrization does not describe well PSFs for different azimuthal coordinates on the frame -- PSFs are not symmetric in respect to the centre of the frame, which was not anticipated at this stage of analysis.

In general, the first parametrization gives better results in the distance from the frame centre tests and provides model PSFs more resembling reconstructed profiles. Therefore this parametrization was chosen for the further use.

\begin{figure}[p!]
\begin{center}
\subfigure[0 pixels from the frame centre]{
	\includegraphics[width=0.32\textwidth]{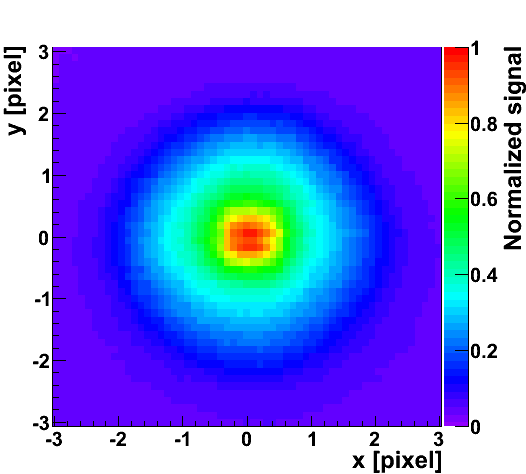}
	\includegraphics[width=0.32\textwidth]{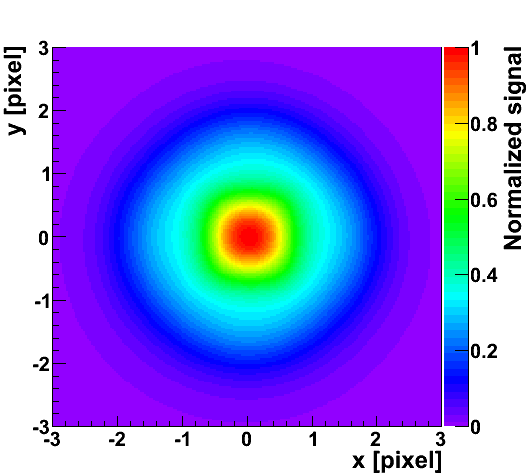}
	\includegraphics[width=0.32\textwidth]{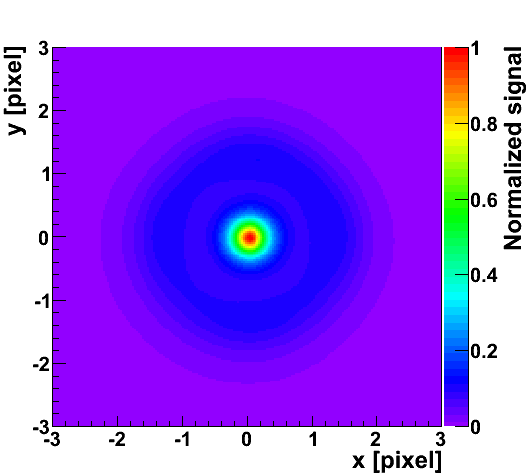}
}
\subfigure[200 pixels from the frame centre]{
	\includegraphics[width=0.32\textwidth]{polynomials/psf1.png}
	\includegraphics[width=0.32\textwidth]{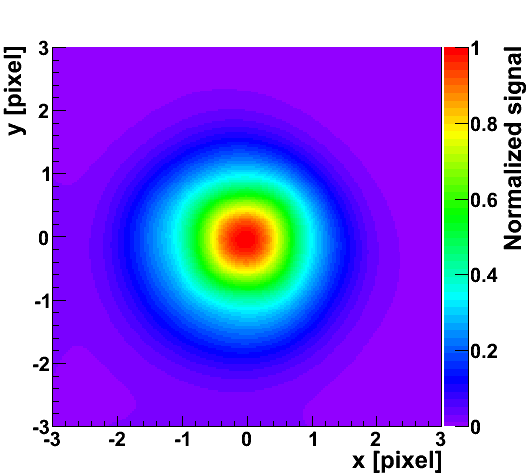}
	\includegraphics[width=0.32\textwidth]{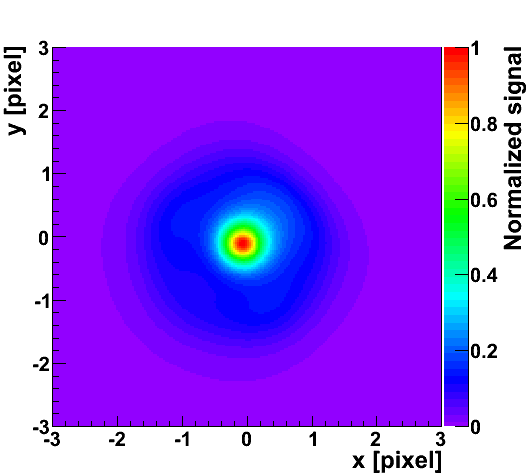}
}
\subfigure[400 pixels from the frame centre]{
	\includegraphics[width=0.32\textwidth]{polynomials/psf2.png}
	\includegraphics[width=0.32\textwidth]{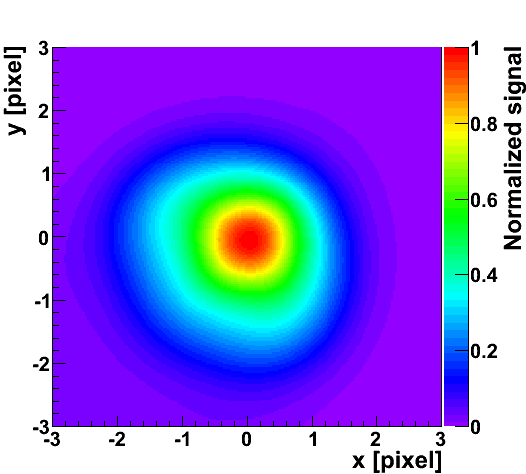}
	\includegraphics[width=0.32\textwidth]{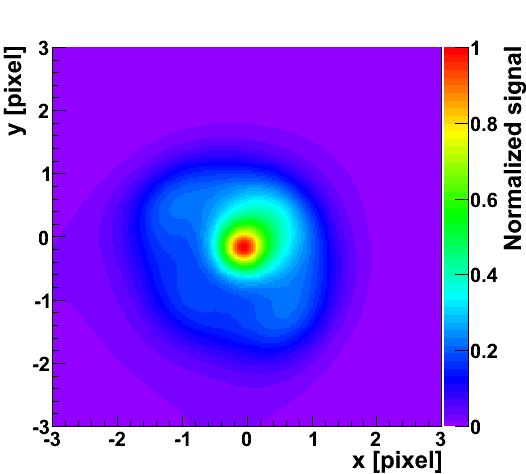}
}
\subfigure[600 pixels from the frame centre]{
	\includegraphics[width=0.32\textwidth]{polynomials/psf3.png}
	\includegraphics[width=0.32\textwidth]{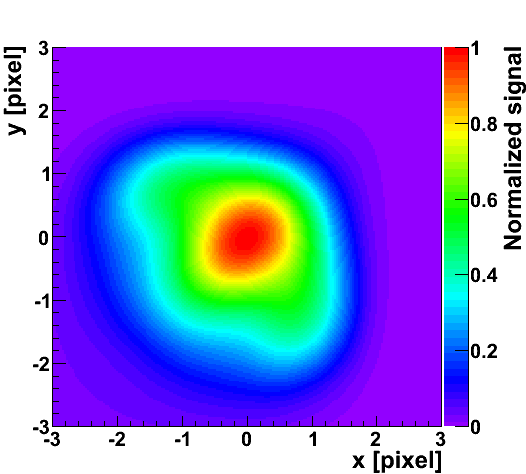}
	\includegraphics[width=0.32\textwidth]{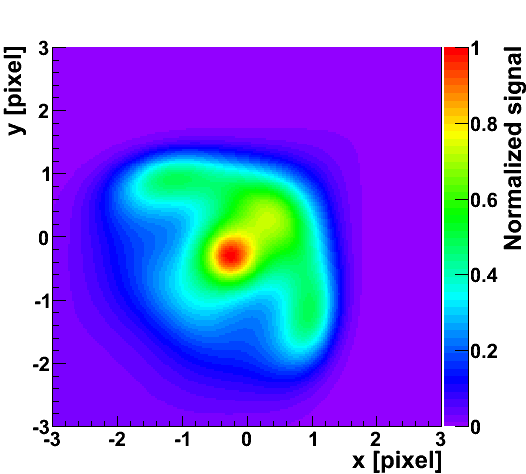}
}
\end{center}
\caption{Final results for the point spread function modelling. Shown for different distances from the frame centre are: measured profile (left), polynomial model of the PSF convoluted with the CCD pixel response (centre) and the obtained shape of the optical PSF before convolution with the CCD pixel structure (right).}
\label{fig_fin_psf}
\end{figure}

\begin{figure}[p!]
\ContinuedFloat
\begin{center}

\subfigure[800 pixels from the frame centre]{
	\includegraphics[width=0.32\textwidth]{polynomials/psf4.png}
	\includegraphics[width=0.32\textwidth]{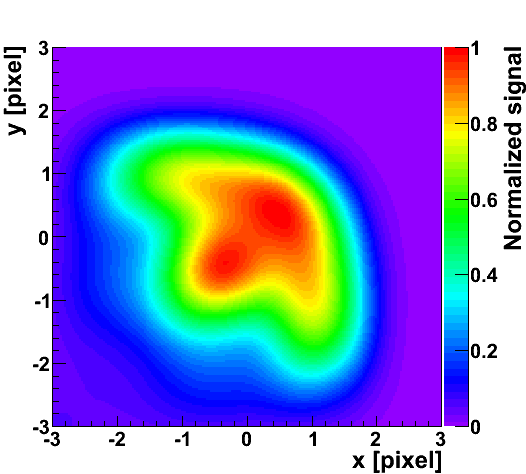}
	\includegraphics[width=0.32\textwidth]{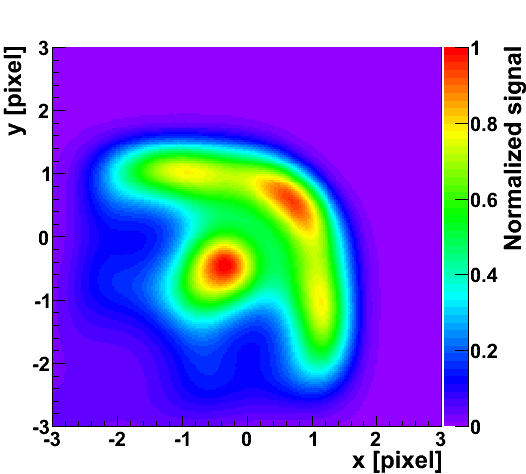}
}
\subfigure[1000 pixels from the frame centre]{
	\includegraphics[width=0.32\textwidth]{polynomials/psf5.png}
	\includegraphics[width=0.32\textwidth]{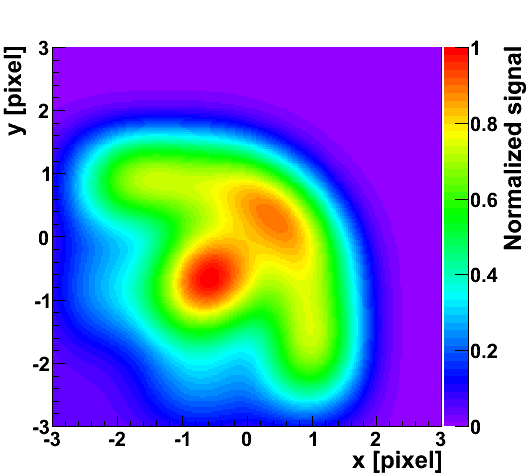}
	\includegraphics[width=0.32\textwidth]{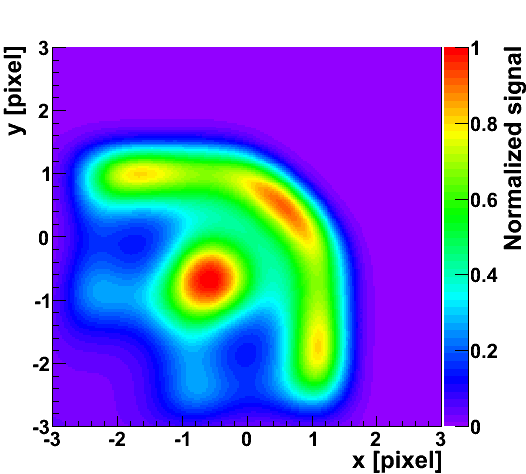}
}
\subfigure[1200 pixels from the frame centre]{
	\includegraphics[width=0.32\textwidth]{polynomials/psf6.png}
	\includegraphics[width=0.32\textwidth]{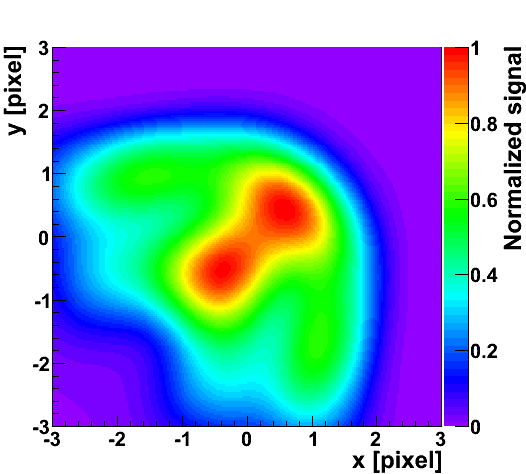}
	\includegraphics[width=0.32\textwidth]{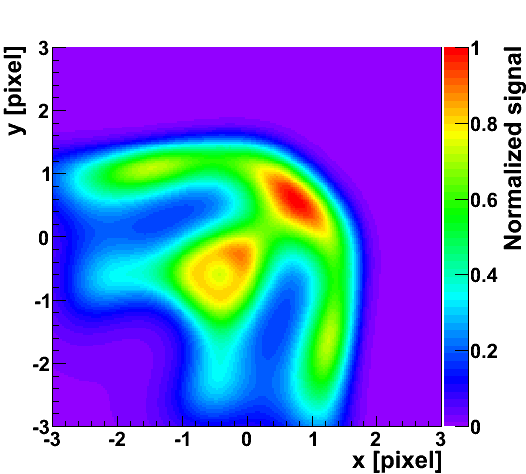}
}
\subfigure[1400 pixels from the frame centre]{
	\includegraphics[width=0.32\textwidth]{polynomials/psf7.png}
	\includegraphics[width=0.32\textwidth]{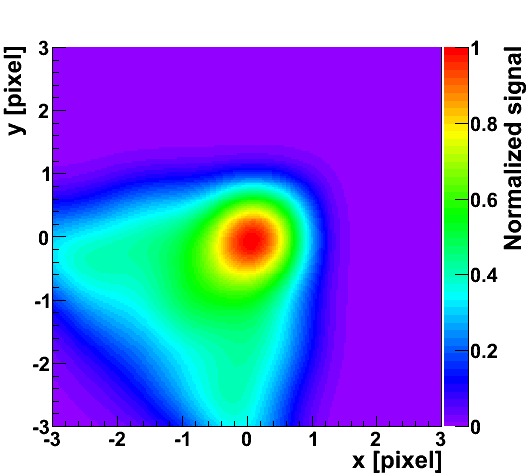}
	\includegraphics[width=0.32\textwidth]{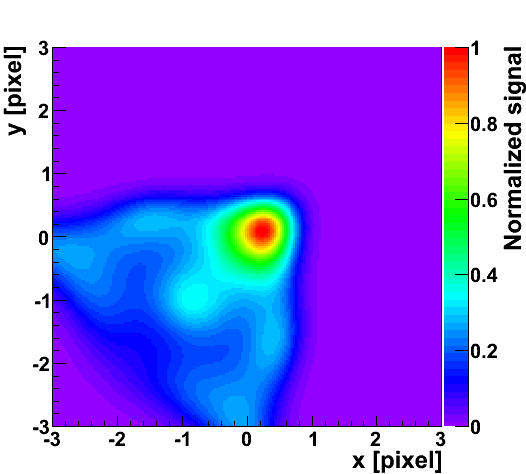}
}
\end{center}
\caption{Continued...}
\end{figure}

\begin{figure}[b!]
\begin{center}
$
\begin{array}{cc}
	\includegraphics[width=0.4\textwidth]{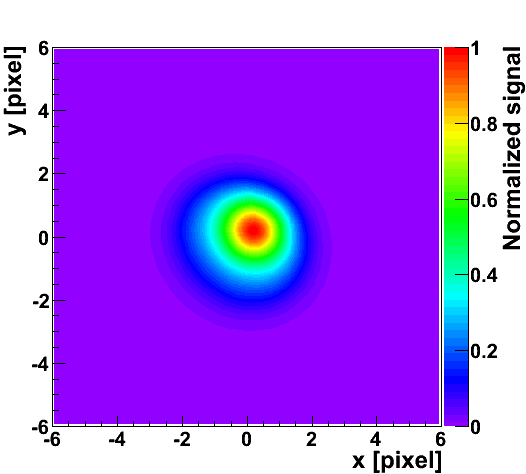} &
	\includegraphics[width=0.4\textwidth]{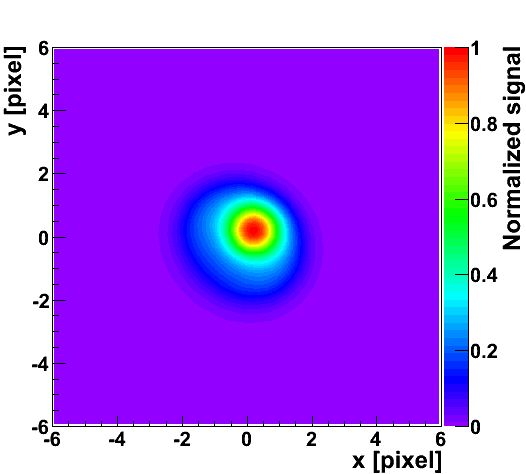} \\
	\includegraphics[width=0.4\textwidth]{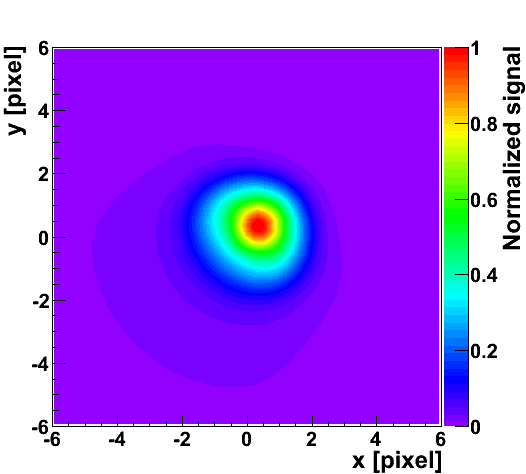} &
	\includegraphics[width=0.4\textwidth]{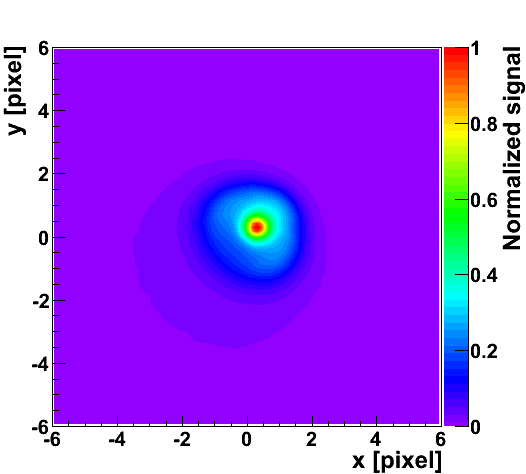} \\
\end{array}
$
\end{center}
\caption{Results of a fit to the PSF measured 400 pixels from the frame centre with different parametrizations: with 3 circular and other best terms for all profiles (top) and the best terms of the least and the most deformed profiles (bottom). Left: resulting model convoluted with $PRF$; right: optical $PSF_L$ only, not convoluted with $PRF$. Within the available data accuracy good results of the final, convoluted fit does not imply similar non convoluted PSFs.}
\label{fig_diff_par_deconv}
\end{figure}

\section{Results of the PSF model}

As shown in fig. \ref{fig_fin_psf}, chosen parametrization describes the central $6\times6$ pixels of the measured PSFs very well, especially up to 800 pixels from the frame centre. For comparison, results for the optical PSF shape, before convolution with the CCD response function, are also shown. For 1000-1400 pixels from the frame centre proportions of the core (red area) seems a little bit different then for measured profiles -- cores are slightly too thick, especially for PSF 1400. Additionally, the more complicated the shape of the PSF, the worse the reproduction of the deep details -- the fact visible most at the distance of 1200 pixels from the centre of the frame. However, we have to realize that the differences are only visible on sub-pixel scales. Therefore we conclude that the obtained model is highly satisfactory.

Deconvoluted profiles smoothly change their shape with the distance from the frame centre, showing that the polynomial base is well suited for interpolation. With such results it is very tempting to assume that the deconvoluted fit results are a good approximation of the real PSF not convoluted with the CCD pixel structure. Such an approximation would be a great base for testing different parametrizations or even the diffractive approach, for the whole time needed for calculation of the convolution could be saved. However, this is not the case -- the available accuracy allows to create different, similarly good models of the measured profile with very different deconvolution results (an example is shown in fig. \ref{fig_diff_par_deconv}).


\subsection{Monochromatic and defocus fits}

\begin{figure}[h!]
\begin{center}
\subfigure[Blue diode with nominal focus $fs=1.4$ m]{
$
\begin{array}{c}
	\includegraphics[width=0.33\textwidth]{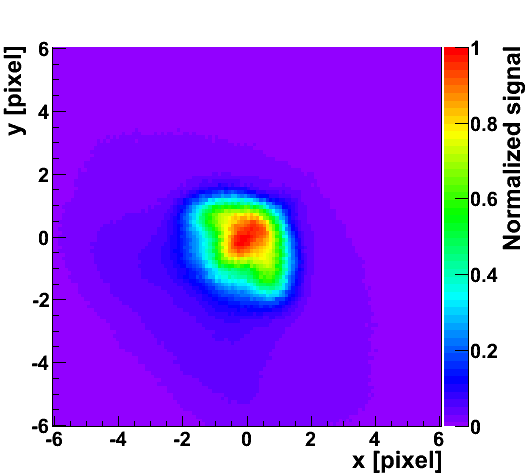}
	\includegraphics[width=0.33\textwidth]{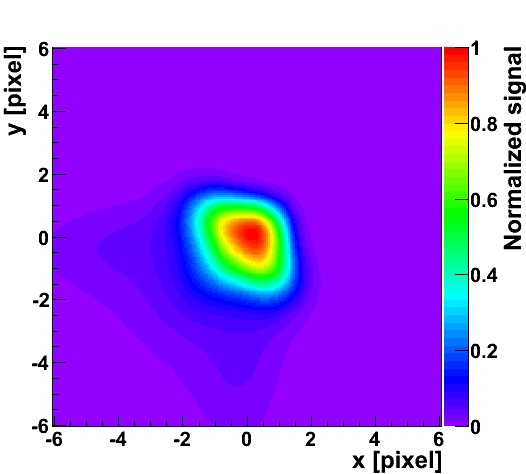}
\end{array}
$
}
\subfigure[Red diode with nominal focus $fs=1.4$ m]{
$
\begin{array}{c}
	\includegraphics[width=0.33\textwidth]{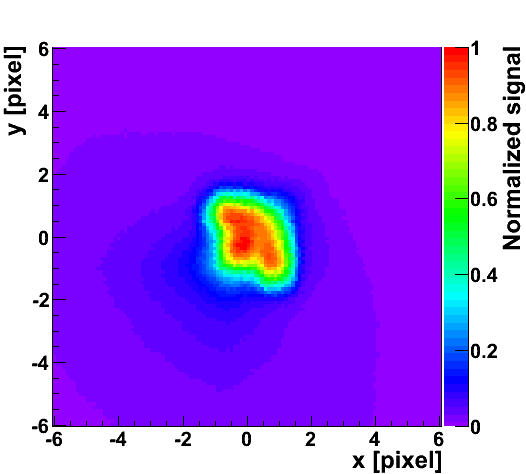}
	\includegraphics[width=0.33\textwidth]{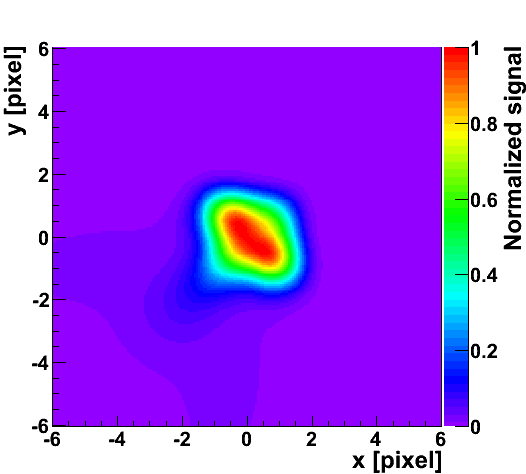}
\end{array}
$
}
\subfigure[White diode with focus changed to $fs=1.2$ m]{
$
\begin{array}{c}
	\includegraphics[width=0.33\textwidth]{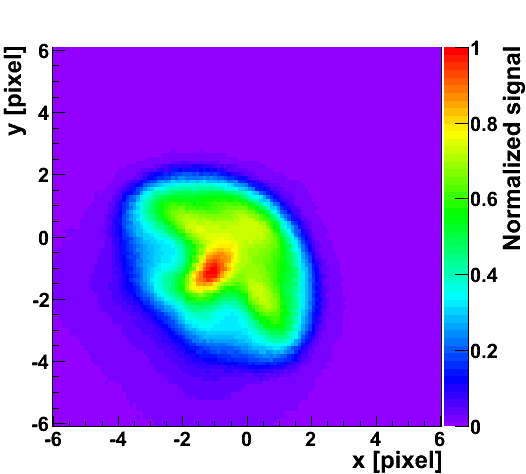}
	\includegraphics[width=0.33\textwidth]{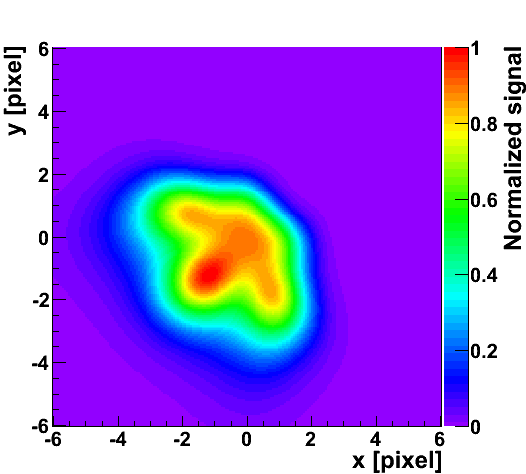}
\end{array}
$
}
\subfigure[White diode with focus changed to $fs=1.6$ m]{
$
\begin{array}{c}
	\includegraphics[width=0.33\textwidth]{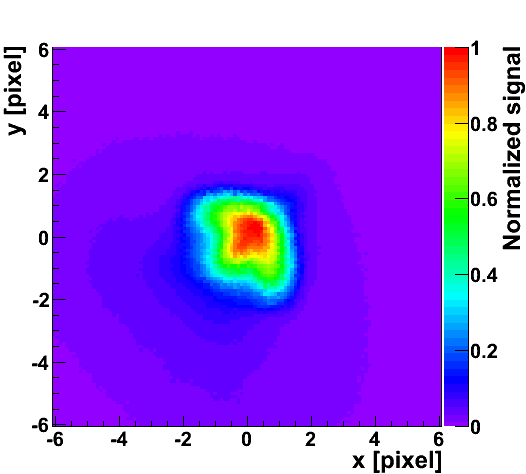}
	\includegraphics[width=0.33\textwidth]{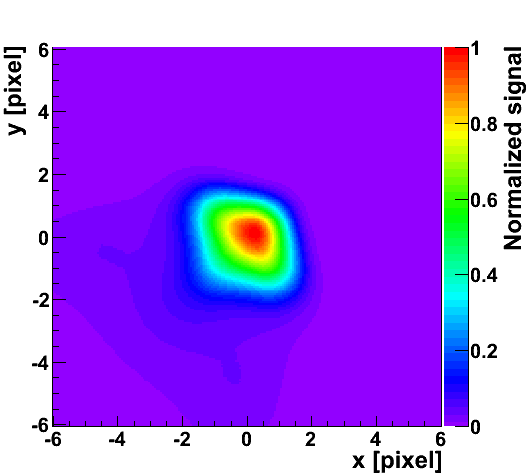}
\end{array}
$
}
\end{center}
\caption{Measured PSFs for colour diodes and for the white diode with changed focusing (left column) compared with their models obtained with the developed parametrization (right column), for distance of 800 pixels from the frame centre (results for other distances are shown in appendix \ref{fig_psf_col_defoc_cont}).}
\label{fig_psf_col_defoc}
\end{figure}

Parametrization used to describe point spread function measured for a white diode can also be used to model monochromatic or defocused profiles. It seems however, that the results of such modelling are not as good as for the white PSFs, as shown in fig. \ref{fig_psf_col_defoc} and \ref{fig_psf_col_defoc_cont}. The size and shape of the profiles (green area) is roughly reproduced, but the details of the core (red part) are similar only for the fit of the white diode with $fs=1.2$ m, although the core in this case seems to be most complicated. That is probably due to the very short tails (blue part) of this PSF. As for other profiles, the fitting algorithm focuses both on the core and tails, failing to reproduce any of them in sufficient detail. 

Worse modelling results despite the shape similarities are another indication that the polynomial approach does not follow real 
physical attributes of the PSF. Thus there were nearly no chances that the focusing or wavelength variations would be followed in any way in the parametrization. Even though, some attempts to check, if this is really the case have been made.

\begin{figure}[h!]
\begin{center}
\subfigure[Measured PSF]{
	\includegraphics[width=0.31\textwidth]{polynomials/red_800.png}
}
\subfigure[Refit of the chosen basis]{
	\includegraphics[width=0.31\textwidth]{polynomials/red_800_ap.png}
}
\subfigure[Fit improve attempt]{
	\includegraphics[width=0.31\textwidth]{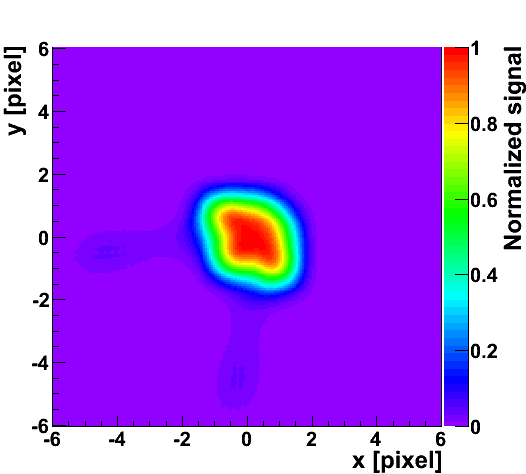}
}
\end{center}
\caption{Comparison between measured PSF for the red diode, 800 pixels from the frame centre (a) and its models resulting from two different fitting approaches (b and c). Model (b) is a fit of the standard polynomial basis, as obtained for the white diode, while model (c) was the standard basis with parameter values taken from the white diode fit, but with refitted $p$ and $\lambda$ parameters of eq. \ref{eq_zer_repar} and \ref{eq_psf_lenses}.}
\label{fig_sample_red_refit}
\end{figure}

In some cases a general shape of the profile for colour or defocused PSF is better fitted with the parameter values obtained for the white, focused profile, but with free $p$ and $\lambda$ parameters in eq. \ref{eq_zer_repar} and \ref{eq_psf_lenses}, as shown in fig. \ref{fig_sample_red_refit}. However, this is not the case for most profiles.

Additionally, a simultaneous fit of 3 focus settings for the same distance from the frame centre has been performed for 800, 1000, 1200 and 1400 pixels. No simple dependence of parameter values on focusing and distance from the frame centre has been observed, even though the analysis was performed for just 3 focusing settings. This study shows, that no simple modelling of chromatic and focusing changes can be obtained with the polynomial approach.

\subsection{Photometry of the ``diode'' frames}

The final test of the obtained PSF model for the white diode was a check of its suitability for photometric purposes. Obtained model was fitted to the single diode images taken for the PSF reconstruction purposes, for all available distances from the frame centre along the frame diagonal. Two fit types were performed -- based on $\chi^2$ and loglikelihood minimisation. Additionally, an ASAS photometry and a simple aperture photometry resembling the fast photometric algorithm used in the \pin were also tested on the same data set. 

\begin{figure}[h!]
\begin{center}
	\includegraphics[width=0.7\textwidth]{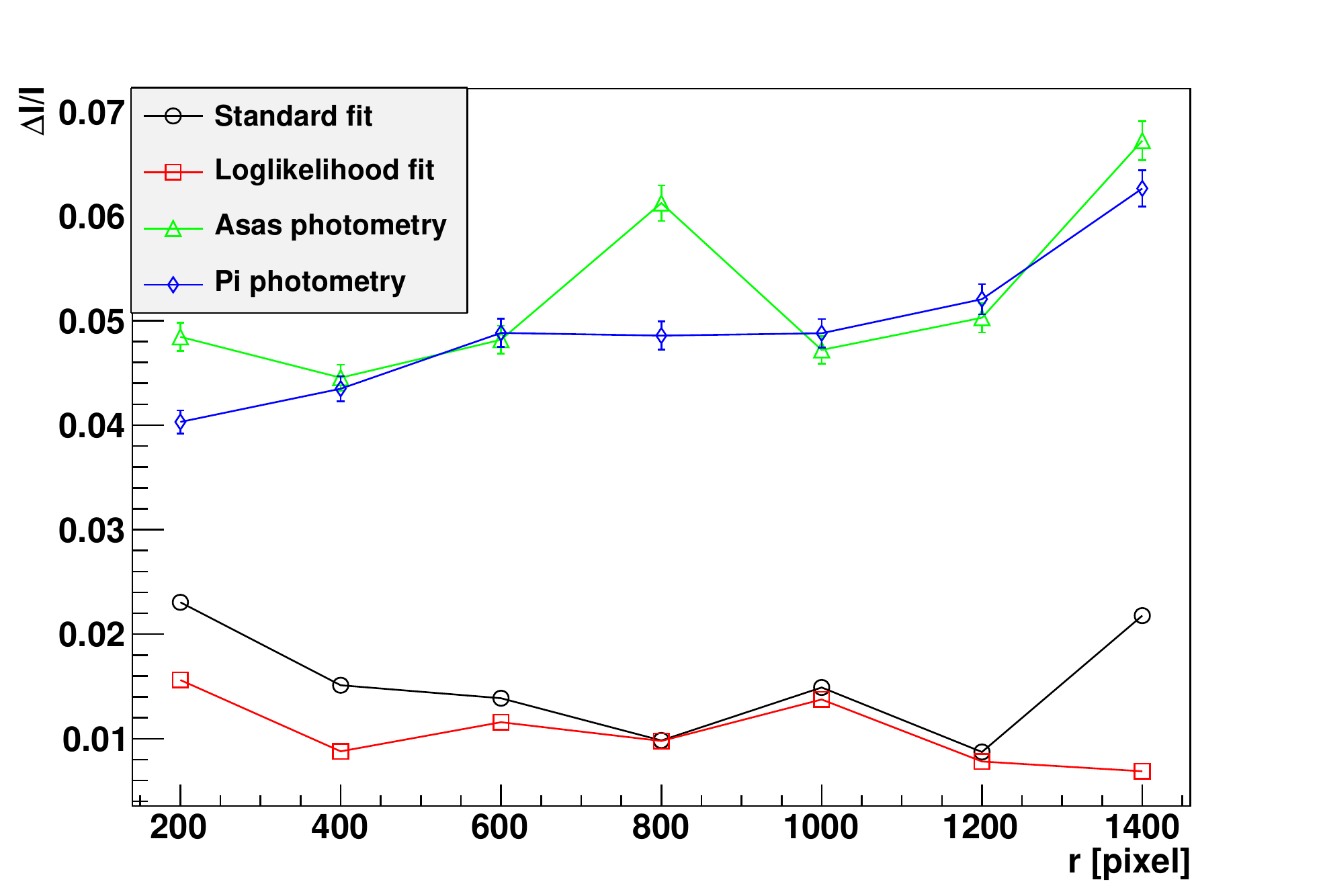}
\end{center}
\caption{Photometry of the diagonal white diode images (used for diode PSF reconstruction). Black and red points stand for the fit of the obtained PSF model (respectively $\chi^2$ and loglikelihood fits), green for the ASAS photometry and blue for the single aperture photometry based on the fast \pin photometric algorithm. Error bars for the fitted model are smaller than markers.}
\label{fig_diode_photometry_compare}
\end{figure}

Results presented in fig. \ref{fig_diode_photometry_compare} show that the obtained model of the PSF can serve the photometric purposes very well. There is no clear dependence of the brightness measurement spread on the distance from the frame centre for the model fits as well as ASAS photometry. Some dependence is visible for the ``Pi photometry'', probably due to the fact, that this method uses a single aperture size, while the diode PSF size changes significantly with its position on the frame (see fig. \ref{fig_fwhm}).

\chapter{Modelling real sky data}
\label{chap_modelling_real_sky_data}

Laboratory measurements of the point spread function were a necessary step for obtaining a model properly reproducing images of the stars seen by \pin cameras. The model described in the previous chapter successfully reproduces the PSF of an artificial point source and gives very promising photometry results. However, it has to be compared to real sky data to find out if it achieves its goal.

A quick glance at data from real sky images taken with the prototype \pin detector in Chile\footnote{Camera used in laboratory experiments and those taking real sky images in Chile were identical in technical specification and were equipped with the same lenses.} shows that the star images are not identical in shape to the laboratory measurements, although have similar general properties. This difference can result from different focusing of lenses used in laboratory and sky measurements. Moreover, star images show that the real PSF does not change with distance from the frame centre only, but also strongly depends on the azimuthal coordinate. This has to be due to mechanical differences appearing in the assemble procedure -- the dependence on azimuthal coordinate is most likely a result of a non-negligible tilt of the optical axis with respect to the the CCD (sensor board) plane.

Different focusing and slight, at a first glance, but non-negligible after deeper study deviations in cameras assembly enforce a recalculation of model parameters for each set of equipment. However, this is an iterative procedure and requires use of the model obtained in the laboratory measurements at the first step.

\section{PSF model recalculation for real star images} 

\subsection{Recalculation procedure}

\begin{figure}[t!]
\begin{center}
	\includegraphics[width=0.49\textwidth]{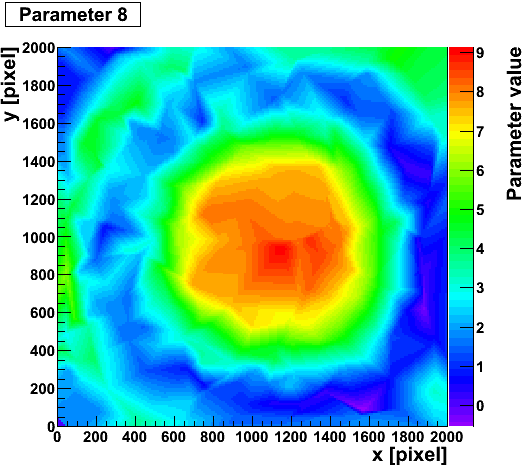}
	\includegraphics[width=0.49\textwidth]{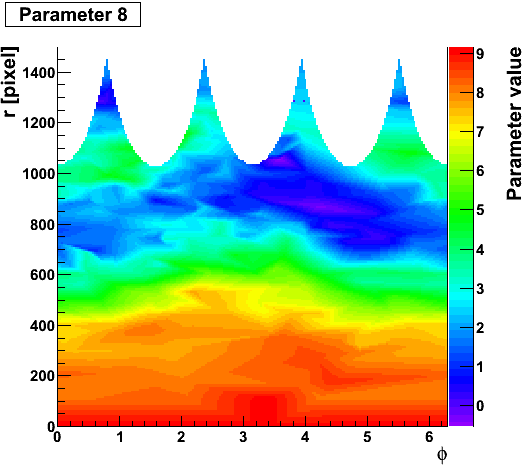}
	\includegraphics[width=0.49\textwidth]{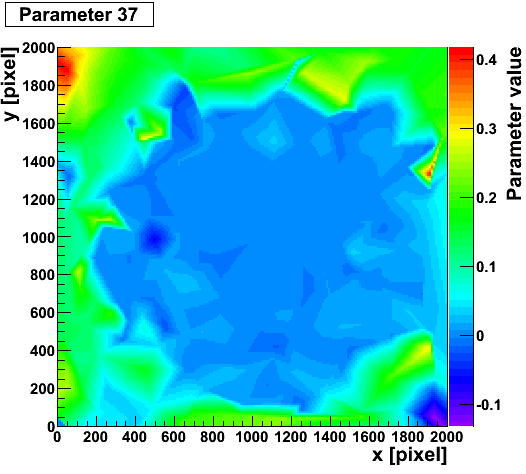}
	\includegraphics[width=0.49\textwidth]{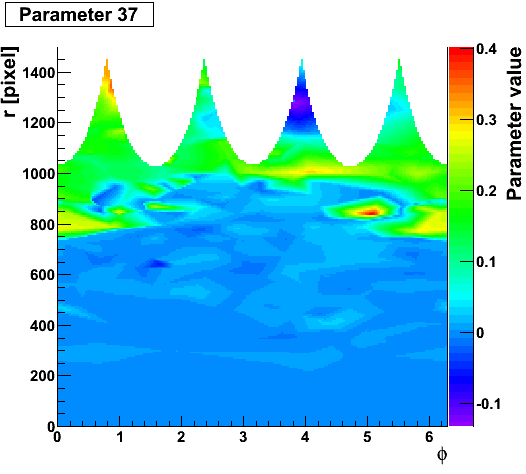}
\end{center}
\caption{Dependence of the selected model parameters on the star position on the frame. On the left: in Cartesian coordinates, on the right: after transformation to polar coordinates. The azimuthal angle $\phi$ is calculated from (0,1) versor, counterclockwise.}
\label{sample_parameters}
\end{figure}

Recalculation of the model parameters requires obtaining high resolution profiles of real stars. This task proved difficult if not impossible for highly deformed PSFs without knowledge of their general shape (as discussed in sec. \ref{ssec_phot_ast_imp}). However, intermediate model obtained in laboratory measurements gives a general shape of stars for different coordinates on the frame and can be used for obtaining the high resolution profile.

A set of 285 bright stars up to $9^\m$, without bright neighbours, scattered all over the frame was chosen. For every star scale (signal), position and background level were fitted with the intermediate model on each of the consecutive 172 frames of a single field. Using obtained positions, a high resolution profile of each star was prepared, similarly as in the chap. \ref{psf_reconstruction}. The results of this reconstruction were successful and, after subtracting neighbouring stars\footnote{For the reconstruction purposes, stars without bright neighbours were chosen. However, dimmer neighbours, often not visible during the procedure of star selection, but visible on the high-resolution profile made from 172 frames remained.} were more than sufficient for model parameters refit.

Similarly as in the intermediate model, PSF has to be described for every position on the CCD, while fits were performed only for a finite set of stars with specific positions on the sensor. Thus each parameter value has been plotted in Cartesian coordinates and a Delaunay triangulation has been used for interpolation between measured points. Results for most of the parameters show general rotational symmetry and it seemed that the polar coordinate system would be the most natural for the interpolation purposes, described in more detail in sec. \ref{ssec_model_quality}.

However, for most parameters the symmetry axis of the plot was shifted from the centre of the CCD, see fig. \ref{sample_parameters}. To estimate the actual position of the optical axis on the CCD sensor we assumed that the radial dependence of the parameter should be described as a $7^{\rm th}$ order polynomial in distance from the axis and looked for axis position which resulted in the best description of given parameter (neglecting azimuthal dependence). The new centre of the frame was obtained from weighted average of centres for all parameters and was used for transformation into polar coordinates.

The model recalculated in described way and interpolated after transformation to polar coordinates could already be used as a general representation of PSF for the discussed data. However, to ensure high quality of the parametrisation an iterative approach should be used -- the procedure of profiles reconstruction, refitting of the model parameters and interpolation should be repeated several times. In our case the final model was obtained after 6 iterations. However, the number of iterations was induced more by the fact, that the procedure was being improved and polished than by the better quality obtained at each next run. A well known practice for such PSF iterative modelling (although for cases with much simpler PSF) is to perform 3 repetitions of the procedure.

\begin{figure}[b!]
\begin{center}
	\includegraphics[width=0.49\textwidth]{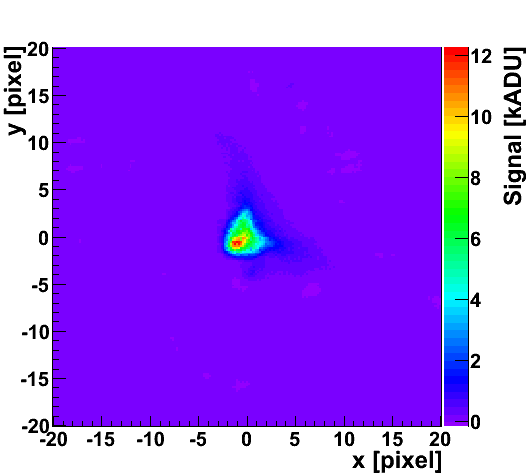}
	\includegraphics[width=0.49\textwidth]{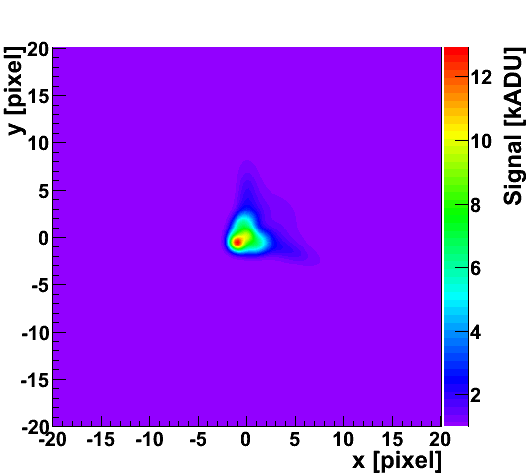}
	\includegraphics[width=0.49\textwidth]{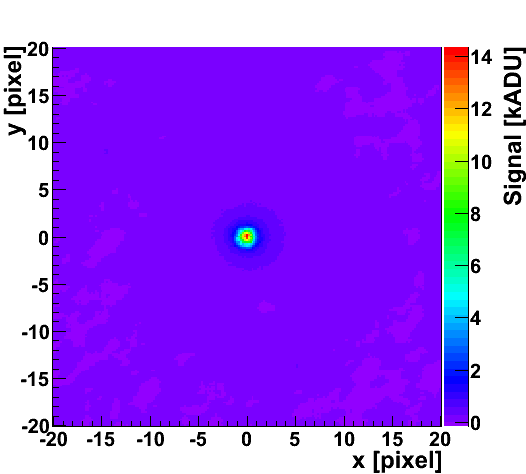}
	\includegraphics[width=0.49\textwidth]{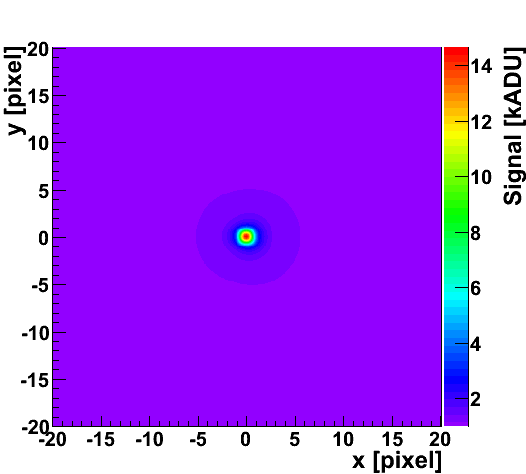}
\end{center}
\caption{Two sample reconstructed star profiles (on the left), used to obtain the final PSF model (on the right), for about 1300 (top) and 100 (bottom) pixels from the frame centre respectively. The model reproduces profiles in high detail, although far tails and asymmetric parts are lost.}
\label{sample_PSF}
\end{figure}

\subsection{Quality of the model}
\label{ssec_model_quality}

The final model recalculated for the real sky data seems to reproduce collected images quite well, for slightly asymmetric profiles close to the frame centre and for very deformed ones close to the frame edge, as shown in fig. \ref{sample_PSF}. However, as in the laboratory measurements, far ``tails'' of deformed PSFs are not properly described. Moreover, one can see that the reconstructed profile is slightly asymmetric -- an effect that was not visible during laboratory measurements and is probably induced by the slight tilt of CCD surface relative to the lenses axes.

\begin{figure}[b!]
\begin{center}
	\includegraphics[width=0.6\textwidth]{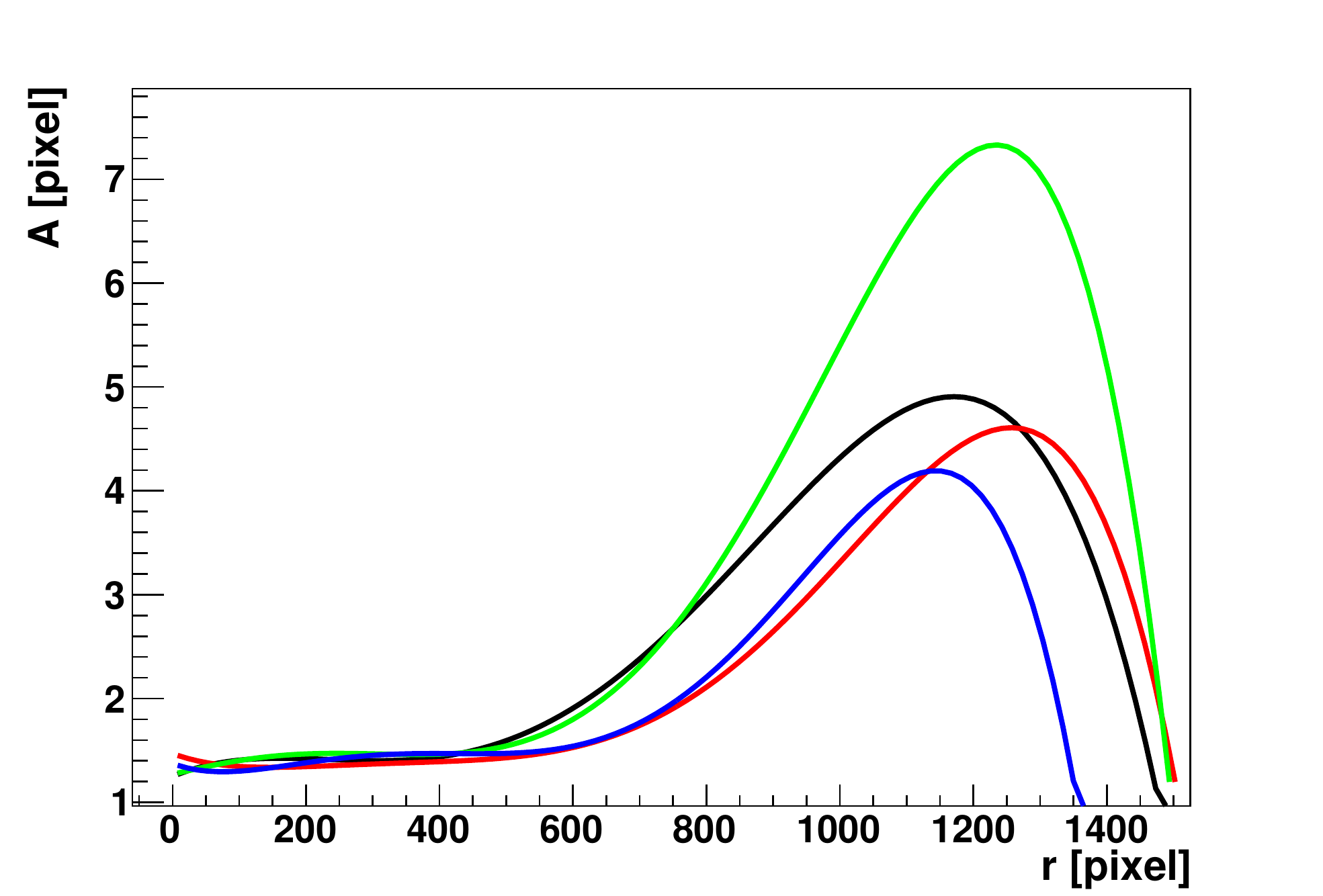}
\end{center}
\caption{Sensor area A covered by the signal larger than $50\%$ of the maximal signal for the final model of PSF, applied along the frame diagonals (in distance from the frame centre r). Four curves represent fits to the results obtained at different quarters of the CCD.}
\label{model_fwhm}
\end{figure}

As mentioned before, the general shape of the PSF and its development on the CCD is similar for the artificial point source (laboratory measurements) and the real sky data. The similarity of the PSF shape evolution is visible on the plots showing area above $50\%$ of the maximal signal for the final model (fig. \ref{model_fwhm}) and the laboratory model (fig. \ref{fig_fwhm}). The plots cannot be directly compared, for in the former case a mathematical function is integrated, while in the latter real measurements are analysed. Still, the behaviour is the same: signal is contained in small number of pixels close to the frame centre and near the frame corner and spread over area even few times larger for intermediate positions. The maximum position here, however, is oscillating around 1200 pixels from the frame centre, while it was about 1000 pixels for the laboratory measurements. The difference between maxima for different quarters of the CCD is probably due to the sensor tilt described previously.

In general, the number of considered star measurements (number of frames) was sufficient for profile reconstruction and model fit. However, the number of bright stars in the very corners of the frame seem to be somewhat scarce compared to the rapid changes of the PSF shape in this area. This obstacle cannot be easily solved, for there are simply very few stars in the corner of the frame compared to the other parts of the frame due to the sensor geometry. Stars from more fields should be gathered but this requires extra time, needed for preparing each field, and introduces some additional uncertainties. In general, the lower density of stars per area in corners of the frame, the lower quality of the model for these coordinates.

Additional systematic uncertainties in the model are due to the selected method of interpolation. Choosing polar coordinates for interpolation seems to be natural for model describing PSF whose development is approximately radially symmetric, although not all parameters show equal radial dependence (fig. \ref{sample_parameters}). Still, a problem of the interpolation on non-equidistant grid remains. The Delaunay triangles interpolation\cite{delaunay_tri} gives quite good general description of the parameter values, but is not smooth neither in Cartesian nor in polar coordinates and suffers from rapid transitions between different areas. Rapid transitions could not be avoided also for popular Shephard's interpolation\cite{shepard_int}. The 2D interpolation in our case seems to be a separate topic requiring sophisticated studies. Fortunately results of photometry based on models using interpolation in Cartesian and polar coordinates, for Delaunay triangles and Shephard's method were very similar, so we conclude that the corresponding uncertainties are negligible.


\section{Photometry and astrometry results}

\subsection{Preparation}

The final model of the point spread function (introducing 2D Delaunay triangles interpolation in polar coordinates) has been used as a basis for simple photometry and astrometry. Each star on a frame has been fitted with a PSF specific for its coordinates, only scale (signal level), background and position on the pixel were treated as free parameters, not given by the model. Position on the pixel in x and y coordinates were the only numerically fitted parameters, while the scale was being calculated analytically, according to the formula:

\begin{equation}
S = \frac{\displaystyle\sum_{x,y}f(x,y)\frac{s(x,y)+b}{\sigma(x,y)^2}-b*\displaystyle\sum_{x,y}\frac{f(x,y)}{\sigma(x,y)^2}}{\displaystyle\sum_{x,y}\frac{f(x,y)^2}{\sigma(x,y)^2}}
\label{scale_fit_formula}
\end{equation} 

where $x,y$ stand for pixels coordinates on the image, $f(x,y)$ denote the PSF model function value for centres of image pixels (for a given PSF position), $b$ is the background and $\sigma(x,y)$ stands for the uncertainty of the signal in pixel $(x,y)$. We assume here that $\sigma \simeq \sqrt{s(x,y)+b}$, where $s(x,y)$ is the signal in pixel $(x,y)$. The eq. \ref{scale_fit_formula} is obtained from an analytic minimization of a $\chi^2$ equation for the model function to real signal and background at given PSF position. Although background can be fitted analytically as well, results turned up to be most stable if we set background as to a trimmed median\footnote{The trimmed median in this case was median of values of all pixels, after removing pixel values outside mean $\pm 3\cdot\mathrm{RMS}$.} of all pixels within the area $40\times 40$ pixels around the star's centre.

The position on the pixel in the fitting procedure was being initialized with a position estimated from the transformation of celestial coordinates of the star to frame coordinates, using general astrometry of the frame. However, this transformation procedure does not take into account different shapes of the PSF and thus different positions of a geometrical centre of the star's image on the frame\footnote{The term ``geometrical centre'' of the star is probably a little bit misused in this case. In truth, it is the centre of the modified radial Zernike polynomials set used for obtaining the model PSF. This point was the best centre of the star's shape according to the numerical procedure fitting the model parameters, but its physical meaning had not been determined and is rather not relevant. In the general case it is different from the ``centre of mass'' position.}. Thus a correction had to be determined -- the systematic shift between position from transformation and fitted position on the pixel all over the sensor.

\begin{figure}[t!]
\begin{center}
	\includegraphics[width=0.6\textwidth]{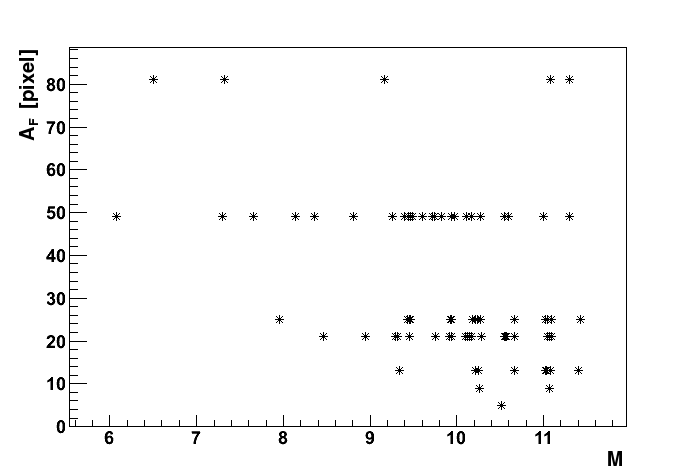}
\end{center}
\caption{Fit area $\mathrm{A_F}$ giving the smallest brightness measurement spread vs. magnitudo M for a test set of stars.}
\label{fitting_radius}
\end{figure}

The other important step for photometry and astrometry preparation was a determination of an optimal area around the star centre for which the fitting procedure should be performed. Photometry have been made on 172 frames for 78 chosen stars and 8 different fitting areas: 5 pixels (a cross), 9 pixels (a $3\times 3$ square), 13 pixels (as for 9, but with 4 additional), 21 pixels (a $5\times 5$ square without corners), 25 pixels (a $5\times 5$ square), 49 pixels (a $7\times 7$ square), 81 pixels (a $9\times 9$ square) and 121 pixels (a $11\times 11$ square). For each star the area with the smallest spread of the fitted signal as calculated from the photometry results on the 172 considered frames was chosen. Results of the best fit area vs. magnitudo are visible on fig. \ref{fitting_radius}. No straightforward dependency of the fit area on the star's brightness is visible on this plot, but it forms an input for an educated guess: for stars brighter than $10^\m$ the 49 pixel area and for dimmer the 25 pixel area have been chosen as resulting in best results for most stars.

It has to be emphasized at this point that the discussed photometry and astrometry procedure is a very simple one, performed rather to test the PSF model than to compete with professional profile algorithms. The main difference is the fact that all the stars were fitted in a one-stage, straightforward manner, without, for example, subtracting neighbours of each fitted star -- a common procedure in many profile algorithms.


\subsection{Photometry results}
\label{sec_phot_res}

\begin{figure}[t!]
\begin{center}
	\includegraphics[width=0.47\textwidth]{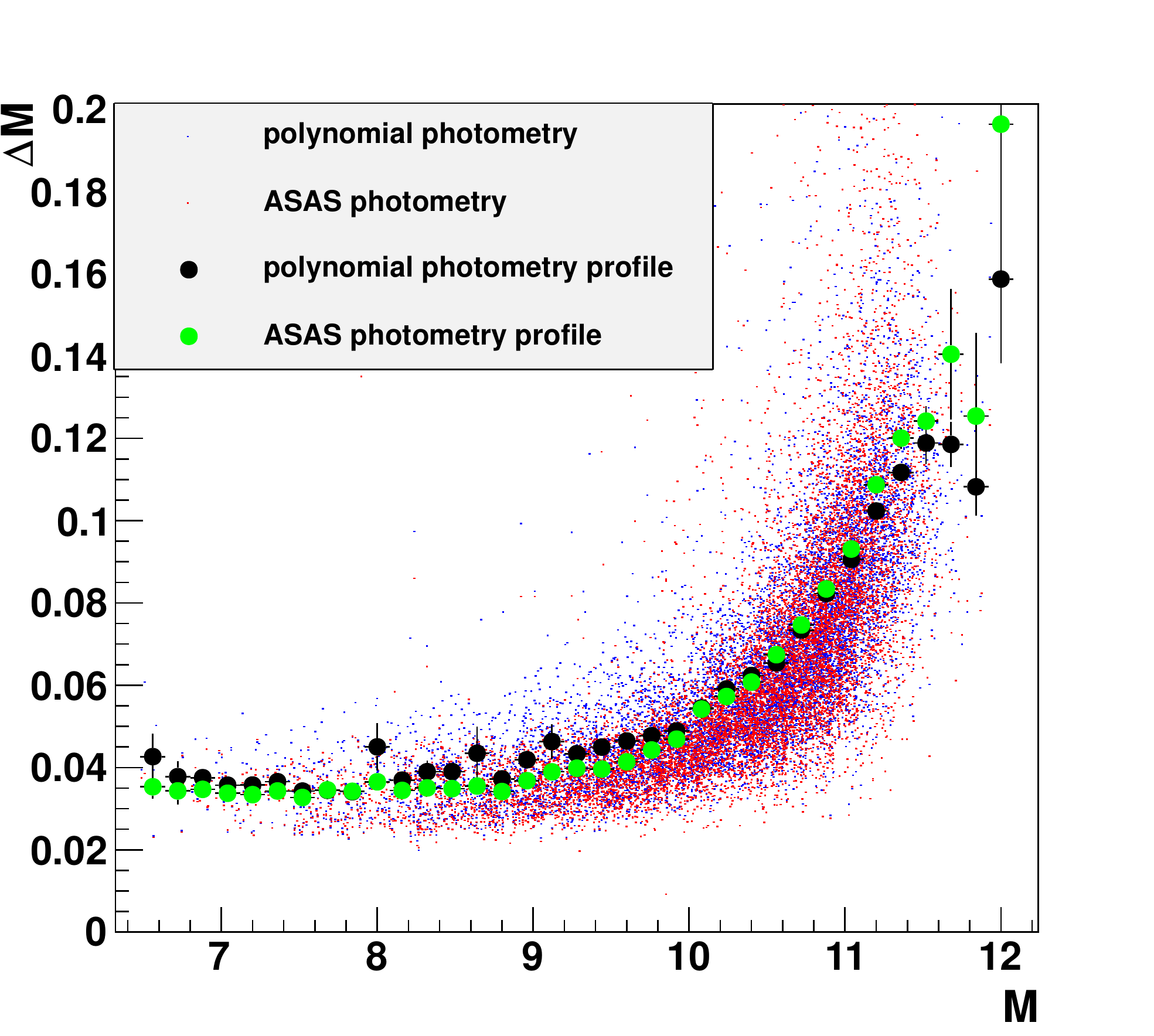}
	\includegraphics[width=0.47\textwidth]{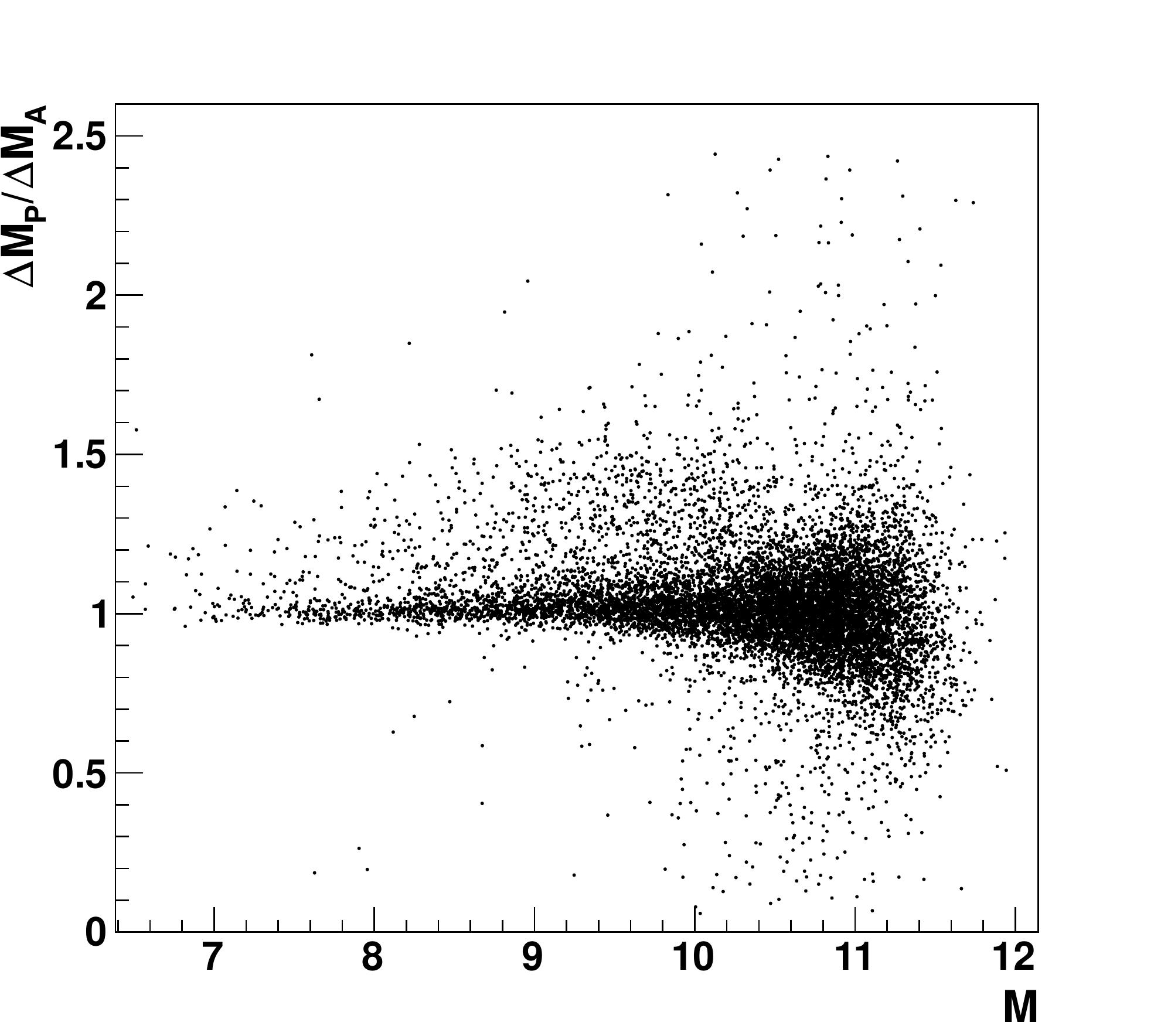}
\end{center}
\caption{Measured magnitudo spread $\mathrm{\Delta M}$ for polynomial (blue points) and ASAS photometry (red points) (left) and the ratio of the polynomial to ASAS photometry spread $\frac{\mathrm{\Delta M_P}}{\mathrm{\Delta M_A}}$ (right) as a function of the magnitudo M of the measured star.}
\label{fig_dmag_mag}
\end{figure}

The detailed photometry studies have been performed on two fields of one night and one field of a different night. All results were very similar, so only the results for 172 frames of one field are shown here. Results of the fitted magnitudo spread\footnote{The magnitudo spread is obtained from the fitted signal, without applying any corrections based on reference stars, due to the correction calculation time and its small relevance for comparison of photometry methods.} vs magnitudo, as obtained from the polynomial profile photometry and compared to ASAS photometry are shown in fig. \ref{fig_dmag_mag}. The smallest obtained brightness spread is about $0.02^\m$ for nearly the whole available brightness range (fig. \ref{fig_dmag_mag}, left). Up to $9^\m$ the spread rarely exceeds $0.05^\m$, but for dimmer stars it steeply ascends to reach close to $0.2^\m$ for $\sim11^\m$.
The best range for polynomial photometry all over the frame is around $7^\m-9^\m$. For brighter stars fits are less stable probably due to more evident ``tails'' of PSF not properly described by the model. For dimmer stars the reason is simply the descending signal to noise ratio and neighbouring brighter stars disturbing the measurement -- the number of these increasing with magnitudo.

The polynomial photometry is comparable to ASAS aperture photometry, similar or slightly worse for most stars up to $\sim8.7^\m$, and slightly improving with magnitudo (fig.~\ref{fig_dmag_mag}, right). However, for the most stars the ratio of brightness measurements uncertainties remain close to 1 all over the magnitudo range. There are probably several reasons of the model giving in general results not better then the aperture photometry. The first is the possible instability of the fit in the profile photometry, especially when there is another star in proximity, while the aperture photometry is always stable. The second is the fact, that ASAS algorithm calculates a mean value of 4 apertures of different sizes. The result of such calculation is shown here, while results of single aperture are in most cases worse than the polynomial photometry. Furthermore, brightness spread for each star was calculated only for those frames accepted by the ASAS photometry, the number of such frames significantly decreasing with magnitudo, thus results shown favour ASAS algorithm slightly.

\begin{figure}[t!]
\begin{center}
	\includegraphics[width=0.47\textwidth]{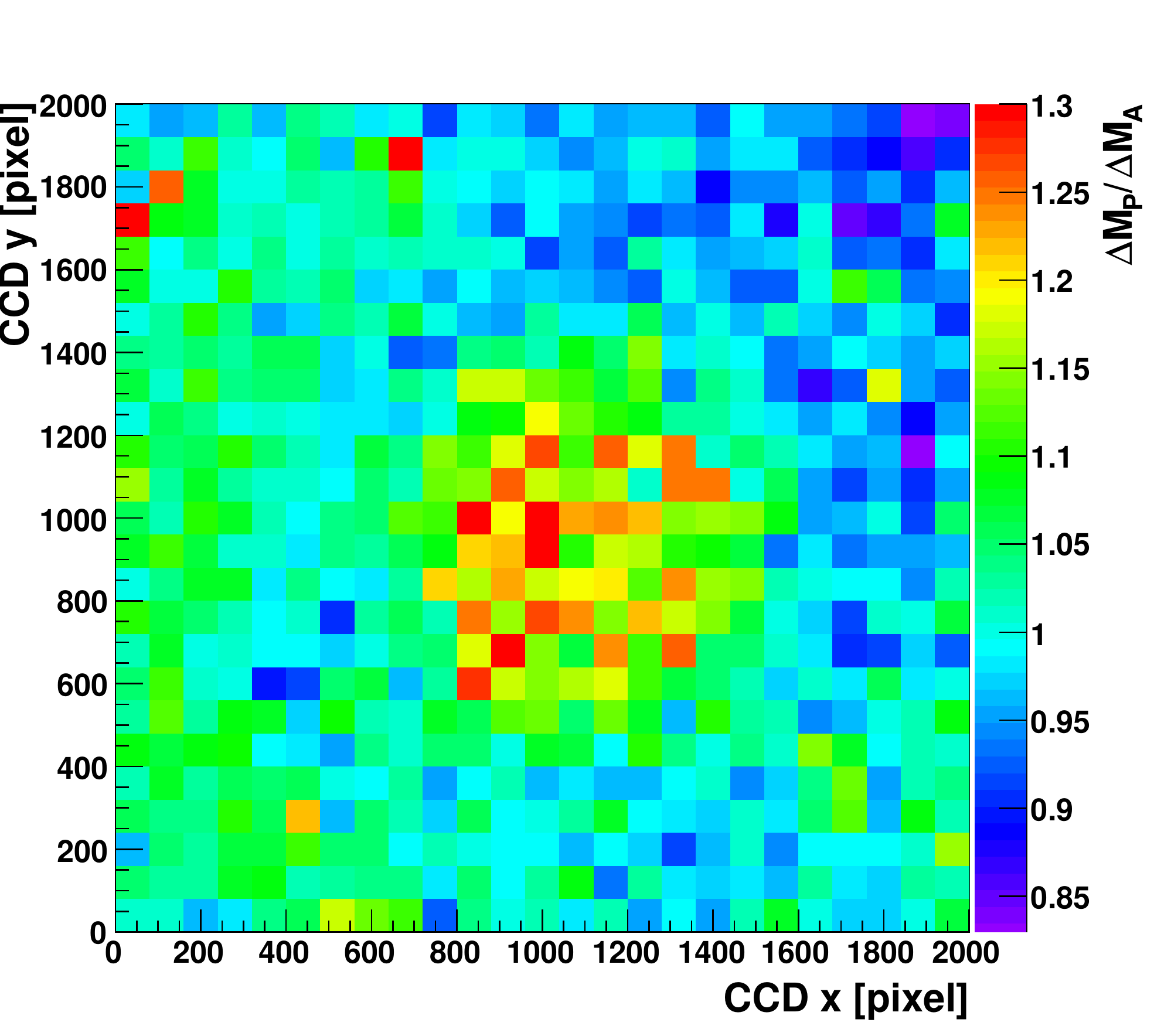}
	\includegraphics[width=0.47\textwidth]{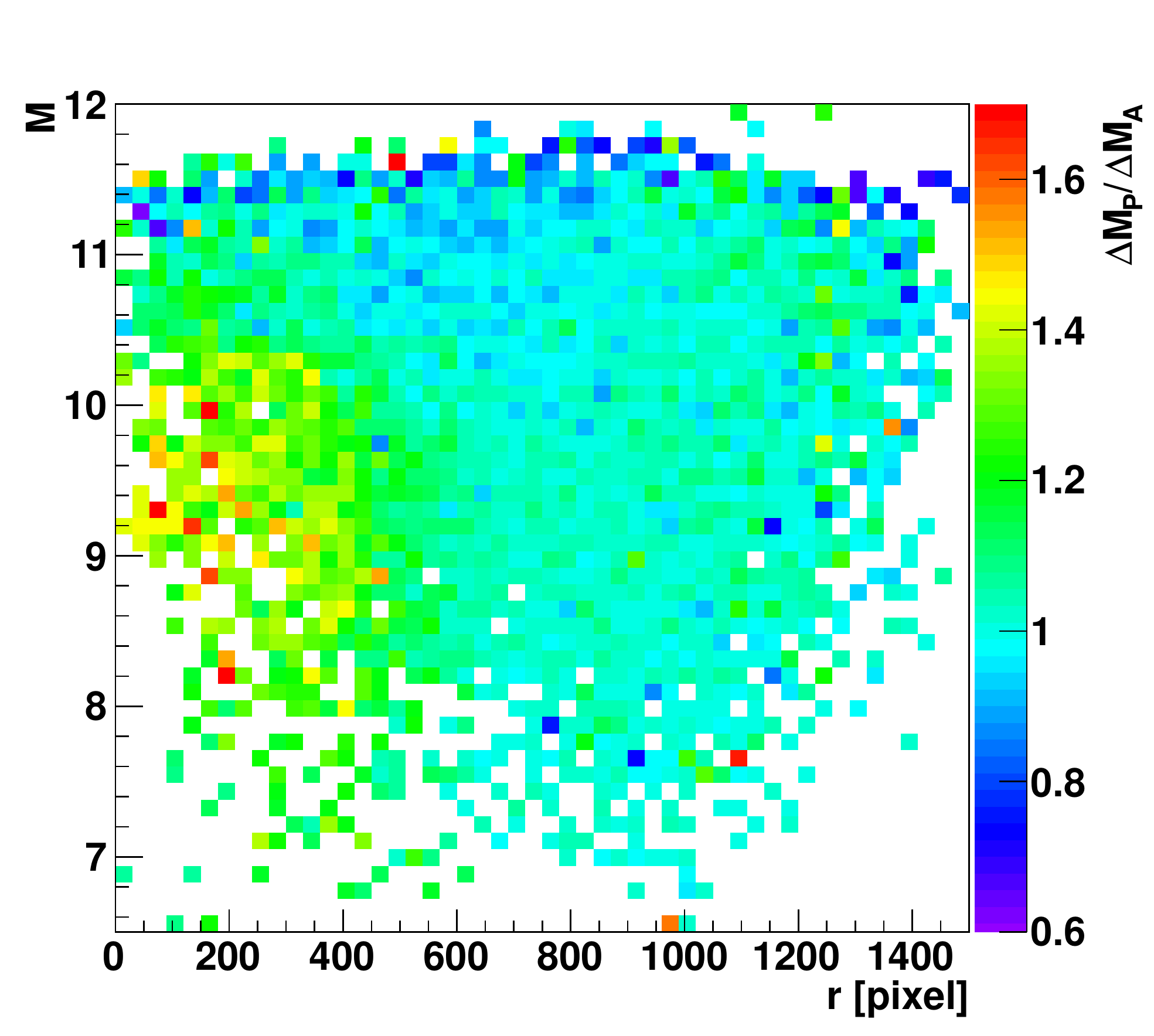}
\end{center}
\caption{Ratio of the polynomial to ASAS photometry magnitudo spread $\frac{\mathrm{\Delta M_P}}{\mathrm{\Delta M_A}}$ as a function of the star position on the chip (left), and as a function of star magnitudo and distance from the frame centre (right).}
\label{fig_dmag_pos_mag}
\end{figure}

Figure \ref{fig_dmag_pos_mag} shows comparison of the two photometry methods as a function of the position on the chip. The discussed ratio of the profile photometry spread to the ASAS photometry spread is smallest for an annulus about 500-1000 pixels around the frame centre (fig. \ref{fig_dmag_pos_mag}, left). This covers the range of an extended PSF size, according to fig. \ref{model_fwhm}. Additionally, the ratio is smallest for the top-right corner of the frame. The reason of this may be related to the change of the PSF shape with the azimuthal coordinate on the frame, but higher statistics in corners is needed to check this hypothesis.

The polynomial photometry is in most cases better than ASAS photometry for stars above $11^\m$ (fig. \ref{fig_dmag_pos_mag}, right). Based on this results, it can be stated that the polynomial photometry is better for dim stars far from the frame centre. However, according to the results shown in fig. \ref{fig_dmag_mag} (right), even for bright stars there is a number of objects for which the model performs better. This shows that it is definitely worth trying to fit ``objects of special interest'' with the polynomial method -- it is possible that the frame-by-frame fit in human controlled condition would eventually give a better result than the aperture photometry.
		
\subsection{Astrometry results}

\begin{figure}[b!]
\begin{center}
\subfigure[Position of a star]{
	\includegraphics[width=0.47\textwidth]{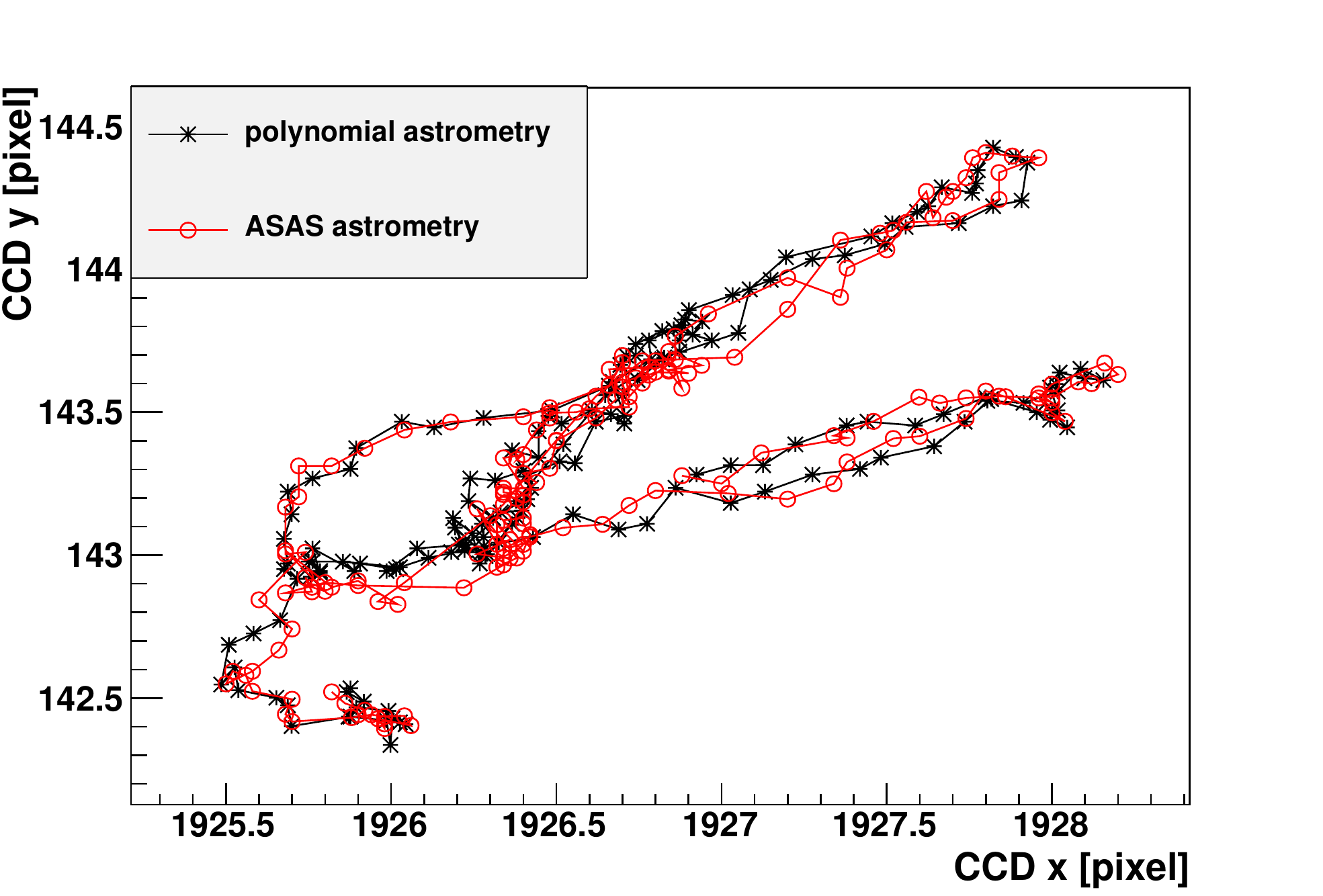}
\label{astro_single_pos}
}
\subfigure[Distance between two stars]{
	\includegraphics[width=0.47\textwidth]{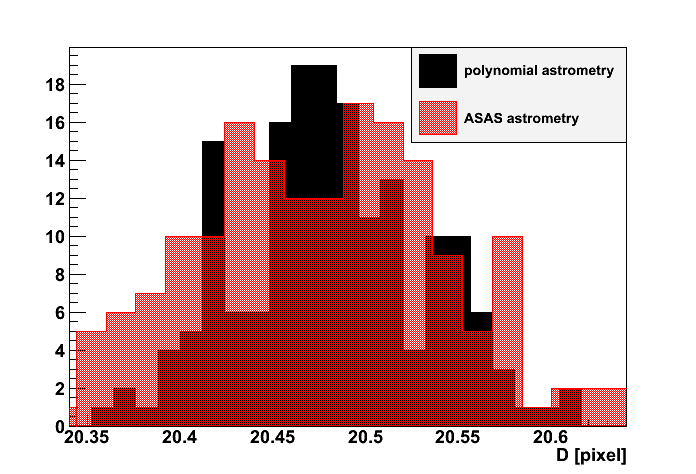}
\label{astro_single_dist}
}
\end{center}
\caption{Changes in the position of the selected bright star on the CCD during 172 exposures (left) and the distribution of the measured distance between two stars D (right). Compared are results from polynomial and ASAS astrometry.}
\end{figure}

The polynomial astrometry is performed in the same manner and in the same time as photometry, but the subject of analysis is not the analytically fitted scale (signal) of a star, but its numerically fitted position on the CCD. The position of a star relative to the pixel edge cannot be accurately determined from a simple centre of mass calculation (as explained in sec. \ref{psf_reconstruction}), but the accuracy for bright stars is much improved for more sophisticated algorithms, like ASAS aperture astrometry and discussed polynomial astrometry performed by fitting the modelled PSF to the star. The star position on the CCD of the \pin camera can be a subject to significant frame-to-frame changes, as seen on fig. \ref{astro_single_pos}, thus the accurate astrometry on a single frame is important. The movement of a bright star ($\sim 8^\m$) during 172 exposures is clearly visible in results from both polynomial and ASAS astrometries and is similar for all the bright stars on the frame. However, slight differences between results of the two algorithms remain and are directly related to the differences in their accuracies, becoming more visible for dimmer stars. The question is, how to compare astrometry accuracies?

Perhaps the simplest idea would be to analyse the difference between a star position obtained from the astrometric algorithm and coordinates from a stars catalogue. However, this method involves transformation of the position on the frame to the celestial coordinates. This introduces significant uncertainty related to the estimation of the frame's centre and orientation in celestial coordinates.

\begin{figure}[t!]
\begin{center}
	\includegraphics[width=0.47\textwidth]{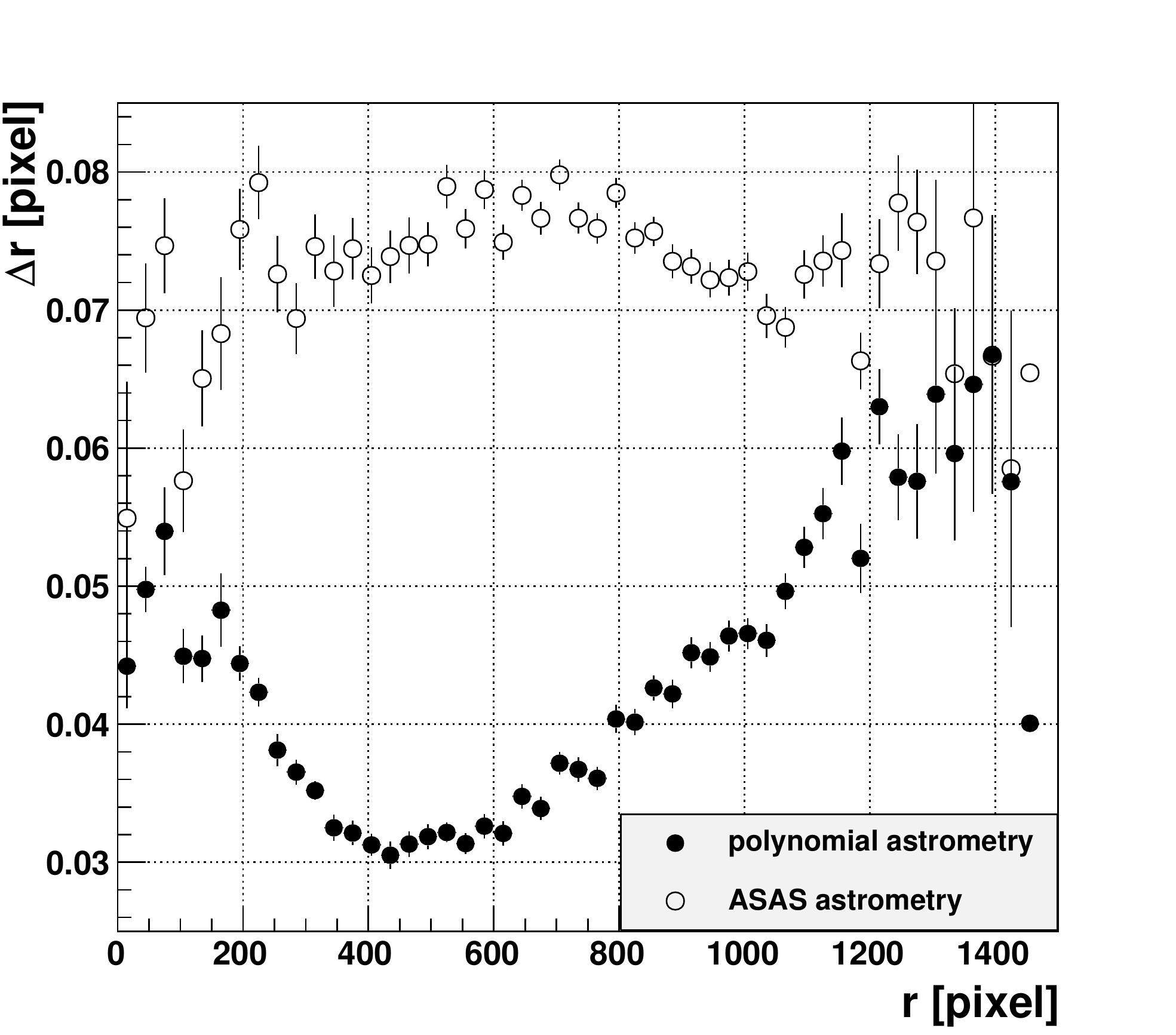}
	\includegraphics[width=0.47\textwidth]{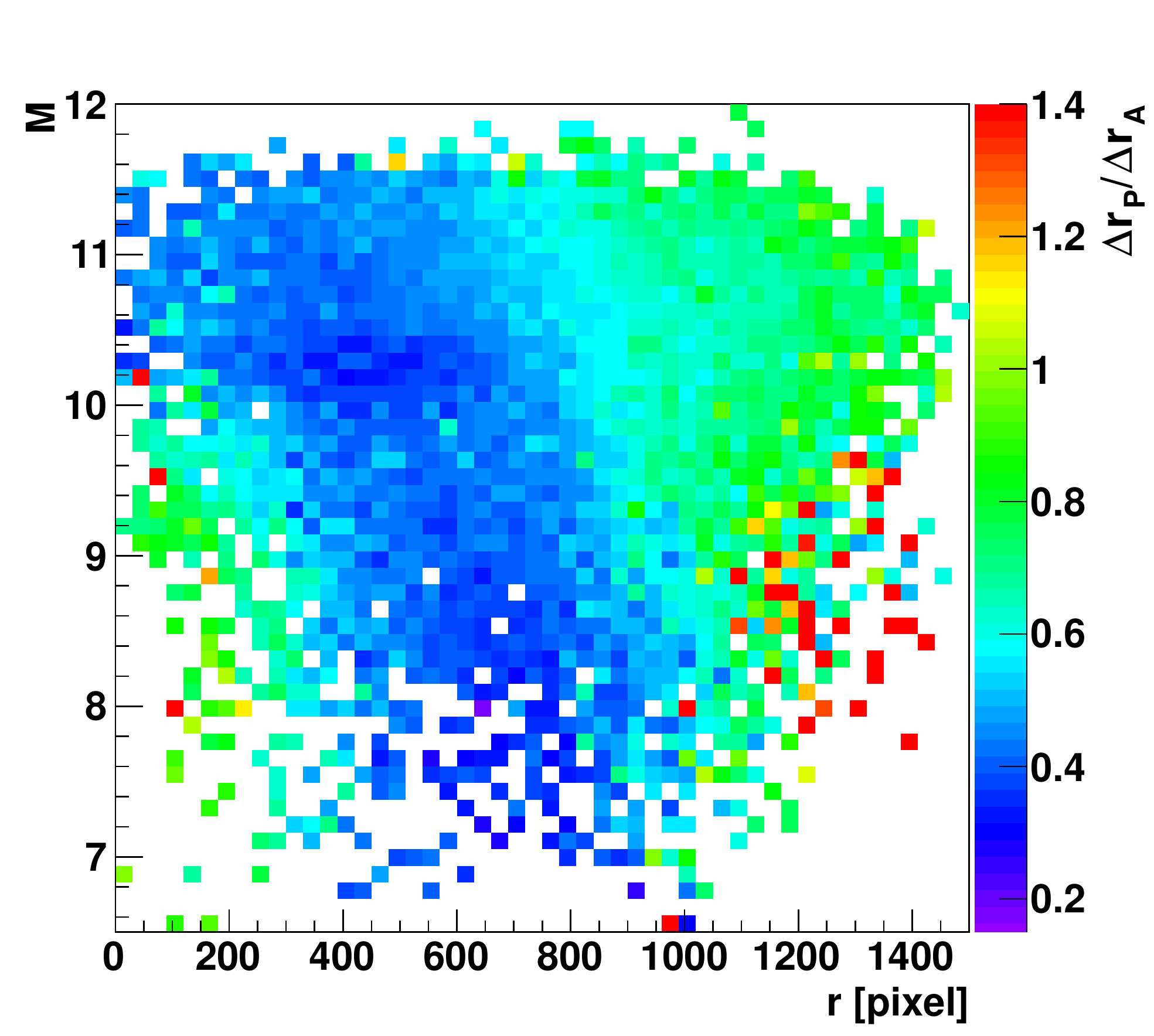}
\end{center}
\caption{Comparison between polynomial and ASAS astrometry results. Left: position spread $\mathrm{\Delta r}$ for stars with magnitudo $M<9^\m$ as a function of the distance from the frame centre r. Right: ratio $\frac{\mathrm{\Delta r_P}}{\mathrm{\Delta r_A}}$ of position spread of polynomial and ASAS astrometry as a function of star brightness and distance from the frame centre.}
\label{astro_multi}
\end{figure}

Nevertheless, all the stars on CCD shift from frame to frame in approximately the same manner\footnote{The general movement direction and range on the considered frames is similar for all stars, however details in tracks are reproduced only for stars in proximity. This fact is probably due to the rotation of the sensor while following the sky movement.}, so the distance between two stars should remain constant on all frames. Thus any spread of measured distance between stars should be a direct result of inaccuracies in position determination and can be used to quantify astrometry quality.

For a given star up to 100 stars in $300\times 300$ pixel proximity and $\pm 0.5^\m$ range are found. For each neighbouring star a histogram of its distance from the given star is calculated for all the frames out of 172 on which ASAS astrometry returned results for both stars. The RMS of such a histogram (example presented in fig. \ref{astro_single_dist}) is used as the spread of the distance between two stars. The spread is calculated for each pair and the average of all results gives the final estimate of position uncertainty for a given star. The number of frames analysed decreases with the magnitudo of a star while the number of neighbouring star in $\pm 0.5^\m$ range increases with magnitudo. As a result the uncertainty of the estimate is the biggest for very bright and very dark stars. This may affect the absolute astrometry uncertainty determination in these areas, but should not affect comparison between polynomial and ASAS astrometries, for measurements based on the same data sample are being compared.

Contrary to the photometry results, the polynomial astrometry performs much better than the ASAS algorithms. In the whole magnitudo range polynomial astrometry is at least $20\%$ better, being as much as 2.5 times better for bright stars $400-600$ pixels from the frame centre, the distance uncertainty being $\sim 0.75$~pixel for the polynomial and $\sim 0.3$~pixel for the ASAS astrometry (see fig. \ref{astro_multi}, left). The ratio of the estimated uncertainties is the smallest for the part of the frame containing the highest number of stars, due to geometrical reasons. The quality of the algorithm drops for the distance very close and far from the frame centre, where the PSF of a star becomes smaller -- probably due to the fit performing better on larger shapes.

Comparison of the two astrometry methods as a function of the distance to frame centre and magnitudo is shown in fig. \ref{astro_multi} (right). In general, for most of the magnitudo-position space polynomial astrometry performs better than ASAS. The ratio best values, reaching 0.3 for bright stars slightly increases with magnitudo, but the range in the distance from the frame centre where the profile astrometry performs much better becomes wider. The brightness dependency shows, that ASAS astrometry gives better results for stars very close and far from the frame centre, but only for magnitudo $\le 9.5$. Moreover, closer examination of these regions reveals that the higher ratio is rather due to instabilities in the fit than due to systematic differences.

Results of the polynomial astrometry show that this algorithm should be used in nearly all cases for which a longer processing time is acceptable. Such cases include position determination for objects of special interest, for which such a precise measurement is important, for example, unidentified artificial satellites of Earth, for which determining orbital parameters would help in creation of catalogues of ``cosmic debris''.

\section{GRB080319B precursor search}

Applications of the obtained PSF model are not limited to photometry and astrometry. One of other possible utilizations is a dedicated search for signal at specific coordinates on the frame. An example presented here is the search for optical precursor\cite{optic_precursor} to ``the naked eye'' burst GRB080319B, which, due to its unprecedented brightness perfectly suits the task. \pin started observing the position of the burst more than 19 minutes before the trigger, providing required data. The position of the GRB prior to the explosion and on the first 3 exposures was close to the edge of the frame, thus the burst image and its possible precursor were a subject to a large PSF deformation. In this circumstances measurement of a possible precursor signal with the dedicated PSF model should give a larger signal to noise ratio than a standard profile or aperture photometry based on circular/elliptical PSFs. 

\subsection{Single camera analysis}

\begin{figure}[tb!]
\begin{center}
	\includegraphics[width=0.555\textwidth]{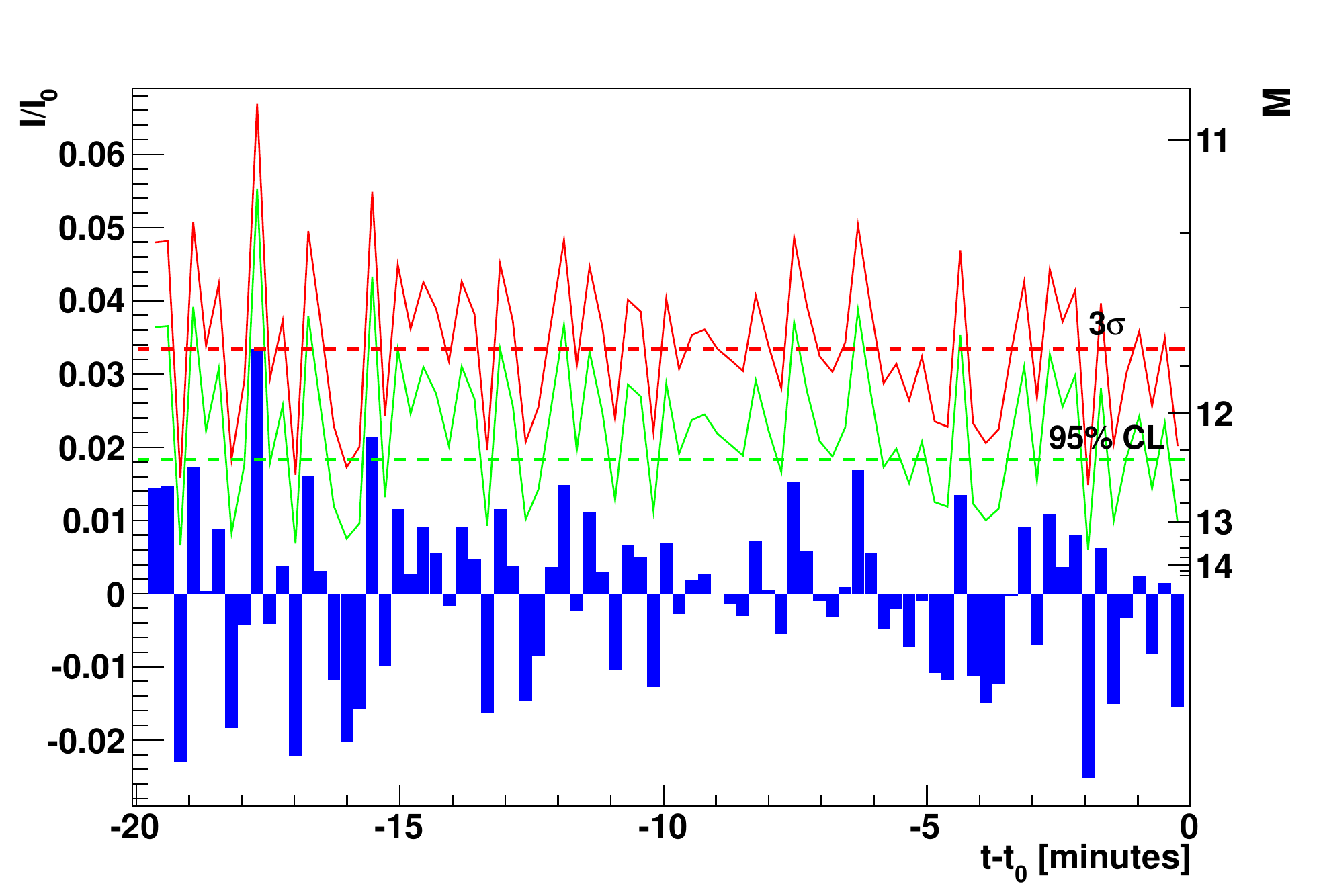}
	\includegraphics[width=0.395\textwidth]{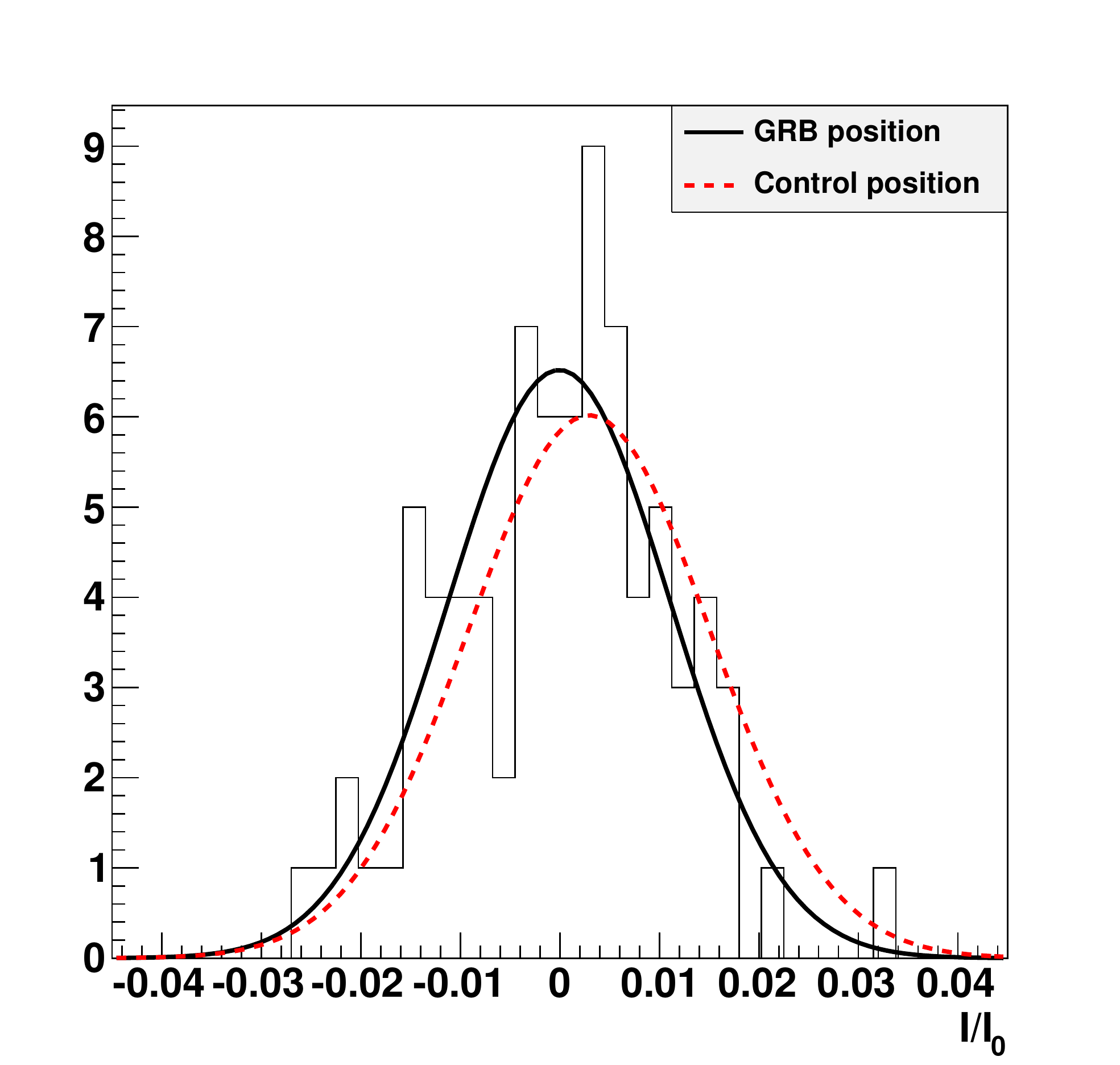}
\end{center}
\caption{Left: the measured signal $\mathrm{\frac{I}{I_0}}$ in the expected place of the optical precursor to the GRB080319B as a function of time before the GRB $\mathrm{t-t_0}$ and the corresponding limits on its brightness for k2a camera. Right: distribution of the measured signal for 84 frames before the burst with fit of normal distribution indicated.}
\label{fig_precursor_single}
\end{figure}

The analysis was performed by fitting a model PSF to GRB coordinates on the frame on all the frames covering 19 minutes prior to the explosion, on two cameras (with internal names: k2a and k2d) of the \pin prototype. Figure \ref{fig_precursor_single} (left) shows the fitted signal value $\rm \frac{I}{I_0}$ for k2a camera, for all considered frames. To suppress systematic uncertainty fitted signal I is referred to the scale $\rm I_0$ of a nearby $7.98^\m$ star. No signal exceeding $3\sigma $ limit has been found on k2a camera. Standard approach in case of ``no signal'', at least in astronomical observations, is to quote $3\sigma$ limit assuming measured value of signal to be zero. Estimation of all the limits is based on an assumption that the measurement error for a small signal that would be emitted by the optical precursor is similar to the fit error for the sky background at this position. The uncertainty $\sigma$ was calculated from a fit of the normal distribution to the histogram of signal ratio $\rm \frac{I}{I_0}$ (shown in fig. \ref{fig_precursor_single}, right). The result is consistent with distributions of signal in 3 empty control areas in the burst proximity which were considered as a cross-check. In this approach the resulting $3\sigma$ limiting magnitudo for k2a camera in the polynomial photometry is $11.67^\m$. The published limit given by the standard \pin photometry\footnote{In this case not the ASAS photometry, but so called ``fast Pi photometry'', a slightly different aperture based algorithm, was used.} was $\sim 11.5^\m$, so a $\sim 0.17^\m$ increase in limit value was obtained by using the new method. On the $95\%$ confidence level ($=1.96\sigma$) the limiting magnitudo for k2a camera is $12.13^\m$.

The advantage of the polynomial photometry is that it also allows setting limits on particular frames based on the signal level in the GRB coordinates (or rather, in this case, background fluctuation in the place where we seek for the signal). The limit was numerically calculated according to the formula given in \cite{stat_limits}, which allows extracting limits with well defined confidence level (CL) all over the measured signal range, including negative signal fluctuations. The $3\sigma$ limits ($99.73\%$ CL) calculated on single frames of k2a camera range from $10.8^\m$ to $12.5^\m$, fluctuating in most cases between $11.5^\m$ and $12.25^\m$. The $95\%$ confidence level limits (which are commonly used in high energy physics and seem more appropriate for single frame analysis) range from $11.1^\m$ to $13.5^\m$, fluctuating in most cases between $11.6^\m$ and $13^\m$. The $3\sigma$ limit on the last frame before the GRB is $12.25^\m$, and this value can be compared to the limit set by standard \pin photometry of $11.5^\m$ (which is actually the limit based on the same frame, but with assumption of no signal). In this case we obtain $0.75^\m$ limit improvement when using polynomial photometry.

\subsection{Combined results for two cameras}

\begin{figure}[tb!]
\begin{center}
	\includegraphics[width=0.555\textwidth]{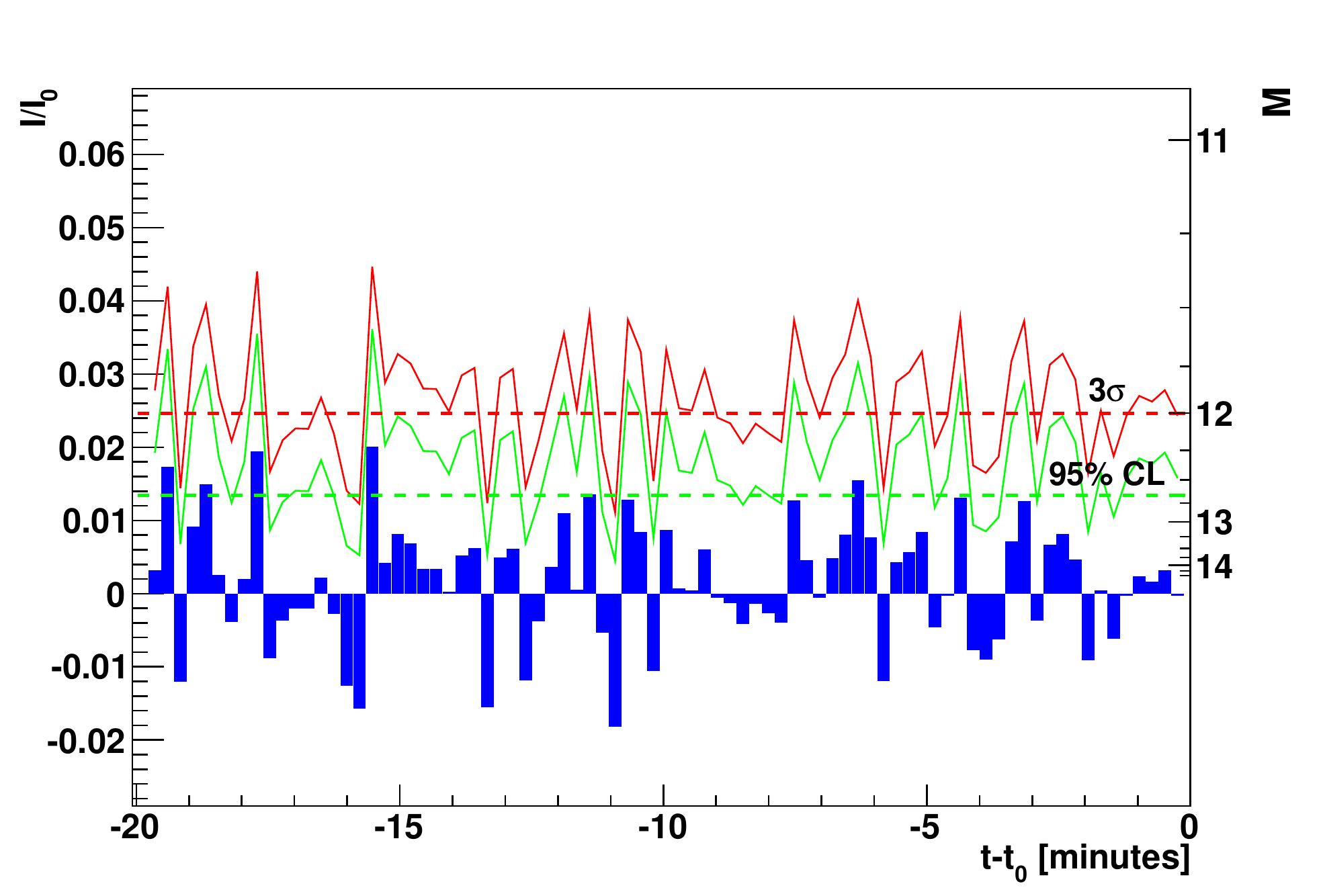}
	\includegraphics[width=0.395\textwidth]{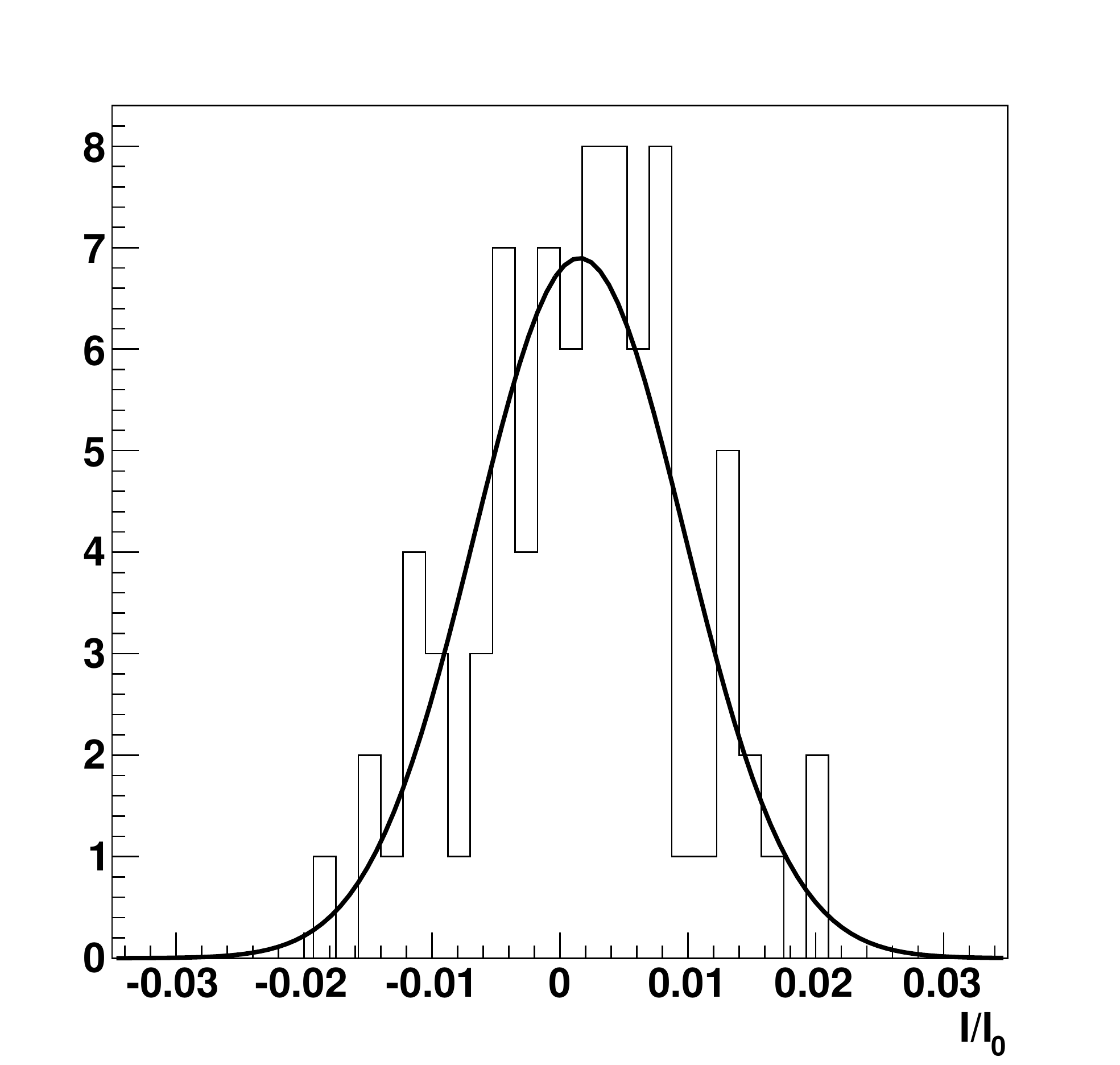}
\end{center}
\caption{Left: the measured signal $\mathrm{\frac{I}{I_0}}$ in the expected place of the optical precursor to the GRB080319B as a function of time before the GRB $\mathrm{t-t_0}$ and the corresponding limits on its brightness combined for two -- k2a and k2d -- cameras. Right: distribution of the combined signal for 84 frames before the burst with fit of normal distribution indicated. }
\label{fig_precursor_combined}
\end{figure}

The polynomial photometry, measuring the signal in the expected position of the precursor, allows also combining measurements from different cameras, increasing the signal to noise ratio and thus giving a possibility for deeper search or setting better limits. Prior to the comparison, the signal from k2a had to be normalized to the level of the signal on k2d camera, due to differences in amplification and other hardware related factors. The normalization constant was taken from the comparison of the signal between the star of reference on two cameras. After the normalization, noise levels (readout errors) for each camera have been calculated as the standard deviation of normal distribution fitted to single-camera $\rm \frac{I}{I_0}$ histograms. The combined scale for precursor I was then computed as an weighted average of scales fitted on both cameras. Similar procedure (but taking into account the star signal error) was used for the scale $\rm I_0$ of the star of reference.

No signal above $3\sigma$ level is visible in the combined signal distribution, as shown in fig. \ref{fig_precursor_combined}, (left). However, most limits greatly improve. Standard $3\sigma$ limits calculated assuming zero signal is $12^\m$, same as limiting magnitude given by standard \pin photometry for 20 coadded frames and higher by $0.33^\m$ than the single camera limit. The $95\%$ CL limit also increases by $0.33^\m$ reaching $12.46^\m$. The $3\sigma$ limit based on measured signal for single frames is between $11.5^\m$ and $12.6^\m$ for most frames. For $95\%$ CL limit is contained between $11.5^\m$ and $13.8^\m$, and for most frames between $11.7^\m$ and $13.3^\m$. The $3\sigma$ limiting magnitudo on the last frame before the GRB is $12^\m$, this time background fluctuation into low values in k2a camera is compensated by the opposite fluctuation in k2d camera, resulting in signal consistent with zero.

In general, the combined limits based on signal measured in two cameras (fig. \ref{fig_precursor_combined}, right) are more stable than limits on the single, k2a camera. The measurement error was reduced by nearly $30\%$, from $0.011$ for k2a to $0.008$ for combined signal. This is the main reason for limit improvement with two cameras.

The example of the precursor search shows that the polynomial model of PSF gives better chances in dedicated searches for a small signal in specific coordinates. In case of no signal found, it allows us to use more advanced limit setting procedures, giving better results. Moreover, it allows combining the signal from multiple cameras of \pin observing the same place, which provided important result in this case and may lead to even better outcomes in the final \pin system, where 4 cameras will observe the same part of the sky (in ``deep'' mode of work).


\chapter{Simulator}
\label{chap_simulator}

The polynomial model of point spread function described in this work can be used in multiple applications, as described in chap. \ref{chap_modelling_real_sky_data}. However, the main idea behind its development was to gain a better understanding of the \pin detector. During the process of the model development we found out how the star image (PSF) in our cameras is formed (in general) and described it in a parametric way along with spacial and energetic pixel response functions. The obtained knowledge allows reconstruction of a \pin frame in high detail. Good description of real data opens the possibility to extend the study and to test frame processing as well as dedicated analysis algorithms in controllable circumstances using simulated data samples. This approach, often referred as Monte Carlo method, is widely used in high energy physics, to test and develop new detectors, predict future results and analyse those already obtained. Inspired by methods from experimental particle physics, dedicated simulator code has been created, its features, tests and possible applications are presented in this chapter.

\section{Frame generation}

The main aim of the simulator is to generate a frame, which would correspond to a frame measured by \pin camera for selected sky coordinates. This task is performed in following steps:

\begin{figure}[tb!]
\begin{center}
	\includegraphics[width=0.47\textwidth]{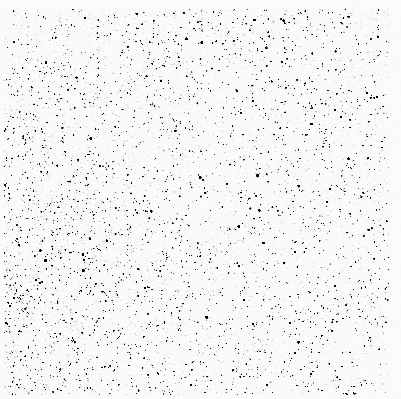}
	\includegraphics[width=0.47\textwidth]{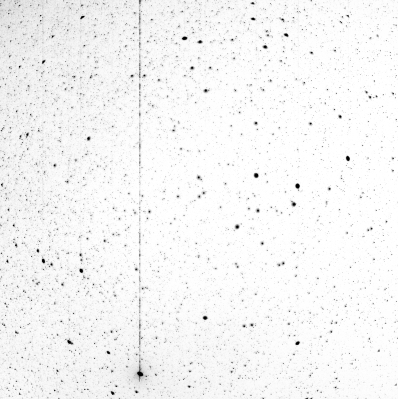}
\end{center}
\caption{Comparison between simulated (left) and real (right) images of the sky. Some non-star objects are visible on the real frame but not included in the simulation, such as a very bright, saturated satellite flash visible near the lower edge of the frame.}
\label{fig_sample_frames}
\end{figure}

\begin{description}
	\item[Sky background generation] \hfill \\
		All pixels on the frame are initialized to a value of sky background, which is one of the simulator parameters.\footnote{Lenses transmission is taken into account by multiplying the background by a transmission function reconstructed from the real flat-field frames. If requested, flat-divided frame (i. e. without correction) can be generated.}
	\item[Star distribution reproduction] \hfill \\
		The frame range on the sky is computed from position of its centre in celestial coordinates. All stars positions and brightnesses in this range are read from TYCHO catalogue. 
	\item[Frame coordinates fluctuations] \hfill \\
		If requested, star positions on the CCD surface are shifted by a value randomly selected from an assumed distribution of pointing fluctuations (flat distribution in a single pixel range is used as default); one shift vector is generated for the whole frame. This reproduces slight changes in frames coordinates due to mount movement. Obtained star positions form an input for each star image creation (fig. \ref{fig_sample_frames}).
	\item[Star brightness fluctuations] \hfill \\
		If requested, each star's brightness is multiplied by a value randomly selected from an assumed distribution (normal distribution centered in 1 and with $1\%$ standard deviation was used in results presented further in this chapter). This step reproduces source brightness fluctuations due to, for example, turbulence of the atmosphere.
	\item[Star image reproduction] \hfill \\
		An image of each star is computed from the model PSF for given coordinates on the frame. After normalizing to the star magnitudo, taking into account fluctuations, if included, it is projected on the frame.
	\item[Poisson statistics application] \hfill \\
		Each pixel value is changed according to Poisson distribution, which describes fluctuations in number of electrons collected by CCD in given period of time. This takes into account fluctuations of both incoming photon flux and CCD quantum efficiency.
	\item[Readout noise] \hfill \\
		CCD readout noise is added to each pixel. The noise is modelled by a Gauss distribution. The default noise level (sigma of the distribution) is 16e -- a value characterizing simulated camera used in laboratory measurements.
	\item[Gain fluctuations] \hfill \\
		Collected charge is converted to the output signal units (Analog-Digital converter Units: ADU) based on the assumed CCD gain (default is 2.7e/ADU). In addition fluctuations of the CCD gain, approximated by a Gauss distribution, can be applied to each pixel.
	\item[Energetic pixel response function application] \hfill \\
		Signal measured in all pixels is recalculated according to the energetic pixel response function, which introduces non-linearity of generated signal for bright light sources.
\end{description}

After the requested number of frames is generated, the result are stored in FITS files -- a standard format for astronomical data. This output is suitable for all software working on FITS frames including most of the \pin analysis algorithms. Additionally, results can be stored in the ROOT framework\cite{root} TTree classes, which were the basic data format for analysis performed in this thesis.

From the point of view of external programs, generated frames are identical to real sky data. Thus the simulation of a real part of the sky as seen by the \pin camera is a good tool for testing photometric, astrometric, sources detection and cataloguing algorithms as well as other analysis programs in well defined conditions. Moreover, the PSF model can be easily changed to describe different apparatus.

\begin{figure}[tb!]
\begin{center}
	\includegraphics[width=0.47\textwidth]{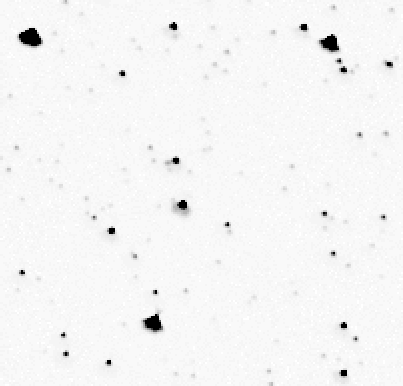}
	\includegraphics[width=0.47\textwidth]{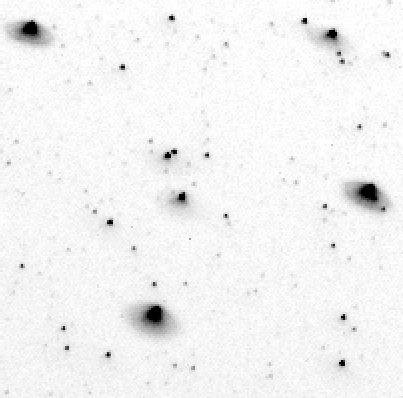}
\end{center}
\caption{Comparison between zoomed parts of simulated (left) and real (right) frames. The core of PSF is similar, while the real data show a second order ``halo'' effect for the brightest stars, not included in the model and thus the simulation. The bright flash visible near the middle of the right border of the real frame part is due to an artificial satellite.}
\label{fig_sample_stars}
\end{figure}

\section{Quality of the simulation}

Fig. \ref{fig_sample_frames} shows the simulated and the real frame of the same part of the sky. The spatial distribution of stars is the same, although the real frame reveals some objects not included in the catalogue of stars, such as satellites or planets. Clearly visible in \ref{fig_sample_frames} is a flash of one of the Earth's artificial satellites: the brightest point near the lower edge of the frame creating a bright belt of charge going up to the upper edge\footnote{The belt o charge is a result of so-called ``blooming''. Charge from saturated pixels floods neighbouring pixels and leaves a trail in the most brightest pixels' columns during the frame readout process.}. The background on the real frame is not as uniform as on the simulated frame due to true background variations or an imperfect flat frame used in the reduction process. Nevertheless, function describing spatial distribution of the background can be easily added to the simulator.

The differences between apparent brightnesses of bright stars on the real frame are bigger than on the simulated frame. This is an illusion caused not by star's image real brightness but its size. As mentioned in chap. \ref{chap_modelling_real_sky_data}, far ``tails'' of the PSF are not reproduced by the model. Additionally, a kind of ``halo'' is visible around the brightest stars (fig. \ref{fig_sample_stars}), which is a second-order effect, not taken into account in the model. The ``halo'' is the main cause behind the brightest stars appearing larger and thus brighter on the real frame compared to the simulated frame. However, the magnification of the frame reveals that dimmer stars are reproduced in high detail.

It has to be noted, that the results of the visual comparison of the real and the simulated frame, although good, are rather of a lower importance. What matters most is the degree of similarity between results of analysis obtained on real and simulated data. Therefore a polynomial photometry has been performed on a series of 25 simulated catalogue frames.

\begin{figure}[t!]
\begin{center}
	\includegraphics[width=0.47\textwidth]{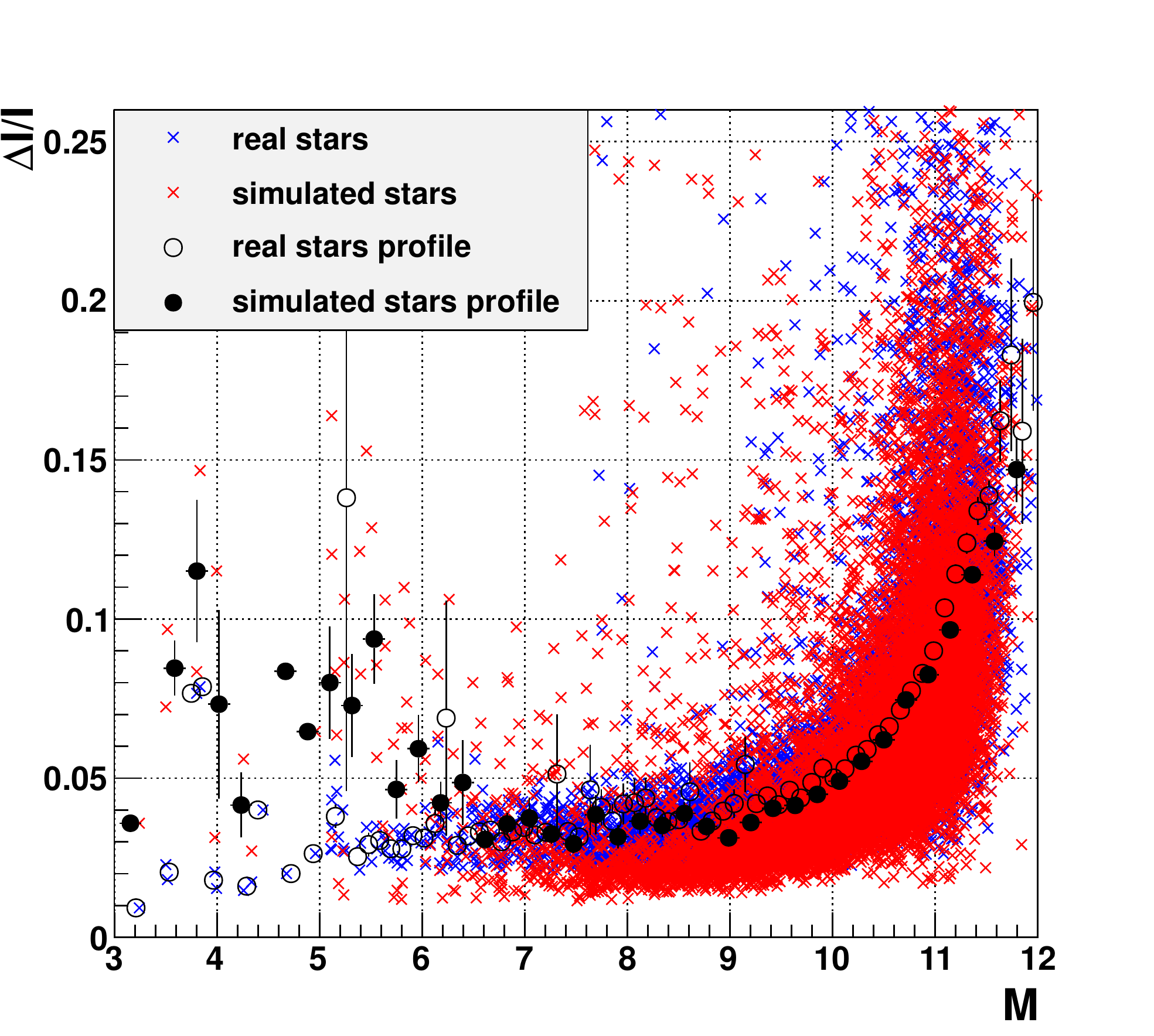}
	\includegraphics[width=0.47\textwidth]{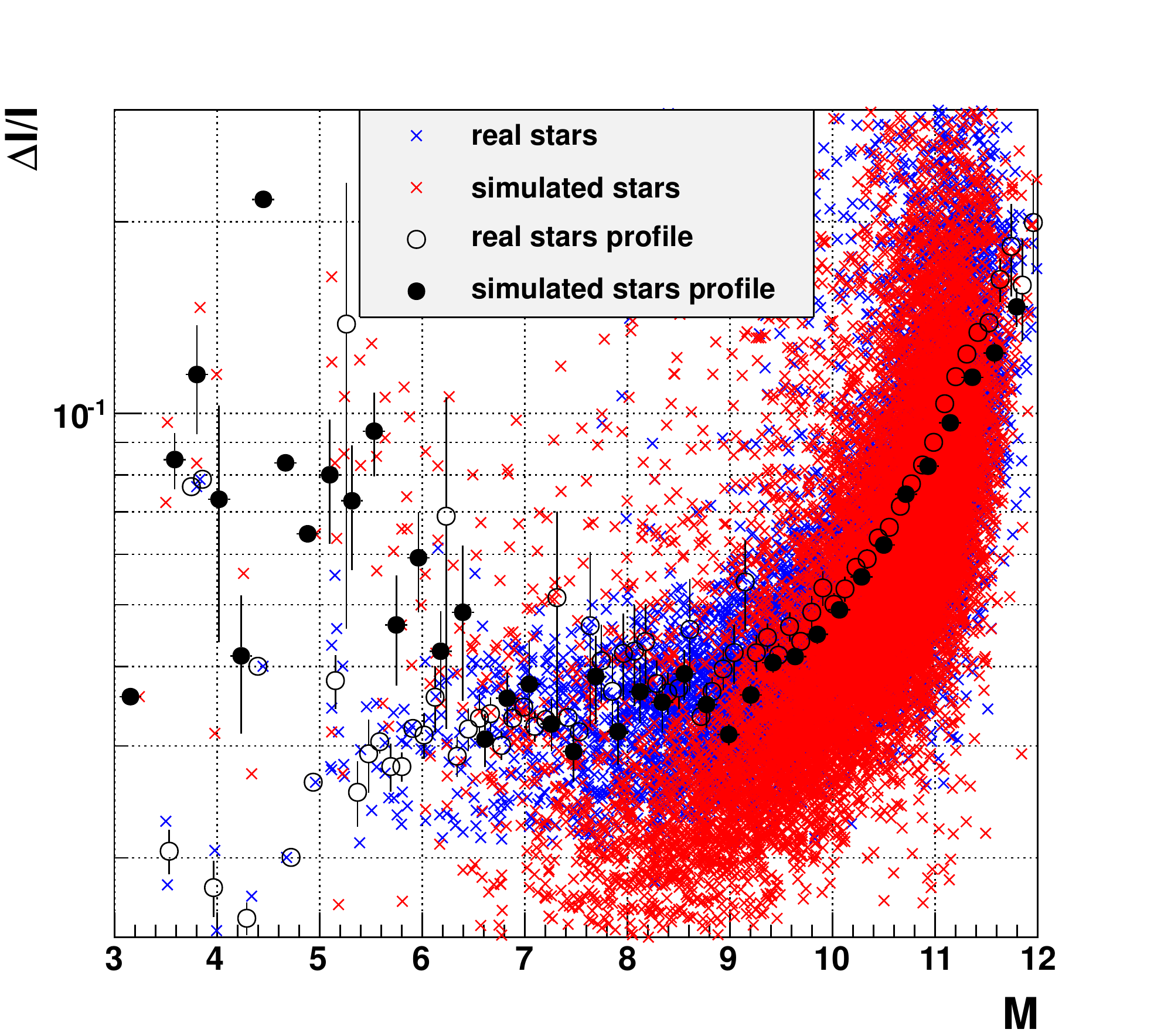}
\end{center}
\caption{Polynomial photometry results for simulated and real sky images. Dependence of the relative brightness measurement error $\mathrm{\frac{\Delta I}{I}}$ on the star magnitude M is shown in linear (left) and logarithmic (right) scale.}
\label{fig_sim_real_sigma_mag}
\end{figure}

The dependence of the stars signal spread on the brightness is very similar for real and simulated data, as shown in fig. \ref{fig_sim_real_sigma_mag}. The slightly better match of results could be obtained by fine-tuning of the simulation parameters describing measurement fluctuations. It is visible that the number of poorly-measured stars is similar all over the magnitudo range. Additionally, other details of the dependence are well reproduced, such as the high instability of the brightest stars, caused probably by the CCD energetic response non-linearity, and the subgroup of dim stars with low brightness measurement uncertainty between $\sim10.5^\m$ and $\sim11.5^\m$. 

\begin{figure}[tb!]
\begin{center}
\subfigure[simulated data]{
	\includegraphics[width=0.47\textwidth]{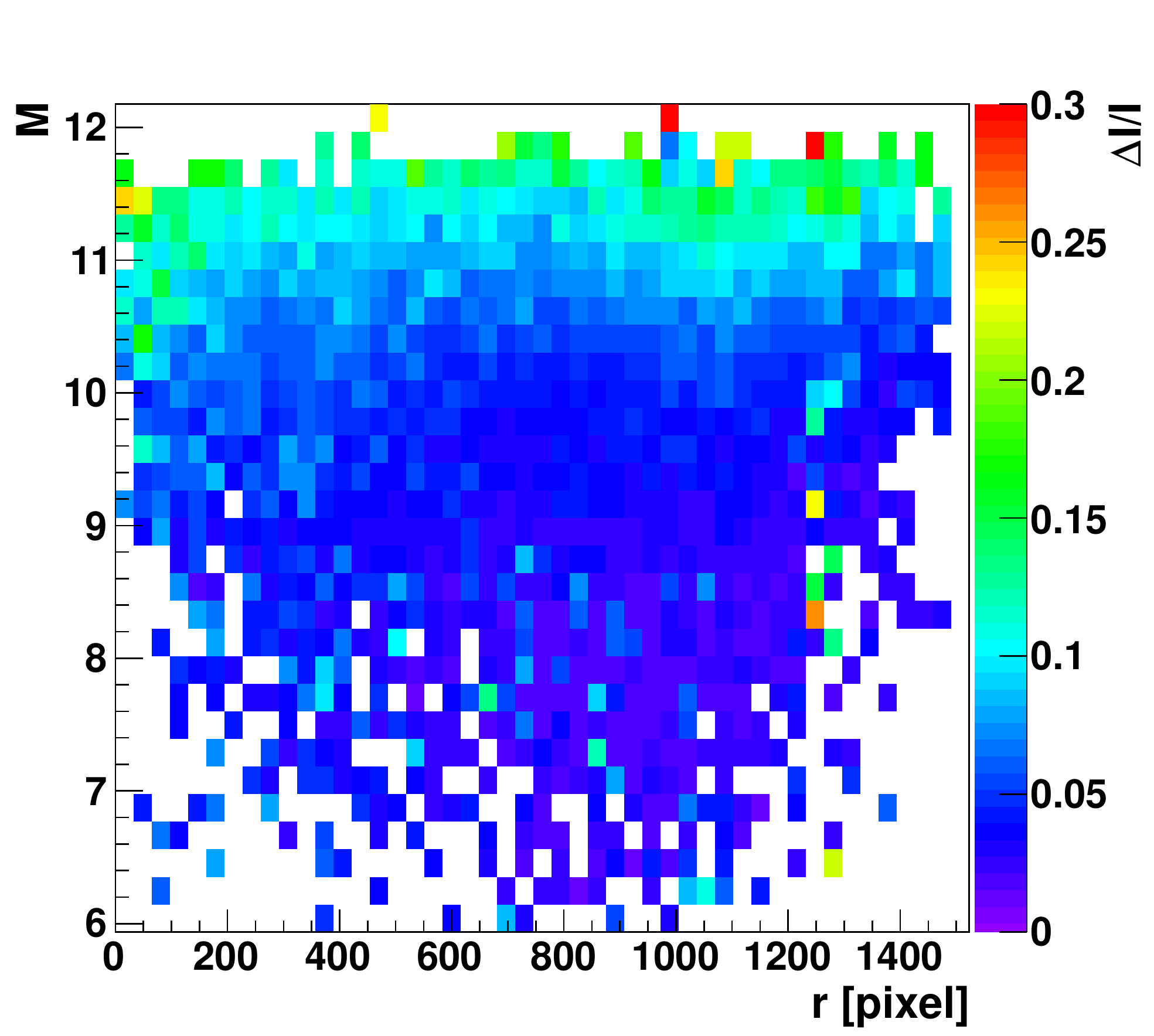}
}
\subfigure[real data]{
	\includegraphics[width=0.47\textwidth]{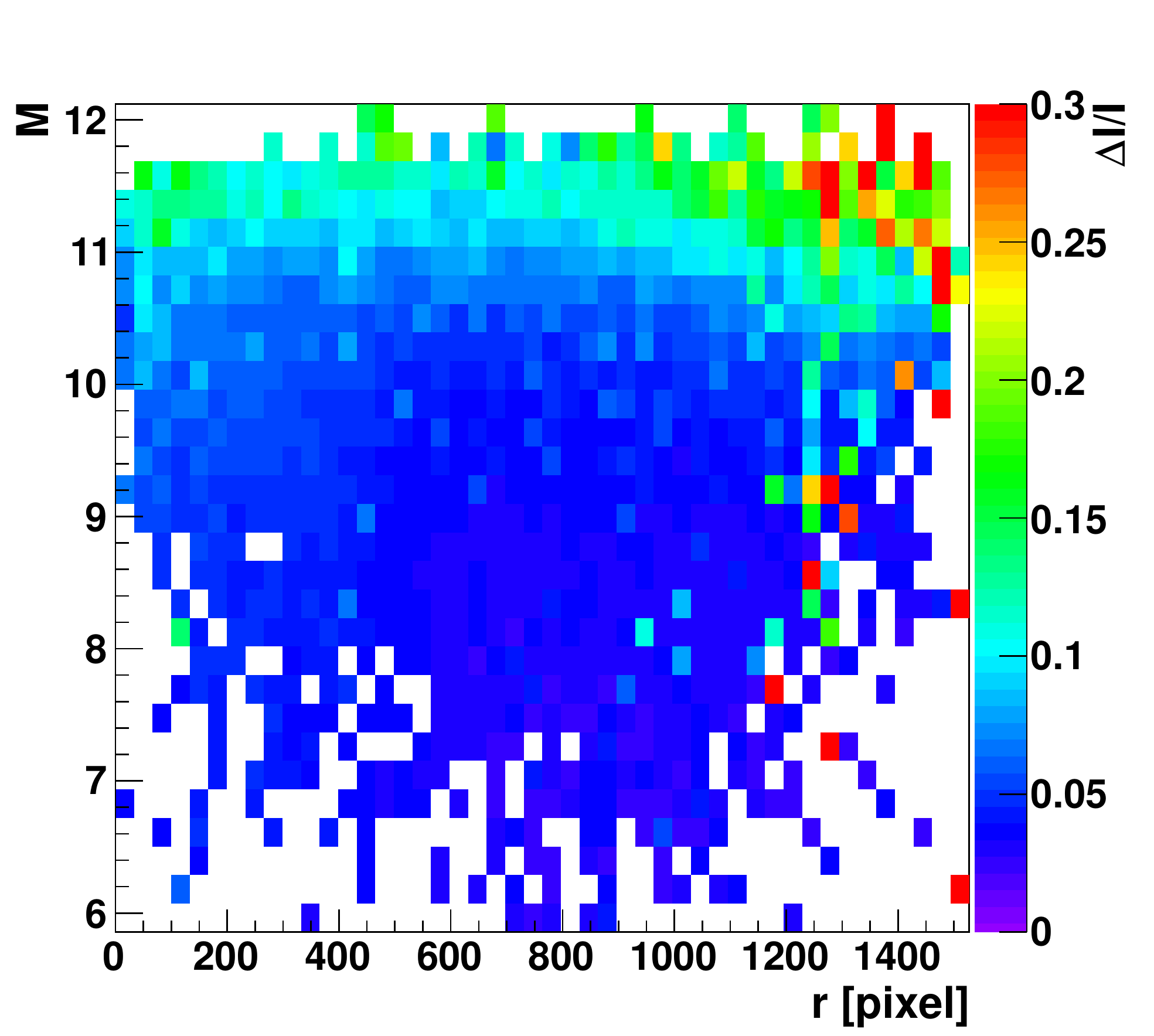}
}
\end{center}
\caption{Dependence of the relative signal measurement uncertainty $\mathrm{\frac{\Delta I}{I}}$ on the star magnitudo M and the distance from the frame centre r, for polynomial photometry performed on simulated sky images (left) and on real data (right).}
\label{fig_sim_real_sigma_mag_r}
\end{figure}

The spatial distribution of the photometry uncertainty on the frame is also very similar for real and simulated data, as shown in fig. \ref{fig_sim_real_sigma_mag_r}. The similarity proves that the PSF is the main factor responsible for varying photometry quality on the frame.

\begin{figure}[b!]
\begin{center}
	\includegraphics[width=0.49\textwidth]{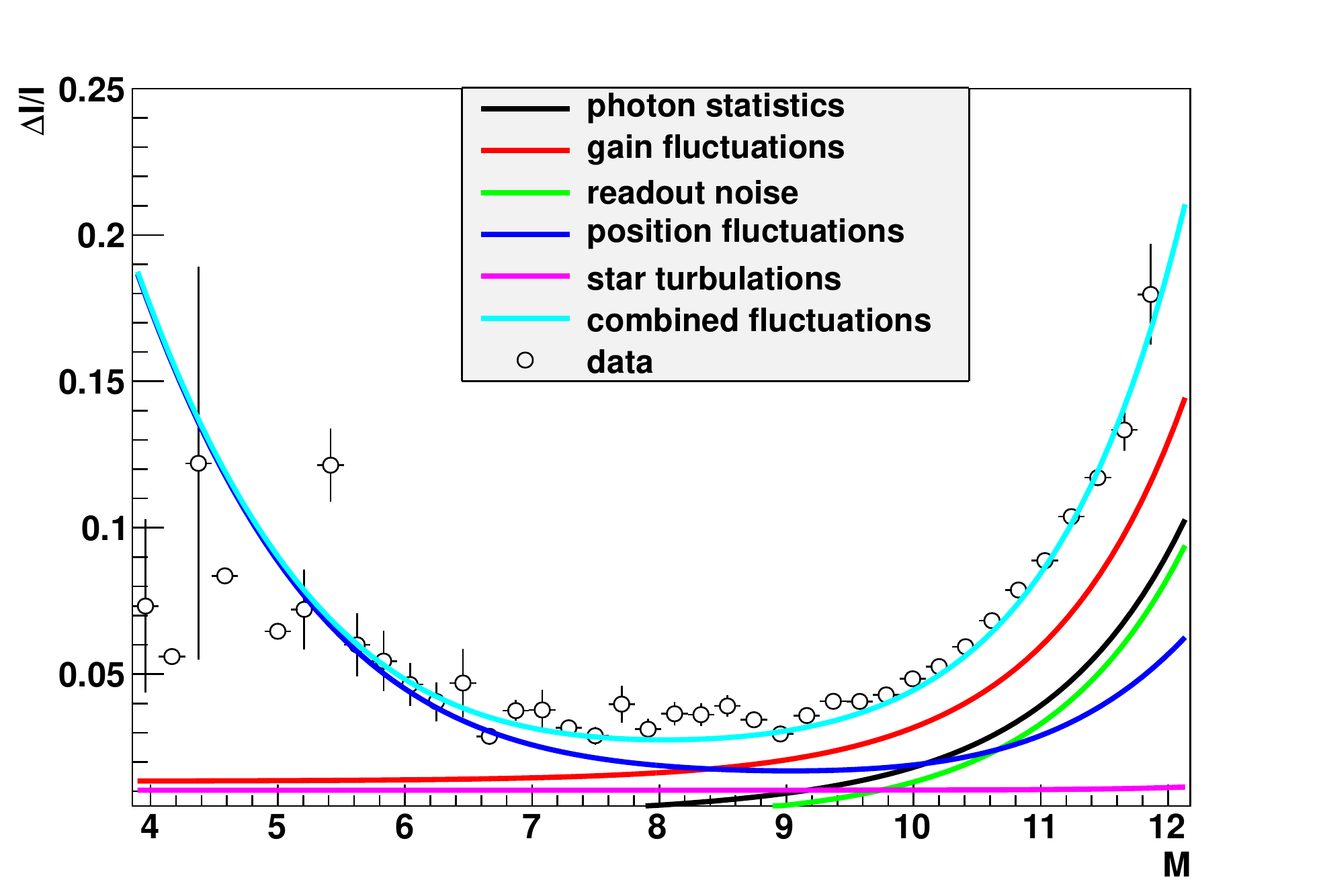}
	\includegraphics[width=0.49\textwidth]{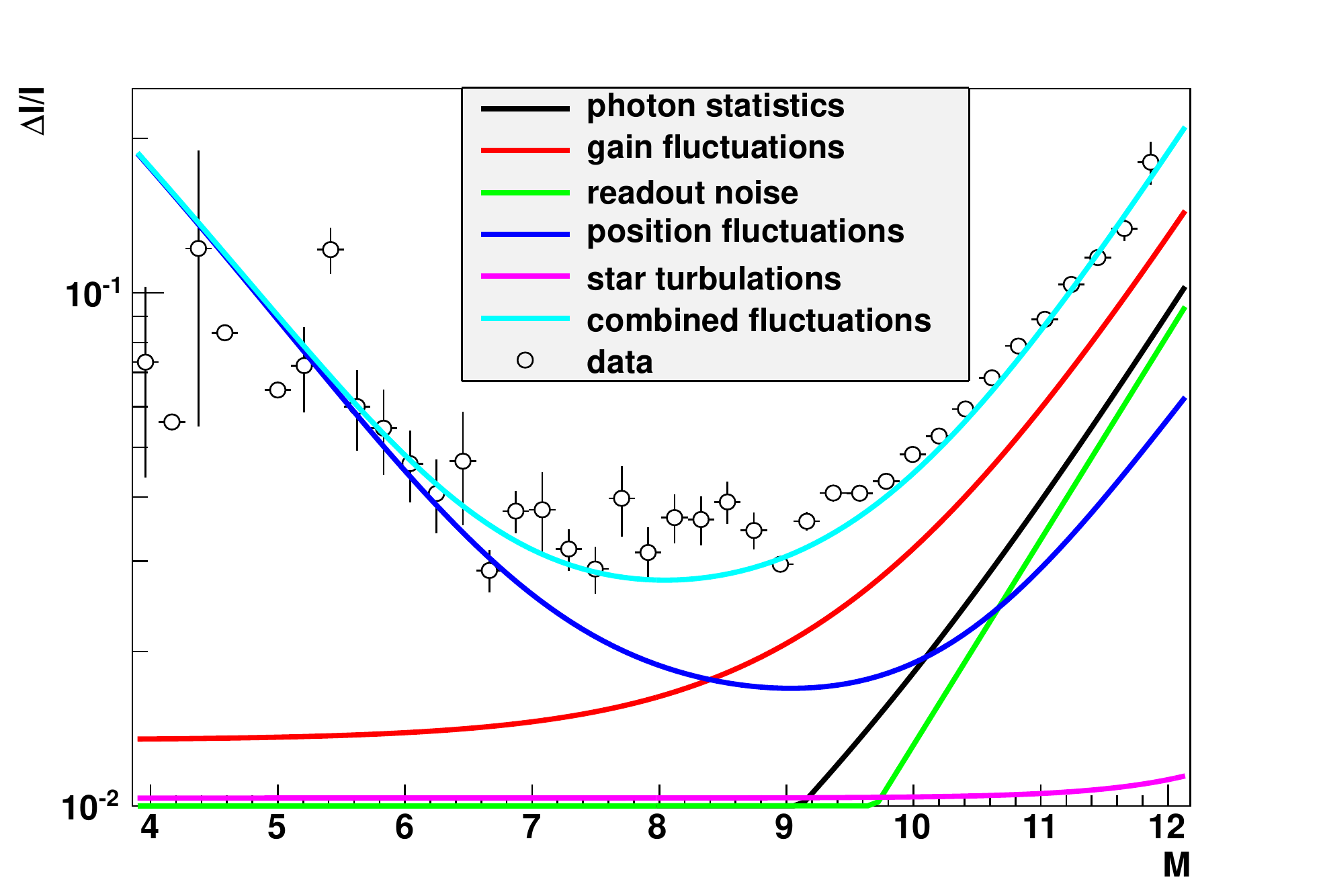}
\end{center}
\caption{Relative error of the photometry performed on the simulated data, as a function of the star brightness (points) and the estimated contribution from different sources of variability (colour lines). The error is shown in linear (left plot) and in logarithmic scale (right plot).}
\label{fig_components}
\end{figure}

The key to the good agreement between results of photometry performed on generated images and real data is not only the PSF, but also proper simulation of the frame variability -- the fluctuations of stars brightness caused by miscellaneous factors. The described simulation takes into account 5 factors contributing to the frame-to-frame variability. The overall signal for each star (mainly due to atmospheric turbulence) is modified by gaussian fluctuations with $\sigma=1\%$, and the star position is shifted by a random value (uniform inside a pixel). Charge deposited in each pixel of the frame is smeared according to the Poisson distribution, readout noise of 32 electrons and gaussian gain fluctuations of $3.7\%$ are also applied\footnote{The readout noise for the described camera as measured in laboratory conditions was 16 electrons, but the value of 32 electrons seems to describe real data better. This can be attributed to the contribution from the dark frame subtraction process, as well as to aging of the electronics.}.

The impact of each variability factor on the photometry quality is shown in fig. \ref{fig_components}. The biggest uncertainty is caused by gain fluctuations, which dominate in nearly the whole region except for the bright stars below $8.4^{\mathrm{m}}$, where energetic pixel response function non-linearity becomes significant. In this region the most important factor is the uncertainty caused by position fluctuations. For stars dimmer than $\sim10^\m$ Poisson fluctuations become more significant than position fluctuations. Star fluctuations (due to turbulence) add a factor constant (compared to the other factors) nearly all over the magnitudo range and were introduced here mainly for the simulator functionality presentation purposes. The readout noise is of a small importance over the whole magnitudo range. Combined contributions from the separate factors follow the shape of the data coming from the simulation with all the factors included.

All the variability parameters, except the photon statistics, were an educated guess based on the photometry of real data. However, the proportion between factors should not change significantly after tuning. The important conclusion coming from the simulation is that the biggest improvement in photometry stability could be obtained by reducing CCD gain fluctuations. Still, the uncertainty caused by position fluctuations is significant and should be a subject of further studies. Reducing readout noise of the detector is of a rather minor importance.

\section{Possible uses}

Generation of series of frames consisting of catalogue stars is a good tool for general software and system tests, and for planning future measurements. It can be used for predicting cameras range in different circumstances, estimating photometry/astrometry efficiency in different parts of the frame and for different star's magnitudo, etc. Electronics noises and mechanical tracking precision can be easily adjusted and thus those calculations can be easily repeated to study the all possible impacts of system electronics and/or mechanics upgrade.

Although this mode of operation allows general testing of algorithms, their deeper analysis and tuning can be performed in a much more efficient way. The key is a generation of a frame dedicated for such tasks, consisting of a small number of stars with specific parameters, like coordinates, brightness, etc., and their changes in time. These features create two possible groups of additional simulator uses.

\subsection{Coordinates-focused analysis}

\begin{figure}[tb!]
\begin{center}
\subfigure[site 1 frame]{
	\includegraphics[width=0.47\textwidth]{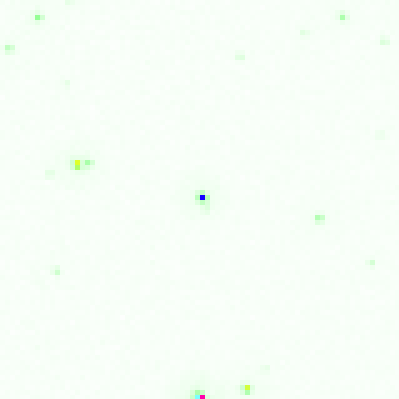}
}
\subfigure[site 2 frame]{
	\includegraphics[width=0.47\textwidth]{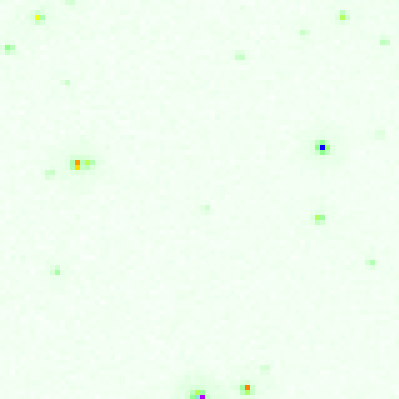}
}
\end{center}
\caption{Simulation of the parallax phenomenon: satellite 23000 km above Earth as seen from two \pin sites distant by 100 km.}
\label{fig_parallax_example}
\end{figure}

Performing analysis on a frame of ``randomly'' scattered stars has a particular drawback -- the number of stars changes with the distance from the frame centre and is especially small in the corners of the frame. Thus corners of the frame always suffer from a small number of measurements. The simulator capacity of generating sets of stars for requested positions allows for a detailed study of algorithms performance on any position on the frame.

However, generating stars in requested positions aids studying more sophisticated tasks. The first among them is the study of the possibility of using parallax phenomenon in the \pin project. Frames reproducing parts of the sky as seen from different sites on the Earth can be generated, with an image of a near-Earth object appearing in different places. Images shown in fig. \ref{fig_parallax_example} were generated assuming altitude of 23000 km for a satellite directly above \pin sites distant by 100 km, resulting in $\sim 15'$ angular separation of the images.

In this case simulated data can help determining the maximal distance of the object from Earth for which it could be stated, that it appears in two cameras at different coordinates. Assuming that the smallest measurable position difference of 0.1 pixel ($3.6'$) allows identifying near-Earth objects maximally $\sim 1432000$ km from Earth surface for sites distant by 100 km and $\sim 430000$ km for sites distant by 30 km, in both cases behind the Moon orbit. However, the real separation distance changes (mainly due to the PSF changes) with coordinates on the frame and can be calculated using the simulated data.

\begin{figure}[tb!]
\begin{center}
	\includegraphics[width=0.47\textwidth]{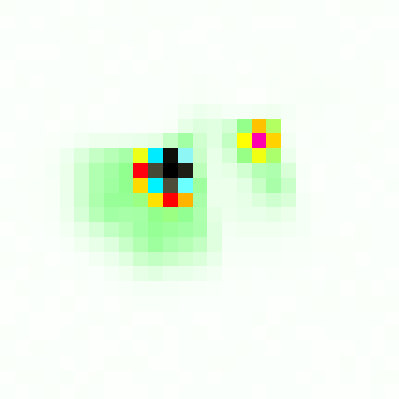}
\end{center}
\caption{Example image from the simulation of a dense field -- three overlapping stars of $5^\m$, $7^\m$ and $9^\m$.}
\label{fig_overlapping_stars}
\end{figure}

A similar task is studying efficiency of analysis of crowded fields. Simulated images of overlapping stars (fig. \ref{fig_overlapping_stars}) of different magnitudos can be used for determining when the stars become indistinguishable, how their position/brightness measurements depends on their distance, magnitudo, position on the frame, etc. Additionally, the simulator allows specifying a function describing a star movement through the frame during specified time. This capacity mimics mount tracking errors and can help in compensating stars variability caused by it.

\subsection{Lighcurve-focused analysis}

In the project such as \pin, dedicated to observing rapidly variable phenomena on the sky, a special stress is put on the algorithms analysing stars variability and detecting new, explosive sources, such as gamma ray bursts. So far optimal parameters and efficiency of the flash detection algorithm have been estimated in a very simple simulation\cite{lwp_mgr}. Similar estimations have been performed for other detection and variability analysis software used in the project.

\begin{figure}[b!]
\begin{center}
	\includegraphics[width=0.6\textwidth]{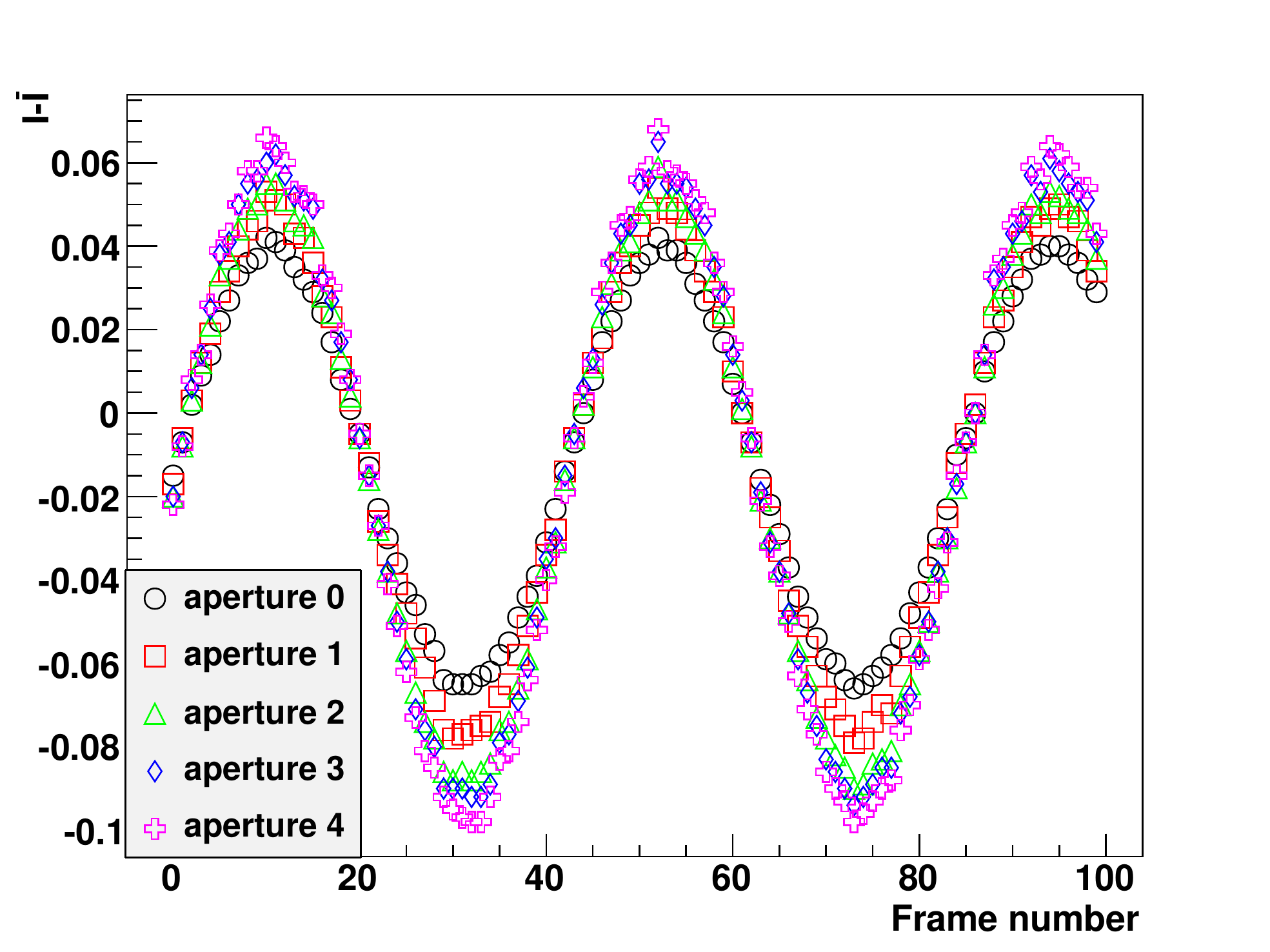}
\end{center}
\caption{ASAS photometry results (signal I shifted by mean signal $\mathrm{\bar{I}}$ to cover the same range) for a simulation of a sinusoidally-variable star of $\sim6.5^\m$. Differences between amplitudes given by apertures of different sizes are probably caused by a non-linearity of energetic pixel response function.}
\label{fig_simul_lightcurve}
\end{figure}

Also in this case, the simulator aids much more sophisticated analysis of such algorithms. It allows creation of star-like objects with specified lightcurve, reproducing requested variability, as shown in fig. \ref{fig_simul_lightcurve}. Thus for periodic variable stars a period and amplitude determination algorithms can be tested. For flare, novae stars or optical transients thresholds for detection can be estimated, etc. Additionally, all other features of the simulator can be used simultaneously, like generation of a neighbouring star with specific parameters, mount tracking instabilities, electronic noises, etc., giving testing possibilities unprecedented in previous simulations.

\section{Possible extensions}

The design of the simulator allows easy creation of extensions, such as introducing new catalogues of objects like planets or stars, which would help replicating the real sky view in more detail. Simulating near-earth objects could help develop algorithms for orbits determination. Additionally, more features related to the electronics performance can be added, such as blooming (signal saturation) effect, hot pixels, nongaussian distribution of noises, etc., which in some conditions affect the data analysis.

Presented features of the simulator make it a very powerful tool for data analysis algorithms. Frames are reproduced in a very high detail and posses the most important attributes of the real sky. That fact, together with highly adjustable stars characteristics and easy extendability should prove simulator a software greatly accelerating future algorithms development and testing.


\chapter{Summary}

\pin is a set of robotic telescopes aiming for continuous observation of a large part of the sky with high temporal and optical resolution using wide field-of-view instruments. Its primary goal is to look for optical afterglows associated with the gamma ray bursts (GRB), but it is also well suited to study any kind of short time scale astrophysical phenomena. The very-wide sky coverage of the final \pin system, about 1.5~sr, requires large field of view of a single camera, $20^{\circ}\times 20^{\circ}$. Lenses allowing such coverage introduce large deformations of the point spread function. This results in increased uncertainties of measured brightness and positions of the observed objects.

The main goal of the presented study was to develop a detailed model of the camera response including a parametrization of the star image shape (point spread function -- PSF). For this purpose the author of this thesis prepared and configured a unique laboratory setup, which allowed precise measurements of the detector response in fully controlled conditions. The point spread function, along with the pixel sensitivity function as well as spatial and energetic pixel response functions, for multiple wavelengths were reconstructed. These results were a starting point for development of the diffraction-based model of the PSF. 

The diffraction-based model describes the observed PSF shape as a result of the light propagation through camera's lenses. In the considered approximation the wavefront aberrations were parametrized as a sum of the so-called Zernike polynomials. In this approach PSF parametrization required integration of the wavefront, which is a rapidly oscillating function, over the aperture. Therefore calculations could not be made efficiently with standard numerical algorithms. Non-standard method of integration was successfully implemented along with numerous numerical simplifications. The diffractive model reproduced measured profiles quantitatively, proving that the approach is proper in general. However, numerical calculations involved are still very time consuming and not precise enough to allow for practical application of this method for modelling or simulation purposes.

As an alternative, an effective model parametrizing the PSF in the image plane has been developed based on laboratory measurements. This required constructing a dedicated mathematical base, derived from Zernike polynomials, then selecting proper model parameters and parameter interpolation methods. After adjusting the model to describe the real sky data from the prototype, it was incorporated into photometric and astrometric algorithms and tested on collection of real sky images. No improvement in photometry compared to standard aperture based algorithm was obtained. However, significant improvement of more than a factor of two (for bright stars) has been reached in astrometry accuracy. Additionally, new limits on the optical precursor to ``the naked-eye burst'' have been set, based on the signal level reconstructed on each frame in the place of GRB and combined from two cameras. This approach was not possible prior to the model development.

The improvement in astrometry convinced us that the effective model describes data properly. Therefore, the model was used as an input for a general purpose package prepared for Monte-Carlo simulation of sky images as seen by the \pin camera. The simulator can be used to generate sky frame for given celestial coordinates properly reproducing the deformed stars shapes. It is also capable of creating frames of special interest, consisting of objects with specified parameters (eg. GRB flashes or satellites), which can then be used for testing analysis algorithms. The description of measurement fluctuations introduced in the frame generation was successful at reproducing the quantitative behaviour of photometry errors in the real data, thus proving the simulator a well suited tool for development of data analysis algorithms.

The development of the CCD camera response model, which was the subject of this thesis, turned out to be a very complicated task. At many stages of the analysis different approaches were possible. Because of time limitation, not all the possibilities were explored in details. Nevertheless some very important results have been reached, mainly a more than significant improvement in astrometry precision and new limits on the GRB080319B optical precursor, improving previous limits by $0.75^\m$. The results of the simulation indicated the most significant factor responsible for stars brightness measurement errors -- the gain fluctuations of the camera's electronics. Up to now, this factor has been underestimated and this new information may help to improve future \pin cameras significantly.

It is clear that further improvements are possible both in the PSF modelling and in the frame generator. A valuable achievement would be to complete the diffractive model of the light passage through lenses. Although this approach would be, in any case, to slow to be incorporated into the data analysis algorithms, it could be used as a reference for tuning of the effective model. Additionally, it could allow for predicting the PSF changes with spectrum of light and different focusing of cameras. However, the complexity of this task calls for a separate, dedicated research.

We hope that the detailed PSF description will be useful in analysis of the data from the final \pin system. The first mount of the full system shows better tracking precision and reduced mechanical vibrations. This should result in reduced shape fluctuations of real star images and allow for much more precise measurements. Reduced readout noise and gain fluctuations of new cameras should also increase the impact of properly parametrized star shape on objects brightness and position determination. We do hope that the brightness measurement accuracy, much better than in the prototype setup can be obtained. First results from the new detector in Spain should be available soon.

We expect that the developed model, after possible improvements, will be used for a numerous new studies not considered in this work. Precise astrometry may result in more accurate summing of star images from consecutive frames, extending the \pin range. Simultaneously, profile photometry of dense fields could be now much more precise. In both cases, tests and optimization of the analysis algorithms can be performed with the developed simulator.

The full \pin system, with its large sky coverage, will have the unprecedented capability of detecting short optical transients on the sky. Moreover, undoubtful identification of flashes of an astrophysical origin, thanks to parallax. This is yet an untouched field in optical astronomy and as such it may lead to many surprises, such as not yet discovered phenomena.

Together with still the biggest optical sky coverage, \pin gives hope for large discoveries. However, these detections will require studies of the parallax determination, as well as, assuming \pin is going to be the source of optical transients alerts distributed to other observatories, most precise measurements of brightness and position. In all these, the developed simulator package should be of great help. Therefore, we are sure that the results presented in this thesis will be helpful for future studies. Full \pin system with its large optical sky coverage and parallax gives us hope for important discoveries in the coming years.

\setcounter{page}{101}
\def\thechapter{\Alph{chapter}} 
\setcounter{chapter}{1}
\chapter*{Appendix A}
\addcontentsline{toc}{chapter}{Appendix A}
\chaptermark{Appendix}
\markboth{Appendix}{Appendix}
\label{app_a}

Number of plots included in the main text of this thesis was reduced for clarity. Shown in this appendix are additional plots complementing results presented in chapter \ref{chap_polynomials}:
\begin{itemize}
	\item dependence of all coefficients of Zernike polynomial on the star position, for the first parameter set, is shown in fig. \ref{fig_app_first_par_val} (refer fig. \ref{fig_all_par_fit_dist})
	\item same dependence for the second parameter set is shown in fig. \ref{fig_app_fin_par_val} (refer fig. \ref{fig_fin_par_val})
	\item comparison of measured PSFs for colour and white diodes with their model description, for different focus settings and distances from the frame centre are shown in fig. \ref{fig_psf_col_defoc_cont} (refer fig. \ref{fig_psf_col_defoc})
\end{itemize}

\begin{figure}[p]
\begin{center}
	\includegraphics[width=0.99\textwidth]{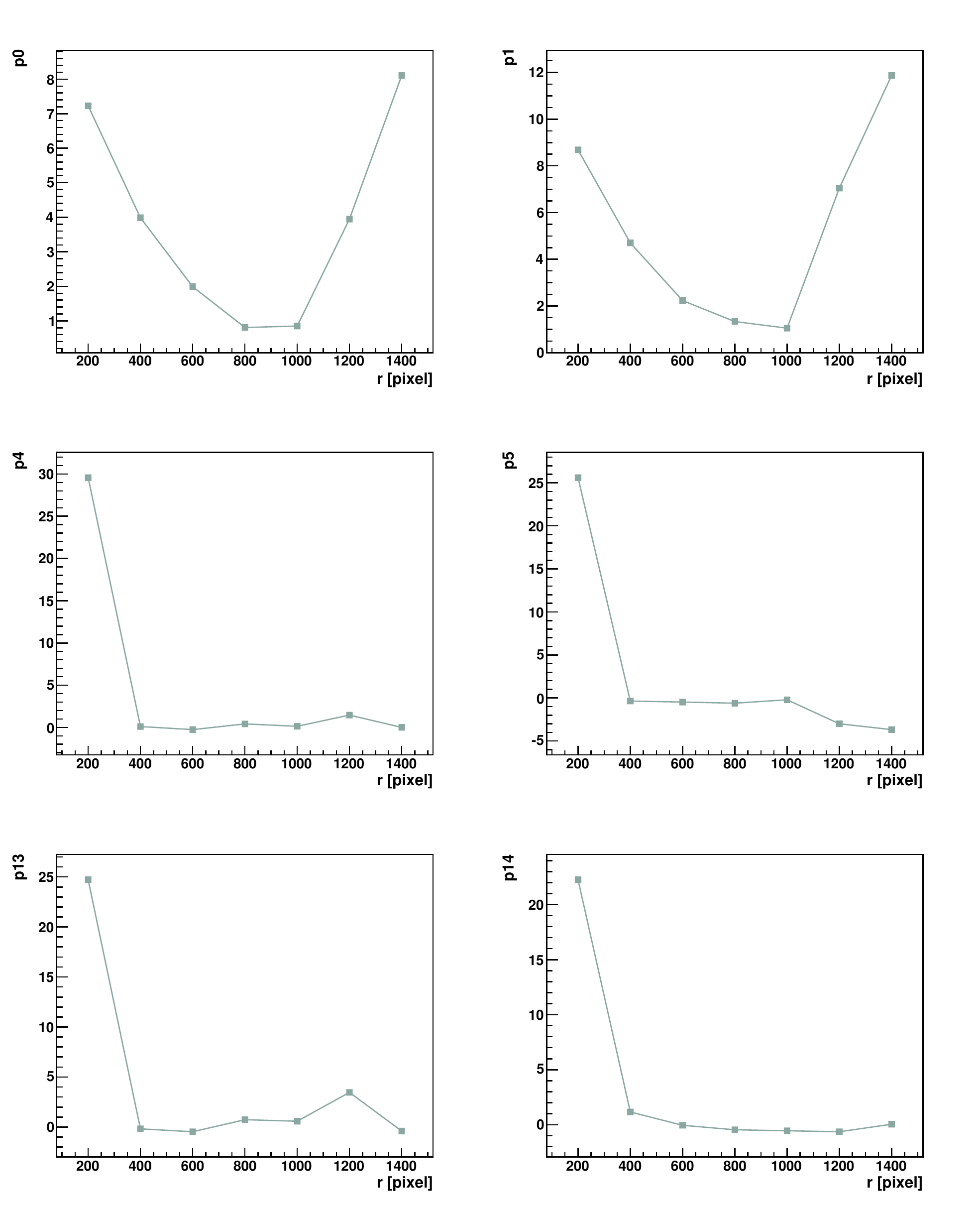}
\end{center}
\caption{Coefficients of Zernike polynomial as a function of the distance from the frame centre r -- results of the best parameters searching and fitting procedure.}
\label{fig_app_first_par_val}
\end{figure}

\begin{figure}[p]
\ContinuedFloat
\begin{center}
	\includegraphics[width=0.99\textwidth]{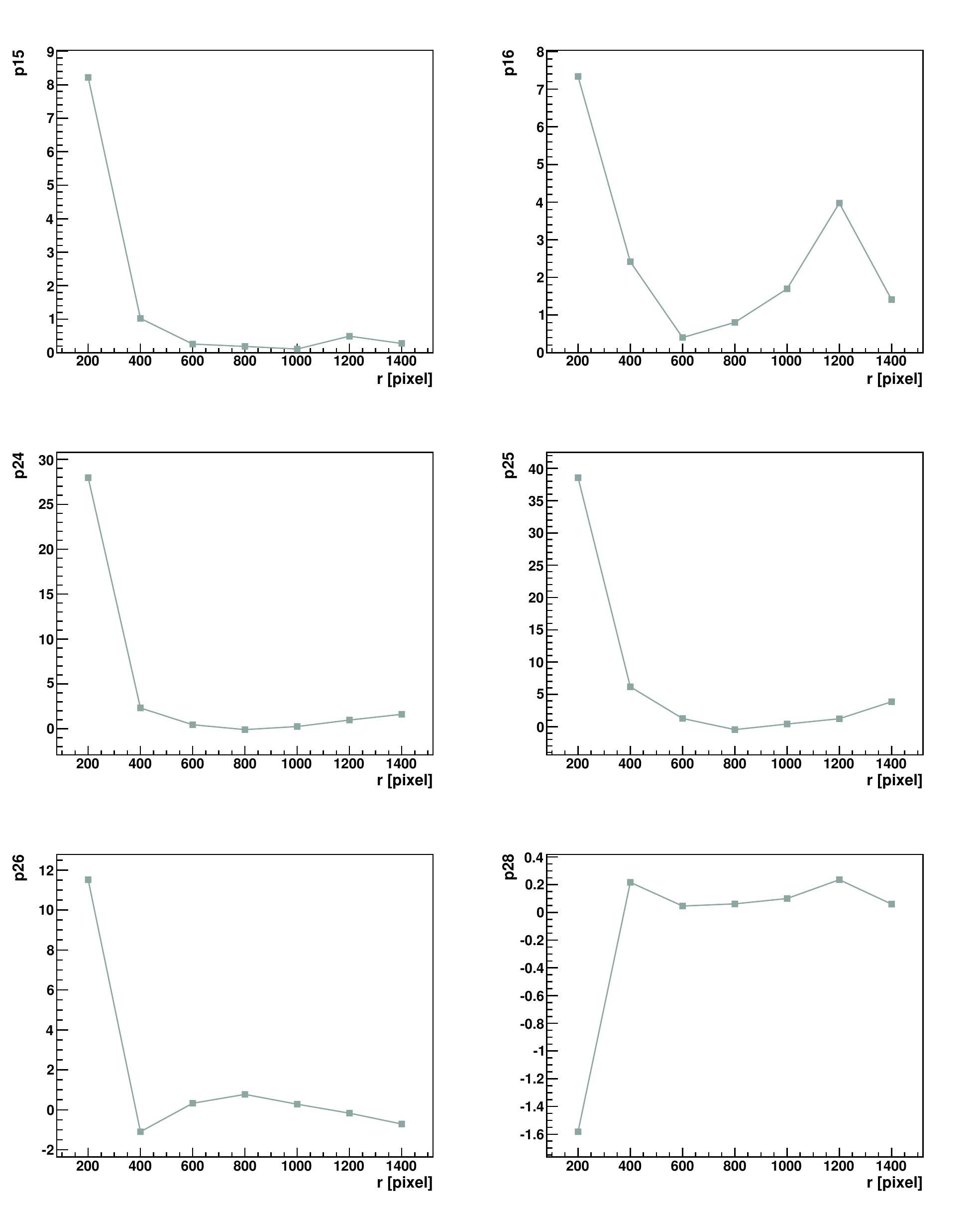}
\end{center}
\caption{Continuation...}
\end{figure}

\begin{figure}[p]
\ContinuedFloat
\begin{center}
	\includegraphics[width=0.99\textwidth]{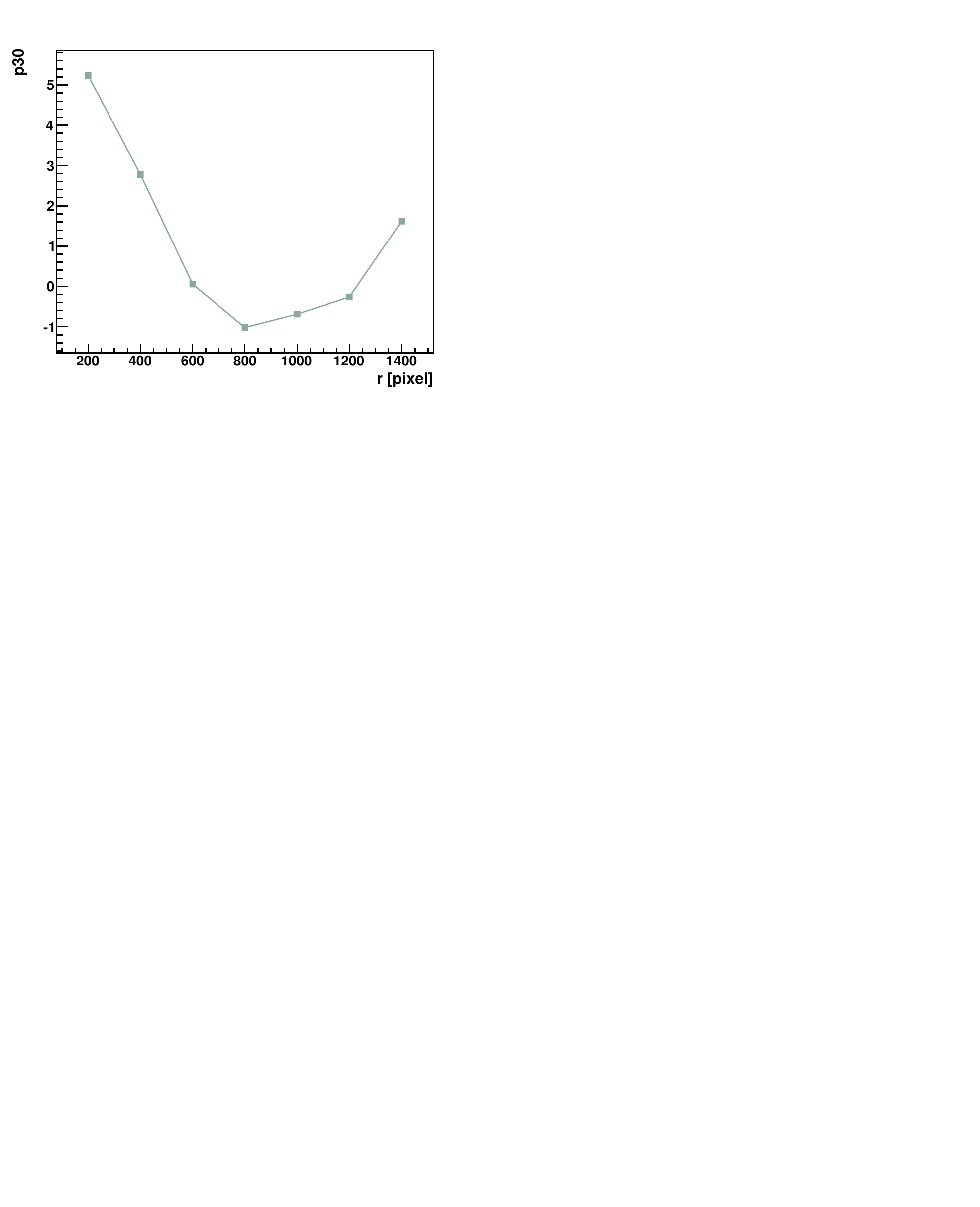}
\end{center}
\caption{Continuation...}
\end{figure}

\begin{figure}[p]
\begin{center}
	\includegraphics[width=0.99\textwidth]{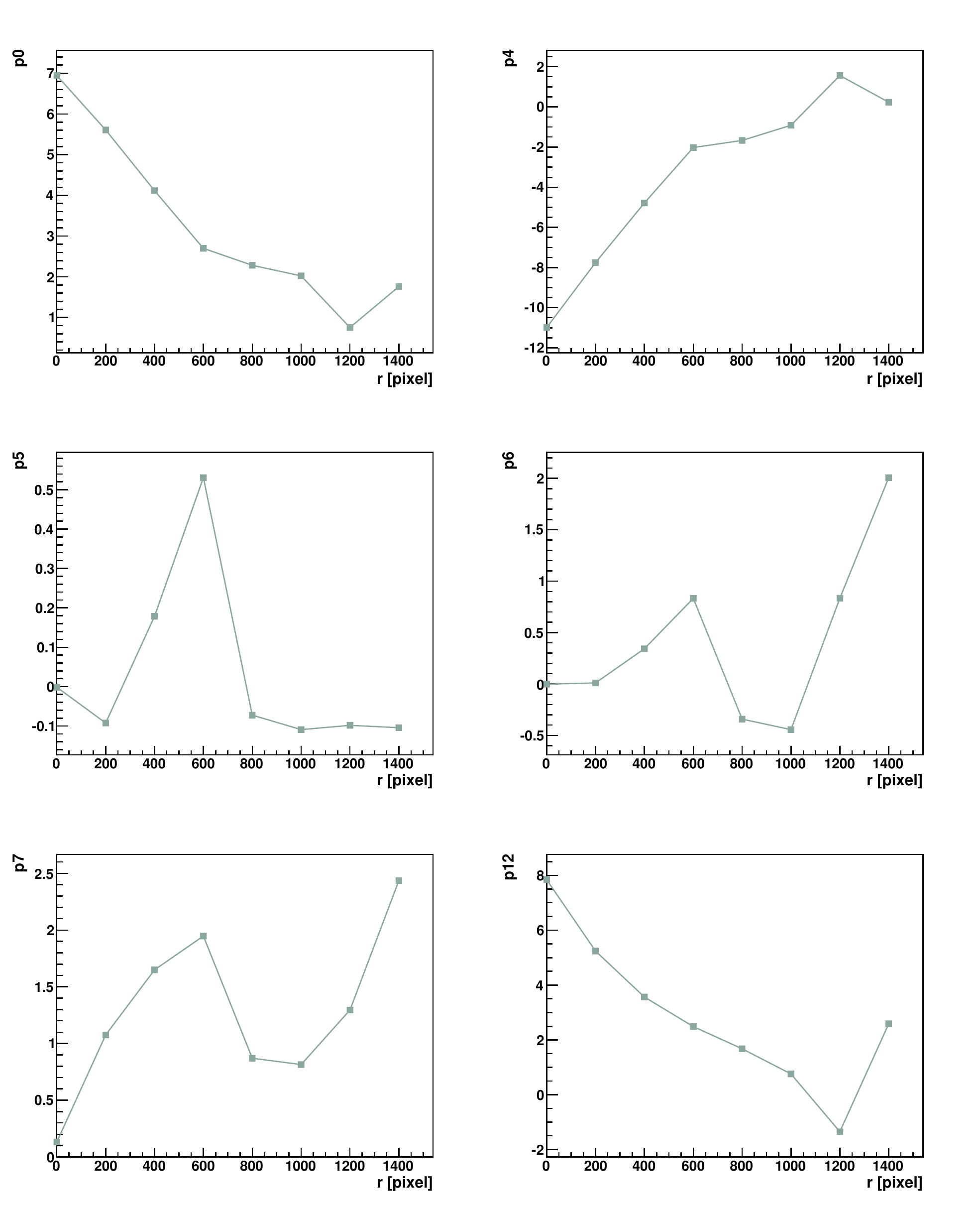}
\end{center}
\caption{Coefficients of Zernike polynomial as a function of the distance from the frame centre r for the finally chosen basis of 17 polynomial terms -- 3 circular and 14 giving best $\chi^2$ for all measured profiles.}
\label{fig_app_fin_par_val}
\end{figure}

\begin{figure}[p]
\ContinuedFloat
\begin{center}
	\includegraphics[width=0.99\textwidth]{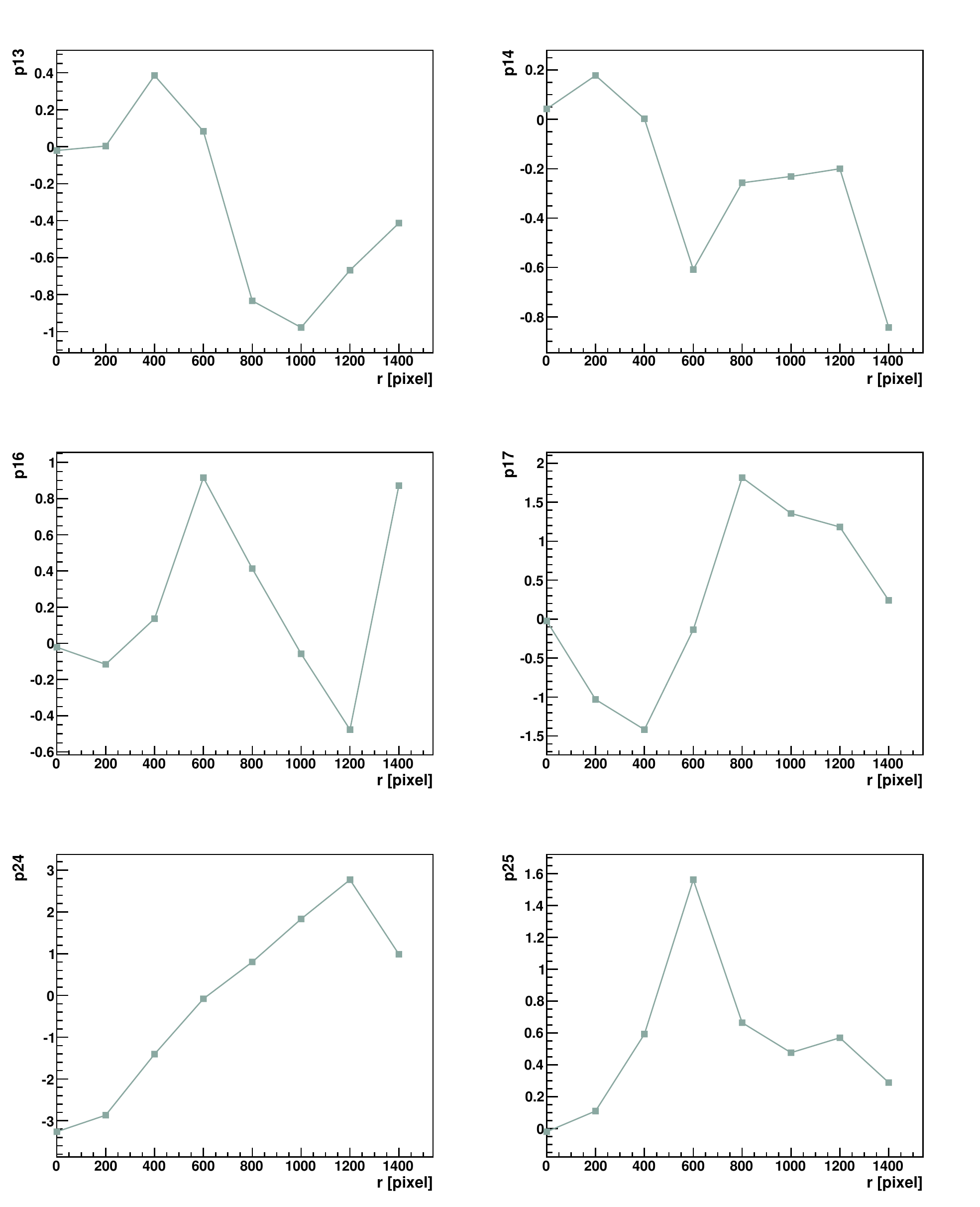}
\end{center}
\caption{Continuation...}
\end{figure}

\begin{figure}[p]
\ContinuedFloat
\begin{center}
	\includegraphics[width=0.99\textwidth]{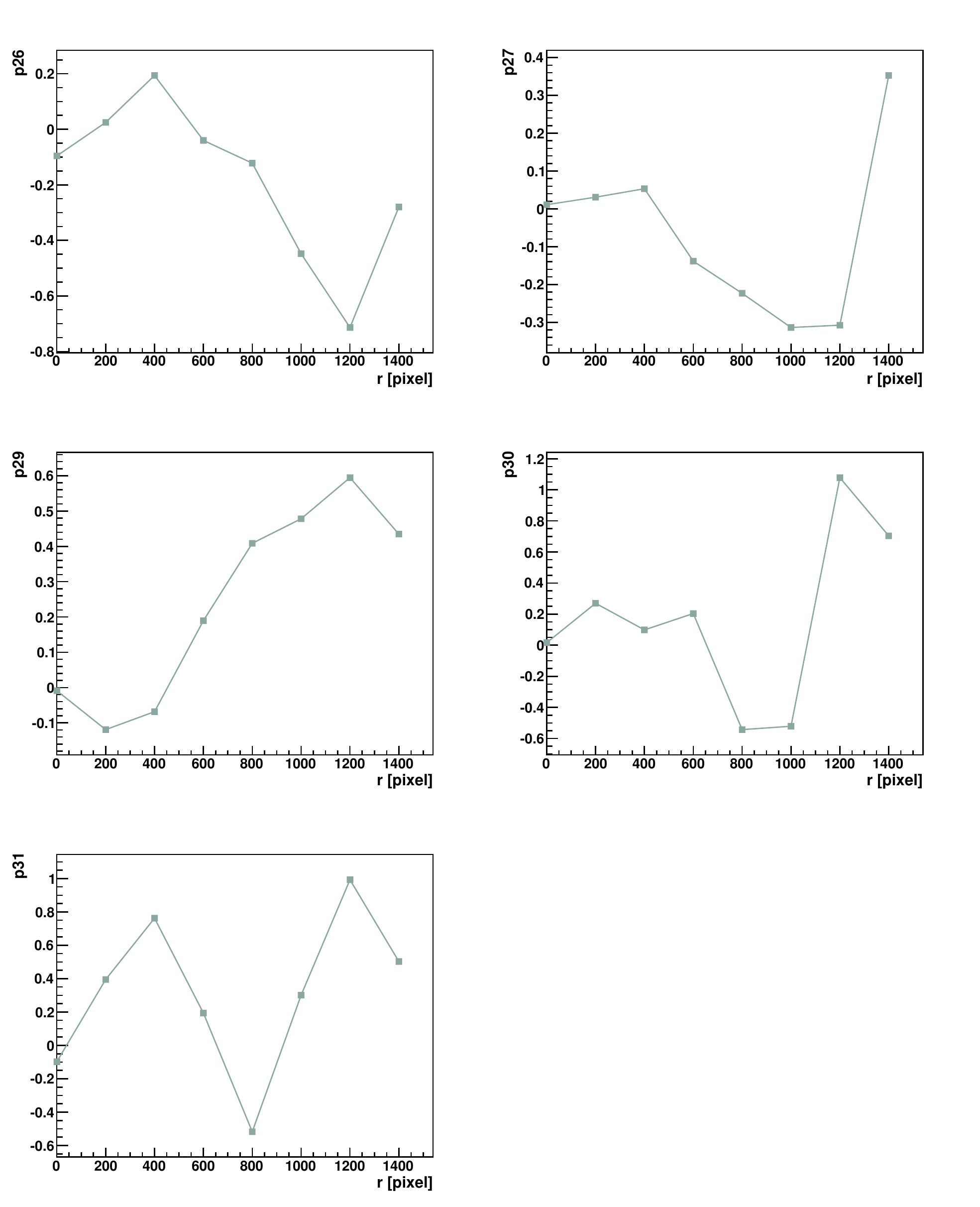}
\end{center}
\caption{Continuation...}
\end{figure}

\begin{figure}[h!]
\begin{center}
\subfigure[Blue diode with nominal focus $fs=1.4$ m]{
$
\begin{array}{c}
	\includegraphics[width=0.31\textwidth]{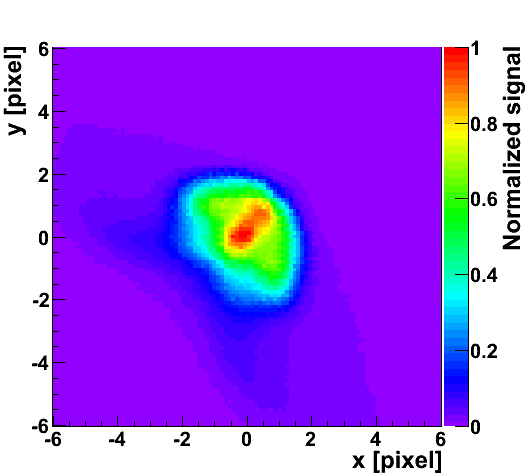}
	\includegraphics[width=0.31\textwidth]{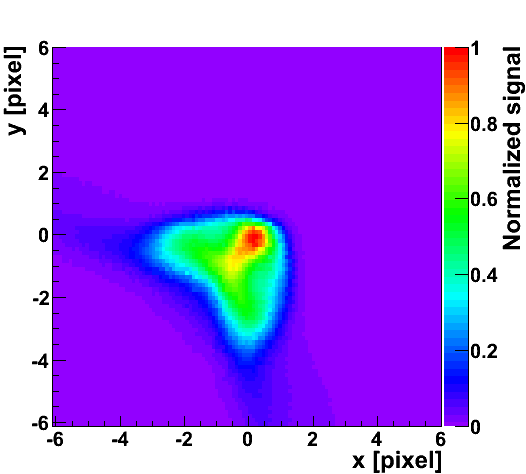}
	\includegraphics[width=0.31\textwidth]{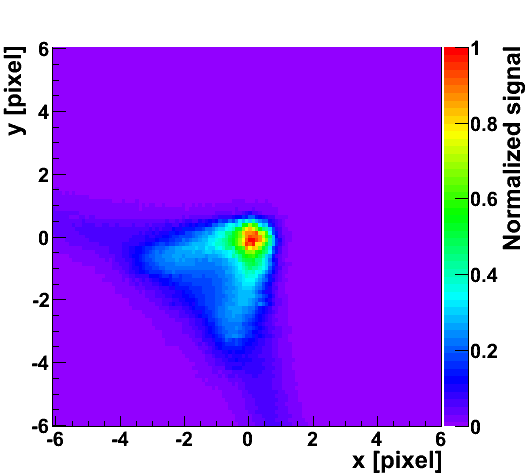}\\
	\includegraphics[width=0.31\textwidth]{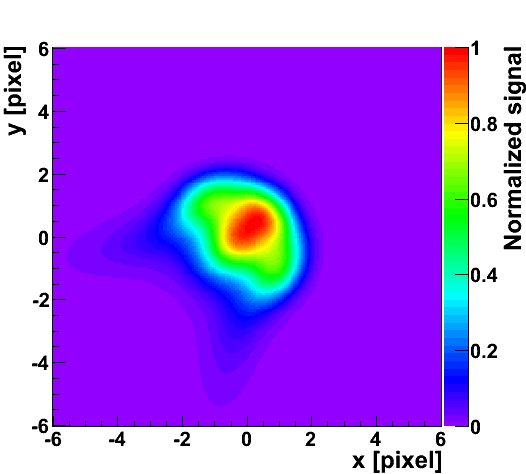}
	\includegraphics[width=0.31\textwidth]{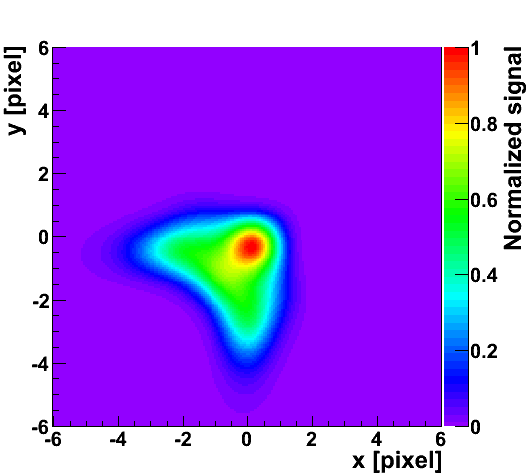}
	\includegraphics[width=0.31\textwidth]{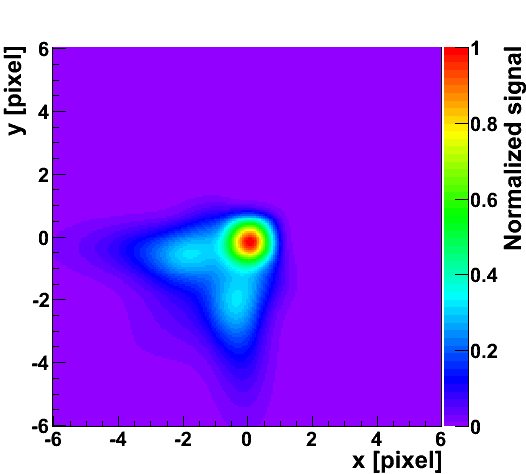}
\end{array}
$
}
\subfigure[Red diode with nominal focus $fs=1.4$ m]{
$
\begin{array}{c}
	\includegraphics[width=0.31\textwidth]{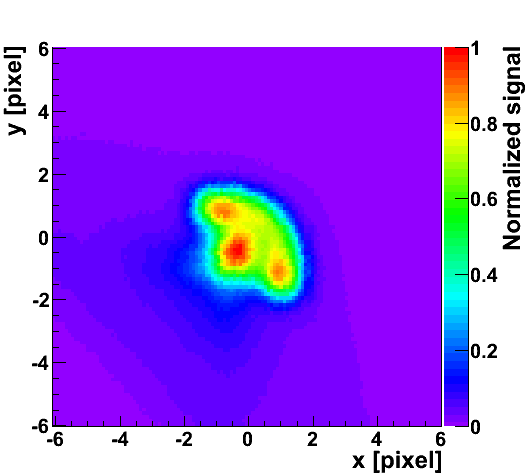}
	\includegraphics[width=0.31\textwidth]{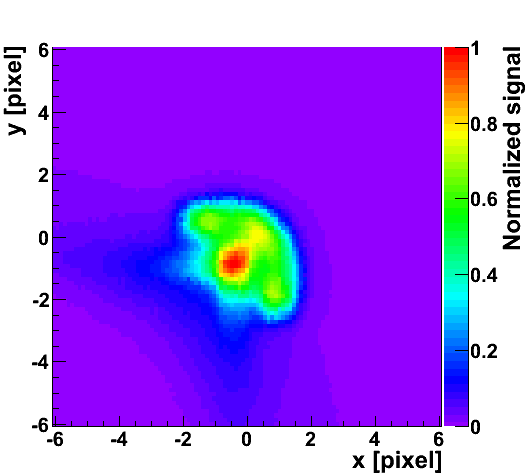}
	\includegraphics[width=0.31\textwidth]{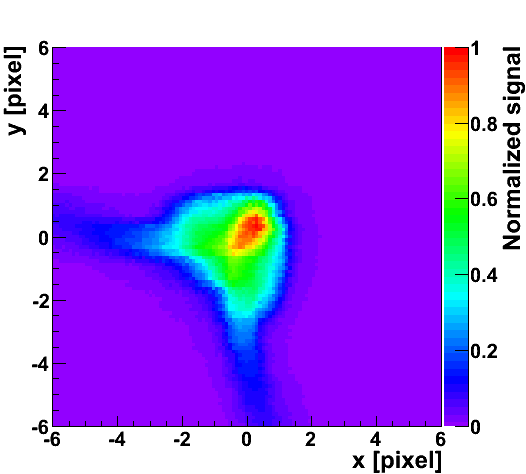}\\
	\includegraphics[width=0.31\textwidth]{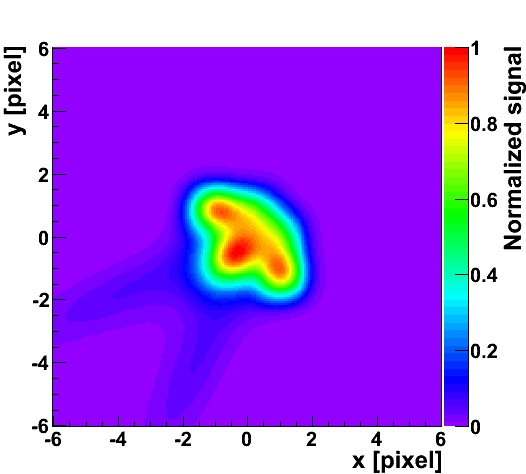}
	\includegraphics[width=0.31\textwidth]{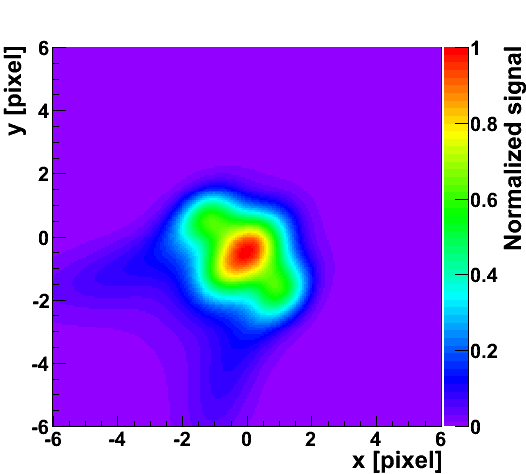}
	\includegraphics[width=0.31\textwidth]{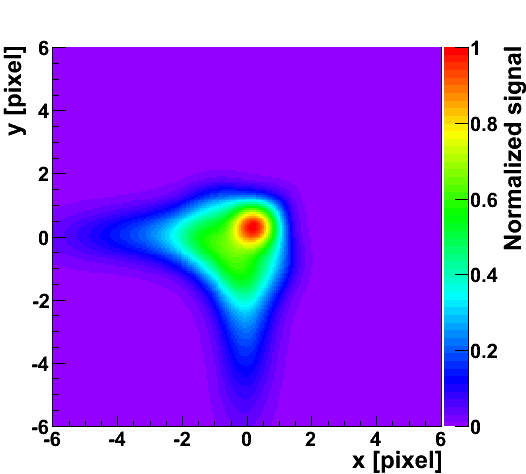}
\end{array}
$
}
\end{center}
\caption{Measured PSFs for colour diodes and for the white diode with changed focusing (first rows) compared with their models obtained with the developed parametrization (second rows), for distance of 1000 (left column), 1200 (centre), 1400 (right column) pixels from the frame centre. Plots for 800 pixels from the frame centre were shown on fig. \ref{fig_psf_col_defoc}.}
\label{fig_psf_col_defoc_cont}
\end{figure}

\begin{figure}[h!]
\ContinuedFloat
\begin{center}
\subfigure[White diode with focus changed to $fs=1.2$ m]{
$
\begin{array}{c}
	\includegraphics[width=0.31\textwidth]{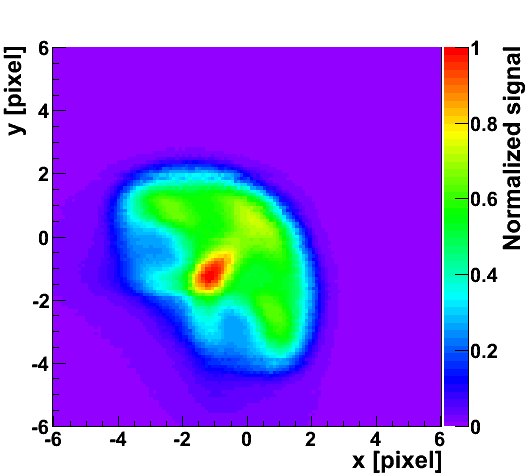}
	\includegraphics[width=0.31\textwidth]{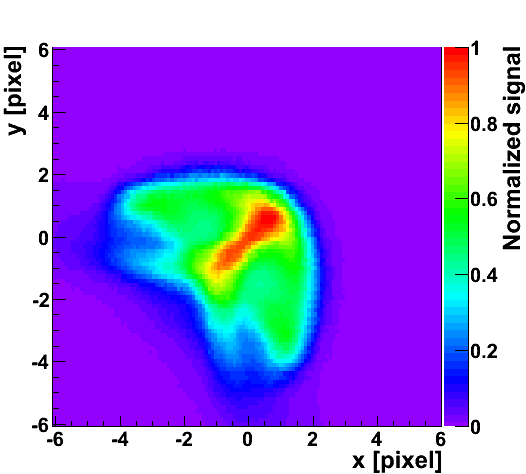}
	\includegraphics[width=0.31\textwidth]{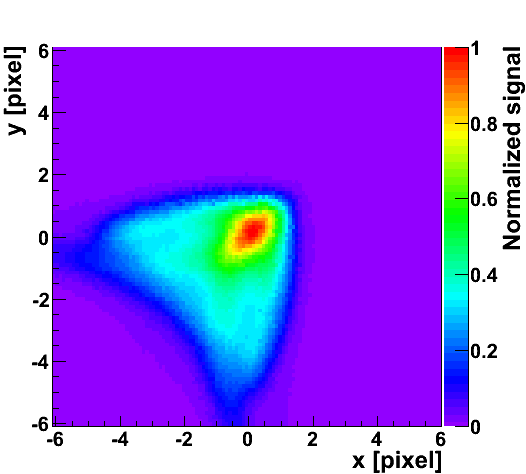}\\
	\includegraphics[width=0.31\textwidth]{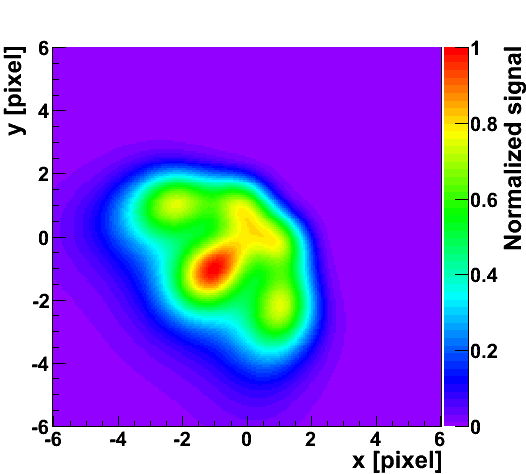}
	\includegraphics[width=0.31\textwidth]{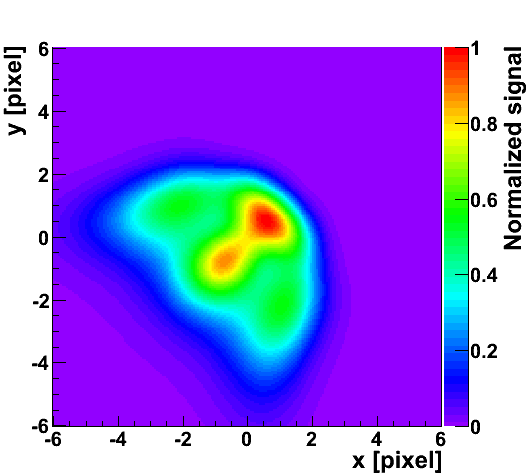}
	\includegraphics[width=0.31\textwidth]{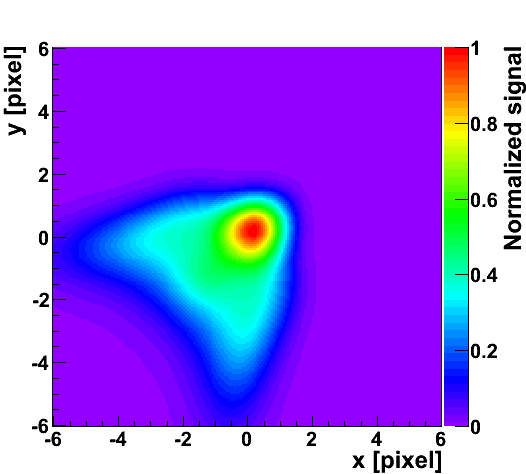}
\end{array}
$
}
\subfigure[White diode with focus changed to $fs=1.6$ m]{
$
\begin{array}{c}

	\includegraphics[width=0.31\textwidth]{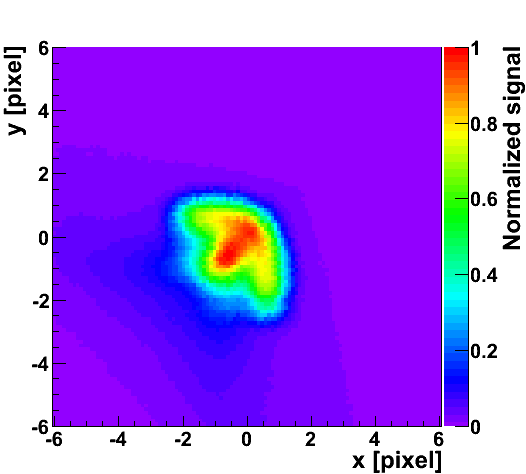}
	\includegraphics[width=0.31\textwidth]{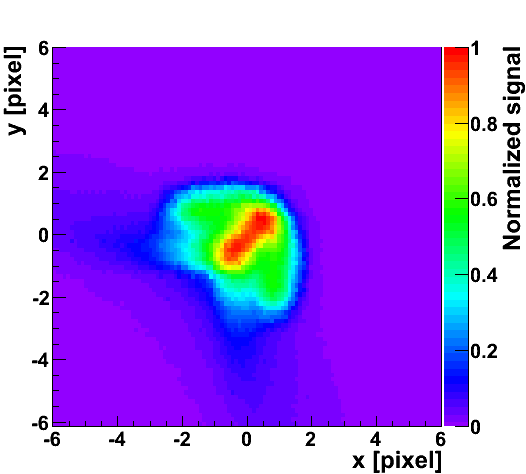}
	\includegraphics[width=0.31\textwidth]{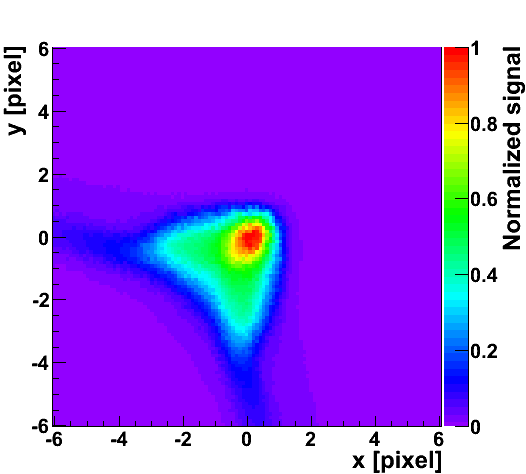}\\
	\includegraphics[width=0.31\textwidth]{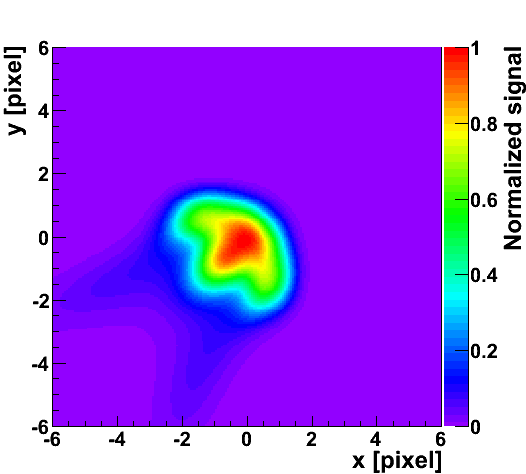}
	\includegraphics[width=0.31\textwidth]{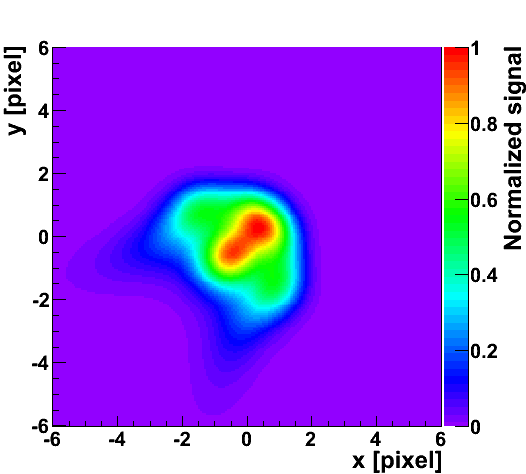}
	\includegraphics[width=0.31\textwidth]{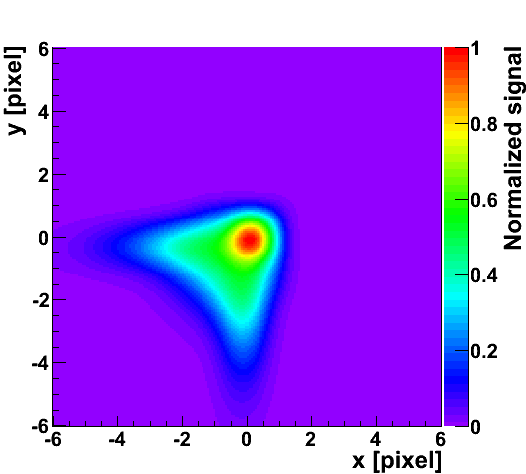}
\end{array}
$
}
\end{center}
\caption{Continued...}
\end{figure}



\thispagestyle{empty}
\cleardoublepage

\include{acknowledgments}



\thispagestyle{empty}
\cleardoublepage

\setcounter{page}{113}

\bibliography{mybib}{}
\bibliographystyle{ieeetr}

\end{document}